\documentclass[]{aastex631}

\shorttitle{MANOS colors}
\shortauthors{Moskovitz et al.}
\graphicspath{{./}{figures/}}
\usepackage{rotating}
\begin{document}

\title{NEO Colors from The Mission Accessible Near-Earth Object Survey (MANOS)}

\correspondingauthor{Nicholas Moskovitz}
\email{nmosko@lowell.edu}

\author[0000-0001-6765-6336]{Nicholas Moskovitz}
\affiliation{Lowell Observatory, 1400 West Mars Hill Road, Flagstaff, AZ, 86001, USA}

\author[0000-0003-1008-7499]{Theodore Kareta}
\affiliation{Department of Astrophysics and Planetary Science, Villanova University, Villanova, PA, USA}
\affiliation{Lowell Observatory, 1400 West Mars Hill Road, Flagstaff, AZ, 86001, USA}

\author[0009-0006-1160-3829]{Samantha Hemmelgarn}
\affiliation{Lowell Observatory, 1400 West Mars Hill Road, Flagstaff, AZ, 86001, USA}
\affiliation{Northern Arizona University, Flagstaff, AZ, USA}

\author[0009-0000-5376-1143]{Hannah Zigo}
\affiliation{Arizona State University, Tempe, AZ, 85281, USA}

\author[0000-0002-6509-6360]{Maxime Devogèle}
\affiliation{ESA NEO Coordination Centre, Largo Galileo Galilei, 1, 00044 Frascati (RM), Italy}

\author[0000-0002-1506-4248]{Audrey Thirouin}
\affiliation{Lowell Observatory, 1400 West Mars Hill Road, Flagstaff, AZ, 86001, USA}

\author[0009-0001-6122-1291]{Katie Breeland-Newcomb}
\affiliation{Northern Arizona University, Flagstaff, AZ, USA}
\affiliation{Lowell Observatory, 1400 West Mars Hill Road, Flagstaff, AZ, 86001, USA}

\author[0000-0002-6423-0716]{Brian Burt}
\affiliation{Lowell Observatory, 1400 West Mars Hill Road, Flagstaff, AZ, 86001, USA}

\author[0000-0002-7600-4652]{Annika Gustaffson}
\affiliation{Northern Arizona University, Flagstaff, AZ, USA}
\affiliation{Lowell Observatory, 1400 West Mars Hill Road, Flagstaff, AZ, 86001, USA}

\author[0009-0006-2183-484X]{Mitchell Magnuson}
\affiliation{Lowell Observatory, 1400 West Mars Hill Road, Flagstaff, AZ, 86001, USA}

\author[0000-0002-8132-778X]{Michael Mommert}
\affiliation{Stuttgart University of Applied Sciences, Stuttgart, Germany}

\author[0000-0002-6977-3146]{David Polishook}
\affiliation{Faculty of Physics, Weizmann Institute of Science, Rehovot 0076100, Israel}

\author{Robert Schottland}
\affiliation{Lowell Observatory, 1400 West Mars Hill Road, Flagstaff, AZ, 86001, USA}

\author[0000-0001-5306-6220]{Brian Skiff}
\affiliation{Lowell Observatory, 1400 West Mars Hill Road, Flagstaff, AZ, 86001, USA}

\author[0000-0003-3091-5757]{Cristina Thomas}
\affiliation{Northern Arizona University, Flagstaff, AZ, USA}

\author{Mark Willman}
\affiliation{Institute for Astronomy, University of Hawaii, Hilo, HI, USA}



\begin{abstract}

We present spectro-photometric $griz$ colors for 189 near-Earth objects (NEOs) collected by the Mission Accessible Near-Earth Object Survey (MANOS). Data acquisition involved non-simultaneous multi-band exposures, thus particular attention was given to the influence of rotational lightcurves on the derived colors. We show that colors measured without accounting for lightcurve variations can significantly influence results for individual objects and potentially have systematic offsets for ensemble studies. Color-based taxonomic classifications were used to investigate the distribution of spectral types. Our results were combined with other visible wavelength surveys to highlight a previously reported change in the observed taxonomic distribution of NEOs as a function of size, namely a decrease in S complex and an increase in X complex objects with increasing absolute magnitude. Plausibility arguments are given to suggest that Main Belt source region, thermal modification, discovery bias, tidal resurfacing, regolith grain size, and impact shock darkening are unlikely explanations for this size-dependent trend. Consistent with recent NEO population models and work on the connection between meteorites and young asteroid families in the Main Belt, this trend is best explained by a compositional gradient in the NEO population. In particular, the observed abundance of S complex or ordinary chondrite-like NEOs decreases by a factor of two from $\sim65\%$ of the population at km-scales down to a third at sizes $\lesssim50$m. This result has implications for understanding the initial pre-impact population of meteorite parent bodies prior to atmospheric filtering. Furthermore, this will have implications for probabilistic impact risk assessment models.

\end{abstract}

\keywords{Asteroids; Near-Earth Objects; Multi-color photometry}


\section{Introduction} \label{sec:intro}

Near-Earth objects (NEOs) are Earth's closest celestial neighbors. As dynamically transient bodies with average lifetimes around 10 Myr \citep{gladman00}, NEOs are replenished from various sources throughout the inner Solar System \citep{granvik18}. As such they are amongst the most accessible population for studying processes of planet formation and evolution specific to the terrestrial planets. Their proximity to Earth also means that NEOs can be ideal targets for manned or robotic spacecraft exploration \citep{elvis11}, which includes the possibility of in situ harvesting of resources such as water \citep{colvin20} and metals \citep{cannon23}. Such utilization of NEOs may ultimately prove necessary for human exploration outside of the Earth-Moon system. On the other hand, NEOs can and do impact the Earth with consequences ranging from infrequent mass extinction events to common and less energetic meteorite dropping impacts \citep{shoemaker83,halliday89,jenniskens09,popova13}. Each of these endeavors -- scientific inquiry, spacecraft exploration, resource utilization, planetary defense -- require an understanding of the compositions of NEOs, both on an individual level and for the population as a whole.

Aside from rare in situ or sample return missions, the primary means of constraining NEO compositions involves spectroscopic or spectro-photometric (imaging) observations. Telescopically, these observations are most commonly conducted at visible (0.4-0.8 $\mu m$), near-infrared (0.8-2.5 $\mu m$), and mid-infrared (2.5-25 $\mu m$) wavelengths where a myriad of electronic transitions and vibrational modes produce absorption and emissivity features that are diagnostic of mineralogy \citep{reddy15,emery19}. Dedicated surveys have produced spectral data for many hundreds of NEOs \citep[e.g.][]{binzel01,mommert16,erasmus17,perna18,popescu19,devogele19,binzel19,sanchez24,navarro-meza24,birlan24}. Classification of these data into taxonomic systems \citep[e.g.][]{bus02,demeo09} has helped to understand the diversity and distribution of NEO spectral types. However, interpretation of remote data is complicated by a number of factors including space weathering \citep{chapman04}, the effects of tidal encounters with the terrestrial planets \citep{binzel10}, thermal modification \citep{delbo14}, and the shock effects of impacts \citep{sanchez25}. A wholistic picture of NEO spectral properties and compositions is further complicated by discovery bias against low albedo objects \citep{stuart04}. While progress has been made in understanding some of the consequences of these evolutionary factors \citep{devogele19,graves19,sergeyev23}, their relative importance to the NEO population remains uncertain.

Recent work \citep{broz24,broz24b,marsset24} has provided fresh insight into the connection between meteorites and their parent asteroids in the Main Belt, which has implications for NEOs. These works suggest that the flux of meteorites on Earth is dominated by a small number of young collisional families in the Main Belt whose size frequency distributions are steep and thus still contain significant numbers of meter- and decameter-scale asteroids, i.e. the immediate parent bodies of meteorites. This implies that large (km-scale) NEOs are derived from the population of collisional families in the Main Belt, both young and old, that are located near secular and mean motion resonances, which act as conduits for transferring objects into near-Earth space. Conversely, small ($<100$ meter) NEOs come from only the youngest of these families and would thus have a different compositional distribution, dominated by a few families with ages $\lesssim100$ Myr.

This model is consistent with a number of lines of evidence including the cosmic ray exposure ages of meteorites, bolide statistics, and the presence of infrared emission from dust bands at specific orbital inclinations. However, the spectroscopic and compositional record of NEOs is biased towards the larger (and easier to observe) members of the population. New data that can further support this model include spectroscopic observations of meter to decameter-sized NEOs. Such data would also  constrain the ``top of the atmosphere" population of the most frequent Earth impactors and meteorite parent bodies. With those constraints, the compositional bias imposed by atmospheric filtering of our meteorite collection \citep{shober25} could be better understood.

The Mission Accessible Near-Earth Object Survey (MANOS) conducts characterization observations of objects on low $\Delta v$ orbits \citep{shoemaker78} and in the sub-km size regime where knowledge of physical properties has traditionally been limited (Section \ref{sec:targets}). The work presented here focuses on the measurement of SDSS $griz$ colors collected between 2014 and 2025 using three 4-meter-class telescopes: the 4.3-m Lowell Discovery Telescope (LDT) in northern Arizona, the 4.1-m Southern Astrophysical Research Telescope (SOAR) on Cerro Pachon in Chile, and the 4.0-m Mayall telescope at Kitt Peak National Observatory in Arizona (Section \ref{sec:obs}).

The derived broad-band colors (Section \ref{sec:colors}) were taxonomically classified in the Bus-DeMeo system \citep{demeo09}. Multiple validation steps were taken to ensure reliability in the derived classifications, for example, comparison to spectroscopic data and comparison of colors independently measured on more than one night. These classifications were used to study the underlying distribution of spectral types and to highlight several noteworthy objects (Section \ref{sec:results}). Individual objects of interest include the potential Moon impactor 2024~YR4, an enigmatic D-type object 2022~BX5, and a set of objects on orbits with very low $\Delta v < 4$ km/s. 

The measurement of colors for small NEOs can be complicated by rotational lightcurve variations that occur at periods comparable to observational cadences. Ideally, one would measure colors for such objects with multiple filters  simultaneously (unfortunately such instruments are rare) or employ long observing runs where any lightcurve variability could be measured independently in each filter, but that would be an expensive use of observing time. We adopted a simpler approach to address lightcurve variations (Section \ref{sec:LC}). All of the colors presented here were obtained with consecutive but non-simultaneous multi-band exposures. Generally, colors were measured with a reference filter interleaved into a sequence of multiple exposures to track lightcurve variability. For a quarter of our sample (N=49), new rotational lightcurves were measured with roughly an hour of dedicated observations outside of the color sequences. This enabled more robust lightcurve corrections. It is shown (Section \ref{sec:LC}) that colors measured without accounting for lightcurve variations can significantly influence the results for individual objects and potentially introduce systematic effects for ensemble studies. More specifically, ignoring lightcurve variations for the set of 49 objects with measured lightcurves results in an artificial enhancement of C complex objects. It remains unclear whether the underlying cause of this enhancement is a systematic effect of not correcting for lightcurve variations or is simply due to random fluctuations in our particular data set.

Our color data were combined with results from other visible wavelength surveys to highlight a previously reported \citep{devogele19} change in the observed taxonomic distribution of NEOs as a function of size from multi-km down to meter-scale objects (Section \ref{sec:size}). In particular, we demonstrate a decrease in S complex and an increase in X complex objects with increasing absolute magnitude $H$. Arguments to explain this trend lead to the plausible conclusion that there is a size-dependent compositional gradient in the NEO population (Section \ref{sec:discussion}). This result has implications for understanding the compositional distribution of meteorite parent bodies in space and for planetary defense in terms of informing models of impact risk.

\section{Target Selection} \label{sec:targets}

Targets for MANOS were typically selected in the hours before the start of each observing night to take advantage of newly discovered objects and potentially short accessibility windows. The following criteria were used to prioritize the set of observable NEOs on a given night.
\begin{itemize}
 \item For relevance to mission accessibility, priority was given to objects on low $\Delta$v ($<$6 km/s) orbits as defined by the formalism of \citet{shoemaker78}. 
 \item To probe an under-represented subset of the NEO population, focus was given to absolute magnitudes $H>20$, which corresponds to diameters smaller than a few hundred meters.
 \item To probe potentially hazardous asteroids (PHAs) and the influence of planetary encounters on physical properties \citep[e.g.][]{binzel10}, we favored minimum orbit intersection distances (MOID) relative to Earth $<$ 0.05 au.
 \item To achieve a requisite signal-to-noise ratio S/N $\gtrsim10$ with available instruments, we targeted objects with predicted apparent magnitude $V \lesssim 22$. 
\end{itemize}

Depending on the specific telescope and available instrument suite, the final set of objects for a given night may have included spectroscopic targets \citep{devogele19}, lightcurve targets \citep{thirouin16,thirouin18}, and targets for multi-band color photometry. The latter were collected to provide an independent assessment (albeit at low spectral resolution) of spectral properties. Practical benefits of color photometry include increased time efficiency relative to spectroscopy -- a typical set of visible wavelength colors required less than 20 minutes to collect versus about an hour for the equivalent spectrum -- and the ability to probe fainter targets than could be accessed by a spectrograph on the same telescope.

\begin{figure}
    \centering
    \includegraphics[]{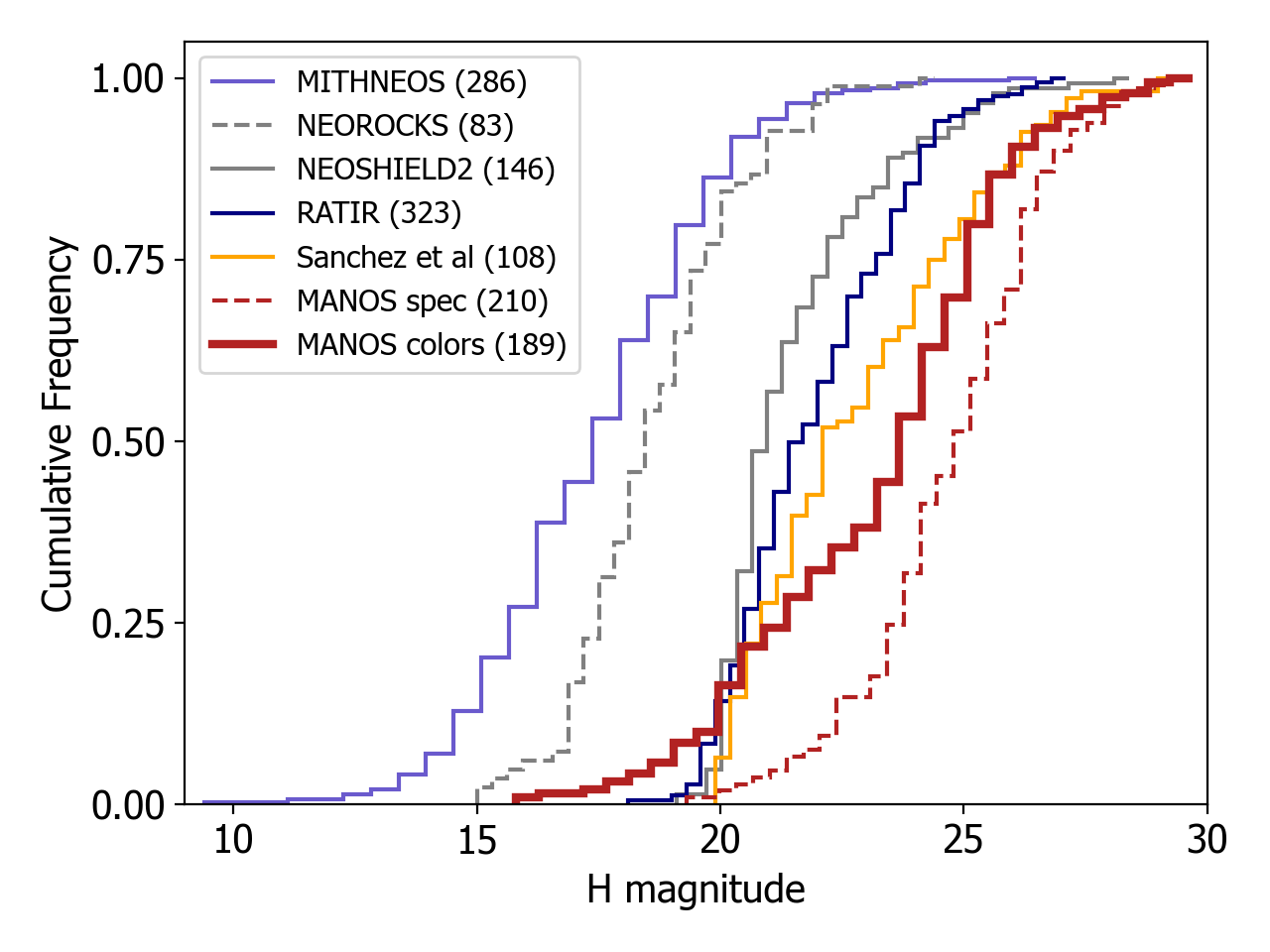}
    \caption{Cumulative frequency distributions for several major NEO spectroscopic and color surveys. The median absolute magnitude $H$ for each survey is: 17.8 for MITHNEOS \citep{binzel19}, 18.6 for NEOROCKS \citep{birlan24}, 21.0 for NEOSHIELD2 \citep{perna18}, 21.8 for RATIR \citep{navarro-meza24}, 22.4 for \citet{sanchez24}, 25.1 for MANOS spectra \citep{devogele19}, and 23.9 for the MANOS colors presented here. The number of objects observed by each of these surveys is given in parentheses in the legend. MANOS has systematically focused on sub-km NEOs ($H>20$), objects that are under-represented in existing datasets.}
    \label{fig:H-Mag}
\end{figure}

Figure \ref{fig:H-Mag} shows the $H$ magnitude cumulative frequency distribution for the MANOS color sample relative to other NEO surveys. The other surveys in Figure \ref{fig:H-Mag} represent a sampling of data from the literature that are good analogs to the MANOS sample in terms of wavelength coverage, sample size, and/or $H$ magnitude coverage. As was done in the spectroscopic study of \citet{devogele19}, we include comparison here to only the visible wavelength spectra from the MITHNEOS survey \citep{binzel19}. There are significantly more spectra presented in that work, however those are primarily at near-infrared wavelengths ($0.8-2.5~\mu m$) and thus not directly comparable to our visible colors. The NEOROCKS survey was a wide-ranging set of investigations into NEO physical properties and dynamics, including a color photometry component that is used for comparison here \citep{birlan24}, again because this provides the most direct comparison. The NEOSHIELD2 spectroscopic survey \citep{perna18} fills an important gap in $H$ magnitude coverage between the MITHNEOS sample and MANOS, and thus is key to analyses presented later (Section \ref{sec:size}). The survey of \citet{navarro-meza24} used a simultaneous multi-band imager called RATIR \citep[the Reionization And Transients InfraRed camera,][]{butler12} and was similar to MANOS in its focus on rapid response to newly discovered objects. The near-IR spectroscopic survey of \citet{sanchez24} was carried out over many years with the SpeX instrument \citep{rayner03} at NASA's Infrared Telescope Facility (IRTF), and most closely matches the MANOS $H$ magnitude distributions. Finally, we include a comparison to the MANOS spectral survey \citep{devogele19}, which is highly complementary to the colors presented here. In particular, the MANOS target selection criteria resulted in sampling $H$ magnitudes that range from 1-5 magnitudes fainter than other dedicated spectral and color surveys (Figure \ref{fig:H-Mag}).

It is interesting to note that on average the MANOS spectroscopic sample probes higher $H$ values than the MANOS color sample. This might seem counterintuitive as practical apparent magnitude limits for colors should generally be fainter than those for spectroscopy. However, the cause of this offset reflects a difference in the style of observation for each of these samples. The MANOS spectroscopic sample was dominated by observations from the Gemini North and South 8-m telescopes, which operate in queue mode. Targets of interest were loaded into the Gemini queues shortly after discovery with observations generally happening within a week or so of that. This ``slow target of opportunity" style of observation enabled access to objects at the smallest end of the NEO size distribution. In contrast, the color sample was built up through time at classically scheduled 4-m facilities, where nights were scheduled months in advance. This effectively reduced the capacity for rapid response, which can be essential for accessing small objects during their short observing windows.

\section{Observations and Data Reduction} \label{sec:obs}

We present multi-band, visible wavelength color photometry for 189 NEOs collected on 83 different nights from UT 2014-04-24 to 2025-02-06. A total of 199 sets of colors were obtained with several objects observed on multiple nights. These data were obtained at the 4.3-m Lowell Discovery Telescope\footnote{The LDT was known as the Discovery Channel Telescope (DCT) until 2019 when it was renamed.} in northern Arizona (LDT; MPC observatory code = G37), the 4.2-m Southern Astrophysical Research (SOAR) telescope on Cerro Pach\'on in central Chile (MPC observatory code = I33), and the Mayall 4-m telescope at Kitt Peak National Observatory in southern Arizona (MPC observatory code = 695 for Kitt Peak). Details for each of these facilities and the instruments employed are given in the following sub-sections. Observing circumstances for each target are presented in Table \ref{tab:observations}, which includes absolute magnitude $H$, and the predicted $V$-band magnitude of the targets, the heliocentric distance $r$, geocentric distance $\Delta$, and solar phase angle $\alpha$ at the time of observation. These data were pulled from a combination of the Lowell Observatory astorb system \citep{moskovitz22} and JPL Horizons.

\startlongtable
\begin{deluxetable}{llclccccc}
\tablecaption{Observational circumstances for the MANOS color sample. \label{tab:observations}}
\tabletypesize{\small}
\tablehead{\colhead{Number} & \colhead{Designation} & \colhead{Facility} & \colhead{UT Obs.} & \colhead{H} & \colhead{$V$ (mag)} & \colhead{$r$ (au)} & \colhead{$\Delta$ (au)} & \colhead{$\alpha$ (deg)}}
\startdata
10302 & 1989 ML & LDT 4.3m & 2022-04-07 & 19.45 & 18.6 & 1.23 & 0.29 & 35 \\
12923 & Zephyr & LDT 4.3m & 2021-10-29 & 15.8 & 15.7 & 1.24 & 0.37 & 41 \\
22099 & 2000 EX106 & LDT 4.3m & 2023-12-29 & 18.09 & 19.3 & 1.09 & 0.57 & 64 \\
52381 & 1993 HA & LDT 4.3m & 2022-05-31 & 20.07 & 20.4 & 1.46 & 0.49 & 21 \\
85628 & 1998 KV2 & LDT 4.3m & 2021-10-16 & 17.28 & 18.6 & 1.26 & 0.60 & 51 \\
89136 & 2001 US16 & LDT 4.3m & 2023-11-05 & 20.28 & 21.8 & 1.56 & 0.73 & 29 \\
141018 & 2001 WC47 & LDT 4.3m & 2021-12-04 & 19.04 & 20.8 & 1.29 & 0.76 & 49 \\
141018 & 2001 WC47 & LDT 4.3m & 2022-02-11 & 19.04 & 19.6 & 1.09 & 0.41 & 65 \\
154555 & 2003 HA & LDT 4.3m & 2021-10-16 & 16.67 & 18.3 & 0.82 & 0.65 & 84 \\
162004 & 1991 VE & LDT 4.3m & 2023-12-29 & 18.29 & 19.8 & 1.48 & 0.70 & 34 \\
264993 & 2003 DX10 & LDT 4.3m & 2024-03-06 & 20.43 & 17.9 & 1.12 & 0.15 & 30 \\
267337 & 2001 VK5 & LDT 4.3m & 2024-01-13 & 18.05 & 21.4 & 1.80 & 1.40 & 33 \\
292220 & 2006 SU49 & LDT 4.3m & 2023-12-29 & 19.39 & 18.3 & 1.02 & 0.18 & 74 \\
303450 & 2005 BY2 & LDT 4.3m & 2025-02-02 & 20.5 & 19.4 & 1.07 & 0.21 & 60 \\
350713 & 2001 XP88 & LDT 4.3m & 2018-05-24 & 20.73 & 21.2 & 1.47 & 0.52 & 23 \\
350751 & 2002 AW & LDT 4.3m & 2023-11-12 & 20.82 & 18.1 & 1.18 & 0.19 & 7 \\
412983 & 1996 FO3 & LDT 4.3m & 2021-11-15 & 20.16 & 19.9 & 1.21 & 0.34 & 42 \\
412983 & 1996 FO3 & LDT 4.3m & 2022-02-11 & 20.16 & 18.6 & 1.03 & 0.16 & 70 \\
439437 & 2013 NK4 & LDT 4.3m & 2024-05-01 & 18.95 & 16.2 & 1.15 & 0.15 & 22 \\
455176 & 1999 VF22 & LDT 4.3m & 2025-02-02 & 20.71 & 19.8 & 1.28 & 0.32 & 20 \\
481965 & 2009 EB1 & LDT 4.3m & 2024-01-13 & 20.69 & 21.5 & 1.43 & 0.55 & 30 \\
510529 & 2012 EY11 & LDT 4.3m & 2022-05-31 & 21.58 & 21.1 & 1.31 & 0.34 & 28 \\
513312 & 2007 DM41 & LDT 4.3m & 2016-02-22 & 22.06 & 20.3 & 1.11 & 0.18 & 43 \\
513312 & 2007 DM41 & LDT 4.3m & 2025-02-02 & 22.06 & 22.3 & 1.24 & 0.41 & 44 \\
523599 & 2003 RM & LDT 4.3m & 2023-11-12 & 19.64 & 22.0 & 1.88 & 1.00 & 19 \\
523813 & 2008 VB1 & LDT 4.3m & 2021-11-15 & 20.66 & 20.3 & 1.44 & 0.46 & 8 \\
523813 & 2008 VB1 & LDT 4.3m & 2021-12-04 & 20.66 & 20.3 & 1.37 & 0.41 & 15 \\
523828 & 1992 BC & LDT 4.3m & 2023-11-12 & 19.96 & 21.8 & 1.80 & 0.87 & 16 \\
533638 & 2014 KT86 & LDT 4.3m & 2023-12-29 & 20.39 & 19.9 & 1.15 & 0.30 & 50 \\
 & 1991 TF3 & LDT 4.3m & 2021-10-16 & 19.29 & 17.9 & 1.25 & 0.27 & 20 \\
 & 1994 WR12 & LDT 4.3m & 2023-12-29 & 22.4 & 19.1 & 1.05 & 0.10 & 43 \\
 & 1998 KY26 & LDT 4.3m & 2024-05-01 & 25.7 & 22.2 & 1.11 & 0.11 & 22 \\
 & 2000 WO148 & LDT 4.3m & 2021-10-29 & 20.53 & 20.8 & 1.41 & 0.48 & 25 \\
 & 2001 QJ142 & LDT 4.3m & 2024-09-07 & 23.98 & 21.2 & 1.08 & 0.11 & 50 \\
 & 2002 JR100 & LDT 4.3m & 2018-05-24 & 24.3 & 21.2 & 1.11 & 0.11 & 33 \\
 & 2002 TP69 & LDT 4.3m & 2021-10-28 & 21.92 & 18.8 & 1.05 & 0.10 & 49 \\
 & 2003 EH1 & LDT 4.3m & 2024-09-07 & 16.15 & 20.9 & 2.42 & 2.13 & 25 \\
 & 2003 SM84 & LDT 4.3m & 2021-12-04 & 23.02 & 22.1 & 1.21 & 0.28 & 32 \\
 & 2005 QA5 & LDT 4.3m & 2018-05-24 & 21.29 & 21.2 & 1.48 & 0.48 & 11 \\
 & 2007 EF & LDT 4.3m & 2022-05-07 & 21.34 & 22.0 & 1.06 & 0.40 & 72 \\
 & 2010 XN & LDT 4.3m & 2024-07-06 & 24.17 & 19.9 & 1.06 & 0.06 & 43 \\
 & 2011 CG2 & LDT 4.3m & 2025-01-07 & 21.41 & 20.4 & 1.24 & 0.29 & 25 \\
 & 2011 CG2 & LDT 4.3m & 2025-02-06 & 21.41 & 21.3 & 1.17 & 0.35 & 51 \\
 & 2011 YQ10 & LDT 4.3m & 2021-10-16 & 19.09 & 17.0 & 1.11 & 0.16 & 43 \\
 & 2012 BA35 & LDT 4.3m & 2022-02-11 & 23.75 & 20.8 & 1.13 & 0.15 & 17 \\
 & 2012 BF86 & LDT 4.3m & 2016-02-22 & 22.69 & 19.1 & 1.04 & 0.08 & 46 \\
 & 2013 PA7 & LDT 4.3m & 2024-04-12 & 22.67 & 21.3 & 1.20 & 0.24 & 31 \\
 & 2013 RB6 & LDT 4.3m & 2024-08-14 & 21.83 & 21.7 & 1.24 & 0.35 & 43 \\
 & 2014 HW & LDT 4.3m & 2014-04-24 & 28.4 & 19.2 & 1.01 & 0.01 & 15 \\
 & 2014 KH39 & LDT 4.3m & 2014-06-03 & 26.2 & 17.3 & 1.02 & 0.01 & 61 \\
 & 2014 SC324 & LDT 4.3m & 2014-10-17 & 24.39 & 18.7 & 1.04 & 0.04 & 21 \\
 & 2014 TR57 & LDT 4.3m & 2014-10-17 & 25.2 & 19.3 & 1.03 & 0.04 & 26 \\
 & 2014 VL6 & LDT 4.3m & 2021-11-15 & 21.38 & 21.0 & 1.22 & 0.33 & 40 \\
 & 2015 CG & LDT 4.3m & 2015-02-10 & 25.6 & 18.6 & 1.00 & 0.02 & 32 \\
 & 2015 CO & LDT 4.3m & 2015-02-10 & 26.2 & 20.4 & 1.04 & 0.05 & 7 \\
 & 2015 EP & LDT 4.3m & 2015-03-10 & 26.0 & 18.9 & 1.01 & 0.02 & 26 \\
 & 2015 TM143 & LDT 4.3m & 2022-05-07 & 23.54 & 20.5 & 1.15 & 0.15 & 16 \\
 & 2015 XE352 & LDT 4.3m & 2021-10-29 & 20.94 & 20.7 & 1.38 & 0.42 & 18 \\
 & 2015 XT129 & LDT 4.3m & 2016-02-22 & 21.98 & 19.7 & 1.17 & 0.19 & 20 \\
 & 2016 AD166 & LDT 4.3m & 2016-01-19 & 23.58 & 19.9 & 1.07 & 0.10 & 26 \\
 & 2016 AG166 & LDT 4.3m & 2016-01-19 & 24.0 & 20.1 & 1.06 & 0.09 & 30 \\
 & 2016 AG193 & LDT 4.3m & 2025-02-02 & 23.5 & 21.0 & 1.12 & 0.15 & 30 \\
 & 2016 AO131 & LDT 4.3m & 2016-01-19 & 24.3 & 20.8 & 1.05 & 0.09 & 40 \\
 & 2016 AO131 & LDT 4.3m & 2022-02-11 & 24.3 & 20.9 & 1.13 & 0.14 & 9 \\
 & 2016 AV164 & LDT 4.3m & 2016-01-19 & 24.9 & 20.2 & 1.06 & 0.08 & 13 \\
 & 2016 DK & LDT 4.3m & 2016-02-22 & 22.4 & 19.6 & 1.05 & 0.11 & 54 \\
 & 2016 GW216 & LDT 4.3m & 2022-05-23 & 22.63 & 20.1 & 1.15 & 0.16 & 27 \\
 & 2016 GW216 & LDT 4.3m & 2022-05-31 & 22.63 & 20.6 & 1.17 & 0.18 & 32 \\
 & 2016 HP3 & LDT 4.3m & 2016-05-22 & 23.7 & 19.2 & 1.04 & 0.05 & 53 \\
 & 2016 LT1 & LDT 4.3m & 2016-06-07 & 29.0 & 17.9 & 1.02 & 0.00 & 21 \\
 & 2016 QJ44 & LDT 4.3m & 2021-10-28 & 20.24 & 19.8 & 1.27 & 0.35 & 32 \\
 & 2016 QJ44 & LDT 4.3m & 2021-10-29 & 20.24 & 19.8 & 1.26 & 0.34 & 33 \\
 & 2017 VR12 & LDT 4.3m & 2017-12-09 & 20.66 & 20.2 & 1.24 & 0.33 & 34 \\
 & 2017 XE & LDT 4.3m & 2017-12-09 & 26.0 & 17.0 & 1.00 & 0.01 & 14 \\
 & 2018 CN2 & LDT 4.3m & 2018-02-08 & 27.8 & 20.3 & 1.00 & 0.02 & 42 \\
 & 2018 HO1 & LDT 4.3m & 2018-04-25 & 25.31 & 19.7 & 1.04 & 0.04 & 36 \\
 & 2018 JJ2 & LDT 4.3m & 2018-05-24 & 23.9 & 19.4 & 1.06 & 0.06 & 36 \\
 & 2018 JJ3 & LDT 4.3m & 2018-05-24 & 24.0 & 19.8 & 1.08 & 0.08 & 26 \\
 & 2018 KH & LDT 4.3m & 2018-05-24 & 24.5 & 20.3 & 1.07 & 0.07 & 35 \\
 & 2018 LB1 & LDT 4.3m & 2018-06-12 & 20.11 & 21.7 & 1.68 & 0.76 & 22 \\
 & 2018 LC1 & LDT 4.3m & 2018-06-12 & 24.9 & 19.8 & 1.04 & 0.04 & 49 \\
 & 2018 LV2 & LDT 4.3m & 2018-06-12 & 25.3 & 21.4 & 1.07 & 0.07 & 39 \\
 & 2018 MB7 & LDT 4.3m & 2018-07-03 & 23.8 & 17.0 & 1.02 & 0.01 & 82 \\
 & 2018 PK21 & LDT 4.3m & 2019-10-18 & 25.88 & 21.1 & 1.05 & 0.06 & 25 \\
 & 2019 AM10 & LDT 4.3m & 2019-01-26 & 25.2 & 19.7 & 1.03 & 0.05 & 18 \\
 & 2019 EJ & LDT 4.3m & 2019-03-15 & 24.6 & 20.9 & 1.06 & 0.09 & 36 \\
 & 2019 GB4 & LDT 4.3m & 2019-05-28 & 23.94 & 20.0 & 1.10 & 0.09 & 21 \\
 & 2019 LV1 & LDT 4.3m & 2019-06-24 & 25.43 & 19.5 & 1.03 & 0.02 & 56 \\
 & 2019 PR2 & LDT 4.3m & 2019-10-18 & 18.68 & 18.7 & 1.17 & 0.35 & 52 \\
 & 2019 QE1 & LDT 4.3m & 2019-09-06 & 25.2 & 20.1 & 1.02 & 0.03 & 62 \\
 & 2019 QR6 & LDT 4.3m & 2019-10-18 & 20.05 & 20.0 & 1.17 & 0.35 & 52 \\
 & 2019 QT2 & LDT 4.3m & 2019-09-06 & 23.94 & 20.3 & 1.08 & 0.09 & 34 \\
 & 2019 QU4 & LDT 4.3m & 2019-09-06 & 24.8 & 21.4 & 1.00 & 0.03 & 110 \\
 & 2019 QV4 & LDT 4.3m & 2019-09-06 & 24.3 & 19.6 & 1.04 & 0.05 & 51 \\
 & 2019 QW4 & LDT 4.3m & 2019-09-06 & 24.6 & 19.8 & 1.07 & 0.06 & 19 \\
 & 2019 RA & LDT 4.3m & 2019-09-06 & 25.46 & 17.0 & 1.02 & 0.01 & 20 \\
 & 2019 SE9 & LDT 4.3m & 2019-10-03 & 25.16 & 19.7 & 1.03 & 0.04 & 43 \\
 & 2019 SF1 & LDT 4.3m & 2019-09-22 & 26.37 & 20.2 & 1.05 & 0.04 & 8 \\
 & 2019 SJ8 & LDT 4.3m & 2019-10-18 & 24.5 & 18.1 & 1.02 & 0.03 & 24 \\
 & 2019 SU3 & LDT 4.3m & 2019-10-03 & 27.27 & 18.7 & 1.02 & 0.01 & 9 \\
 & 2019 SY4 & LDT 4.3m & 2019-10-03 & 24.7 & 20.0 & 1.06 & 0.07 & 21 \\
 & 2019 VW & LDT 4.3m & 2019-11-06 & 26.75 & 18.9 & 1.01 & 0.02 & 18 \\
 & 2019 WR4 & LDT 4.3m & 2020-01-04 & 26.04 & 20.1 & 1.01 & 0.03 & 41 \\
 & 2019 XX & LDT 4.3m & 2020-01-03 & 23.58 & 19.5 & 1.07 & 0.09 & 19 \\
 & 2019 YD3 & LDT 4.3m & 2020-01-04 & 25.9 & 20.0 & 1.02 & 0.04 & 23 \\
 & 2019 YF4 & LDT 4.3m & 2020-01-04 & 27.0 & 18.9 & 1.00 & 0.01 & 26 \\
 & 2019 YM3 & LDT 4.3m & 2020-01-03 & 23.3 & 18.0 & 1.04 & 0.06 & 17 \\
 & 2019 YT2 & LDT 4.3m & 2020-01-03 & 24.77 & 19.3 & 1.02 & 0.04 & 35 \\
 & 2020 AE & LDT 4.3m & 2020-01-04 & 25.7 & 20.4 & 1.02 & 0.05 & 28 \\
 & 2020 BE8 & LDT 4.3m & 2020-02-27 & 25.18 & 20.9 & 1.09 & 0.10 & 7 \\
 & 2020 BP13 & LDT 4.3m & 2022-04-07 & 21.06 & 20.3 & 1.03 & 0.20 & 76 \\
 & 2020 CA3 & LDT 4.3m & 2020-02-27 & 25.2 & 19.8 & 1.05 & 0.06 & 11 \\
 & 2020 DJ & LDT 4.3m & 2020-02-27 & 26.2 & 19.2 & 1.01 & 0.03 & 18 \\
 & 2020 FK3 & LDT 4.3m & 2020-03-30 & 25.3 & 19.6 & 1.03 & 0.04 & 30 \\
 & 2021 GK1 & LDT 4.3m & 2021-04-30 & 27.02 & 19.4 & 1.02 & 0.01 & 51 \\
 & 2021 GY1 & LDT 4.3m & 2024-09-26 & 23.97 & 21.2 & 1.17 & 0.17 & 12 \\
 & 2021 PB15 & LDT 4.3m & 2022-02-11 & 21.92 & 20.3 & 1.09 & 0.18 & 52 \\
 & 2021 PS16 & LDT 4.3m & 2022-02-11 & 23.12 & 19.3 & 1.06 & 0.08 & 35 \\
 & 2021 UH2 & LDT 4.3m & 2025-02-02 & 25.13 & 21.0 & 1.10 & 0.12 & 4 \\
 & 2021 VX7 & LDT 4.3m & 2021-11-15 & 24.67 & 20.7 & 1.06 & 0.09 & 26 \\
 & 2022 BX5 & LDT 4.3m & 2022-02-08 & 29.04 & 21.8 & 1.01 & 0.02 & 17 \\
 & 2022 CJ1 & LDT 4.3m & 2022-02-08 & 25.56 & 21.4 & 1.05 & 0.07 & 35 \\
 & 2022 DC5 & LDT 4.3m & 2022-03-09 & 22.94 & 20.2 & 1.13 & 0.15 & 24 \\
 & 2022 DX & LDT 4.3m & 2022-03-09 & 28.3 & 21.5 & 1.01 & 0.02 & 28 \\
 & 2022 FC3 & LDT 4.3m & 2022-04-02 & 25.5 & 22.0 & 0.93 & 0.96 & 63 \\
 & 2022 NX1 & LDT 4.3m & 2022-08-30 & 28.07 & 22.3 & 1.02 & 0.02 & 66 \\
 & 2022 OB5 & LDT 4.3m & 2025-02-02 & 28.96 & 22.8 & 1.00 & 0.03 & 48 \\
 & 2023 BU & LDT 4.3m & 2023-01-25 & 29.69 & 18.7 & 0.99 & 0.00 & 28 \\
 & 2023 OP2 & LDT 4.3m & 2023-11-12 & 22.99 & 20.0 & 1.07 & 0.11 & 41 \\
 & 2024 NH & LDT 4.3m & 2024-07-06 & 25.41 & 21.0 & 1.07 & 0.06 & 31 \\
 & 2024 OY2 & LDT 4.3m & 2024-08-14 & 25.14 & 18.4 & 1.02 & 0.02 & 65 \\
 & 2024 PJ1 & LDT 4.3m & 2024-08-14 & 24.56 & 20.6 & 1.11 & 0.10 & 16 \\
 & 2024 PT5 & LDT 4.3m & 2024-08-14 & 27.45 & 18.0 & 1.02 & 0.00 & 56 \\
 & 2024 RZ3 & LDT 4.3m & 2024-09-07 & 26.88 & 21.4 & 1.04 & 0.04 & 32 \\
 & 2024 YR4 & LDT 4.3m & 2025-01-07 & 23.92 & 20.0 & 1.08 & 0.10 & 16 \\
 & 2024 YR11 & LDT 4.3m & 2025-01-07 & 25.45 & 20.2 & 1.02 & 0.04 & 36 \\
138175 & 2000 EE104 & Kitt Peak Mayall 4m & 2017-01-19 & 20.49 & 19.5 & 1.29 & 0.33 & 18 \\
226554 & 2003 WR21 & Kitt Peak Mayall 4m & 2017-01-19 & 19.62 & 20.7 & 1.11 & 0.53 & 62 \\
334412 & 2002 EZ2 & Kitt Peak Mayall 4m & 2016-04-23 & 20.32 & 20.0 & 1.30 & 0.36 & 30 \\
363505 & 2003 UC20 & Kitt Peak Mayall 4m & 2017-01-19 & 18.47 & 20.3 & 1.04 & 0.79 & 63 \\
382875 & 2004 KE1 & Kitt Peak Mayall 4m & 2017-01-07 & 21.73 & 21.4 & 1.45 & 0.47 & 7 \\
403247 & 2008 XO2 & Kitt Peak Mayall 4m & 2016-04-23 & 19.25 & 20.0 & 1.24 & 0.47 & 51 \\
467917 & 2011 OP24 & Kitt Peak Mayall 4m & 2016-04-23 & 19.91 & 19.7 & 1.05 & 0.27 & 74 \\
 & 2006 QA31 & Kitt Peak Mayall 4m & 2016-04-23 & 20.05 & 19.4 & 1.23 & 0.30 & 37 \\
 & 2009 DL46 & Kitt Peak Mayall 4m & 2016-05-07 & 21.99 & 19.2 & 1.01 & 0.07 & 86 \\
 & 2013 WT45 & Kitt Peak Mayall 4m & 2016-04-23 & 20.36 & 19.9 & 0.97 & 0.17 & 96 \\
 & 2014 OV3 & Kitt Peak Mayall 4m & 2015-02-10 & 23.2 & 20.3 & 1.15 & 0.17 & 10 \\
 & 2014 UD57 & Kitt Peak Mayall 4m & 2014-10-27 & 25.8 & 18.4 & 1.02 & 0.02 & 14 \\
 & 2014 UX7 & Kitt Peak Mayall 4m & 2014-10-27 & 25.6 & 20.7 & 1.07 & 0.07 & 9 \\
 & 2015 AA44 & Kitt Peak Mayall 4m & 2015-02-10 & 23.9 & 19.7 & 1.01 & 0.05 & 61 \\
 & 2015 GC14 & Kitt Peak Mayall 4m & 2015-04-24 & 24.8 & 20.9 & 1.10 & 0.10 & 19 \\
 & 2015 HU9 & Kitt Peak Mayall 4m & 2015-05-08 & 23.36 & 20.8 & 1.10 & 0.13 & 43 \\
 & 2015 JD & Kitt Peak Mayall 4m & 2015-05-08 & 25.5 & 18.3 & 1.02 & 0.02 & 38 \\
 & 2015 KM120 & Kitt Peak Mayall 4m & 2015-05-26 & 24.7 & 20.3 & 1.04 & 0.05 & 54 \\
 & 2015 RF36 & Kitt Peak Mayall 4m & 2015-09-14 & 23.4 & 18.8 & 1.05 & 0.06 & 35 \\
 & 2016 CF194 & Kitt Peak Mayall 4m & 2016-06-08 & 24.18 & 19.7 & 1.07 & 0.06 & 33 \\
 & 2016 CL29 & Kitt Peak Mayall 4m & 2016-02-08 & 24.6 & 19.9 & 1.04 & 0.06 & 24 \\
 & 2016 FU13 & Kitt Peak Mayall 4m & 2016-04-23 & 22.3 & 19.5 & 1.10 & 0.13 & 37 \\
 & 2016 LR1 & Kitt Peak Mayall 4m & 2016-06-08 & 23.6 & 20.6 & 1.09 & 0.11 & 43 \\
 & 2017 AT4 & Kitt Peak Mayall 4m & 2017-01-07 & 26.6 & 19.1 & 1.01 & 0.03 & 1 \\
141018 & 2001 WC47 & SOAR 4.2m & 2017-01-13 & 19.04 & 18.8 & 1.20 & 0.35 & 44 \\
487583 & 2015 FJ36 & SOAR 4.2m & 2016-02-22 & 22.19 & 19.5 & 1.13 & 0.16 & 24 \\
 & 2009 CR4 & SOAR 4.2m & 2016-02-22 & 21.67 & 19.5 & 1.14 & 0.18 & 28 \\
 & 2012 KT12 & SOAR 4.2m & 2019-05-14 & 26.8 & 18.3 & 1.02 & 0.01 & 15 \\
 & 2015 FO124 & SOAR 4.2m & 2018-01-26 & 21.34 & 19.5 & 1.26 & 0.28 & 4 \\
 & 2015 TY178 & SOAR 4.2m & 2016-02-22 & 21.8 & 19.3 & 1.14 & 0.17 & 23 \\
 & 2016 CS247 & SOAR 4.2m & 2016-02-22 & 25.66 & 19.5 & 1.02 & 0.03 & 27 \\
 & 2016 NK39 & SOAR 4.2m & 2016-08-16 & 23.86 & 20.4 & 1.07 & 0.09 & 46 \\
 & 2016 UJ101 & SOAR 4.2m & 2016-11-14 & 20.52 & 19.5 & 1.12 & 0.23 & 51 \\
 & 2017 QK & SOAR 4.2m & 2017-08-21 & 23.86 & 20.9 & 1.15 & 0.15 & 18 \\
 & 2017 XG1 & SOAR 4.2m & 2017-12-21 & 24.88 & 20.7 & 1.03 & 0.07 & 45 \\
 & 2017 XJ1 & SOAR 4.2m & 2017-12-21 & 23.71 & 20.5 & 1.04 & 0.09 & 50 \\
 & 2018 BQ & SOAR 4.2m & 2018-01-26 & 25.7 & 19.5 & 1.00 & 0.02 & 51 \\
 & 2018 BU1 & SOAR 4.2m & 2018-01-26 & 24.6 & 16.3 & 0.99 & 0.01 & 36 \\
 & 2018 PU23 & SOAR 4.2m & 2018-08-20 & 28.1 & 20.6 & 1.03 & 0.02 & 18 \\
 & 2018 RA4 & SOAR 4.2m & 2018-09-18 & 24.33 & 20.2 & 1.10 & 0.09 & 15 \\
 & 2018 RS1 & SOAR 4.2m & 2018-10-20 & 23.8 & 20.6 & 1.05 & 0.09 & 53 \\
 & 2018 TO1 & SOAR 4.2m & 2018-10-20 & 21.7 & 19.6 & 1.19 & 0.21 & 17 \\
 & 2018 TZ5 & SOAR 4.2m & 2018-10-20 & 24.4 & 19.6 & 1.06 & 0.07 & 17 \\
 & 2018 YS2 & SOAR 4.2m & 2019-01-17 & 24.5 & 20.8 & 1.09 & 0.11 & 18 \\
 & 2019 BJ2 & SOAR 4.2m & 2019-03-26 & 23.33 & 20.1 & 1.12 & 0.13 & 18 \\
 & 2019 CC3 & SOAR 4.2m & 2019-02-15 & 25.8 & 20.8 & 1.06 & 0.07 & 9 \\
 & 2019 CE3 & SOAR 4.2m & 2019-02-15 & 25.4 & 20.4 & 1.04 & 0.06 & 27 \\
 & 2019 CE5 & SOAR 4.2m & 2019-03-16 & 23.43 & 19.8 & 1.08 & 0.10 & 29 \\
 & 2019 CY1 & SOAR 4.2m & 2019-02-15 & 25.7 & 19.3 & 1.03 & 0.04 & 4 \\
 & 2019 CZ2 & SOAR 4.2m & 2019-02-15 & 24.6 & 19.3 & 1.03 & 0.05 & 24 \\
 & 2019 EH & SOAR 4.2m & 2019-03-16 & 24.92 & 20.6 & 1.05 & 0.07 & 38 \\
 & 2019 EO2 & SOAR 4.2m & 2019-03-26 & 23.36 & 20.4 & 1.05 & 0.10 & 53 \\
 & 2019 EP1 & SOAR 4.2m & 2019-03-26 & 25.6 & 21.0 & 1.08 & 0.09 & 9 \\
 & 2019 EP2 & SOAR 4.2m & 2019-03-26 & 23.5 & 19.7 & 1.08 & 0.09 & 28 \\
 & 2019 JL3 & SOAR 4.2m & 2019-05-14 & 24.92 & 19.2 & 1.03 & 0.03 & 41 \\
 & 2019 RH1 & SOAR 4.2m & 2019-10-09 & 24.8 & 20.0 & 1.06 & 0.07 & 19 \\
 & 2019 UN12 & SOAR 4.2m & 2019-11-07 & 21.86 & 18.3 & 1.09 & 0.11 & 20 \\
 & 2019 VK & SOAR 4.2m & 2019-11-07 & 24.63 & 20.4 & 1.04 & 0.06 & 43 \\
 & 2019 YD1 & SOAR 4.2m & 2020-01-03 & 25.07 & 20.0 & 1.03 & 0.05 & 32 \\
 & 2019 YE2 & SOAR 4.2m & 2020-01-03 & 24.67 & 20.4 & 1.06 & 0.08 & 19 \\
 & 2019 YF4 & SOAR 4.2m & 2020-01-03 & 27.0 & 19.2 & 1.00 & 0.02 & 28 \\
 & 2019 YK & SOAR 4.2m & 2020-01-03 & 24.1 & 19.7 & 0.99 & 0.04 & 79 \\
 & 2020 AC & SOAR 4.2m & 2020-01-03 & 26.7 & 18.6 & 0.99 & 0.01 & 59 \\
\enddata
\end{deluxetable}

The observing strategy for each of the facility/instrument combinations was similar. In all cases, we used the SDSS $griz$ filter set, with associated band centers at 0.47, 0.62, 0.75, and 0.89 $\mu m$ respectively \citep{fukugita96}. For all three of the instruments, filter changes were required to collect the consecutive multi-band images. This non-simultaneous approach to collecting images thus demanded careful consideration of several timescales critical to the derivation of accurate colors. The most important timescale was the rotation period of the asteroid. If the brightness of the asteroid changed significantly during the collection of the multi-band images, the derived colors could be wildly incorrect. This is demonstrated and explored in greater detail in Section \ref{sec:LC}. In short, if the rotation period of an asteroid were comparable to the time required to collect a full set of $griz$ images (typically $\sim10's$ of minutes), the derived colors could be erroneous. We can consider previously studied objects analogous to the MANOS sample to assess the possible extent of this issue. From the Asteroid Lightcurve Database \citep{warner09}, 570 out of 1230 NEOs with $H>20$ have rotation periods less than 1 hour, suggesting that lightcurve variations could be an issue for up to about 50\% of objects in the MANOS color sample. Of course, very fast rotation (periods $<<1$ minute) would not be a problem as brightness variations would average out during typical exposures, but such rapid rotators only represent a few percent of the measured periods for objects with $H>20$ \citep{warner09}.

Unfortunately, rotation periods and lightcurve amplitudes for the large majority of our targets were not known in advance. Several strategies were thus implemented to mitigate the effects of rotation on the derived colors. First, we checked on existing observations cataloged at the Minor Planet Center (MPC). Generally the focus for data submitted to the MPC is astrometry, however surveys such as ATLAS \citep{tonry18} and PanSTARRS \citep{chambers16} employ reduction pipelines that produce consistent photometry that can be used to check for indications of lightcurve variability. Intra- and sometimes inter-night photometric variations from these surveys can be an indication that the lightcurve of a given object may need to be accounted for in color calibration.

A second strategy to mitigate lightcurve effects involved an observing sequence with an interleaved reference filter to track brightness variations. For example, a typical color sequence would involve exposures in the following filter order: $r-g-r-i-r-z-r$. In this case, the $r$-band served as a reference to track underlying lightcurve variability, which was then compensated for when deriving colors (Section \ref{sec:colors}). The $r$-band was most commonly used as the reference filter as it typically produced the highest S/N of the four filters. Exceptions were made when the temporal sampling of the $r$ band was less complete than other filters, when the S/N in $r$ was lower than other filters, or when a greater number of individual exposures in another band provided a better constraint on the lightcurve variability.

A final strategy that was used for a quarter of our sample (49 objects) involved actually measuring the lightcurve before collecting colors. In these cases, the color photometry was rotationally corrected using a Fourier fit to the lightcurve photometry (Section \ref{sec:colors}).

While these observing approaches were similar for all three facilities, there were notable differences related to the specific capabilities of each telescope and instrument.

\subsection{LDT}

At LDT we employed the Large Monolithic Imager \cite[LMI,][]{bida14} for all of our observations. LMI consists of a single 6k $\times$ 6k e2v charge-couple device (CCD) with 15 $\mu m$ pixels. It images a 12.3 arcminute square with an unbinned pixel scale of 0.12 arcsec/pixel. The observations were almost exclusively carried out with the instrument binned 3$\times$3.

In addition to the $griz$ filter set, we used a broad $VR$ filter with high throughput from about 0.5-0.7 $\mu m$. This $VR$ filter was used exclusively for measuring lightcurves and was not used to directly contribute to the determination of colors.

For the large majority of LDT observations, the telescope was tracked at sidereal rates, letting the objects move through the LMI field of view. Individual exposure times were set to minimize trailing to less than two times the local seeing based on the target's non-sidereal rates of motion. 

\subsection{Mayall}

Two instruments were used at the Kitt Peak Mayall 4-m: the Mosaic-1.1 and Mosaic-3 imagers. The Mosaic-1.1 camera consisted of eight 2k $\times$ 4k e2v CCDs that covered a 36 arcminute square field of view at a spatial resolution of 0.26 arcsec/pixel. In November of 2015 Mosaic-1.1 was replaced with the Mosaic-3 camera, which consisted of four 4k $\times$ 4k CCDs from Lawrence Berkeley National Labs (LBNL) that covered the same field of view with the same spatial sampling as its predecessor.

In addition to the $griz$ filter set, we used a broad $wh$ (white) filter, which was essentially a long-pass filter with high throughput longwards of 0.35 $\mu m$. As with the $VR$ filter at LDT, this $wh$ filter was used for measuring lightcurves.

\subsection{SOAR}

At SOAR we used the Goodman Spectrograph and Imager \citep{clemens04}, which in imaging mode captured a 7.2 arcminute circular field imaged onto the red-sensitive 4k square e2v CCD. Goodman is a flexible instrument with both imaging and spectroscopic modes, but it has a limitation of only four available filter slots for imaging. Thus, a typical setup for the instrument involved selecting the three $giz$ filters and a custom $VR$ filter designed after the one at LDT. These specific filters were chosen to capture the visible spectral slope with the $g-i$ color, probe for the presence of a 1 $\mu m$ absorption band with the $i-z$ color, and capture a high S/N lightcurve with the $VR$ filter.

\subsection{Photometry Data Reduction}

All data were reduced using standard bias subtraction and flat field correction techniques. Initial processing of MOSAIC data from the Mayall 4-m was handled by the default NOAO (re-branded as NOIRLab in 2019) pipeline \citep{swaters07}. For both Mosaic-1.1 and Mosaic-3, the data were packaged in multi-extension FITS files. Only the image extensions that contained the targets were extracted and used for downstream processing. The data from SOAR and LDT were all bias and flat corrected using custom Python scripts. The only noteworthy difference was a z-band fringe correction applied to LDT/LMI data taken in that filter. A master fringe frame was constructed from a median combination of 18 z-band images taken at different locations on sky and across multiple nights from November 2019 to January 2020. These images had exposure times ranging from 3 to 40 seconds. This master fringe frame was normalized to its mean and, because fringing is an additive effect, was then multiplied by the mean value of individual frames before being subtracted off.

Photometry measurements for all instruments were performed with the \texttt{Photometry Pipeline} \citep[\texttt{PP},][]{Mommert17}. Use of the \texttt{PP} followed the descriptions for LDT and SOAR in \citet{moskovitz24}. In short, the \texttt{PP} is a Python package with a set of automated routines for deriving calibrated photometry based on catalog values of on-chip field stars. The \texttt{PP} source identification and astrometric registration are based on \texttt{SExtractor} \citep{Bertin96} and \texttt{SCAMP} \citep{Bertin06}. Gaia Data Release 2 \citep[VizieR catalog ID I/345,][]{gaia18} was used to derive an astrometric solution for each frame. A curve-of-growth analysis to determine an optimal photometric aperture was run to optimize S/N on both the asteroid target and calibration field stars. The photometric zero point for each frame was typically solved based on field star magnitudes in the PanSTARRS Data Release 1 catalog \citep[VizieR catalog ID II/349,][]{flewelling20}, though for pointings at southern latitudes that were outside of PanSTARRS sky coverage, we calibrated frames against the SkyMapper Data Release 1 catalog \citep[VizieR catalog ID II/358,][]{wolf18}. To minimize zero-point calibration errors, only field stars with roughly solar colors ($g-r$ and $r-i$ within 0.2 mag of the Sun) were used to calibrate each image. Generally, this resulted in at least half a dozen, and sometimes many tens of field stars for calibration. In rare cases no solar-like stars were present, for example at low galactic latitude where the effects of galactic dust extinction reddened all field stars. In these cases no solar filter was applied to the catalog stars used for calibration. For two reasons, this was unlikely to have had significant consequences for individual objects or the sample as a whole. First, we intentionally avoided targets at low galactic latitude (only 14 of the 199 sets of colors were obtained at latitudes $<10^\circ$). Second, checking the difference in colors derived for objects with and without the solar color filter applied resulted in differences at the level of a few hundredths of a magnitude, much less than typical photometric error bars.

For a small number of targets, the S/N on source in individual frames was too low ($<2$) to derive reliable photometry. This was usually for either the $g$ or $z$ bands where instrumental sensitivity was lowest. When possible, multiple frames in these filters were stacked to extract a single measurement. This image stacking leveraged the {\it pp\_stackedphotometry} routine in the \texttt{PP}. This routine was validated by comparing colors derived from stacked and unstacked images for a selection of high S/N objects; this comparison produced essentially identical results. Of course stacking images precluded any correction for lightcurve variability between exposures, but it did provide a way to recover useful data for a small subset of objects. 

\subsection{Spectroscopic Observations} \label{subsec:spectra}

In addition to the primary data set of colors, we collected new spectra of 6 objects that serve as an independent check on the color data (Section \ref{subsec:consistency}). The observational circumstances for these spectroscopic observations are summarized in Table \ref{tab:spec_obs}. The GMOS instruments at the twin Gemini North and South 8-m telescopes, and the Deveny spectrograph at the LDT were used to collect visible spectra. The instrumental setups, reduction steps, and process of taxonomic assignment for these visible spectra were identical to those described in \citet{devogele19}. One object, 2023~BU, was a target for near-IR spectroscopy from NASA's Infrared Telescope Facility (IRTF) with the SpeX instrument \citep{rayner03}. Reduction of those data leveraged the Spextool package \citep{cushing04} and followed the procedures described in \citet{moskovitz17}.

\begin{deluxetable}{llclccccc}
\tablecaption{Observational circumstances for spectroscopic targets. \label{tab:spec_obs}}
\tabletypesize{\small}
\tablehead{\colhead{Number} & \colhead{Designation} & \colhead{Facility/Instrument} & \colhead{Obs. Date} & \colhead{H} & \colhead{$V$ (mag)} & \colhead{$r$ (au)} & \colhead{$\Delta$ (au)} & \colhead{$\alpha$ (deg)}}
\startdata
 439437 & 2013 NK4 & LDT/Deveny & 2024-05-01 & 18.95 & 18.3 & 1.04 & 0.21 & 76 \\
  & 2019 CE3 & Gemini-S/GMOS & 2019-02-14 & 25.3 & 20.2 & 1.04 & 0.05 & 25 \\
  & 2019 QE1 & Gemini-S/GMOS & 2019-08-30 & 25.2 & 19.8 & 1.05 & 0.04 & 34 \\
  & 2019 VW & Gemini-N/GMOS & 2019-11-09 & 26.8 & 19.9 & 1.01 & 0.02 & 32 \\
  & 2019 YK & Gemini-S/GMOS & 2019-12-25 & 24.1 & 22.3 & 1.23 & 0.24 & 15 \\
  & 2023 BU & IRTF/SpeX & 2023-01-25 & 28.7 & 17.4 & 0.99 & 0.003 & 28 \\
\enddata
\end{deluxetable}

\section{Derivation of Colors} \label{sec:colors}

\begin{figure}
    \centering
    \includegraphics[]{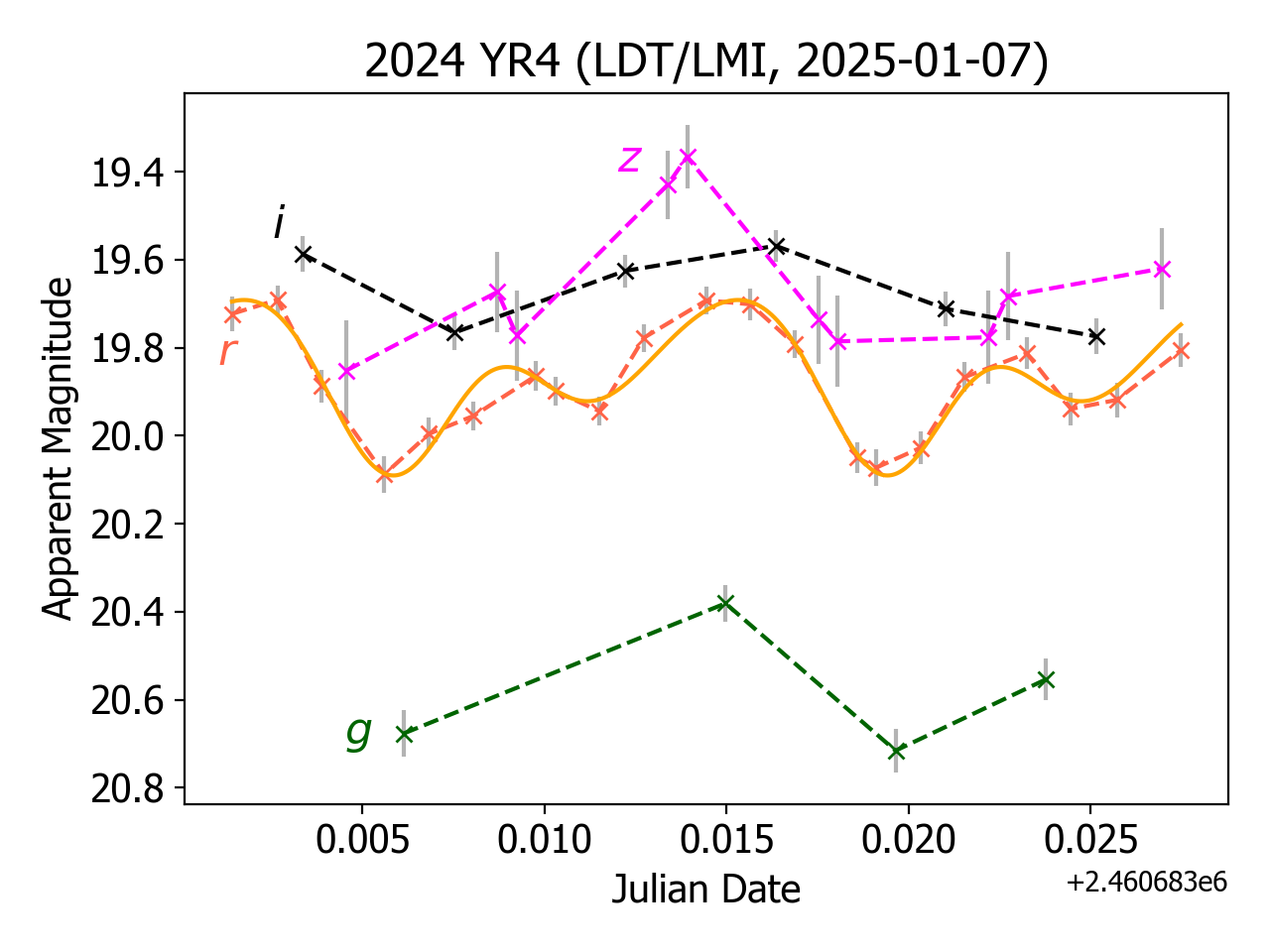}
    \caption{Time series $griz$ photometry for 2024~YR4 obtained from the LDT on 7 January 2025. Sets of exposures in the individual filters are color coded and annotated. The $r$ band data were obtained to track lightcurve variations over the 30 minute color sequence. Fitting just those data with a 3rd order Fourier series (orange line) produced a reasonable period solution (P $\sim$ 0.33 hr) and amplitude (0.4 mag) consistent with other studies of this object \citep{bolin25}. This fitted lightcurve was used to compute the $r$ magnitude at specific times throughout the imaging sequence so that $g-r$, $r-i$, and $r-z$ colors could be retrieved.}
    \label{fig:yr4}
\end{figure}

Following the data reduction steps in the previous section, we now describe the process of deriving colors. The full set of measured colors is presented in Table \ref{tab:results}. As noted above, a typical color sequence would be captured with a single reference filter interleaved to track lightcurve variations. In some cases the reference filter data were used to constrain a lightcurve solution (Figure \ref{fig:yr4}). For cases where the reference filter indicated slower rotation, e.g. a monotonic change in brightness, lightcurve variability was approximated with a low-order polynomial fit. The order of the polynomial was set to minimize the reduced chi-squared statistic associated with the fit to the reference filter data. For a subset of targets, dedicated lightcurve observations were obtained before the color sequence. These lightcurve data differed from the reference filter sequence in several ways: the lightcurve observations typically involved an uninterrupted set spanning $\sim1$ hour, they were captured in one of the broad $VR$ or $wh$ filters, and the Fourier fits to these data were then extrapolated to the times of the color sequence. The Fourier fitting process followed the standard formalism of \citet{harris89}. More details on the measured lightcurves are given in Section \ref{sec:LC}.

\startlongtable
\begin{deluxetable}{llclccccccccclc}
\tablecaption{Set of measured colors and uncertainties for our full sample of 189 objects. The method of lightcurve  correction (LC Corr.) is indicated in the third column with the fitting function either a Fourier series or a polynomial of Nth order, e.g. a second order polynomial is indicated with poly=2. The mean reference magnitude (Ref. mag) used to track LC variations is indicated in the corresponding column. The final two columns are the best fit taxonomic classification in the Bus-DeMeo system and the RMS associated with that fit. More details and notes on specific objects are given in the text.  \label{tab:results}}
\tabletypesize{\scriptsize}
\tablehead{\colhead{Number} & \colhead{Designation} & \colhead{LC Corr.} & \colhead{Ref. mag} & \colhead{Ref. mag err} & \colhead{$g-r$} & \colhead{$\sigma_{g-r}$} & \colhead{$r-i$} & \colhead{$\sigma_{r-i}$} & \colhead{$g-i$} & \colhead{$\sigma_{g-i}$} & \colhead{$i-z$} & \colhead{$\sigma_{i-z}$} & \colhead{Type} & \colhead{RMS}}
\startdata
10302 & 1989 ML&poly=2&r=18.087&0.023&0.542&0.006&0.144&0.013&0.687&0.015&0.024&0.025&Xk&0.009 \\
12923 & Zephyr&poly=2&r=15.685&0.016&0.637&0.009&0.181&0.003&0.818&0.01&-0.104&0.022&Sq&0.013 \\
22099 & 2000 EX106&poly=1&r=19.037&0.045&0.807&0.087&0.109&0.032&0.916&0.093&-0.183&0.049&Sa&0.045 \\
52381 & 1993 HA&poly=0&r=20.281&0.035&0.580&0.063&-&-&-&-&-&-&Xe&0.013 \\
85628 & 1998 KV2&poly=0&g=18.970&0.02&0.699&0.001&0.192&0.022&0.891&0.021&-0.164&0.042&Sr&0.018 \\
89136 & 2001 US16&poly=2&r=21.590&0.082&0.681&0.043&0.143&0.067&0.823&0.079&-0.187&0.112&Q&0.029 \\
138175 & 2000 EE104&poly=1&g=19.813&0.037&0.482&0.024&0.199&0.031&0.681&0.02&-0.023&0.044&Xk&0.026 \\
141018 & 2001 WC47&poly=2&r=20.692&0.035&0.701&0.023&0.258&0.023&0.959&0.032&-0.036&0.034&L&0.034 \\
141018 & 2001 WC47&poly=0&r=19.401&0.032&0.789&0.006&0.232&0.01&1.021&0.012&-0.113&0.01&A&0.025 \\
141018 & 2001 WC47&poly=1&i=18.321&0.071&-&-&-&-&1.095&0.084&-0.095&0.038&A&0.081 \\
154555 & 2003 HA&poly=1&g=18.864&0.03&0.647&0.005&0.117&0.017&0.764&0.016&-0.146&0.017&Q&0.021 \\
226554 & 2003 WR21&poly=1&r=20.993&0.053&0.648&0.025&0.242&0.055&0.89&0.06&-0.388&0.114&V&0.039 \\
264993 & 2003 DX10&poly=0&r=17.606&0.027&0.542&0.024&0.205&0.022&0.747&0.033&-0.034&0.042&K&0.023 \\
267337 & 2001 VK5&poly=0&r=21.640&0.048&0.671&0.083&0.203&0.046&0.874&0.095&-0.724&0.113&V&0.094 \\
292220 & 2006 SU49&poly=0&r=17.699&0.016&0.77&0.056&0.206&0.054&0.976&0.078&-0.216&0.058&Sa&0.024 \\
303450 & 2005 BY2&poly=1&r=19.177&0.093&0.64&0.05&0.173&0.039&0.813&0.063&-0.104&0.093&Sq&0.013 \\
334412 & 2002 EZ2&poly=2&r=19.642&0.058&0.825&0.116&0.161&0.006&0.987&0.116&-0.171&0.039&Sa&0.024 \\
350713 & 2001 XP88&poly=3&g=21.514&0.077&0.641&0.072&0.161&0.074&0.802&0.018&-0.135&0.098&Sq&0.025 \\
350751 & 2002 AW&poly=1&i=17.790&0.017&0.482&0.032&0.082&0.029&0.564&0.015&0.001&0.018&B&0.014 \\
363505 & 2003 UC20&poly=1&r=19.997&0.031&0.503&0.009&0.155&0.007&0.659&0.011&0.073&0.045&X&0.004 \\
382875 & 2004 KE1&poly=2&i=21.128&0.070&-&-&0.251&0.133&-&-&-0.295&0.205&Q&0.052 \\
403247 & 2008 XO2&poly=1&r=19.940&0.032&0.74&0.039&0.161&0.021&0.902&0.044&-0.198&0.053&R&0.028 \\
412983 & 1996 FO3&poly=1&r=19.753&0.02&0.628&0.014&0.168&0.017&0.796&0.022&0.02&0.024&Xe&0.016 \\
412983 & 1996 FO3&poly=7&r=18.816&0.03&0.536&0.01&0.117&0.022&0.652&0.024&0.085&0.044&Xc&0.015 \\
439437 & 2013 NK4&poly=1&i=16.511&0.023&0.695&0.007&0.048&0.006&0.742&0.004&-0.493&0.001&O&0.074 \\
455176 & 1999 VF22&poly=0&r=19.477&0.019&0.464&0.007&0.1&0.019&0.564&0.02&0.025&0.03&C&0.011 \\
467917 & 2011 OP24&poly=1&r=19.416&0.048&0.499&0.029&0.087&0.029&0.587&0.041&-0.215&0.058&O&0.034 \\
481965 & 2009 EB1&poly=1&r=21.287&0.047&0.497&0.07&0.139&0.046&0.636&0.084&0.055&0.074&Xk&0.017 \\
487583 & 2015 FJ36&poly=3&r=19.266&0.039&0.649&0.026&0.174&0.026&0.823&0.037&-0.232&0.037&Q&0.030 \\
510529 & 2012 EY11&poly=1&r=21.098&0.039&0.511&0.065&0.138&0.061&0.649&0.089&0.056&0.098&Xc&0.012 \\
513312 & 2007 DM41&poly=4&r=21.801&0.075&0.723&0.093&-0.035&0.136&0.689&0.165&-0.48&0.228&O&0.097 \\
523599 & 2003 RM&poly=0&r=21.450&0.038&0.516&0.03&0.156&0.001&0.672&0.03&-0.012&0.098&Xk&0.024 \\
523813 & 2008 VB1&poly=0&r=20.131&0.02&0.593&0.037&0.165&0.007&0.758&0.037&-0.07&0.031&Sq&0.030 \\
523813 & 2008 VB1&poly=3&r=20.283&0.026&0.549&0.02&0.192&0.032&0.74&0.038&-0.017&0.057&K&0.021 \\
523828 & 1992 BC&poly=0&r=21.455&0.047&0.584&0.065&0.242&0.016&0.826&0.067&0.066&0.105&D&0.024 \\
533638 & 2014 KT86&poly=0&r=19.527&0.02&0.697&0.017&0.144&0.026&0.841&0.031&-0.163&0.048&Sq&0.034 \\
& 1991 TF3&poly=1&g=18.400&0.025&0.561&0.008&0.168&0.024&0.729&0.023&-0.078&0.037&Sq&0.041 \\
& 1994 WR12&poly=1&r=19.080&0.054&0.501&0.055&0.178&0.026&0.679&0.061&0.021&0.065&Xk&0.013 \\
& 1998 KY26&poly=0&r=21.734&0.026&0.534&0.05&0.244&0.051&0.778&0.072&0.223&0.112&D&0.068 \\
& 2000 WO148&poly=1&r=20.630&0.032&0.682&0.026&0.188&0.015&0.87&0.03&-0.014&0.097&S&0.028 \\
& 2001 QJ142&fourier&r=20.647&0.045&0.526&0.046&0.126&0.029&0.652&0.055&0.021&0.062&Xk&0.024 \\
& 2002 JR100&poly=2&g=21.448&0.078&0.611&0.012&0.044&0.067&0.654&0.066&0.053&0.069&Cg&0.025 \\
& 2002 TP69&poly=1&g=19.293&0.025&0.708&0.01&0.203&0.024&0.911&0.021&0.012&0.024&L&0.014 \\
& 2003 EH1&poly=3&r=20.568&0.022&0.46&0.011&0.14&0.008&0.6&0.014&-0.001&0.035&C&0.014 \\
& 2003 SM84&poly=4&r=22.217&0.075&0.723&0.037&0.21&0.035&0.933&0.052&-0.111&0.036&S&0.023 \\
& 2005 QA5&poly=0&g=21.763&0.098&0.622&0.035&0.192&0.062&0.814&0.051&0.018&0.064&Xe&0.027 \\
& 2006 QA31&poly=1&r=19.030&0.044&0.62&0.024&0.165&0.067&0.785&0.071&-0.168&0.086&Q&0.031 \\
& 2007 DM41&poly=1&r=19.916&0.068&0.527&0.086&0.136&0.049&0.663&0.099&-0.624&0.102&O&0.089 \\
& 2007 EF&poly=2&r=21.567&0.086&0.61&0.095&0.36&0.056&0.97&0.11&-0.366&0.138&R&0.077 \\
& 2009 CR4&poly=1&i=19.184&0.039&0.728&0.069&0.183&0.032&0.911&0.062&-0.082&0.08&S&0.019 \\
& 2009 DL46&poly=2&r=19.236&0.03&0.562&0.074&0.254&0.052&0.816&0.091&-0.013&0.054&T&0.037 \\
& 2010 XN&poly=2&r=19.655&0.052&0.539&0.03&0.09&0.042&0.629&0.052&0.009&0.072&Cg&0.016 \\
& 2011 CG2&poly=3&r=20.731&0.04&0.533&0.01&0.069&0.02&0.602&0.022&0.06&0.045&Cgh&0.020 \\
& 2011 CG2&poly=2&r=20.953&0.066&0.613&0.041&0.057&0.035&0.67&0.054&-0.127&0.058&Q&0.044 \\
& 2011 YQ10&poly=0&g=17.355&0.016&0.637&0.004&0.122&0.007&0.76&0.006&-0.395&0.011&O&0.044 \\
& 2012 BA35&poly=3&r=21.494&0.131&0.646&0.208&0.219&0.099&0.865&0.23&-0.152&0.183&Sr&0.021 \\
& 2012 BF86&fourier&r=19.332&0.06&0.839&0.138&0.198&0.035&1.037&0.142&-0.092&0.053&A&0.025 \\
& 2012 KT12&poly=4&g=18.930&0.071&-&-&-&-&0.756&0.028&-0.002&0.057&Xe&0.019 \\
& 2013 PA7&poly=0&i=16.511&0.023&0.695&0.007&0.048&0.006&0.742&0.004&-0.493&0.001&Sq&0.017 \\
& 2013 RB6&poly=2&r=21.430&0.046&0.548&0.034&0.203&0.054&0.751&0.064&0.022&0.066&Xe&0.018 \\
& 2013 WT45&poly=0&r=19.782&0.064&-&-&0.112&0.065&-&-&-0.288&0.135&O&0.024 \\
& 2014 HW&fourier&r=18.974&0.044&0.554&0.082&0.177&0.103&0.732&0.132&-0.016&0.127&K&0.022 \\
& 2014 KH39&fourier&i=17.508&0.06&0.638&0.133&0.182&0.068&0.82&0.114&-0.087&0.104&Sq&0.017 \\
& 2014 OV3&fourier&i=20.145&0.488&0.643&0.084&0.33&0.034&0.973&0.077&-&-&Sv&0.056 \\
& 2014 SC324&fourier&g=18.935&0.028&0.736&0.033&0.059&0.047&0.795&0.034&-0.413&0.056&O&0.067 \\
& 2014 TR57&poly=0&z=19.064&0.195&0.644&0.06&0.163&0.095&0.807&0.112&-0.057&0.095&K&0.020 \\
& 2014 UD57&fourier&g=19.508&0.137&0.772&0.115&0.12&0.137&0.892&0.075&-&-&Sa&0.037 \\
& 2014 UX7&fourier&g=21.542&0.616&0.491&0.133&0.159&0.146&0.65&0.062&-&-&X&0.007 \\
& 2014 VL6&fourier&r=20.858&0.024&0.608&0.033&0.224&0.033&0.832&0.047&-0.113&0.102&Sq&0.027 \\
& 2015 AA44&fourier&r=19.578&0.117&0.480&0.034&0.197&0.006&0.677&0.035&-&-&X&0.020 \\
& 2015 CG&poly=0&r=18.367&0.034&0.653&0.411&0.263&0.341&0.916&0.534&-0.476&0.62&V&0.034 \\
& 2015 CO&poly=0&r=20.788&0.054&0.504&0.038&0.147&0.035&0.652&0.052&0.021&0.234&Xk&0.013 \\
& 2015 EP &poly=0&r=18.573&0.042&0.636&0.001&0.181&0.027&0.817&0.027&-0.429&0.027&V&0.041 \\
& 2015 FO124&poly=0&z=19.010&0.052&-&-&-&-&0.776&0.031&-0.053&0.024&K&0.023 \\
& 2015 GC14&poly=0&i=20.601&0.469&0.332&0.1&0.223&0.057&0.555&0.082&-0.004&0.053&Cb&0.040 \\
& 2015 HU9&poly=0&r=20.149&0.193&0.661&0.222&0.306&0.278&0.967&0.356&-0.41&0.401&R&0.056 \\
& 2015 JD&fourier&i=17.340&0.07&0.543&0.046&0.161&0.033&0.704&0.032&0.012&0.03&Xk&0.011 \\
& 2015 KM120&fourier&r=19.952&0.226&0.592&0.108&0.082&0.056&0.674&0.122&-0.118&0.163&Q&0.039 \\
& 2015 RF36&poly=0&g=18.961&0.125&0.708&0.055&0.114&0.057&0.821&0.015&-0.187&0.057&Q&0.027 \\
& 2015 TM143&poly=1&r=20.258&0.026&0.417&0.008&0.16&0.03&0.577&0.032&0.037&0.073&Cb&0.015 \\
& 2015 TY178&poly=3&r=19.108&0.032&0.614&0.033&0.173&0.011&0.787&0.035&-0.093&0.04&Sq&0.020 \\
& 2015 XE352&poly=1&r=20.888&0.047&0.607&0.042&0.036&0.039&0.643&0.057&0.099&0.093&Cg&0.037 \\
& 2015 XT129&poly=3&r=19.530&0.049&0.518&0.024&0.207&0.033&0.725&0.041&-0.074&0.049&Xk&0.041 \\
& 2016 AD166&fourier&r=19.898&0.029&0.586&0.046&0.176&0.014&0.762&0.048&-0.192&0.035&Q&0.029 \\
& 2016 AG166&poly=0&r=20.173&0.036&0.46&0.008&0.158&0.016&0.618&0.018&0.078&0.02&X&0.018 \\
& 2016 AG193&poly=2&r=20.756&0.047&0.561&0.029&0.141&0.048&0.702&0.056&0.128&0.129&T&0.032 \\
& 2016 AO131&poly=0&r=20.503&0.057&0.552&0.096&0.088&0.044&0.64&0.105&-0.316&0.209&O&0.015 \\
& 2016 AO131&poly=0&r=20.934&0.049&0.633&0.001&0.06&0.023&0.693&0.023&-0.334&0.074&O&0.032 \\
& 2016 AV164&fourier&r=20.245&0.045&0.592&0.017&0.24&0.018&0.833&0.025&-0.274&0.064&R&0.042 \\
& 2016 CF194&fourier&g=19.999&0.046&0.694&0.022&-&-&-&-&-&-&S&0.021 \\
& 2016 CL29&poly=0&r=19.547&0.03&0.649&0.075&0.16&0.024&0.808&0.078&-0.225&0.042&Q&0.024 \\
& 2016 CS247&poly=0&r=19.167&0.041&0.651&0.068&0.203&0.041&0.855&0.079&-0.082&0.064&S&0.015\\
& 2016 DK&fourier&r=19.673&0.055&0.6&0.086&0.172&0.018&0.772&0.088&-0.114&0.041&Sq&0.028 \\
& 2016 FU13&fourier&r=19.716&0.085&0.681&0.039&0.176&0.05&0.857&0.064&-0.487&0.094&V&0.037 \\
& 2016 GW216&poly=3&r=19.852&0.052&0.604&0.041&0.184&0.024&0.788&0.047&-0.119&0.048&Sq&0.023 \\
& 2016 GW216&poly=0&r=20.225&0.021&0.622&0.009&0.184&0.03&0.805&0.031&-0.148&0.059&Sq&0.025 \\
& 2016 HP3&poly=2&r=19.178&0.05&0.663&0.072&0.204&0.047&0.868&0.086&-0.154&0.074&Sr&0.016 \\
& 2016 LR1&poly=0&i=20.433&0.073&0.73&0.025&0.159&0.023&0.889&0.009&-0.141&0.034&Sr&0.020 \\
& 2016 LT1&fourier&g=17.219&0.047&0.474&0.056&0.102&0.251&0.575&0.244&0.154&0.256&X&0.035 \\
& 2016 NK39&fourier&r=19.710&0.045&0.627&0.099&0.126&0.086&0.753&0.131&-0.03&0.098&K&0.031 \\
& 2016 QJ44&poly=0&g=20.181&0.017&0.668&0.002&0.229&0.026&0.897&0.026&-0.053&0.063&Sv&0.018 \\
& 2016 QJ44&poly=1&r=19.612&0.018&0.706&0.017&0.216&0.014&0.922&0.022&-0.087&0.021&Sv&0.015 \\
& 2016 UJ101&poly=2&g=19.835&0.139&-&-&-&-&0.742&0.012&0.049&0.081&Xe&0.022 \\
& 2017 AT4&fourier&z=19.133&0.049&0.517&0.091&0.093&0.118&0.61&0.148&-0.001&0.118&C&0.016 \\
& 2017 QK&fourier&g=21.408&0.054&-&-&-&-&0.810&0.050&-0.109&0.091&Sq&0.035 \\
& 2017 VR12&poly=0&r=19.799&0.024&0.859&0.046&0.139&0.03&0.998&0.055&-0.375&0.056&V&0.049 \\
& 2017 XE&poly=1&r=17.127&0.025&0.651&0.08&0.13&0.08&0.782&0.113&-0.004&0.089&K&0.020 \\
& 2017 XG1&poly=1&i=20.207&0.064&-&-&-&-&0.816&0.130&-0.195&0.216&Sq&0.045 \\
& 2017 XJ1&fourier&i=19.995&0.053&-&-&-&-&0.810&0.118&0.017&0.072&T&0.029 \\
& 2018 BQ&poly=1&g=19.751&0.052&-&-&-&-&0.719&0.018&0.037&0.062&Xe&0.016 \\
& 2018 BU1&poly=1&g=16.741&0.053&-&-&-&-&0.976&0.055&-0.176&0.104&Sv&0.042 \\
& 2018 CN2 &fourier&r=19.217&0.036&0.59&0.065&0.158&0.027&0.747&0.071&0.087&0.039&T&0.024 \\
& 2018 HO1&fourier&r=19.625&0.105&0.616&0.15&0.146&0.134&0.763&0.201&0.003&0.199&K&0.014 \\
& 2018 JJ2&fourier&g=20.054&0.047&0.598&0.02&0.252&0.036&0.85&0.03&-0.057&0.049&Sv&0.026 \\
& 2018 JJ3&fourier&g=20.030&0.082&0.735&0.021&0.16&0.035&0.895&0.028&-0.272&0.055&R&0.027 \\
& 2018 KH&poly=0&g=21.086&0.141&0.719&0.067&0.154&0.077&0.874&0.039&-0.212&0.073&R&0.029 \\
& 2018 LB1&poly=2&r=21.613&0.048&0.664&0.071&0.118&0.07&0.782&0.099&-0.012&0.121&K&0.027 \\
& 2018 LC1&poly=5&r=19.995&0.04&0.551&0.032&0.164&0.038&0.715&0.05&0.075&0.079&Xe&0.020 \\
& 2018 LV2&poly=0&r=21.273&0.067&0.664&0.05&0.169&0.067&0.833&0.083&-0.039&0.099&K&0.022 \\
& 2018 MB7&poly=1&r=16.228&0.077&0.605&0.037&0.087&0.034&0.692&0.05&0.126&0.071&Xe&0.037 \\
& 2018 PK21&fourier&g=21.156&0.05&0.609&0.044&0.196&0.074&0.804&0.06&-0.149&0.112&Sq&0.026 \\
& 2018 PU23&fourier&g=20.835&0.124&-&-&-&-&0.734&0.068&-0.049&0.068&K&0.032 \\
& 2018 RA4&poly=2&g=20.566&0.102&-&-&-&-&0.667&0.075&-0.290&0.087&O&0.043 \\
& 2018 RS1&poly=0&g=18.145&0.088&-&-&-&-&0.850&0.026&0.107&0.027&D&0.026 \\
& 2018 TO1&poly=1&g=20.188&0.150&-&-&-&-&0.774&0.039&-0.033&0.059&K&0.022 \\
& 2018 TZ5&fourier&g=20.273&0.168&-&-&-&-&0.875&0.100&-0.149&0.169&Sv&0.038 \\
& 2018 YS2&poly=0&i=20.077&0.033&-&-&-&-&0.795&0.034&-0.493&0.103&V&0.063 \\
& 2019 AM10&poly=2&r=19.588&0.044&0.698&0.106&0.121&0.032&0.818&0.111&0.009&0.082&K&0.031 \\
& 2019 BJ2&poly=0&i=19.603&0.053&-&-&-&-&0.493&0.036&0.127&0.026&Cb&0.035 \\
& 2019 CC3&poly=0&z=20.699&0.149&-&-&-&-&0.771&0.179&-0.107&0.087&Sq&0.036 \\
& 2019 CE3&poly=2&i=20.752&0.125&-&-&-&-&0.656&0.053&0.043&0.167&Xk&0.012 \\
& 2019 CE5&fourier&g=20.552&0.106&-&-&-&-&0.777&0.047&-0.012&0.053&Xe&0.024 \\
& 2019 CY1&poly=0&g=19.502&0.061&-&-&-&-&0.780&0.047&0.074&0.070&T&0.010 \\
& 2019 CZ2&fourier&g=19.619&0.049&-&-&-&-&0.724&0.018&-0.033&0.025&Xk&0.029 \\
& 2019 EH&poly=1&i=19.980&0.045&-&-&-&-&0.696&0.054&0.072&0.018&X&0.023 \\
& 2019 EJ&poly=2&g=21.093&0.041&0.668&0.018&0.166&0.03&0.834&0.024&-0.25&0.123&Q&0.032 \\
& 2019 EO2&poly=2&g=20.207&0.068&-&-&-&-&0.547&0.040&0.037&0.101&Cb&0.015 \\
& 2019 EP1&poly=0&i=20.874&0.094&-&-&-&-&0.622&0.201&0.017&0.013&Cb&0.025 \\
& 2019 EP2&poly=0&g=19.966&0.067&-&-&-&-&0.708&0.003&0.009&0.012&Xk&0.018 \\
& 2019 GB4&poly=3&r=19.763&0.033&0.605&0.06&0.177&0.039&0.781&0.072&-0.105&0.146&Sq&0.023 \\
& 2019 JL3&poly=4&g=19.489&0.077&-&-&-&-&0.816&0.005&0.025&0.078&T&0.027 \\
& 2019 LV1&poly=6&r=19.123&0.031&0.727&0.055&0.217&0.02&0.944&0.059&-0.121&0.041&Sv&0.024 \\
& 2019 PR2&poly=1&g=19.396&0.047&0.692&0.014&0.372&0.019&1.064&0.012&0.201&0.024&D&0.222 \\
& 2019 QE1&poly=5&r=20.325&0.053&0.521&0.014&0.121&0.038&0.642&0.04&-0.032&0.078&C&0.023 \\
& 2019 QR6&poly=2&g=20.620&0.081&0.582&0.045&0.295&0.079&0.877&0.065&0.2&0.084&D&0.107 \\
& 2019 QT2&poly=3&r=20.124&0.049&0.506&0.035&0.089&0.063&0.595&0.072&0.147&0.103&X&0.032 \\
& 2019 QU4&poly=1&r=21.407&0.049&0.938&0.139&0.145&0.083&1.083&0.162&-0.046&0.18&A&0.044 \\
& 2019 QV4&poly=2&r=19.460&0.032&0.606&0.027&0.195&0.042&0.801&0.049&0.035&0.069&T&0.026 \\
& 2019 QW4&poly=1&r=19.565&0.025&0.494&0.019&0.176&0.034&0.67&0.039&0.064&0.047&X&0.010 \\
& 2019 RA&fourier&r=16.645&0.025&0.701&0.042&-0.015&0.128&0.686&0.135&-0.077&0.149&Q&0.061 \\
& 2019 RH1&fourier&g=20.237&0.100&-&-&-&-&0.732&0.008&0.120&0.099&Xe&0.017 \\
& 2019 SE9&poly=3&r=19.509&0.027&0.588&0.012&0.172&0.035&0.76&0.037&0.043&0.044&Xe&0.009 \\
& 2019 SF1&poly=4&r=19.971&0.022&0.573&0.013&0.167&0.028&0.74&0.031&-0.141&0.037&Q&0.037 \\
& 2019 SJ8&poly=3&g=18.516&0.053&0.536&0.027&0.14&0.097&0.676&0.093&0.127&0.126&X&0.030 \\
& 2019 SU3&fourier&r=18.656&0.02&0.563&0.006&0.144&0.006&0.707&0.009&0.023&0.019&Xk&0.013 \\
& 2019 SY4&poly=0&r=19.929&0.037&0.678&0.038&0.159&0.046&0.837&0.059&-0.103&0.07&Sq&0.012 \\
& 2019 UN12&poly=1&g=18.594&0.071&-&-&-&-&0.938&0.144&-0.544&0.167&V&0.049 \\
& 2019 VK&poly=0&g=20.721&0.128&-&-&-&-&0.604&0.023&-0.519&0.195&O&0.082 \\
& 2019 VW&fourier&g=19.274&0.063&0.715&0.039&0.093&0.098&0.808&0.09&-0.108&0.168&Sq&0.042 \\
& 2019 WR4&fourier&g=20.252&0.059&0.588&0.003&0.135&0.023&0.723&0.023&-0.009&0.073&Xk&0.026 \\
& 2019 XX&poly=1&g=20.080&0.081&0.568&0.076&0.073&0.124&0.641&0.098&-0.142&0.236&Q&0.052 \\
& 2019 YD1&poly=3&g=20.767&0.083&-&-&-&-&0.956&0.077&-0.478&0.116&V&0.053 \\
& 2019 YD3&poly=2&g=20.095&0.081&0.693&0.093&0.096&0.124&0.789&0.082&-0.214&0.185&Q&0.027 \\
& 2019 YE2&poly=1&g=20.642&0.075&-&-&-&-&0.709&0.09&-0.046&0.16&Xk&0.034 \\
& 2019 YF4&poly=1&g=19.270&0.112&-&-&-&-&0.807&0.027&0.051&0.031&T&0.020 \\
& 2019 YF4&fourier&g=19.133&0.036&0.538&0.035&0.231&0.064&0.769&0.054&-0.08&0.081&K&0.032 \\
& 2019 YK&poly=1&g=19.863&0.083&-&-&-&-&0.704&0.070&-0.002&0.113&Xk&0.018 \\
& 2019 YM3&poly=3&g=18.591&0.027&0.688&0.011&0.06&0.039&0.749&0.037&-0.315&0.042&O&0.048 \\
& 2019 YT2&poly=0&g=19.814&0.046&0.552&0.096&0.051&0.146&0.604&0.11&0.013&0.164&Cg&0.028 \\
& 2020 AC&poly=0&g=19.774&0.296&-&-&-&-&0.829&0.043&-0.564&0.336&V&0.068 \\
& 2020 AE&poly=2&g=21.410&0.138&0.747&0.104&-&-&-&-&-&-&V&0.025 \\
& 2020 BE8&poly=0&g=21.283&0.035&0.519&0.009&0.09&0.03&0.609&0.028&0.074&0.035&Cgh&0.028 \\
& 2020 BP13&poly=2&r=20.282&0.068&-&-&0.209&0.092&-&-&-0.018&0.128&Xk&0.029 \\
& 2020 CA3&fourier&g=20.065&0.023&0.544&0.005&0.156&0.03&0.7&0.03&0.002&0.032&Xk&0.014 \\
& 2020 DJ&fourier&g=20.073&0.061&0.683&0.011&0.116&0.035&0.799&0.033&-0.045&0.057&Sq&0.028 \\
& 2020 FK3&fourier&g=20.073&0.038&0.56&0.016&0.165&0.047&0.725&0.045&-0.046&0.048&Xk&0.032 \\
& 2021 GK1&fourier&r=18.924&0.028&0.537&0.009&0.103&0.024&0.64&0.025&-0.077&0.038&B&0.034 \\
& 2021 GY1&poly=2&r=20.915&0.055&0.683&0.029&0.175&0.067&0.858&0.073&-0.133&0.126&Sr&0.013 \\
& 2021 PB15&poly=1&r=20.113&0.028&0.664&0.014&0.182&0.039&0.846&0.041&-0.197&0.052&R&0.034 \\
& 2021 PS16&poly=2&r=18.901&0.036&0.747&0.05&0.156&0.023&0.904&0.055&-0.121&0.061&Sa&0.022 \\
& 2021 UH2&poly=3&r=20.942&0.051&0.522&0.06&0.157&0.03&0.679&0.067&0.081&0.169&X&0.013 \\
& 2021 VX7&fourier&r=20.509&0.048&0.584&0.105&0.139&0.025&0.723&0.108&-0.129&0.039&Q&0.031 \\
& 2022 BX5&poly=1&r=21.645&0.056&0.893&0.039&0.401&0.013&1.295&0.041&0.316&0.042&D&0.425 \\
& 2022 CJ1&poly=4&r=21.112&0.081&0.398&0.08&0.093&0.01&0.491&0.08&-0.045&0.03&B&0.030 \\
& 2022 DC5&poly=2&r=19.908&0.025&0.444&0.006&0.118&0.025&0.562&0.026&-0.023&0.031&B&0.018 \\
& 2022 DX&fourier&r=21.343&0.062&0.485&0.014&0.102&0.122&0.587&0.122&-0.021&0.185&B&0.019 \\
& 2022 FC3&poly=3&r=21.947&0.108&0.693&0.059&0.181&0.09&0.874&0.108&-0.065&0.159&S&0.013 \\
& 2022 NX1&fourier&g=22.418&0.037&0.534&0.136&0.186&0.137&0.721&0.019&-0.139&0.202&Q&0.047 \\
& 2022 OB5&poly=0&r=22.421&0.053&0.582&0.112&0.165&0.13&0.747&0.171&-&-&Xe&0.003 \\
& 2023 BU&fourier&r=17.552&0.018&0.689&0.162&0.132&0.182&0.82&0.243&-0.204&0.407&Q&0.025 \\
& 2023 OP2&fourier&r=20.155&0.028&0.616&0.053&0.072&0.018&0.688&0.056&0.007&0.026&Cg&0.023 \\
& 2024 NH&fourier&r=20.585&0.059&0.615&0.174&0.195&0.226&0.810&0.286&-0.023&0.240&Sv&0.020 \\
& 2024 OY2&poly=5&r=17.948&0.022&0.523&0.004&0.139&0.034&0.661&0.034&0.026&0.07&Xk&0.015 \\
& 2024 PJ1&poly=1&r=20.266&0.032&0.523&0.051&0.123&0.074&0.645&0.09&0.006&0.091&Cg&0.022 \\
& 2024 PT5&poly=3&r=17.751&0.035&0.626&0.02&0.195&0.049&0.821&0.053&-0.132&0.09&Sq&0.017 \\
& 2024 RZ3&fourier&r=20.546&0.038&0.558&0.065&0.112&0.04&0.671&0.076&0.063&0.09&Xk&0.016 \\
& 2024 YR4&fourier&r=19.872&0.036&0.642&0.035&0.2&0.042&0.842&0.055&-0.014&0.108&S&0.028 \\
& 2024 YR11&poly=1&r=19.988&0.032&0.663&0.035&0.26&0.02&0.923&0.04&-0.322&0.103&R&0.029 \\
\enddata
\end{deluxetable}

Ultimately, three quarters (151 objects) of our sample required some form of lightcurve correction (Table \ref{tab:results}), whether that was a full Fourier fit to the underlying lightcurve (49 objects) or simply applying a polynomial fit to the measurements made in the reference filter (102 objects). Objects for which no lightcurve correction was needed or possible are indicated by the label {\it poly=0} in the LC Correction column of Table \ref{tab:results}. In these cases the derived colors were just the difference in the mean magnitude of each filter. Fourier or polynomial fits to the underlying lightcurve were used to interpolate the reference filter magnitude to specific times in the imaging sequence when the other bands were being measured. The errors on the interpolated reference magnitudes were uniformly set to the standard deviation of the fit residuals (e.g. standard deviation of the difference between the lightcurve model and the measured reference magnitudes). Individual colors could then be computed from the interpolated reference magnitude and the measured non-reference band. Errors on these colors were computed by adding in quadrature the errors on the interpolated reference and measured non-reference magnitudes.

For each asteroid, a single value for each color was computed by taking an unweighted average of the individual measurements. For example, in Figure \ref{fig:yr4}, six individual $r-i$ determinations were possible from the six $i$ band exposures and the interpolated $r$ band magnitudes at those times. An unweighted average was used to avoid biasing the final colors towards the highest S/N measurements. This bias could originate from individual filters accessing different parts of the rotational lightcurve or simply because certain filters systematically produce higher S/N (e.g. $r$ band data generally has higher S/N than $z$ band). Of course contamination of the asteroid flux by background sources could also bias individual measurements. This was generally avoided by only using measurements where source extraction was free of any warning flags (\url{https://sextractor.readthedocs.io/en/latest/Flagging.html}).

The specific reference filter used for each object is indicated in Table \ref{tab:results} and was dictated by details of the data including S/N, number of valid exposures per filter, and temporal sampling. Valid exposures were those that produced S/N$>2$ and for which the object was not contaminated by background field stars. In total, 115 objects used the $r$ band as the reference filter, 61 used $g$, 19 used $i$, and 4 used $z$. The different reference filters for different objects resulted in different sets of derived colors. For example, when $r$ was used as reference the resulting colors were $g-r$, $r-i$, and $r-z$, but when $g$ was the reference the resulting colors would be $g-r$, $g-i$, and $g-z$. To produce a uniform dataset for analysis, all colors were converted to a standard set consisting of $g-r$, $r-i$, $g-i$, and $i-z$ colors (Table \ref{tab:results}).

To derive taxonomic assignments, the colors were converted to solar-corrected, normalized reflectance:
\begin{equation}
    R_\lambda = \frac{f_\lambda}{f_{0.55}} = \frac{10^{-0.4(m_\lambda-M_{\odot,\lambda})}}{f_{0.55}} 
\end{equation}
where $R_\lambda$ is the normalized reflectance, $f_\lambda$ is the relative flux, $f_{0.55}$ is a normalization factor equal to the relative flux at 0.55 $\mu m$, $m_\lambda$ is a normalized magnitude, and $M_{\odot,\lambda}$ is the AB magnitude of the Sun, all at a given wavelength or filter band center $\lambda$. The adopted AB magnitudes of the Sun in the $griz$ filter set were $M_{\odot,g}=5.12$, $M_{\odot,r}=4.68$, $M_{\odot,i}=4.57$, $M_{\odot,z}=4.54$, which came from transforms \citep{rodgers06} based on data in the 8th release of the SDSS-III catalog \citep{aihara11}. There are more recent or different versions of the solar magnitudes \citep[e.g.][]{willmer18}, but the differences were sufficiently small, typically at the 0.01 magnitude level or less, that they had no significant bearing on the results presented here. The normalized magnitude $m_\lambda$ was simply calculated by setting the reference magnitude to unity and then using the colors to compute relative magnitudes in the other filters. The normalization factor $f_{0.55}$ was set equal to a linear interpolation of the flux at 0.55 $\mu m$. This interpolation was based on the closest available bands to 0.55 $\mu m$, typically $g$ and $r$, but in some cases $g$ and $i$. A linear interpolation may not have been ideal for spectral types where the underlying spectrum may have had significant curvature around 0.55 $\mu m$, but for the sake of consistency, we applied the same normalization procedure to all objects.

Best fit taxonomic assignments were based on the type templates in the Bus-DeMeo system \citep{demeo09}. The templates for the Bus-DeMeo types were linearly interpolated to the $griz$ central wavelengths. A root-mean-square (rms) value was computed for each type based on the difference between the template and the measured normalized reflectance. The best fit type was selected based on the lowest rms, both of which are given in Table \ref{tab:results}. This procedure yielded a single classification for each object that did not account for errors on the colors, nor did it provide a probabilistic ranking of taxonomic types for each object \citep[c.f.][]{mommert16}. As an example to highlight this process, consider 1991 TF3, observed with LDT on 2021-10-16 (Table \ref{tab:results}). This was a relatively bright target ($g\sim18.4$), displayed only minor lightcurve variations in the reference filter ($\Delta g\sim0.04$), and had low errors on the resulting colors ($\sigma\sim0.02$). The best-fit taxonomic type for this object was Sq with an associated rms=0.041. Its negative $i-z$ color is consistent with the presence of a $1~\mu m$ absorption band and thus classification in the S complex. This classification seems to be correct, because it is consistent with a near-IR spectrum of this object from MITHNEOS \cite{binzel19}. However, from an rms fit perspective, the second and third most likely taxonomic assignments were a Cg type and a K type, both with rms=0.043. So an Sq type is only slightly favored over alternatives that have very different taxonomic and compositional interpretations. In this case we were able to confirm accurate assignment thanks to the MITHNEOS data, however that is not possible for the majority of objects in our sample. To mitigate this issue, we generally rely on the ensemble of results rather than emphasizing specific classifications for any one object. Our full suite of measured colors are included in Table \ref{tab:results} for future analysis that might consider probabilistic approaches to taxonomic classification or the influence of error bars on the assigned types.

Settling on uncommon or rare classifications is a risk of relying solely on rms to identify a single best-fit taxonomic type. For example, 2021 PB15 was classified here as an R type, which make up only about 1\% of spectroscopic datasets \citep{binzel19}. The rms for this R type classification was 0.033908, whereas the second and third most likely classifications were Sr and Sq with rms values of 0.033919 and 0.03596 respectively. The differences between these rms values are unlikely to be significant. And given the rarity of R types, it seems more likely that 2021 PB15 could be an edge member of the S complex. However, to avoid introducing bias in favor of taxonomic types deemed more likely based on relative abundance in prior surveys, we do not adjust classifications based on such judgment calls.

Subtle differences between some of the Bus-DeMeo types are not easily resolved given the low effective spectral resolution of 4 color channels and our use of only visible wavelength data. To mitigate this issue, some {\it types} were combined into a single taxonomic {\it class}. This combination followed the procedure of \citet{devogele19} in which the C class incorporates C, Cg, Cgh, Ch, Cb, and B types, the K class combines K and L types, the Q-class includes Q and Sq types, the S class includes S, Sa, Sr, and Sv types, and the X class includes X, Xc, Xe, Xk, and Xn types. This scheme resulted in 11 taxonomic classes (A, C, D, K, O, Q, R, S, T, V, X) that are the foundation of all analyses presented below. Even with these class definitions, the differences between some classes, e.g. K and S, are not easily resolved with colors, particularly when photometric errors (typically $\sim0.05$ mag) exceed the few percent differences that can differentiate classes. Again, we must simply rely on the overall distribution of types rather than individual objects. Like NEO color surveys in the Johnson-Cousins $UBVR_cI_c$ filter set \citep[][]{dandy03,hromakina23}, we summarize in Table \ref{tab:mean} the mean and standard deviation of our $griz$ colors for each of these taxonomic types. These values are consistent with mean colors in the Johnson-Cousins system when transforms to the $griz$ fitler set are applied \citep{jester05}. The number of objects used to compute these values are shown in this table and are inclusive of multiple observations of individual objects.

\begin{deluxetable}{lrccc}
\tablecaption{Mean and standard deviation of $griz$ colors for each taxonomic class. The number of objects for each class is given in the second column. \label{tab:mean}}
\tabletypesize{\small}
\tablehead{\colhead{Taxon} & \colhead{\#} & \colhead{$g-r$} & \colhead{$r-i$} & \colhead{$i-z$}}
\startdata
A	& 4 & $0.855 \pm 0.076$	& $0.192 \pm 0.044$	& $-0.086 \pm 0.029$ \\
C 	& 23 & $0.503 \pm 0.072$	& $0.100 \pm 0.042$	& $0.016 \pm 0.046$ \\
D 	& 6 & $0.657 \pm 0.144$	& $0.311 \pm 0.073$	& $0.186 \pm 0.089$ \\
K 	& 17 & $0.626 \pm 0.060$	& $0.174 \pm 0.043$	& $-0.027 \pm 0.025$ \\
O 	& 12 & $0.632 \pm 0.088$	& $0.074 \pm 0.048$	& $-0.390 \pm 0.119$ \\
Q 	& 40 & $0.634 \pm 0.047$	& $0.148 \pm 0.052$	& $-0.151 \pm 0.077$ \\
R 	& 8 & $0.673 \pm 0.055$	& $0.228 \pm 0.077$ 	& $-0.281 \pm 0.077$ \\
S 	& 24 & $0.704 \pm 0.057$	& $0.196 \pm 0.046$	& $-0.116 \pm 0.056$ \\
T 	& 8 & $0.580 \pm 0.022$	& $0.187 \pm 0.050$	& $0.050 \pm 0.045$ \\
V 	& 11 & $0.699 \pm 0.079$	& $0.200 \pm 0.046$	& $-0.496 \pm 0.100$ \\
X 	& 46 & $0.535 \pm 0.042$	& $0.155 \pm 0.031$	& $0.035 \pm 0.052$ \\
\enddata
\end{deluxetable}

\subsection{Consistency Checks} \label{subsec:consistency}

Given this complicated set of procedures for deriving colors and taxonomic classifications, we performed a number of consistency checks to help validate our methodology. This involved internally comparing colors for objects that were observed on multiple nights (Table \ref{tab:multiple}) and comparing our colors to spectral data from both new observations and archival sources (Table \ref{tab:spec_comp}). The new spectral observations from LDT, Gemini, and IRTF were introduced in Section \ref{subsec:spectra}.

In total 28 individual objects were assessed for consistency. There may be other objects in our data set that were observed by other programs and not included in this consistency check, but we are simply aiming for a representative sample to demonstrate the relative accuracy of the derived colors/taxa.

The comparison of multiple-night colors shows that 4 out of 9 objects changed from one taxonomic class to another. Most of these are relatively benign shifts that are to be expected based on typical S/N. For example, 2008~VB1 was seen as a K type on 15 November 2021, but then as an Sq type on 4 December 2021. The colors on these two nights were consistent within error bars, but the nominal values were just different enough to cause a taxonomic shift. This is an unavoidable consequence of distilling  four spectral channels into a single discrete classification. The object 2011~CG2 was the one exception in the color comparison where significantly different answers were obtained on two nights: Cgh on 7 January 2025 and Q type on 6 February 2025. This change was well outside of error bars. The image frames from these nights were individually inspected, but no field stars were seen in the vicinity of the asteroid, thus ruling out background contamination as a cause for the spectral change. The calibrated zero points across the image sequences were stable, suggesting no significant change in atmospheric extinction. The solar phase angles on the two dates were significantly different: 25$^\circ$ on 7 January and 51$^\circ$ on 6 February. It is possible that the effects of phase reddening caused up to $\sim$10\% shifts in spectral slope \citep{sanchez12} and thus contributed to the observed change in spectral type. Another possibility is that this object has a heterogeneous surface with markedly different surface properties at different rotational phases and/or viewing geometries. Additional observations would be needed to further distinguish these possibilities.

\begin{deluxetable}{ll | clc | clc}
\tablecaption{Objects with colors obtained on multiple nights. Entries in {\bf bold} are  preferred classifications based on S/N. \label{tab:multiple}}
\tabletypesize{\small}
\tablehead{\colhead{Number} & \colhead{Designation} & \colhead{Facility \#1} & \colhead{Obs. Date \#1} & \colhead{Type \#1} & \colhead{Facility \#2} & \colhead{Obs. Date \#2} & \colhead{Type \#2}}
\startdata
141018 & 2001 WC47 & SOAR 4.2m & 2017-01-13 & A & LDT 4.3m & 2022-02-11 & \bf A \\
141018 & 2001 WC47 & LDT 4.3m & 2021-12-04 & L & LDT 4.3m & 2022-02-11 & \bf A \\
412983 & 1996 FO3 & LDT 4.3m & 2021-11-15 & \bf Xe & LDT 4.3m & 2022-02-11 & Xc \\
513312 & 2007 DM41 & LDT 4.3m & 2016-02-22 & O & LDT 4.3m & 2025-02-02 & \bf O \\
523813 & 2008 VB1 & LDT 4.3m & 2021-12-04 & Sq & LDT 4.3m & 2021-11-15 & \bf K \\
 & 2011 CG2 & LDT 4.3m & 2025-01-07 & \bf Cgh & LDT 4.3m & 2025-02-06 & Q \\
 & 2016 AO131 & LDT 4.3m & 2016-01-19 & O & LDT 4.3m & 2022-02-11 & \bf O \\
 & 2016 GW216 & LDT 4.3m & 2022-05-23 & Sq & LDT 4.3m & 2022-05-31 & \bf Sq \\
 & 2016 QJ44 & LDT 4.3m & 2021-10-28 & Sv & LDT 4.3m & 2021-10-29 & \bf Sv \\
 & 2019 YF4 & LDT 4.3m & 2020-01-04 & \bf K & SOAR 4.2m & 2020-01-03 & T \\
\enddata
\end{deluxetable}

\begin{figure}
    \centering
    \includegraphics[width=0.49\textwidth]{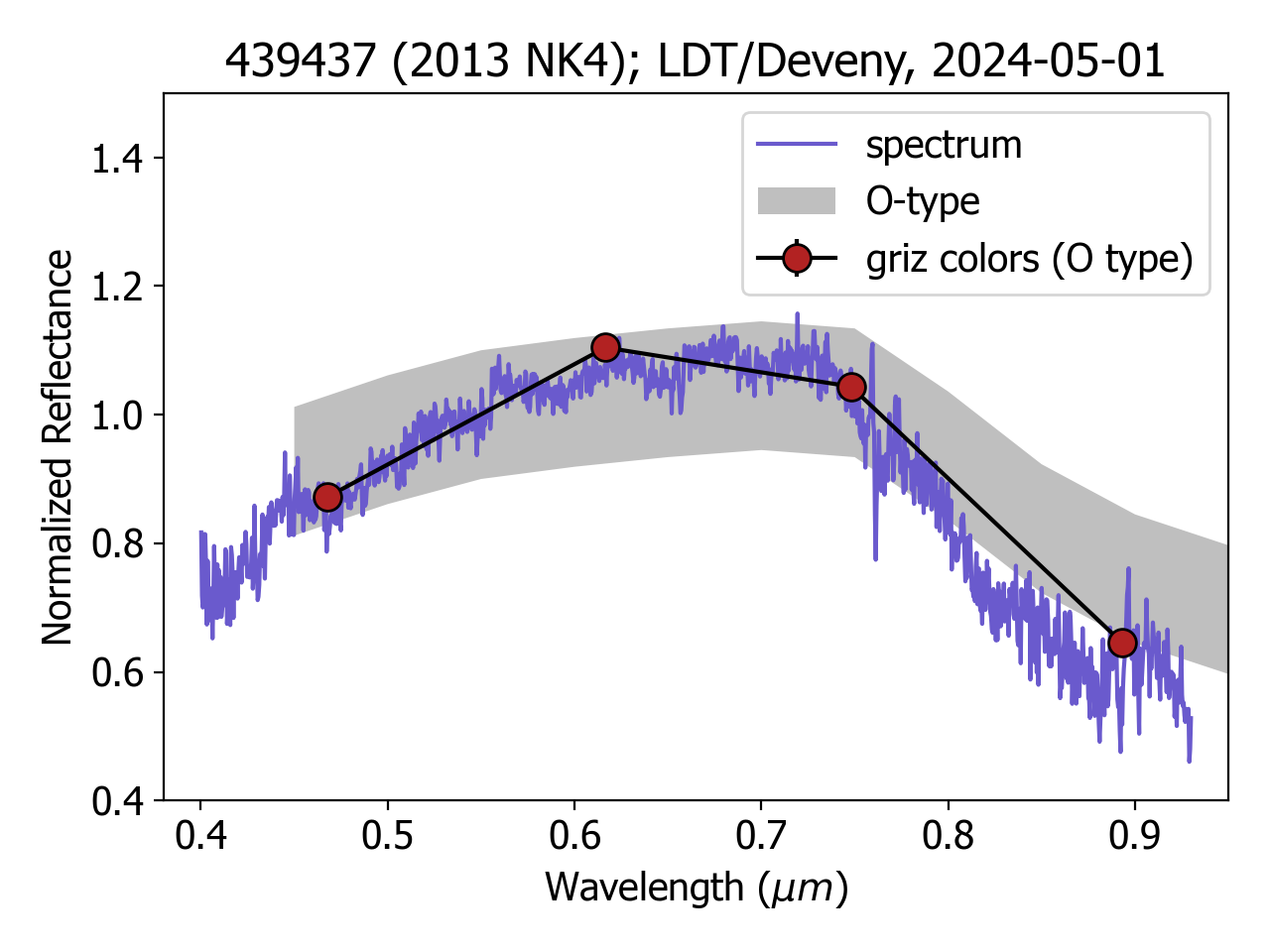}
    \includegraphics[width=0.49\textwidth]{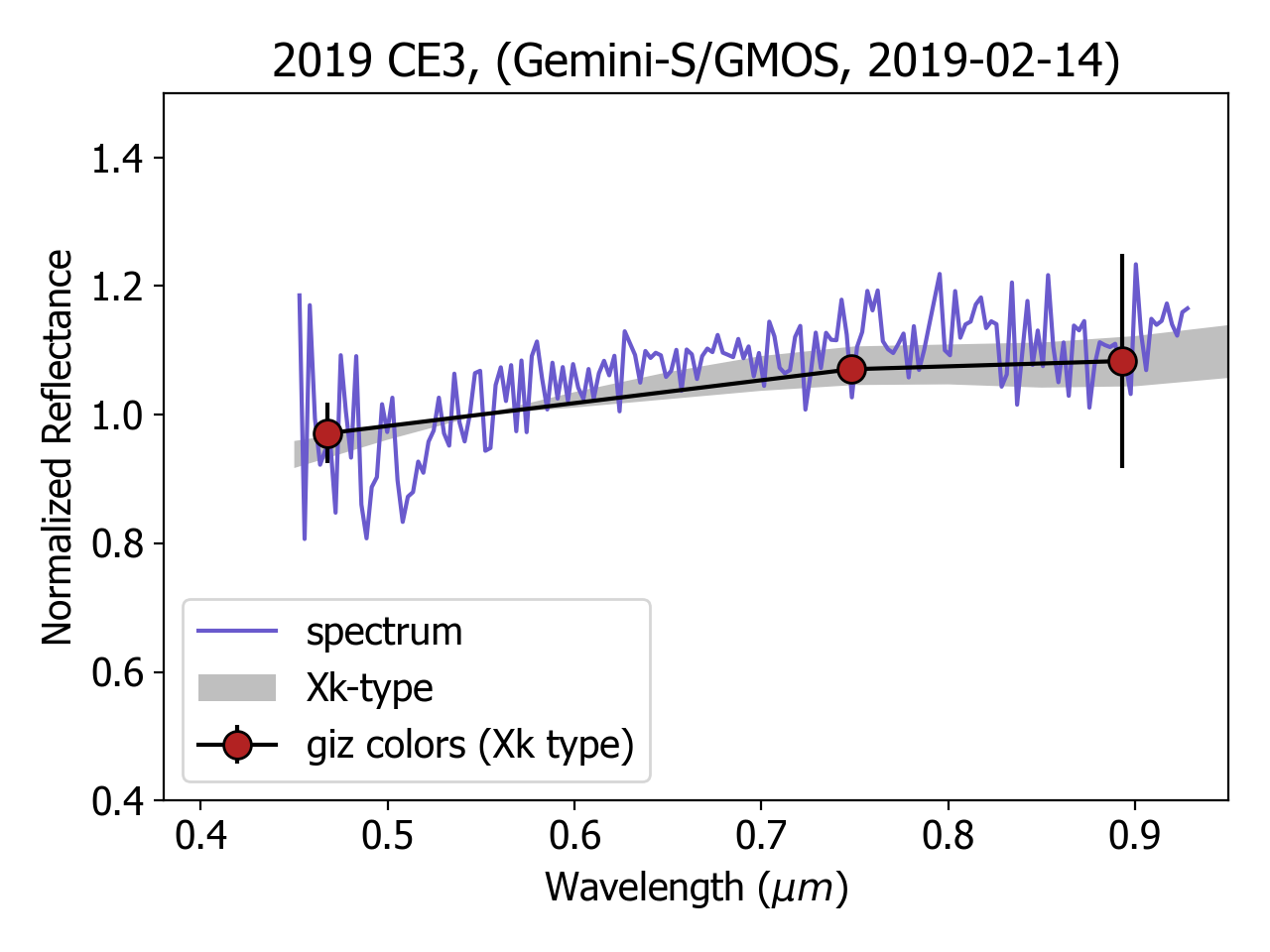}
    \includegraphics[width=0.49\textwidth]{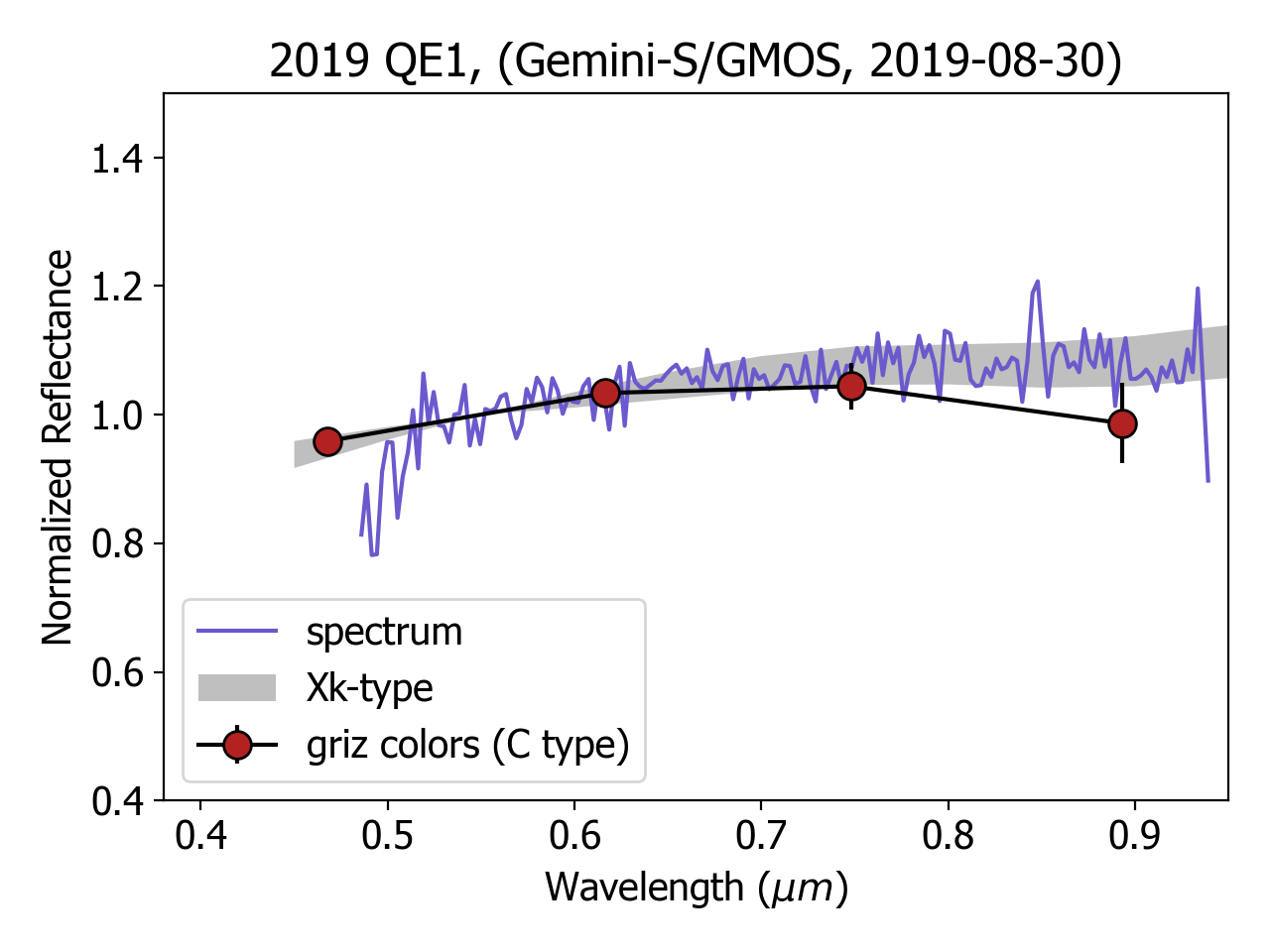}
    \includegraphics[width=0.49\textwidth]{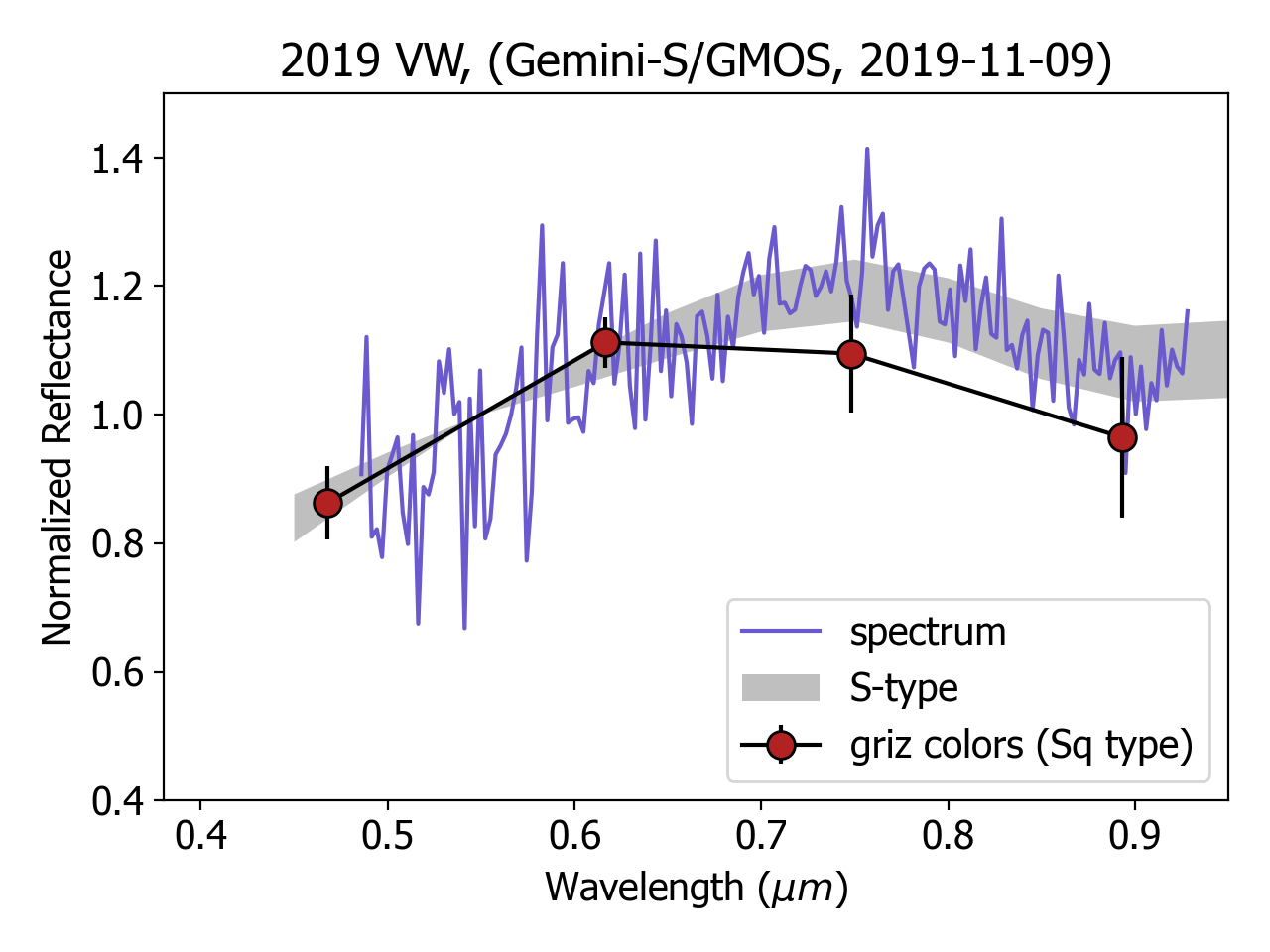}
    \includegraphics[width=0.49\textwidth]{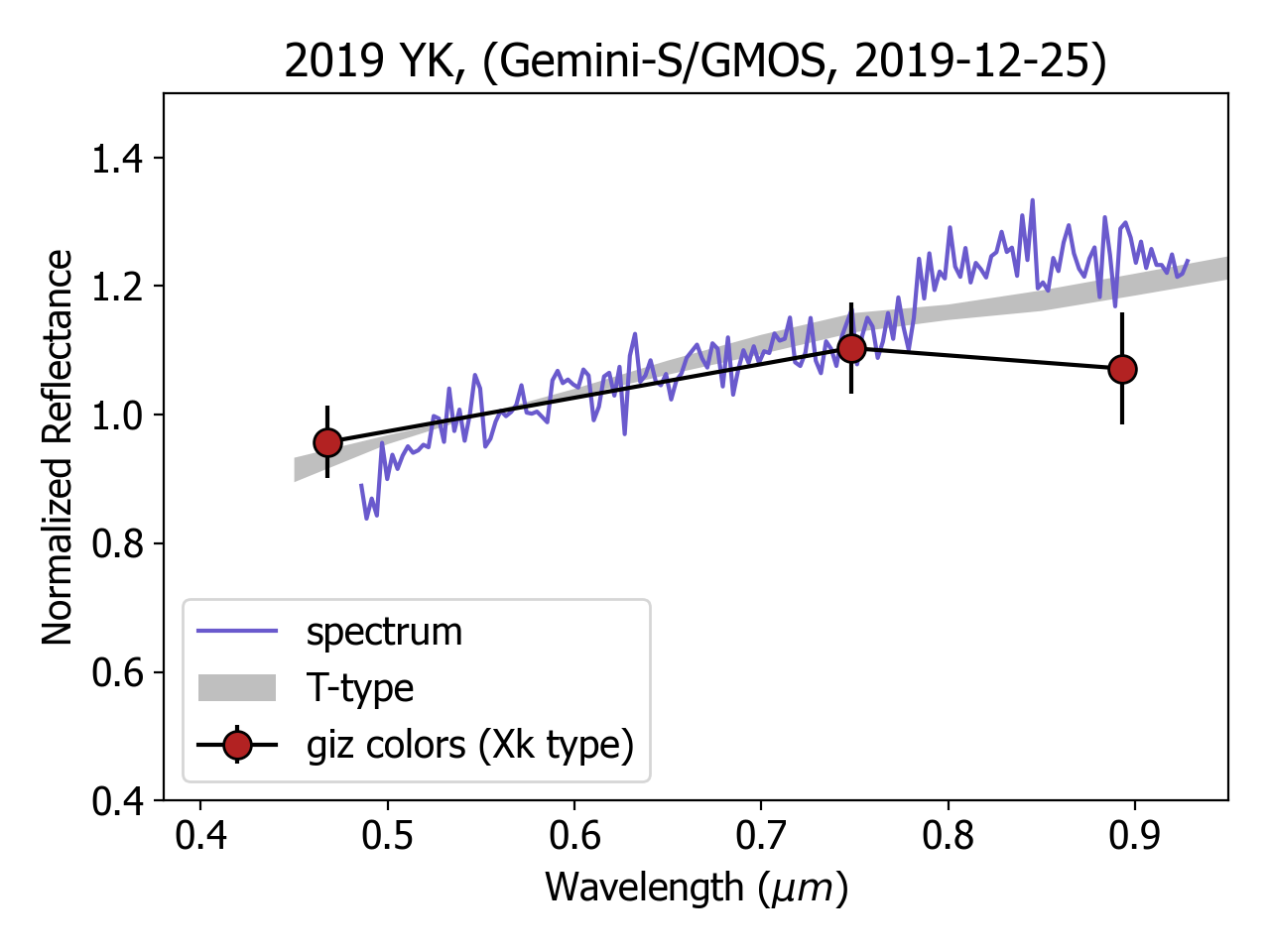}
    \includegraphics[width=0.49\textwidth]{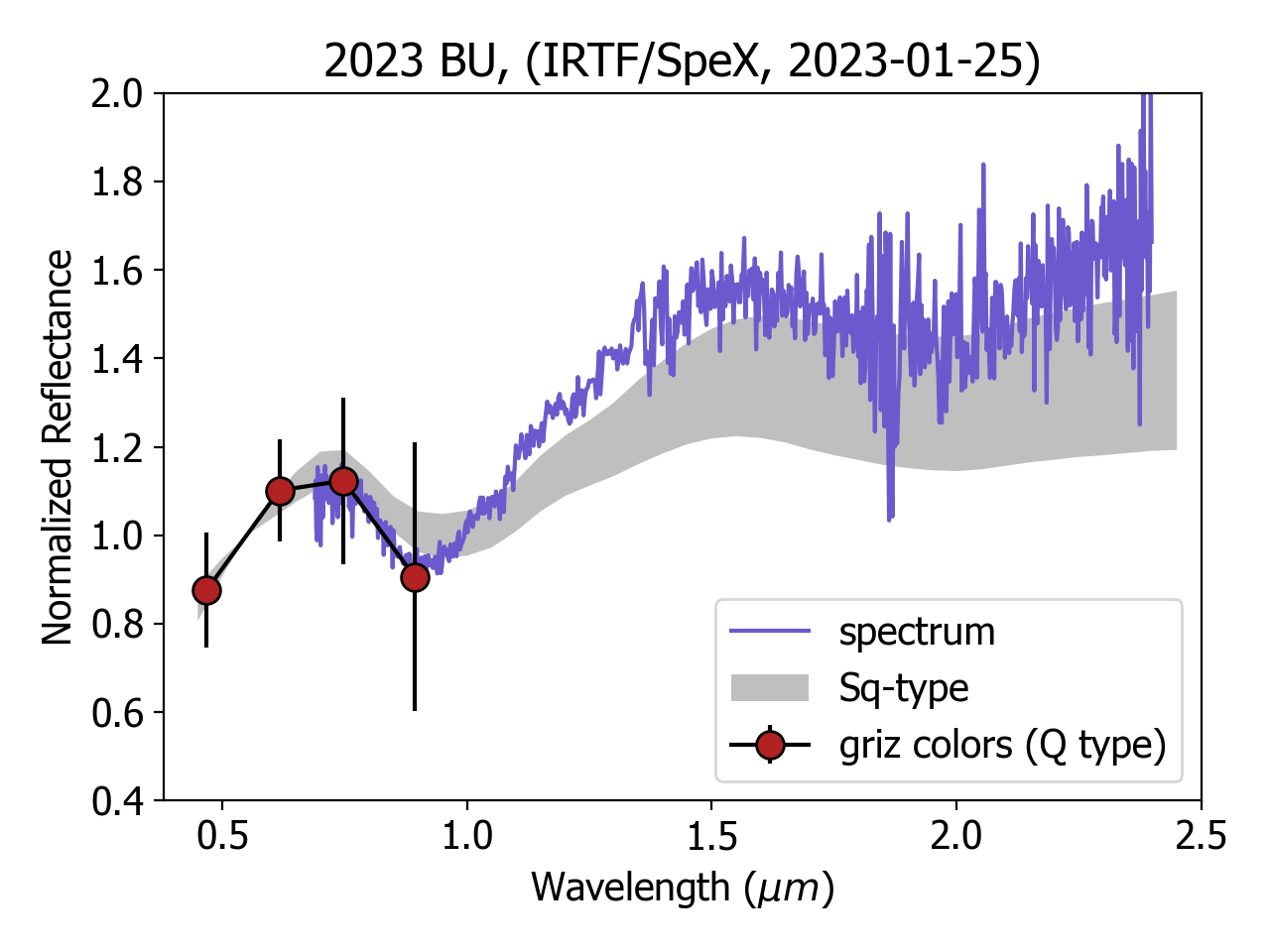}
    \caption{Spectral results with color photometry over-plotted. The spectrograph and date of observation are in the title of each panel. The color data and their best fit taxonomic classification are in the legend. For the spectral data, the best fit taxonomic types and associated 1-sigma envelopes are indicated by the grey filled regions in the background. In all cases, the colors are consistent to within $\sim10\%$ in reflectance relative to the spectra.}
    \label{fig:spec}
\end{figure}

The comparison to spectra includes both new observations and results retrieved from the literature (Table \ref{tab:spec_comp}). The new spectral results are shown in Figure \ref{fig:spec}. Though there are subtle differences between the colors and spectra, they are generally within the reported error bars and at a level of $\sim10\%$ or less in reflectance, which is a reasonable limit to the reproducibility of asteroid spectral data \citep{marsset20}. When differences are seen, they can be attributed to one of several possibilities. First, it is common for differences in spectral slope to cause a shift in taxonomic type. For example, our L-type classification of 2022~TP69 is different from the S-type assignment in \citet{lazzarin05}, which is entirely due to an offset in slope. Second, noisy data can result in different classifications. For example, indications of a tumbling rotation state for 2023~BU (Section \ref{sec:LC}) made it difficult to correct the colors for lightcurve variability and thus resulted in large error bars. Nevertheless, the spectrum and colors for this object are consistent (Figure \ref{fig:spec}). Lastly, spectral data can be susceptible to calibration issues either tied to the specific star used for solar correction or poor removal of telluric features. These issues can manifest as unexpected inflection points in the reduced spectra. For example, the sudden jump in reflectance around 0.78 $\mu m$ for 2019~YK (Figure \ref{fig:spec}) is likely due to imperfect telluric correction as a prominent molecular oxygen absorption band in the atmosphere exists at that wavelength. 

\begin{deluxetable}{llccc}
\tablecaption{Objects with both colors and spectral data. \label{tab:spec_comp}}
\tabletypesize{\small}
\tablehead{\colhead{Number} & \colhead{Designation} & \colhead{Color Taxon} & \colhead{Spectral Taxon} & \colhead{Spectrum Reference}}
\startdata
22099 & 2000 EX106 & Sa & S or Q & \citet{binzel04} \\
89136 & 2001 US16 & Q & Sq/Sr & \citet{kuroda14} \\
141018 & 2001 WC47 & A & Sw & \citet{binzel19} \\
162004 & 1991 VE & C & X or C & \citet{marsset22} \\
350713 & 2001 XP88 & Sq & Xc & \citet{perna18} \\
439437 & 2013 NK4 & O & O & This work \\
 & 1991 TF3 & Sq & S complex & \citet{binzel19} \\
 & 2002 TP69 & L & S & \citet{lazzarin05} \\
 & 2003 HA & Q & C or Q & \citet{carvano10} \\
 & 2014 TR57 & K & S & \citet{devogele19} \\
 & 2015 TM143 & Cb & Cgh & \citet{devogele19} \\
 & 2016 CL29 & Q & Q & \citet{devogele19} \\
 & 2016 CS247 & S & S & \citet{devogele19} \\
 & 2017 AT4 & C & Xc & \citet{devogele19} \\
 & 2017 VR12 & V & V & \citet{devogele19} \\
 & 2019 CE3 & Xk & Xk & This work \\
 & 2019 QE1 & C & Xk & This work \\
 & 2019 VW & Sq & S & This work \\
 & 2019 YK & Xk & T & This work \\
 & 2023 BU & Q & Sq & This work \\ 
\enddata
\end{deluxetable}

Of the 28 objects considered for these consistency checks, 12 switched taxonomic class. Only one of those, 2011~CG2 as mentioned above, was well outside of error bars with colors that changed significantly across nights. Most of the others represented subtle shifts, for example from S to Q-type (e.g. 2023~BU) or X to C-type (e.g. 2019~QE1). It is expected that any differences in these consistency checks are simply random fluctuations that will cancel out in the ensemble analysis presented below. 

\section{Results} \label{sec:results}

\begin{figure}
    \centering
    \includegraphics[]{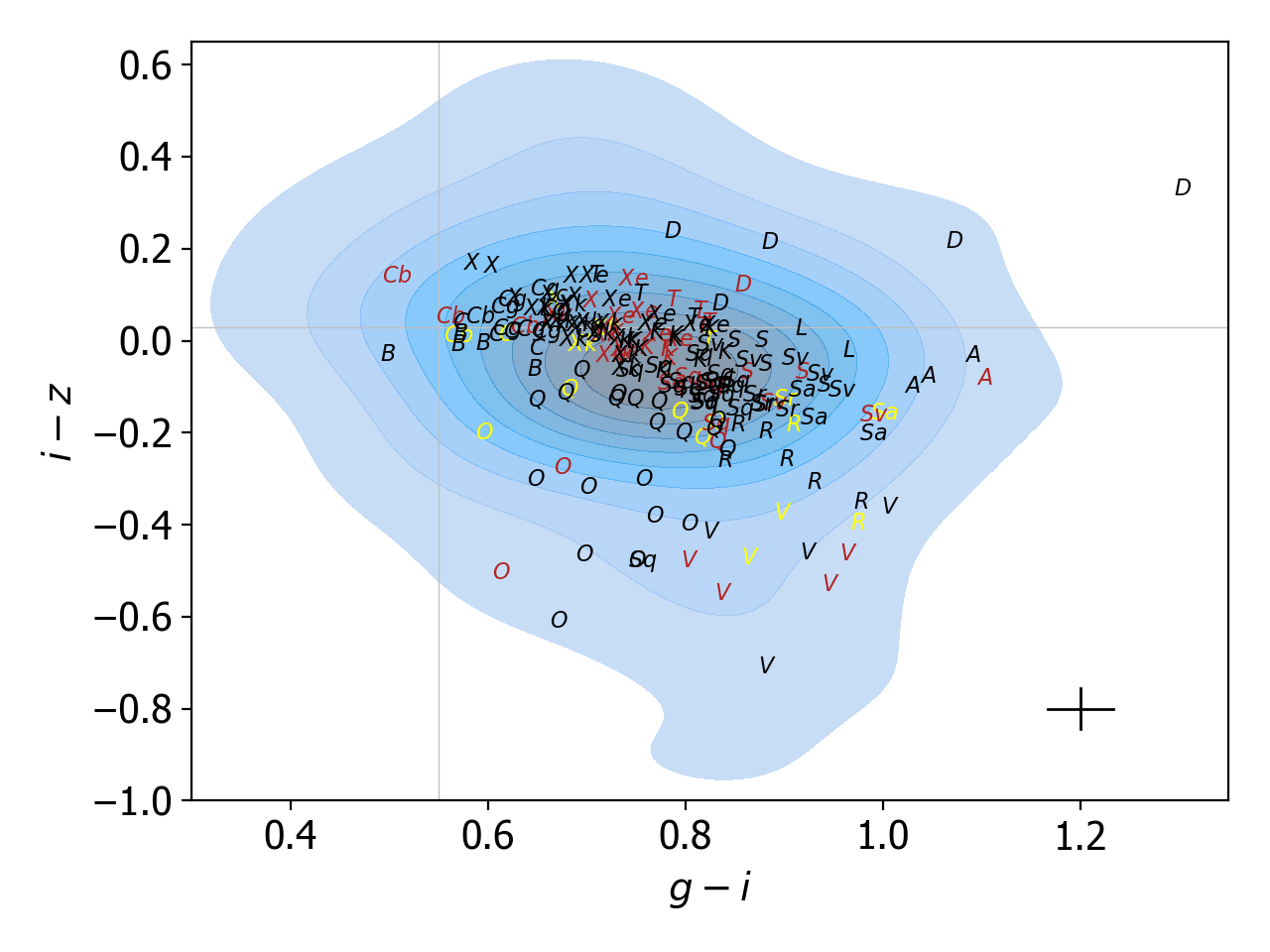}
    \caption{Color-color plot of $g-i$ versus $i-z$ colors of object in the MANOS sample. The assigned taxonomic classification for each object is indicated by the corresponding letters. The text is color coded by facility with the Kitt Peak Mayall in yellow, SOAR in red, and LDT in black. The background contours are a KDE (kernel density estimate) of 233 NEOs from the 3rd release of the SDSS Moving Object Catalog \citep[MOC,][]{ivezic10}. The overlap of our colors with that of the MOC serves as a consistency check to our methodology. Mean error bars are indicated in the lower right. The solar colors are shown as the thin grey lines.}
    \label{fig:colors}
\end{figure}

In total 199 sets of colors for 189 individual objects were obtained in the SDSS $griz$ filter set. A color-color plot of $i-z$ versus $g-i$ for these objects is shown in Figure \ref{fig:colors}. These particular axes were chosen because they serve as a proxy to visible slope ($g-i$) and depth of any 1 $\mu m$ absorption band that may be present ($i-z$), and thus are well suited to distinguish taxonomic types. This works particularly well for end-member classes (e.g. D, O, V, A) as they occupy distinct regions in this parameter space. In Figure \ref{fig:colors}, the overlap of our colors with those of NEOs in the 3rd release of the SDSS Moving Object Catalog \citep{ivezic10} serves as an independent check on the validity of our methods. We opt not to use the composite $a^*$ color from \citet{ivezic01}, which represents a coordinate transform in the $g-r$ versus $r-i$ color space highlighting slope differences between S and C complex objects, because a significant fraction of our sample (34 objects from SOAR) do not have measured $r$ band magnitudes.

The classifications assigned to each object were generally based on a full $griz$ color set. However, there were exceptions when images in a single filter were not obtained or were affected by background star contamination. These are apparent in Table \ref{tab:results} as rows with missing entries. The most obvious of these is the block of objects observed at SOAR where no $r$ band images were obtained due to the limited number of slots in the Goodman filter wheel. Objects 2020~BP13 and 2004~KE1 are missing $g$ band images due to instrumental issues on the night of observation. Objects 1993~HA, 2020~AE, and 2016~CF194 are missing $i$ band data due to either contamination by nearby field stars or issues with the instrument filter wheel. Objects 2022~OB5, 2014~UX7, 2014~UD57, 2015~OV3, and 2015~AA44 are missing $z$ band images because the S/N was too low to yield a reliable detection.

\subsection{Distribution of Taxonomic Types}

\begin{figure}
    \centering
    \includegraphics[]{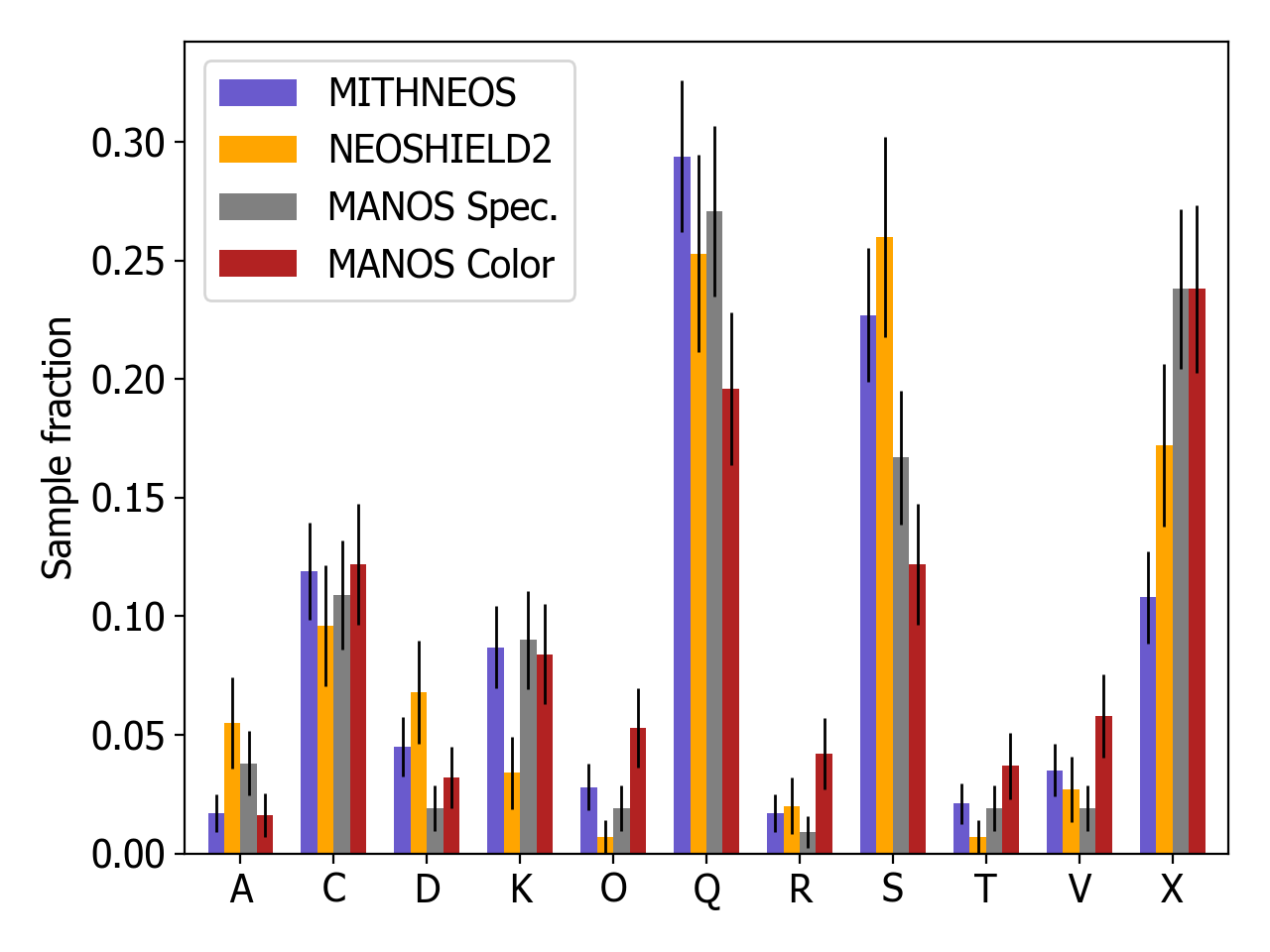}
    \caption{Comparison of the relative fraction of taxonomic types from surveys operating at visible wavelengths. The visible MITHNEOS data \citep{binzel19}, NEOSHIELD2 \citep{perna18}, and MANOS spectroscopy \citep{devogele19} provide sample sizes of 286, 146, and 210 objects respectively. The fractions from the MANOS color sample (N=189) for taxonomic types A, C, D, K, O, Q, R, S, T, V and X are 0.016, 0.122, 0.032, 0.085, 0.053, 0.196, 0.042, 0.122, 0.037, 0.058, and 0.238 respectively. The error bars on each bin are $\sqrt{N}$ Poisson errors.}
    \label{fig:types}
\end{figure}

Our subsequent analysis is primarily focused on using colors to investigate the distribution of taxonomic types. To facilitate that analysis we compare our results to other visible wavelength surveys (Figure \ref{fig:types}). This exclusive focus on visible wavelength data is meant to establish a consistent foundation for comparison. Taxonomic distributions derived from near-IR spectra \citep[e.g.][]{binzel19,sanchez24}, from color data that do not provide similar wavelength coverage \citep[e.g.][]{birlan24,navarro-meza24}, and/or multi-parameter classification schemes \citep[e.g.][]{sergeyev23} are hence excluded from this comparison. We also avoid comparison to data sets with taxonomic assignments that do not readily map to the 11 types used here \citep[e.g.][]{erasmus17,navarro-meza24}. Despite these exclusions, focusing on results from four surveys (Figure \ref{fig:types}) provides a robust set of over 800 combined objects for meaningful statistical analysis. For the MANOS colors, only one taxonomic type was counted when multiple nights of data were obtained (Table \ref{tab:multiple}). In these cases, preference was given to the color set with smallest error bars.

To first order, there is agreement across the four visible wavelength data sets. In all samples, the observed population is dominated Q, S, X and C-types. End-member types A, D, O, R, and V tend to contribute less than 5\% each. However, in detail, we start to notice differences. The reported over-abundance of A and D-types in the NEOSHIELD2 data set for objects with $H>20$ \citep{perna18} is not confirmed by the MANOS colors. However, the MANOS color set does seem to indicate an enhancement in types with the deepest 1 $\mu m$ absorption bands, namely O, R, and V types. For example, R types represent 4\% of our color sample, but no more than 2\% in the other data sets. Granted, the number statistics for these minor types are limited; no more than 10 objects of these types are present in any of the surveys. The cause of this enhancement remains unclear. One possibility is that the higher number of R types may simply be misclassified S types, which are adjacent to one another in color space (Figure \ref{fig:colors}) and could easily get confused. If that is the case, the small deficit of S type colors relative to MANOS spectra could be due to a handful of misclassified R types.

The most prominent difference across the surveys is with the Q, S, and X types. MANOS tends to show an under-abundance of Q and S types relative to the other surveys. This is most pronounced for S types where MANOS colors suggest that these objects represent 12\% of the population, whereas MITHNEOS and NEOSHIELD2 find occurrence rates about twice that. This can be reasonably interpreted given that Q and S types span a continuum of weathering states associated with ordinary chondrite meteorites \citep{binzel19} and that there is an overall bias in the MANOS sample towards small objects (Figure \ref{fig:H-Mag}), which should have younger surfaces due to size-dependent collision rates \citep{bottke05}. In other words, old, weathered ordinary chondrite surfaces would be classified as S-types; MANOS sees fewer of those because we systematically target smaller objects with younger surfaces. As the MANOS sample exhibits a decrease in S and Q types, this is partly compensated by an increase in X types. These trends are explored in greater detail in Section \ref{sec:size}, though we note here that the MANOS data are relatively uniform in quality (S/N) across all values of $H$. This is attributable to an observing strategy that involved real time target selection so that newly discovered, small objects were accessed before they had faded significantly. This is clear in the distribution of apparent $V$ magnitudes in our sample (Table \ref{tab:observations}), which show no correlation with $H$. Thus taxonomic assignments can be confidently compared across the full $H$-range of the data set.

\subsection{Specific Objects of Interest}

A number of objects in our data set stand out as particularly interesting. Our focus on mission accessibility produced data on some of the lowest $\Delta v$ objects in the NEO population. Only 71 NEOs (as of 28 September 2025) have $\Delta v<4$ km/s; we report new physical properties for 7 of those (Table \ref{tab:lowdv}). Highlighting these objects does not necessarily suggest the viability of any particular mission profile. For example, some of these objects have relatively short observation arcs (e.g. the orbit for 2022~BX5 is based on an 18-day arc), which suggests low quality orbit solutions that could pose viability concerns for spacecraft accessibility. Furthermore, the simple \citet{shoemaker78} formalism for computing $\Delta v$ does not provide any insight into specific launch windows. That said, these objects are all on the JPL Near-Earth Object Human Space Flight Accessible Target Study (NHATS) list \citep{abell12}, indicating favorable conditions for future rendezvous missions.

When available, the lightcurves for these objects and their associated period solutions are included in Figure \ref{fig:A1}. Objects with photometry that did not reveal clear lightcurve variability are labeled in Table \ref{tab:lowdv} as ``flat". These objects are either very fast rotators with periods much less than the image exposure times, very slow rotators with periods much longer than a typical exposure sequence ($\sim10-20$ minutes), or had invariant projected area on sky, e.g. due to a pole-on viewing geometry or a spheroidal shape. It can not be determined which of these possibilities is the case with the available data. The diameters of these objects were calculated based on their $H$ values and the mean albedo for each taxonomic type \citep{thomas11,alilagoa13}: Xk type albedo = 0.31, B type albedo = 0.07, D type albedo = 0.02, and Q/Sq type albedo = 0.29.

\begin{deluxetable}{l c c c c c}
\tablecaption{Summary of results for objects on orbits with $\Delta v<4$ km/s. \label{tab:lowdv}}
\tablehead{\colhead{Designation} & \colhead{$\Delta v$ (km/s)} & \colhead{Taxon} & \colhead{Lightcurve Period} & \colhead{Diameter (m)} & \colhead{Next $V<25$ Apparition} }
\startdata
2019 SU3 & 3.88 & Xk & 9.4 min & 8 & Mar 2032 \\
2021 GK1 & 3.81 & B & 1.0 min & 19 & Oct 2029 \\
2022 BX5 & 3.78 & D & flat & 15 & Nov 2049 \\
2022 DX & 3.90 & B & 3.2 min & 11 & Nov 2028 \\
2022 NX1 & 3.83 & Q & 17.5 min & 6 & Dec 2051 \\
2024 PT5 & 3.96 & Sq & flat & 7 & May 2055 \\
2024 RZ3 & 3.97 & Xk & 0.94 min & 10 & Apr 2035
\enddata
\end{deluxetable}

Table \ref{tab:lowdv} indicates the date when these objects will next be observable at $V<25$ magnitude. This is a reasonable approximation for when new physical characterization data of any kind (e.g. orbit refinement, rotation state, absolute magnitude and size, albedo, spectral properties and composition) may be obtained. Until those dates, the current state of knowledge on these objects is unlikely to change appreciably.

A question of provenance is often raised for NEOs discovered at the low end of the $\Delta v$ distribution. Possible origins for these objects can be traced to man-made objects, lunar ejecta, or the Main Belt. Given enough time, generally weeks to months, astrometric monitoring can reveal non-gravitational orbit perturbations from solar radiation pressure that would be inconsistent with the high densities expected for natural objects \citep[e.g.][]{battle24}. A lunar origin can be inferred from dynamical arguments including the $v_\infty$ of the object relative to the Earth and spectroscopic links to lunar-like compositions. Such considerations led \citet{kareta25} to suggest that 2024~PT5 is likely ejecta from an impact on the Moon. Absent evidence of either an artificial or lunar origin, the presumption is that these low $\Delta v$ NEOs have origins that trace back to the Main Belt. That could be the case for the other 6 objects in Table \ref{tab:lowdv}, but future work is needed to better establish the efficiency at which objects can dynamically evolve out of the Main Belt and end up in these very Earth-like orbits.

\begin{figure}
    \centering
    \includegraphics[]{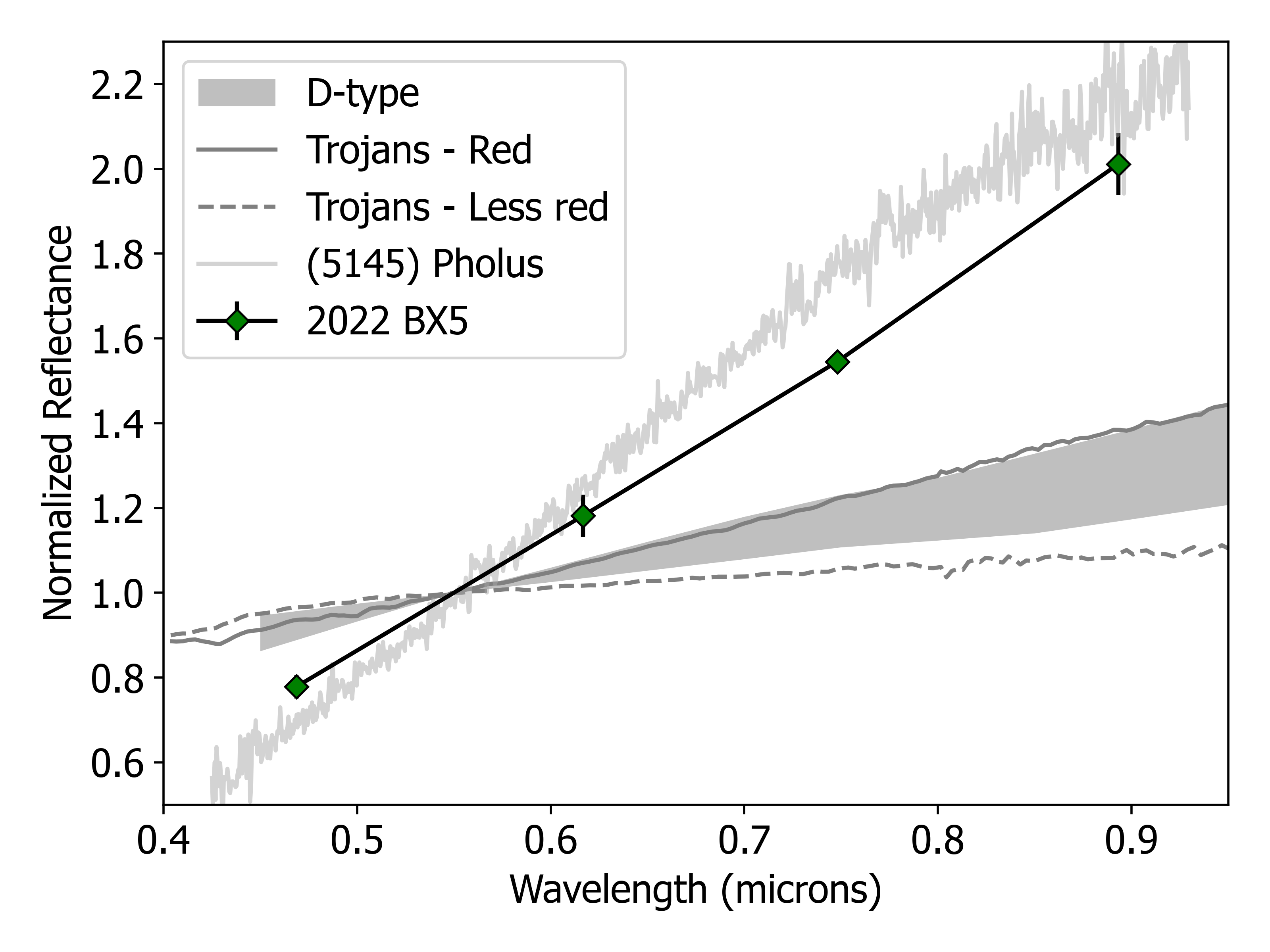}
    \caption{Comparison of spectral properties of 2022~BX5 to the D-type taxonomic class, red and less red Jupiter Trojan spectra from \citet{emery11}, and the extremely red Centaur (5145) Pholus from \citet{fornasier09}. BX5 has the reddest colors of any known NEO, which implies a composition more commonly seen in the outer Solar System.}
    \label{fig:bx5}
\end{figure}

Perhaps the most intriguing of these low $\Delta v$ objects is 2022~BX5 (hereafter BX5). This object stands out in Figure \ref{fig:colors} as the isolated D-type in the upper right corner. The S/N for these observations was excellent, the locations of BX5 in its images were free of any background field stars, and the time series photometry in the reference filter showed no appreciable lightcurve variability over a time span of about 17 minutes. In short, these observations and resulting colors are highly reliable. The colors of BX5 are extremely red (Figure \ref{fig:bx5}), in fact the reddest of any known NEO and are more akin to spectra of objects in the outer Solar System. Such red colors strongly suggest a natural origin as man-made hardware generally does not display such extreme reflectance properties \citep{battle24b}. Unfortunately the 18-day orbital arc for BX5 is not quite long enough to reject an artificial origin based on non-detection of solar radiation pressure. A stack of 10 $\times$ 32 second $r$-band exposures displayed no clear signs of cometary activity. Interestingly, the most likely Main Belt escape route for BX5 is the $\nu_6$ resonance \citep{granvik18}, thus it could have a connection to a population of D type interlopers in the inner Main Belt \citep{demeo14}.

In an attempt to extend the arc of observations for BX5 we searched for any pre-discovery detections with the Canadian Astronomy Data Center (CADC) Solar System Object Image Search (SSOIS) tool. This revealed a promising suggestion that 10$\times$60s images from the 2.6-m VLT Survey Telescope (VST) taken with the OMEGACAM instrument on 2021-11-14 around 02:00 UTC should have contained BX5 at a predicted magnitude $V\sim21.6$. This would have extended the orbital arc by over two months, a factor of about four longer than currently exists. However, BX5 was not found in these images, even with all ten images stacked at the object's ephemeris rate. These images covered the formal uncertainty region for the predicted ephemeris as well as an augmented search region that accounted for upper bounds on any non-gravitational Yarkovsky effect or solar radiation pressure. The Main Belt asteroid 2019~GJ154 happened to be in these images and was easily recovered in an image stack at an apparent Gaia magnitude $G=23.4$. While the predicted magnitude for BX5 could have been off, it would have had to have been 2 magnitudes or more off the nominal prediction to fall below the detection threshold of these images. A change in solar phase angle $\alpha$ from November 2021 ($\alpha=20^\circ$) to the discovery epoch in January 2022 ($\alpha=7^\circ$) was not large enough to cause significant inaccuracies in the predicted magnitudes. Unfortunately it remains unclear why BX5 was not found in these VST images. The next opportunity for recovery could be in mid-2031, but at a predicted magnitude of $V=25$ and given that we couldn't find the object in the pre-discovery 2021 images, this recovery could prove challenging.

In addition to BX5, several other D-types were found outside of the low $\Delta v$ set. The D-type NEO pair 2019~PR2 and 2019~QR6 are interesting objects that were discussed in detail in \citet{fatka22}. The remaining three D-types in our sample were 2018~RS1, 1992~BC, and 1998~KY26. This last object is of particular interest as the extended mission target of the Hayabusa2 spacecraft \citep{kikuchi23} and is an object that exhibits high non-gravitational orbit accelerations \citep{seligman23} that required detailed treatment of radiation forces to fully explain \cite{farnocchia25}. However, our classification of a D-type for 1998 KY26 has large error bars due to both variable seeing and zero point throughout its imaging sequence. This volatility required that the images for each filter be stacked to yield sufficient S/N. As such, there is low confidence in this particular result, which is further highlighted by the Xe type classification recently found by \citet{santanaros25}. The D-type classification for 1992~BC is also surprising given its relatively high albedo of 33\% measured by the Spitzer NEOLegacy project \citep{trilling16}. This object has a moderately high error bar in the $z$ band that makes its reflectance values also formally consistent with a T or L type classification, the latter of which could be consistent with a moderately high albedo \citep{devogele18}. The taxonomic ambiguity associated with these putative D-types and their relative rarity in our sample is suggestive of inevitable uncertainty due to reliance on just four spectral channels, sometimes at low S/N. Statistical fluctuations that can cause such taxonomic ambiguity are at least mitigated in ensemble analyses, if not for these individual objects.

A final object of note is 2024~YR4 (Figure \ref{fig:yr4}). This object gained attention in early 2025 when it reached a 3\% impact probability with the Earth for December of 2032. Our data confirm previous reports of a $\sim20$ minute rotation period and classification in the S-complex \citep{bolin25}. Although an impact with the Earth in 2032 is now ruled out, 2024~YR4 remains of interest to planetary defense, as there currently remains a 4\% impact probability with the Moon \citep{wiegert25}.

\section{Lightcurve Considerations} \label{sec:LC}

A total of 49 objects are presented with newly measured lightcurves (Table \ref{tab:LC}). Figures associated with each of these are presented in Appendix \ref{appendix}. The photometry data were fit with a Fourier series of variable order. The best fit was based on the order that produced the lowest reduced $\chi^2$. A range of periods were scanned to determine a best fit solution. Informed by Nyquist sampling theory, the minimum search period was set to twice the image exposure time and the maximum period was set to twice the length of the observing sequence. The periods and amplitudes (peak-to-trough calculated from the fit) are reported in Table \ref{tab:LC}. The reported errors on the lightcurve period ($P_{err}$) are the offset from the best fit corresponding to an increase of 10\% in the $\chi^2$ statistic relative to the minimum. These lightcurve properties are best estimates based on the available data, however these observations were not necessarily optimized for determining robust lightcurve solutions. Incomplete rotational coverage, non-optimal temporal sampling, tumbling rotation states, or low S/N may have led to inaccurate solutions. However, for our primary objective of providing a means to lightcurve correct the color sequences, this approach was adequate for most cases and clearly helped to produce more reliable final colors. 

\startlongtable
\begin{deluxetable}{lll c c c c c l}
\tablecaption{Summary of measured lightcurve properties. Column descriptions and additional details are in the text. \label{tab:LC}}
\tabletypesize{\scriptsize}
\tablewidth{0pt}
\tablehead{\colhead{} & \colhead{Period} & \colhead{P$_{err}$} & \colhead{Amplitude} & \colhead{Fourier} & \colhead{LC Corrected} & \colhead{Uncorrected} & \colhead{} & \colhead{} \\[-0.3cm] \colhead{Object} & \colhead{(hr)} & \colhead{(hr)} & \colhead{(mag)} & \colhead{Fit Order} & \colhead{Type} & \colhead{Type} & \colhead{poly\_N} & \colhead{Notes}}
\startdata
2001 QJ142 & 0.1577 & 0.0006 & 0.32 & 3 & Xk & Xk & 0 & \\
2012 BF86 & 0.0491 & 0.0001 & 0.38 & 6 & A & Sq & 1 & \parbox[t]{4.5cm}{Consistent with T18} \\
2014 HW & 0.0641 & 0.0001 & 1.13 & 7 & K & Sq & 0 & \parbox[t]{4.5cm}{Actual period could be 0.032 hr} \\
2014 KH39 & 0.0440 & 0.0001 & 3.15 & 4 & Sq & A & 0 & \parbox[t]{4.5cm}{Consistent with T18} \\
2014 OV3 & 0.3471 & 0.0025 & 0.52 & 6 & Sv & Sv & 0 & \parbox[t]{4.5cm}{Consistent with T18} \\
2014 SC324 & 0.3550 & 0.0009 & 0.77 & 5 & O & V & 1 & \\
2014 UD57 & 0.0959 & 0.0001 & 0.94 & 5 & Sa & Sv & 0 & \parbox[t]{4.5cm}{Forced to T18 period, best fit = 0.479 hr} \\
2014 UX7 & 0.0367 & 0.0001 & 0.35 & 3 & X & Ch & 1 & \parbox[t]{4.5cm}{Consistent with T18} \\
2014 VL6 & 0.0723 & 0.0001 & 0.81 & 3 & Sq & A & 1 & \\
2015 AA44 & 1.7249 & 0.0605 & 0.63 & 8 & X & Ch & 1 & \parbox[t]{4.5cm}{Period forced to two-peak lightcurve} \\
2015 JD & 0.0339 & 0.0001 & 0.12 & 6 & Xk & Xk & 0 & \\
2015 KM120 & 2.1386 & 1.8153 & 0.18 & 4 & Q & Q & 0 & \parbox[t]{4.5cm}{Inconsistent with T18, low quality solution} \\
2016 AD166 & 0.0595 & 0.0001 & 0.3 & 4 & Q & C & 0 & \parbox[t]{4.5cm}{Inconsistent with T18, period here yields more reasonable color correction} \\
2016 AV164 & 0.0216 & 0.0001 & 0.49 & 6 & R & Sq & 0 & \parbox[t]{4.5cm}{2x period from T18} \\
2016 CF194 & 0.0346 & 0.0001 & 0.57 & 5 & S & B & 1 & \\
2016 DK & 1.2716 & 0.0095 & 0.79 & 7 & Sq & Sq & 2 & \parbox[t]{4.5cm}{Similar to 1.30 hr period from T18} \\
2016 FU13 & 1.2444 & 0.0800 & 0.26 & 4 & V & V & 1 & \\
2016 LT1 & 0.0525 & 0.0178 & 0.71 & 3 & X & D & 0 & \parbox[t]{4.5cm}{Low quality solution} \\
2016 NK39 & 2.4381 & 0.0303 & 0.52 & 9 & K & Cg & 2 & \parbox[t]{4.5cm}{Inconsistent with T18, period here yields more reasonable color correction} \\
2017 AT4 & 0.0500 & 0.0001 & 0.3 & 6 & C & B & 1 & \parbox[t]{4.5cm}{Period forced to two-peak lightcurve} \\
2017 QK & 0.1578 & 0.0020 & 0.24 & 3 & Sq & Sq & 0 & \parbox[t]{4.5cm}{Consistent with T18} \\
2017 XJ1 & 1.2313 & 0.0080 & 1.57 & 5 & T & T & 2 & \\
2018 CN2 & 0.2028 & 0.0001 & 1.3 & 8 & T & Xe & 2 & \\
2018 HO1 & 0.2152 & 0.0005 & 1.62 & 7 & K & Cg & 0 & \\
2018 JJ2 & 0.0803 & 0.0408 & 0.26 & 4 & Sv & Cg & 1 & \parbox[t]{4.5cm}{Low quality solution, possible NPA rotation} \\
2018 JJ3 & 0.0557 & 0.0001 & 0.44 & 5 & R & Q & 0 & \\
2018 PK21 & 1.2431 & 0.0210 & 0.82 & 4 & Sq & Sq & 2 & \\
2018 PU23 & 0.6198 & 0.0275 & 0.27 & 3 & K & C & 0 & \\
2018 TZ5 & 0.1298 & 0.0039 & 0.27 & 3 & Sv & R & 1 & \\
2019 CE5 & 0.2318 & 0.0070 & 0.39 & 3 & Xe & Xe & 2 & \parbox[t]{4.5cm}{NPA from LCDB} \\
2019 CZ2 & 0.0271 & 0.0001 & 0.39 & 3 & Xk & Q & 1 & \\
2019 RA & 0.0307 & 0.0001 & 0.81 & 5 & Q & Q & 0 & \\
2019 RH1 & 0.6453 & 0.0545 & 0.35 & 6 & Xe & Xe & 1 & \\
2019 SU3 & 0.1565 & 0.0001 & 0.12 & 5 & Xk & Xk & 1 & \\
2019 VW & 0.0867 & 0.0005 & 0.96 & 3 & Sq & Q & 2 & \parbox[t]{4.5cm}{Possible NPA rotation} \\
2019 WR4 & 0.1493 & 0.0001 & 0.72 & 3 & Xk & Q & 2 & \\
2019 YF4 & 0.0262 & 0.0001 & 0.32 & 3 & K & C & 1 & \\
2020 CA3 & 0.7510 & 0.0210 & 0.29 & 2 & Xk & Xk & 2 & \parbox[t]{4.5cm}{Period forced to two-peak lightcurve} \\
2020 DJ & 0.1454 & 0.0005 & 1.06 & 7 & Sq & V & 3 & \\
2020 FK3 & 0.0236 & 0.0059 & 0.29 & 4 & Xk & K & 2 & \\
2021 GK1 & 0.0173 & 0.0001 & 0.39 & 5 & B & Cgh & 1 & \\
2021 VX7 & 0.0182 & 0.0091 & 0.2 & 3 & Q & Q & 1 & \\
2022 DX & 0.0527 & 0.0001 & 0.38 & 4 & B & B & 2 & \\
2022 NX1 & 0.2930 & 0.0530 & 0.66 & 4 & Q & Sq & 0 & \parbox[t]{4.5cm}{LC from two nights of data (JD = 2459817 and 2459821)} \\
2023 BU & 0.1289 & 0.0568 & 1.61 & 3 & Q & Cgh & 0 & \parbox[t]{4.5cm}{Low quality solution} \\
2023 OP2 & 0.2651 & 0.0005 & 0.65 & 6 & Cg & Sq & 0 & \\
2024 NH & 0.1016 & 0.0513 & 1.16 & 4 & K & T & 0 & \parbox[t]{4.5cm}{LC from combined VR and r-band data} \\
2024 RZ3 & 0.0157 & 0.0001 & 0.27 & 3 & Xk & K & 0 & \\
2024 YR4 & 0.3257 & 0.0060 & 0.4 & 3 & S & L & 1 & \\
\enddata
\end{deluxetable}

The final column in Table \ref{tab:LC} provides object-specific notes. In some cases, lightcurve solutions from MANOS were previously published in \citet[][]{thirouin18}, abbreviated as T18. That independent analysis, sometimes on the same data, generally yielded compatible solutions. However, in some cases, slightly different solutions that better corrected the colors are presented here. For objects 2014~UD57, 2015~AA44, 2017~AT4, and 2020~CA3 the solutions were forced to twice the period of the minimum $\chi^2$ value. This adjustment was made because the reference filter magnitudes during the color sequences were better represented by these longer periods and/or because a symmetric two-peak lightcurve was more likely based on the measured amplitude \citep{harris14}.

In the case of two objects, 2022~NX1 and 2024~NH, hybrid lightcurves were constructed. For 2022~NX1 this was a combination of lightcurve data across two separate nights, four days apart. This resulted in a marked improvement to the color corrections. For 2024~NH, its combination was the usual $VR$ band lightcurve augmented with the $r$ band reference filter sequence. Both sets of images were calibrated to the PanSTARRS $r$ band and yielded a clean solution (Figure \ref{fig:A1}).

Several of the lightcurves indicated possible non-principal axis (NPA) rotation states. Generally this would be evident by residual scatter around the best-fit Fourier series that was well outside of the photometric error bars. Examples of this in Figure \ref{fig:A1} include 2018~JJ2, 2018~TZ5, 2019~VW, and 2024~RZ3. We do not attempt to provide an NPA or tumbling lightcurve solution for these objects. The object 2019 CE5 is noted as an NPA rotator in the asteroid lightcurve database \citep[LCDB,][]{warner09}, but our data do not obviously indicate as such.

\subsection{Implications of Lightcurve Variability}

We use these 49 objects to investigate the influence that rotational brightness variations have on colors derived from sequential exposures. This is relevant for any survey that aims to collect non-simultaneous color information for time-variable Solar System objects.

\begin{figure}
    \centering
    \includegraphics[]{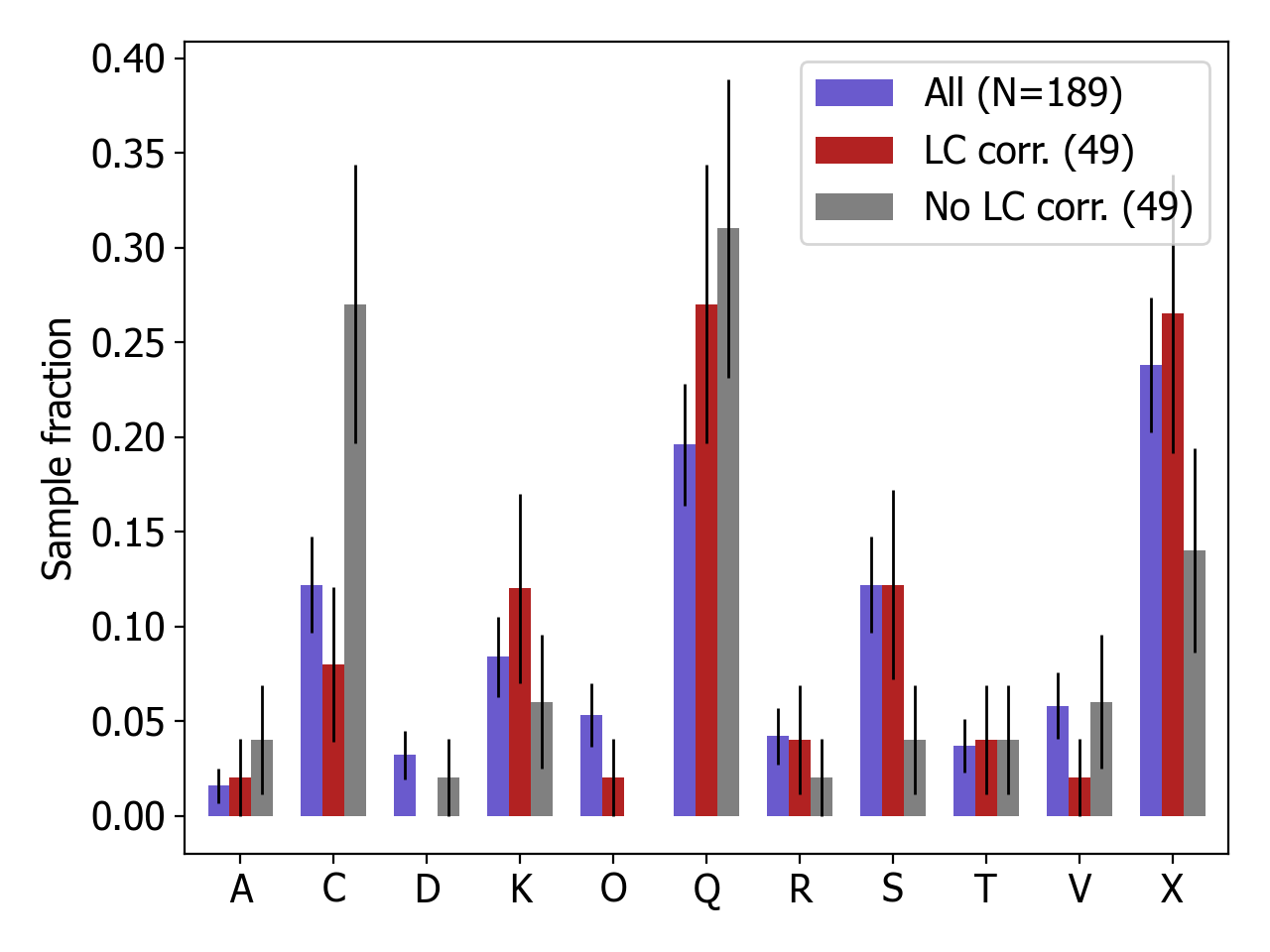}
    \caption{Taxonomic type distributions for the full color sample (blue), the 49 objects with measured lightcurves that were used for rotational corrections (red), and those same 49 objects without proper lightcurve (LC) correction (grey). The full sample and the LC corrected sample are highly consistent within the $\sqrt{N}$ Poison errors on each bin. The non-LC corrected sample shows significant differences, most prominent is a marked over-abundance of C types.}
    \label{fig:lc_correct}
\end{figure}

The lightcurve-corrected taxonomic assignments for our 49 objects are given in Table \ref{tab:LC}. To probe the influence of lightcurves on these results, we assumed an absence of lightcurve information and instead used a simple polynomial fit to the reference filter as a proxy for rotational variability. As with the majority of our sample, the order of this polynomial was set to minimize the reduced chi-squared statistic between the reference filter data and the polynomial fit. The order of these polynomials is indicated in the ``poly\_N" column of Table \ref{tab:LC}. Again a polynomial of zeroth order indicates no lightcurve correction so that the final colors would simply be the difference in mean magnitudes for each filter. Of the 49 objects analyzed in this way, 32 of them (65\%) switched taxonomic type. This included a mix of relatively minor changes, e.g. 2019~VW switching from an Sq to Q type, as well as more significant changes, e.g. 2019~YF4 going from a K type to a C type. The cumulative changes from this exercise are shown in Figure \ref{fig:lc_correct}.

Lightcurve variations affecting taxonomic assignments for two thirds of these 49 objects is roughly consistent with expectations. To first order we would expect any object with significant lightcurve variability across the time span of observation to potentially switch taxonomic types. So we looked at the rotation periods of NEOs in the LCDB \citep{warner09} with $H>20$, i.e.~consistent with the MANOS sample. A bit less than two thirds, about 60\%, of these LCDB objects have lightcurve periods between 36 seconds and 3 hours. This lower limit corresponds to 2$\times$ our mean exposure time, thus periods shorter than 36 seconds would not be resolved with our sampling cadence and would not affect the measurement of colors. The upper limit of 3 hours is an estimate for the time below which lightcurve variability would become apparent in the $\sim10-30$ minutes used to collect a color sequence. Longer periods would not appreciably affect the derived colors. In short, about two thirds of the MANOS-like NEOs in the LCDB have lightcurve periods that could lead to ambiguous taxonomic classification if rotational variability were uncorrected. We find about the same fraction for our test sample of 49 objects.

Surprisingly, we find that the lack of proper lightcurve correction did not result in random re-distribution of taxonomic types. This might have been expected. For example, considering the influence of lightcurve correction on spectral slope, any given object could be just as likely to appear redder as it could bluer. Individual objects might swap taxonomic types, but averaged over the full sample, this would leave the underlying distribution unchanged. However, this is not what we see. Figure \ref{fig:lc_correct} shows an enhancement in the number of C class classifications in the uncorrected set. This enhancement came from 9 objects or 19\% of the set moving into the C class. This represents a larger shift than expected from random fluctuations. To probe this further, we developed a simple model for the influence of lightcurve variability on the measurement of colors.

\subsection{Modeling the Influence of Lightcurves on Colors}

Applying lightcurve corrections when measuring colors can have significant effects both on the resulting distribution of taxonomic types as well for individual objects. For especially complex lightcurves and/or low S/N data, these adjustments can move objects from one taxonomic type to another -- but how frequently are the changes significant (e.g., from an S complex object to an X or a C complex object)? In order to assess what systematic effects lightcurve correction might have on our derived colors and thus taxonomic distribution, we set out to simulate as much of our observational dataset as possible -- from the properties of the target asteroids to the specific choices made by observers at the telescopes.

We first generated a synthetic sinusoidal lightcurve shape drawn from the amplitudes and periods of asteroids in the LCDB \citep{warner09} at sizes similar to what is seen in the MANOS color sample ($H>20$) and then adjusted the lightcurve to have an average apparent magnitude similar to a typical target's apparent magnitude (randomly selected from $19 < m_V < 23$). We then ``sampled" the lightcurve by selecting an exposure time based on a representative, randomly generated non-sidereal rate for the synthetic asteroid  (from $0.8-3.5$ arcseconds per minute) and then set the synthetic exposure to the amount of time that the asteroid would take to cross a typical $1.0"$ seeing disk. We incorporated a six-second delay after every exposure to mimic the changing of filter wheels, the speed of CCD read-out, and other small amounts of dead time. The average magnitude of the asteroid in that filter during that exposure was calculated from the original lightcurve model and then an error was assigned according to the exposure time calculator for the LDT (\url{www2.lowell.edu/users/massey/LMI/etc_calc.php}) in the specified filter under typical conditions. The synthetic uncertainties are filter-specific are based on those from real observations of asteroids with LMI. A critical aspect of determining an object's taxonomy is whether or not it has a $1-\mu{m}$ absorption feature. The $z$ filter is both the filter most sensitive to it and the one in which we typically encountered our highest uncertainties (modeled as $\sim2\times$ that in the $r$ filter), so inclusion of filter-dependent uncertainties was critical to faithfully modeling this part of our observing procedure. In essence, we took a smooth model lightcurve and digitized it with a cadence and filter-dependent SNR that was a good proxy for our typical observations.

This full lightcurve was then split up such that constituent single-filter lightcurves were portioned out non-sequentially to mimic a real observing sequence as described in Section \ref{sec:colors}  (e.g., $r-g-r-i-r-z$, such that the first observation is transformed to $r$, the second to $g$, and so on). We selected a random time offset (between $0$ and half the underlying period of the asteroid) after the `start' of the lightcurve to begin sampling and then selected an integer number of filter cycles (typically three) to record for the next step in the analysis. The relative brightness of the synthetic asteroid in each filter was set by the chosen taxonomy of the asteroid in the Bus-Demeo system \citep{demeo09}, which could either be assigned at random or chosen explicitly. We also selected an approximate hour of the pre-offset portion of the digitized lightcurve to stand in for our $VR$-filter dedicated lightcurve sequence done for the 49 objects described above.

These synthetic multi-filter lightcurves and errors were then treated identically to the real data. Colors were derived with and without lightcurve correction. The corrections were done by fitting the reference filter in the color sequence or by utilizing a Fourier fit to the $VR$-filter dedicated lightcurve collected before the color sequences. Best-fit taxonomic types were then assigned. We could then repeat this procedure to either study the systematic effects on fitting individual lightcurves (e.g., `are different filter sequences while observing an asteroid more reliable than others?') or to study population-level effects on larger groups of objects (e.g., `does lightcurve correction result in a statistically significant change to the derived taxonomic distribution?').

\begin{figure}
    \centering
    \includegraphics[]{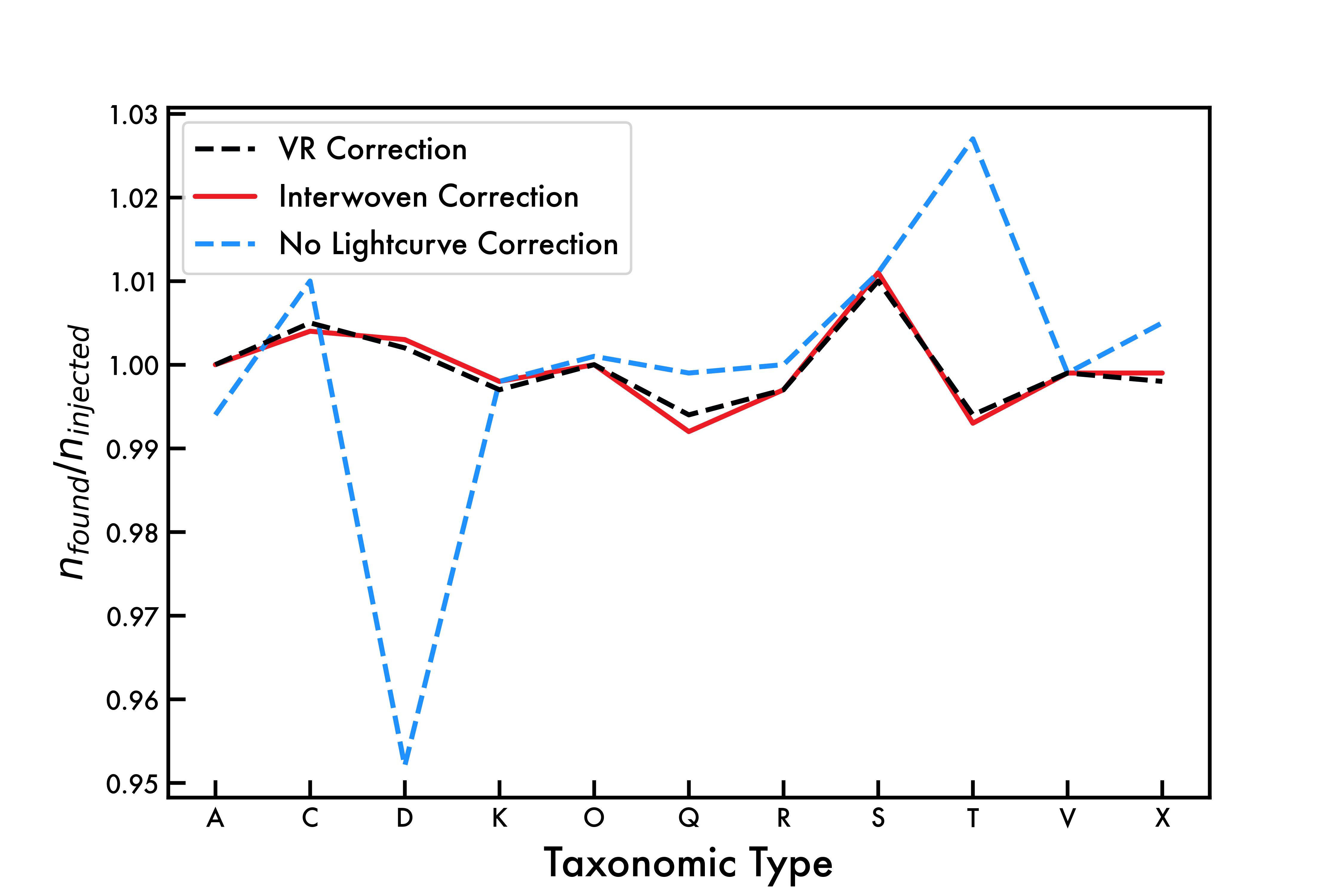}
    \caption{Model output in which synthetic multi-filter asteroid observations are converted to taxonomic types with lightcurve correction based on a dedicated $VR$-band lightcurve sequence (black), on a polynomial fit to an interwoven reference filter during the color sequence (red), and without any correction (blue). If lightcurves did not influence taxonomic assignment, the number of returned objects in each type ($n_{found}$) would equal to the number injected ($n_{injected}$). Lightcurve correction clearly moves the derived taxonomic distribution closer to the true (injected) population. See text for more details.}
    \label{fig:lc_modeling}
\end{figure}

To specifically address the population scale question, we generated $10^3$ synthetic lightcurves in each of our 11 taxonomic types and then taxonomically assigned them utilizing the $VR$-filter based lightcurve correction, a third-order polynomial lightcurve correction on the interwoven Sloan $r'$ filter observations during the color sequences, and with no lightcurve correction at all. We also ran tests with $(1-3)\times10^2$ synthetic lightcurves to mimic the variability one might see in a sample the size of our real one. Each sub-type (e.g., the Cb type inside the C class) was randomly sampled with equal likelihood. The outputs of this approach can be seen in Figure \ref{fig:lc_modeling}. If the bulk taxonomic distribution were unchanged (about as many objects moved into each category as moved out), the curves in this plot would be horizontal lines. Even without lightcurve correction, none of these taxonomic types become over- or under-represented by more than about $\sim5\%$ unless a very small ($<50$) sample size is chosen. Misclassifications between complexes (e.g., observing a C-type and finding it to be best-fit as a S-type) is uncommon and seems to primarily occur when asteroids with only partially-correctable rapid lightcurves are observed in the $z$ filter at their lightcurve minima and thus with higher uncertainties. As noted previously, the $1-\mu{m}$ absorption feature characteristic of stony asteroids is really only detectable with the $z$ filter which makes its typically lower SNR given typical detector sensitivites and asteroid brightnesses an impediment to taxonomic classification. Without any lightcurve correction,  this model predicts slightly more C, S, and T types, and fewer D types, all at the $1-5\%$ level (Figure \ref{fig:lc_modeling}). While these changes align with some of the trends seen in Figure \ref{fig:lc_correct}, they are too small by perhaps an order of magnitude to explain the differences seen in that Figure. In particular, the origin of the systematic overabundance of C types in the non-lightcurve-corrected sample remains unclear, and thus could be a consequence of low number statistics.

There were several aspects of the real dataset not captured by our model, but the most important seems to be lightcurve shape complexity. While our primary simulations assumed sinusoidal lightcurves, experiments with more complex lightcurve shapes generally began to separate the differences between the lightcurve correction techniques. While for most asteroids with rotation periods longer than our typical color sequence the $VR$-filter lightcurve and interwoven Sloan $r'$ correction method worked equivalently (the associated curves in Fig. \ref{fig:lc_correct} overlapped very closely), for shorter periods or more complex lightcurve morphologies the $VR$-filter approach worked better. As surveys similar to our own push to smaller and smaller asteroids, dedicated lightcurve correction sequences might be increasingly critical to deriving robust constraints on the underlying color and taxonomic distributions.

\section{Size-Dependent Trends} \label{sec:size}

Given that the MANOS data characterize the smallest observable NEOs (Figure \ref{fig:H-Mag}), and that MANOS spectra \citep{devogele19} and colors both show a different taxonomic distribution relative to other surveys (Figure \ref{fig:types}), we consider the extent to which the taxonomic distribution of NEOs is size dependent. To fully probe any size dependencies we compiled results across a wide range of H magnitudes. This involved combining NEOSHIELD2 \citep{perna18}, MANOS (spectra and colors), and MITHNEOS \citep{binzel19} visible data. As noted in Section \ref{sec:results} these data sets are consistent in terms of their wavelength coverage and taxonomic classification schemes. Combining the results of these surveys yields classifications for a total of 831 objects with robust sampling from $H=13$ to $H=29$, corresponding to diameters from about 10 km down to a few meters. To ensure a true ``apples-to-apples" comparison, we have intentionally avoided comparison to data sets that provided different wavelength coverage, e.g. the near-infrared component of MITHNEOS \citep{binzel19}, the near-IR spectra of \citet{sanchez24},  or the $BVRI$ colors (lacking the important $z$ band measurement around 0.9 $\mu m$) from the NEOROCKS survey \citep{hromakina23,birlan24}. We also do not compare to the results of hybrid taxonomic schemes that combine diverse data products such as albedos and spectra \citep[e.g.][]{sergeyev23}. 

\begin{figure}
    \centering
    \includegraphics[]{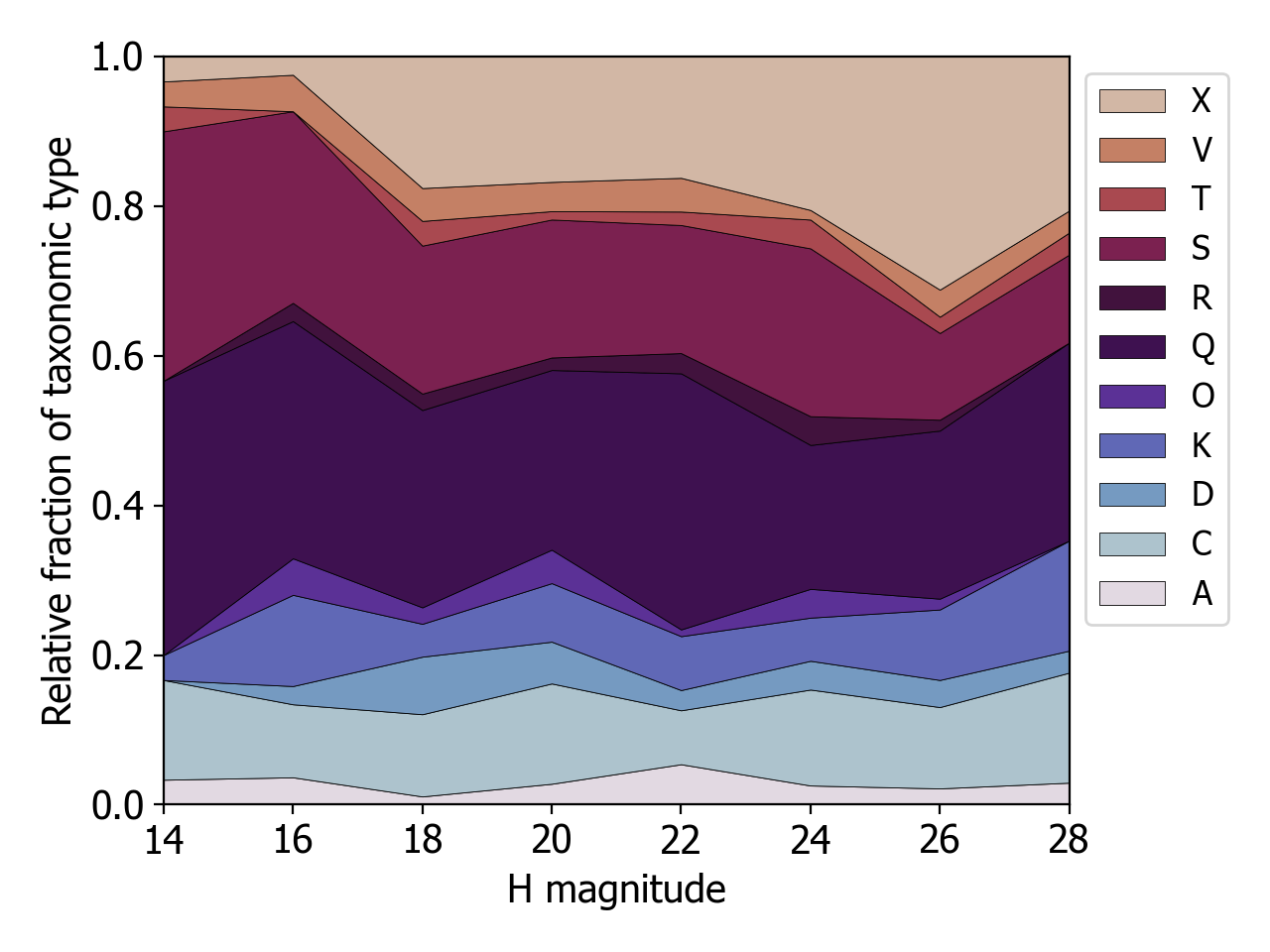}
    \caption{The relative fraction of each taxonomic type as a function of absolute magnitude $H$ derived from a combination of visible-wavelength results from MANOS, NEOSHIELD2, and MITHNEOS. A prominent feature in this figure is the decrease in S and Q types with increasing $H$, which is compensated by an increase in X types.}
    \label{fig:taxH}
\end{figure}

The taxonomic distribution from the combined MANOS, MITHNEOS, and NEOSHIELD2 data is shown as a function of $H$ magnitude in Figure \ref{fig:taxH}. This figure shows the relative fraction of each taxonomic type in $H$ bins that are 2 magnitudes wide from $H=13$ to $H=29$. This covers a significant fraction of the known NEO population. Only 6 NEOs have $H<13$, while the upper limit of $H<29$ is set to ensure a sufficient number of objects in the final bin (N=36). The number statistics for bins at higher $H$ get too low to be meaningfully diagnostic. The 2 magnitude width was set to provide reasonable sampling in each bin. The number of objects per bin ranges from 30 at the lowest $H$ bin to 156 objects in the $H=24$ bin. These counts result in $1\sigma$ Poisson error bars on the relative fraction for each type that range from about 1\% up to 6\%. Thus it is safe to conclude that differences in relative fractions at the few percent level are not significant; differences at the 10\% level or more are meaningful.

While the majority of types (V, T, R, O, K, D, C, A) show no appreciable change in abundance across this $H$ range, there is a prominent change in the relative fraction of X, S, and Q types. Specifically, S and Q types decrease with increasing $H$, while X types increase with $H$. If we combine the S and Q types, which would be consistent with their interpretation as a continuum of ordinary chondrite-like compositions \citep{chapman96}, their combined fraction goes from 70\% of the sample at $H=14$, to 50\% at $H=22$, and to 35\% at $H=28$. This factor of two decrease is consistent with the findings of \citet{devogele19}. The addition of the color data serve to enhance the significance of this trend.

\begin{figure}
    \centering
    \includegraphics[width=5in]{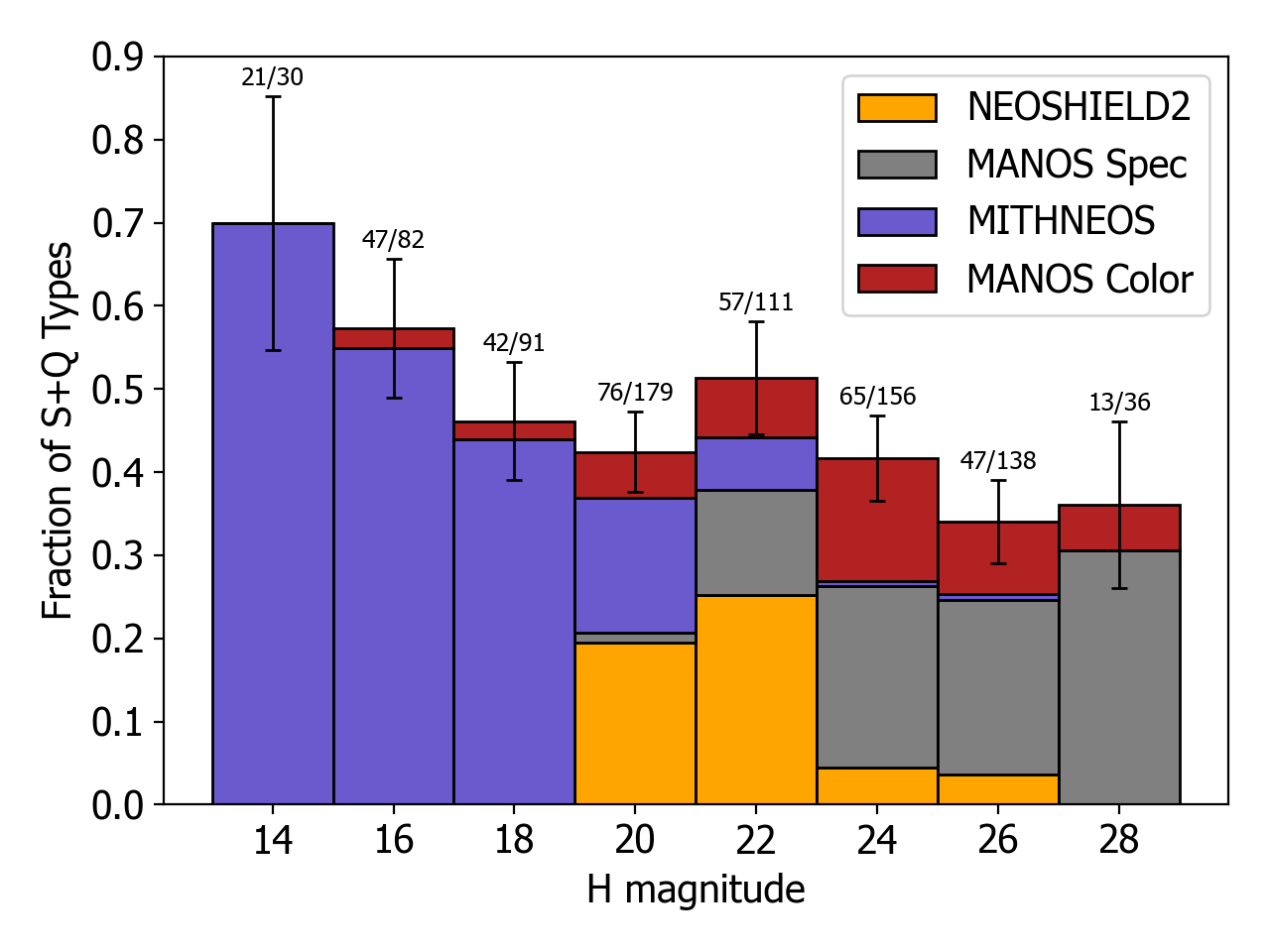}
    \includegraphics[width=5in]{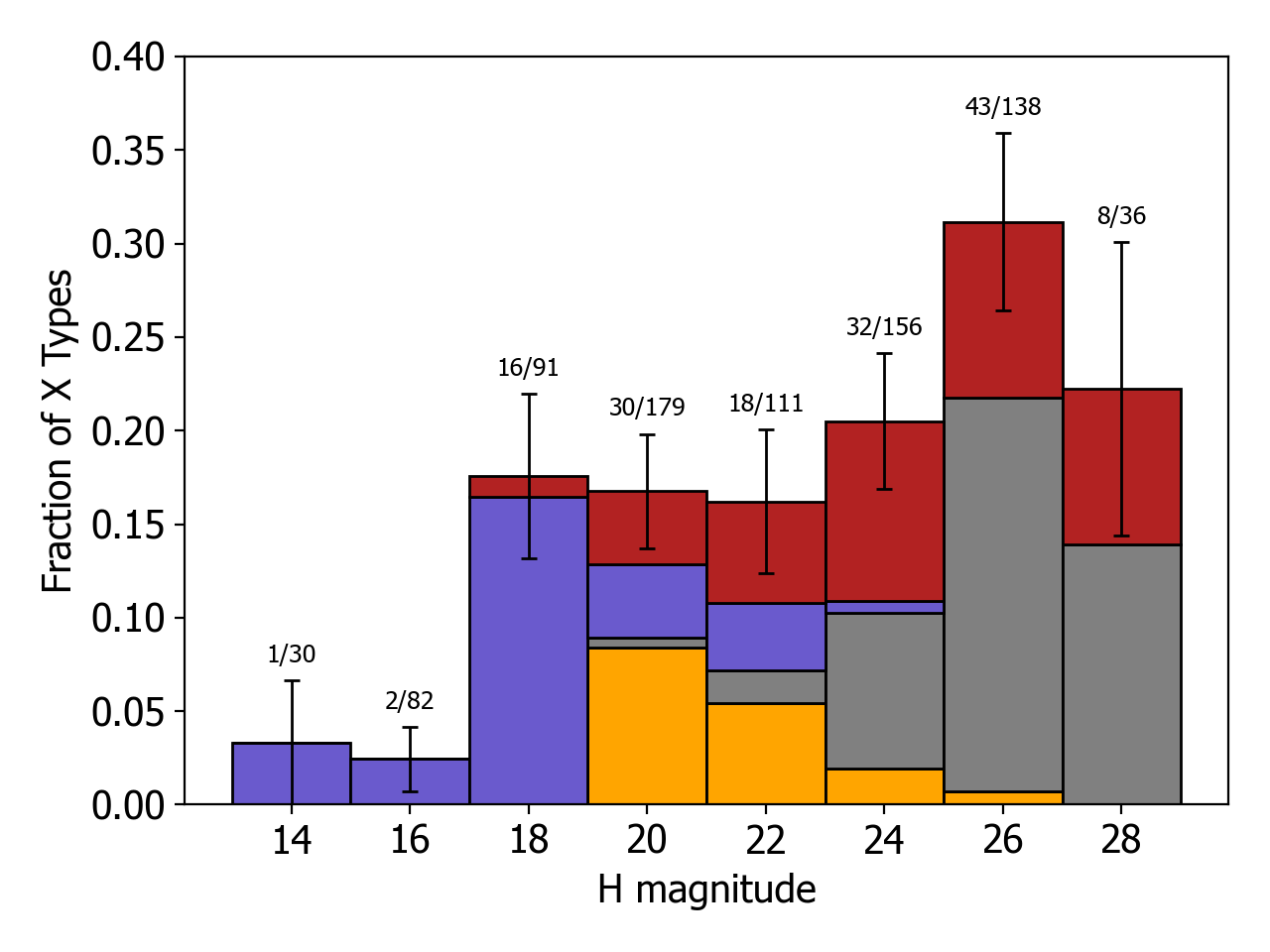}
    \caption{The fraction of S+Q complex (top) and X complex objects (bottom) in a combined set of NEOSHIELD2, MANOS, and MITHNEOS data, shown as a function of absolute magnitude $H$ in 2-magnitude-wide bins. The relative number of objects contributed by each survey is indicated in each bar by the color coding defined in the top panel. The annotated fractions above each bin represent the number of that spectral complex in the numerator versus the total number of objects for that H bin in the denominator. The error bars represent 1-sigma Poisson errors. Both panels show a trend in the distribution of these taxonomic types as a function of $H$.}
    \label{fig:fractions}
\end{figure}

A more detailed view of this change in relative fraction of S+Q and X types is shown in Figure \ref{fig:fractions}. This figure depicts the relative contributions from NEOSHIELD2, MANOS, and MITHNEOS in each $H$ bin. For example, MITHNEOS contributed 100\% of the S and Q types at $H=14$, whereas MANOS spectra and colors made up 100\% of the $H=28$ bin. In general, the MITHNEOS data dominate the $H<20$ population, NEOSHIELD2 in the range $20 < H < 22$, and MANOS for $H>22$. This combination of surveys is important to provide adequate sampling across the full range of $H$ values. Any one survey by itself would not be as well-suited to probe size-dependent trends. It is also clear that the combination of surveys helps to bolster the sample in each bin, as no bin ends up with fewer than 30 objects. 

Performing linear regressions on the relative fractions versus $H$ provides a simple quantification for the significance of these size-dependent changes. For the S+Q fractions, the $p$ value associated with a linear regression is 0.004, corresponding to a correlation significance of $2.9\sigma$. For the X type fractions, the equivalent $p$ value is 0.005 for a correlation significance of $2.8\sigma$. In both cases, the standard error on the fitted slopes suggests these trends are inconsistent with flat distributions at a significance level of $4.2\sigma$.

These size-dependent trends are robust against S/N considerations. As noted above (Section \ref{sec:results}) the MANOS sample is of relatively uniform quality all the way down to the smallest objects in the sample. Thus, as an ensemble, we do not expect systematic biases in the taxonomic distribution due to S/N. This is supported by considering the 2nd and 3rd best (based on rms) taxonomic fits to the color data. If data quality issues, for example low S/N in $z$-band, were systematically biasing taxonomic assignments, we would expect to see shifts in the overall distribution when alternatively adopting the 2nd or 3rd best fits. This does not happen. The distribution of taxonomic types (Figures \ref{fig:types} and \ref{fig:taxH}) does not change at more than the 10\% level for any individual type when considering the 2nd or 3rd best assignments. This is an indication that the colors provide reliable taxonomic assignments and that variations as a function of rms are due to random fluctuations within the 10\% limit for meaningful differences.

We emphasize that the distributions presented in Figures \ref{fig:taxH} and \ref{fig:fractions} are purely observational and do not account for issues of sample bias. However, as we discuss in the following section, formal bias correction like that implemented by \citet{marsset22}, would only exacerbate the size dependent trends highlighted here. Furthermore, through several plausibility arguments in the following section, the cause(s) behind these trends can be constrained to yield novel interpretations even without bias correction.

Lastly, we note that the presentation of results here does not consider a separation of the S and Q class. Canonically, Q types represent fresher, unweathered ordinary chondrites whereas S types have older, more weathered surfaces \citep{chapman04}. In previous works, the ratio of the number of Q to S types as a function of MOID, $H$, and perihelion distance has been used to argue for various mechanisms that may be responsible for modifying the extent of weathered regolith on ordinary chondrite-like surfaces \citep[e.g.][]{binzel10,devogele19,graves19,sergeyev23}. Our results do not necessarily contradict any of these previous studies, but the low effective spectral resolution of our color data makes it difficult to strongly constrain  subtle differences between S and Q class objects in a way that would advance understanding beyond these previous studies. Thus we do not pursue further investigations into the role of space weathering and the mechanism(s) responsible for refreshing ordinary chondrite surfaces.

\section{Discussion} \label{sec:discussion}

We have presented the results of a spectro-photometric color survey of 189 NEOs (Table \ref{tab:results}, Figure \ref{fig:colors}) collected as part of the Mission Accessible Near-Earth Object Survey. The systematic targeting by MANOS of NEOs with $H>20$ produced a unique data set that probes some of the smallest telescopically-accessible objects in near-Earth space (Figure \ref{fig:H-Mag}). 

The color data were carefully curated and validated against several sources including intra-survey comparisons for objects observed on multiple nights (Table \ref{tab:multiple}) and comparisons to spectral data (Figure \ref{fig:spec}, Table \ref{tab:spec_comp}). In general, the colors were fully consistent within error bars and were assigned the same or very similar taxonomic classifications across multiple observations. Differences that did occur were generally attributable to the low spectral resolution of the colors and/or error bars on the photometry that spanned the boundaries between taxonomic types. Most objects that displayed different spectral properties across multiple epochs were assigned similar taxonomic types, e.g. objects were seen to switch from a Q type classification to an Sq type, or from a C to an Xc. A small subset of the objects showed more pronounced shifts. For example, 2011~CG2 displayed colors consistent with a Cgh classification in one epoch and a Q type classification in another. Without additional data it is not clear whether these shifts were due to the influence of rotational lightcurve variability during the measurement of colors, viewing geometry effects tied to phase reddening \citep{sanchez12}, or a spectrally heterogeneous surface. However, such large shifts in taxonomic type were rare, implying that our colors provide a robust assessment of spectral properties.

Specific attention was given to the influence that rotational lightcurves can have on colors derived from non-simultaneous measurements (Section \ref{sec:LC}). This included measuring new rotational lightcurve periods and amplitudes for 49 objects (Table \ref{tab:LC}). It was shown that lightcurve correction, whether that was a Fourier solution to the underlying rotational brightness variations or a simple polynomial fit to parameterize that variability, was needed for 3/4 of our sample to ensure accurate colors and taxonomic assignments. These lightcurve corrections were clearly important for individual objects and unexpectedly indicated systematic implications when left unaddressed. In particular, our data suggest that NEO color surveys that do not correct for lightcurve variations could find a higher fraction of C type classifications (Figure \ref{fig:lc_correct}). The baseline C type fraction found by all four surveys in Figure \ref{fig:types}, including the lightcurve corrected MANOS colors, is about 10-15\%. This is about the same as the fraction of C types found at near-IR wavelengths by \citet{sanchez24} and at the low end of the 10-20\% found by \citet{binzel19}. Unlike the MANOS color sample, these spectroscopic surveys would not be affected by lightcurve variability and thus should provide a solid foundation for comparison. These results thus suggest that the observed fraction of C types is relatively constant at $\sim15\%$ across a wide range of sizes, from kilometer down to meter scale objects (Figure \ref{fig:taxH}). If lightcurve variability can indeed artificially enhance the fraction of C types in color-based data sets, this would have implications for taxonomic assignments derived from surveys like the Vera Rubin Observatory/Legacy Survey of Space and Time (VRO/LSST) that collect color data with non-simultaneous exposures in different filters \citep{kurlander25}. In short, lightcurve variations must be accounted for when collecting colors through traditional methods.

In an attempt to better understand whether lightcurve variability could systematically enhance the C type fraction in the color data, we developed a simple model to account for realistic observational effects connected to exposure cadence, instrumental sensitivity, and resulting S/N. This model was run specifically with lightcurve properties (periods, amplitudes) representative of the MANOS sample (e.g. $H>20$). Unfortunately, the outcome of this did not explain the enhancement of C types in the non-lightcurve corrected set. The root cause of this enhancement remains unclear. However, this modeling exercise did serve to highlight the importance of treating lightcurve variability when measuring colors (Figure \ref{fig:lc_modeling}). These models suggested that there was no appreciable difference in corrections based on a Fourier solution to a dedicated sequence of lightcurve observations versus a simpler polynomial-based correction based on an interleaved reference filter. Both approaches however led to more accurate results ($<1\%$ false classification) than assuming no lightcurve correction (up to 5\% false classification, most pronounced for D and T types). This model assumed a simple sinusoidal lightcurve morphology; simulating more complex lightcurves could yield different results, but is beyond the scope of this work.

From our set of derived colors we highlighted individual objects of interest, for example those on unusually low $\Delta v$ orbits (Table \ref{tab:lowdv}), as well as the ensemble of taxonomic types (Figure \ref{fig:types}). These colors and the results from the MANOS visible spectroscopic survey \citep{devogele19} are broadly consistent. The observed fractions for most taxonomic types (A, C, D, K, S, T, X) from these two studies are the same within error bars. In addition, both works show a deficit of S class asteroids relative to Q class. This could be suggesting that the surfaces of small NEOs are generally fresher and less space weathered (Q class) than their larger counterparts. This would be consistent with the expectation of shorter collisional lifetimes for smaller objects \citep{bottke05}. Other possible explanations such as the perihelia or MOID distributions of the MANOS sample are discussed below.

Slight differences between the colors and the spectra are noted for some end member taxonomic types. For example, a higher percentage of objects with deep 1 $\mu m$ bands, i.e. O, R, and V types, is seen in the color sample. These types represent 16\% of the color sample (31 objects) versus 5\% of the spectroscopic sample (10 objects). The origin of this offset is unclear.  It could be that our procedure of normalizing the spectro-photometry based on a linear interpolation between the $g$ and $r$ bands has consequences for the perceived depth of the 1 $\mu m$ absorption band. The effects of lightcurve variability on colors did not significantly influence the relative abundance of types with deep 1 $\mu m$ bands (Figure \ref{fig:lc_correct}). Despite the overall increase of these spectral types in the color sample, we do see a fraction of V types (6\%) that is consistent with several other surveys \citep[e.g.][]{sergeyev23,sanchez24} and, perhaps coincidentally, the fraction (5\%) of HEDs (Howardite-Eucrite-Diogenite meteorites are traditionally linked to V type asteroids, \citep{burbine01}) from meteorite fall statistics cataloged in the Meteoritical Bulletin Database\footnote{\url{https://www.lpi.usra.edu/meteor/metbull.cfm}}.

The MANOS spectra and color results were combined with taxonomic assignments from the MITHNEOS \citep{binzel19} and NEOSHIELD2 \citep{perna18} surveys (Figure \ref{fig:taxH}). The most notable feature in this figure is the two-fold decrease in S and Q types as a function of $H$ magnitude. Potential causes of this trend are now discussed.

\begin{figure}
    \centering
    \includegraphics[]{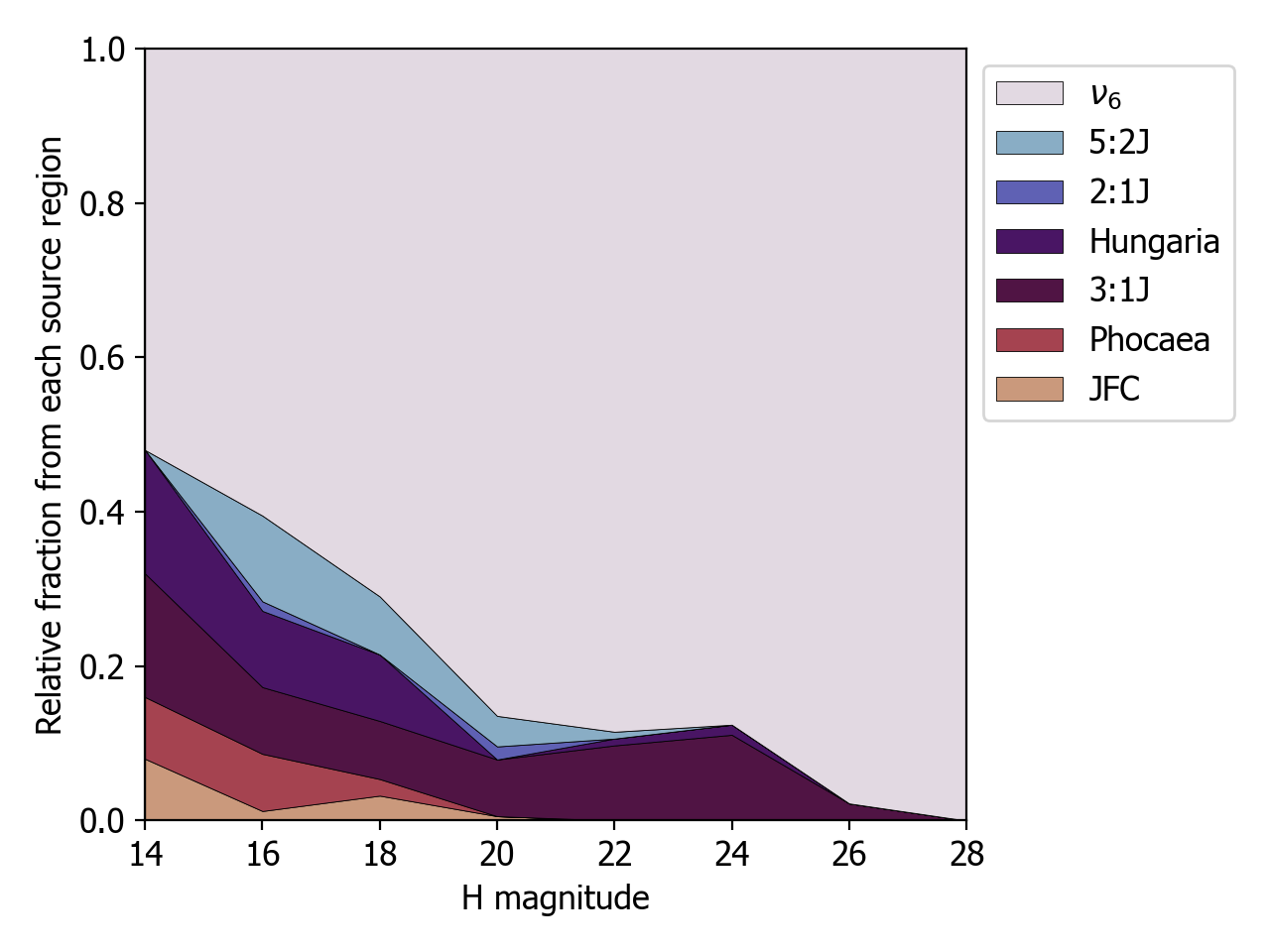}
    \caption{The relative fraction of NEOs in the combined sample from each of the \citet{granvik18} source regions as a function of $H$ magnitude. The prevalence of an inner belt $\nu_6$ origin for small objects suggests they should be dominated by S class objects from the Flora family, however our results find the opposite. This may be because small objects can sample a wider swath of the Main Belt due to their increased mobility under the Yarkovsky effect.}
    \label{fig:source}
\end{figure}

This particular change in the taxonomic distribution is unlikely to be related to size-dependent changes in Main Belt source region \citep{granvik18,nesvorny24a}. Using the astorb database \citep{moskovitz22}, we retrieved the single most probable Main Belt source region for each of the 831 objects in the combined dataset. The relative fractions from each source region -- the $\nu_6$ secular resonance with Saturn, the 5:2, 2:1 and 3:1 mean motion resonances with Jupiter, the Hungaria region, the Phocaea population, and Jupiter Family Comets (JFC) -- are shown in Figure \ref{fig:source}. Since the \citet{granvik18} model only extends to $H=25$, objects with larger $H$ were assigned source region probabilities extrapolated from the $H=25$ bin. The combined set of objects is exclusively sourced from the $\nu_6$ resonance at small sizes. This is consistent with NEO orbital distribution models that predcit a majority of small ($H>25$) NEOs, particularly those on Earth-impacting orbits, come from the $\nu_6$ resonance \citep{granvik18,nesvorny23}. At the inner edge of the Main Belt, the $\nu_6$ should be predominantly contributing S class objects from the Flora family, the largest familiy adjacent to the $\nu_6$. This appears to be supported by \citet{marsset22} who show that the $\nu_6$ is biased in favor of S types (as are all Main Belt source regions due to the higher average albedo of S complex objects relative other major taxonomic types). However, this estimate for the debiased distribution of NEO compositions does not consider any size dependency and, since it is based solely on MITHNEOS spectra, is dominated by objects $>100$m in size. The fact that we see a decrease in S and Q class objects with an increasing likelihood of a $\nu_6$ origin suggests that source region is not necessarily the dominant contributing factor. In fact, the increased spectral diversity at small sizes may be an indication of the increased mobility of small objects under the influence of the Yarkovsky force \citep{bottke06,granvik17}. In other words, these small objects may be coming exclusively from the $\nu_6$ resonance, but due to their mobility, they sample a wider swath of the inner Main Belt before escaping into near-Earth space. This suggests that the spectral types of NEOs $<100$m in size reflect the taxonomic diversity of major families in the inner Main Belt between the $\nu_6$ and 3:1 mean-motion resonances, namely the Flora (S-type), Vesta (V-type), Erigone (C-type), Nysa-Polana (E/X and B-types), and Massalia families (S-type).

\begin{figure}
    \centering
    \includegraphics[]{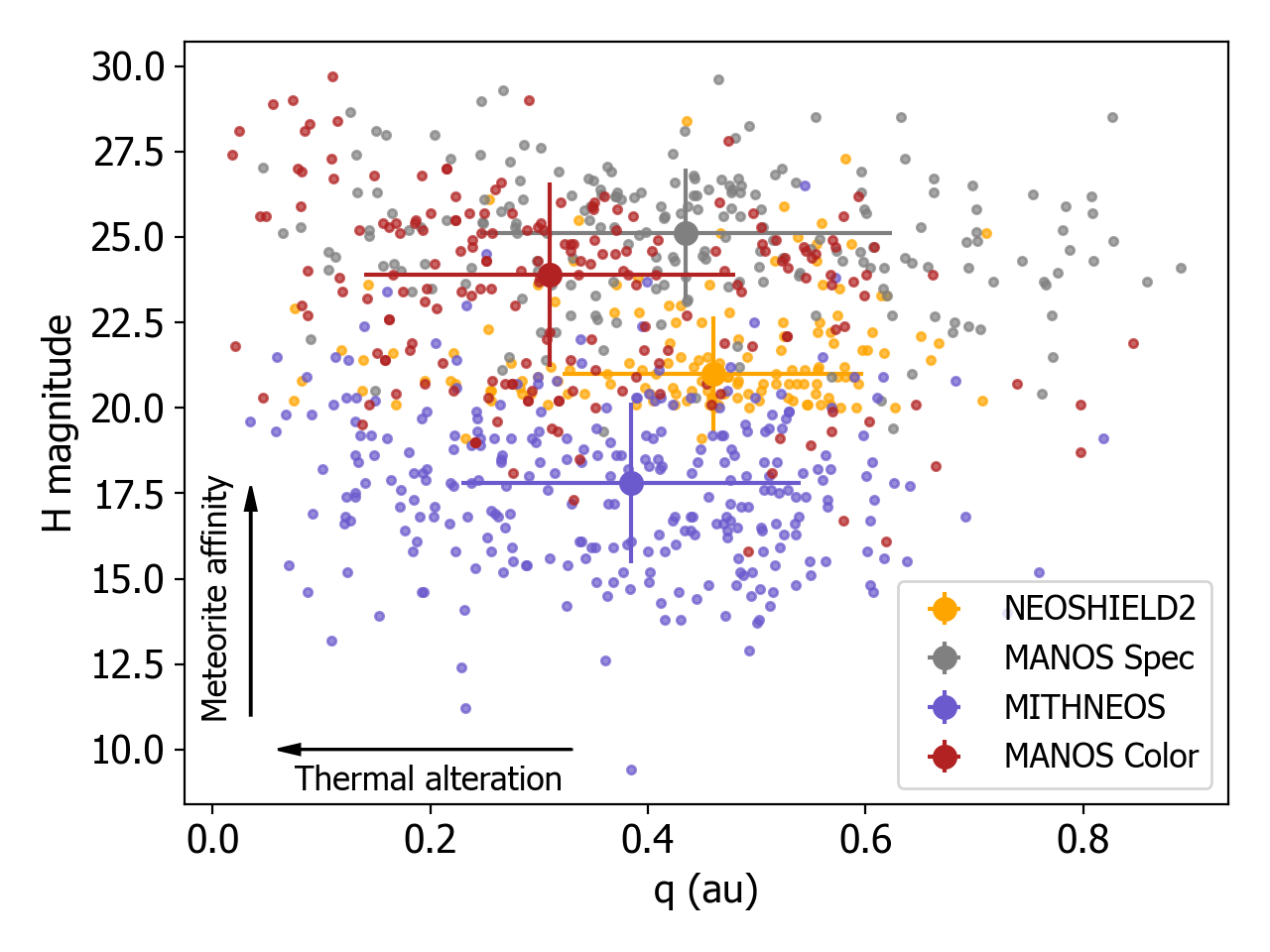}
    \caption{Perihelion distance $q$ versus absolute magnitude $H$. The average values and $1\sigma$ standard deviations are shown for each survey. Objects with smaller $q$ would be more susceptible to thermal alteration. Smaller bodies (larger $H$) should be more closely related to meteorites given their increased frequency of impact.}
    \label{fig:q}
\end{figure}

Thermal modification is another unlikely cause for the $H$-dependent change in S and Q class objects. Thermal fatigue driving the breakdown of regolith as a mechanism for refreshing asteroid surfaces has been demonstrated in the laboratory \citep{delbo14} and supported observationally \citep{graves19,sergeyev23}. However, this is specifically relevant to the S:Q ratio, not necessarily the S+Q fraction of the population. If thermal fatigue were contributing to the abundance trend of S+Q objects, that would suggest smaller objects from the MANOS data sets should be experiencing higher average surface temperatures and thus have lower average perihelion ($q$) distances. This is not the case (Figure \ref{fig:q}). Within 1 standard deviation, all four surveys have the same mean perihelion distance, and span similar ranges. The MANOS color survey sampled slightly lower perihelia, but the MANOS spectral survey did not. Both would have had to systematically sample objects that experienced higher average surface temperatures to explain the size-dependent trend that was observed. Furthermore, the slight offset towards smaller perihelion for the color sample would only correspond to a difference in equilibrium sub-solar temperature of about 80 K relative to the other surveys. This seems insufficient to explain the pronounced shift in S+Q frequency at high $H$.

Discovery or albedo bias is also unlikely to cause the observed drop in S+Q fraction. The sample here is intrinsically affected by discovery bias, because all of these objects were discovered by ground-based telescopes at visible wavelengths. However, that discovery bias would be in favor of mid and high albedo objects, causing them to increase in frequency at small sizes (faint magnitudes) where discovery rates are most biased. For S and Q types we see the opposite: namely a size-dependent decrease for these mid-albedo objects. Interpreting the associated increase in X complex objects is complicated by the mixed albedo of objects in this taxonomic class. The X complex was originally defined to encompass E, M, and P types, which are distinguished by different albedo ranges of $>0.3$, $0.1 - 0.3$, and $<0.1$ respectively \citep{tholen84}. Without knowledge of the underlying albedo distribution in our sample and the relative breakdown of EMP types, it is difficult to know exactly how the relative abundance of X types would change with $H$ in a biased sample. We can gain some insight by looking at existing data sets that combine spectral and albedo information. Perhaps the most complete data set for this purposes is from \cite{sergeyev23}. These authors identify 71 NEOs in the X complex with measured albedo, which break down as 15\% E type, 31\% M type, and 53\% P type. This is similar to the percentages (10\%, 32\%, and 57\%) for a combined Main Belt and NEO sample of 270 objects based on NEOWISE data \citep{mainzer11}. Both of these results suggest that low albedo P types are a majority of X complex objects. If this is the case for our sample, proper de-biasing would further increase the relative fraction of X types at small sizes, again suggesting that discovery bias can not fully account for the trends shown here (Figure \ref{fig:taxH}).

It has been suggested that tidal encounters with the terrestrial planets can induce seismic shaking and subsequent refreshing of NEO surfaces \citep{nesvorny05,binzel10,devogele19}. The small end of the NEO population in the MANOS samples is biased towards objects with low Earth MOID \citep{devogele19}, and thus are more likely to experience close encounters with the Earth. However, the importance of these tidal effects relative to thermal fatigue remains unclear \citep{graves19,sergeyev23}. Furthermore, these effects may ultimately be more important for explaining the S:Q ratio and not as relevant to the observed frequency of S+Q types. Fortunately, a real world test of surface refreshing during a close planetary encounter is coming in 2029 with the flyby of asteroid 99942 Apophis at a distance of about 32,000 km ($\sim5$ Earth radii) above the Earth's surface. This will serve as an important test of the predicted geophysical response of asteroids to a strong tidal impulse. Models of such encounters suggest that the movement of material may be limited to small regions on the surface of Apophis \citep{scheeres05,kim23,ballouz24}, but details of possible outcomes require verification. In-situ spacecraft observations of this event, for example by NASA's OSIRIS-APEX \citep{dellagiustina25} and ESA's RAMSES \citep{michel25} missions, will provide critical data on the role that planetary encounters play in modifying asteroid physical properties. Regardless of the information gleaned from the Apophis flyby, it seems unlikely that subtle surface changes induced by Apophis-like planetary encounters are a major contributor to the overall distribution of S+Q types in the NEO population. 

Tidal encounters closer than the 2029 Apophis flyby, namely inside of $\sim3$ Earth radii, may have significant consequences including catastrophic disruption \citep{richardson98}. Evidence for such disruption events include crater chains on the Moon \citep{bottke97}, doublet craters on the terrestrial planets and the Moon \citep{melosh91}, and the distorted morphologies of some NEOs imaged with delay-Doppler radar \citep{ostro02}. Recent NEO population models \citep[e.g.][]{nesvorny24a} have suggested that a detected excess of small ($25 < H < 28$) NEOs on Earth-crossing orbits ($1 < a < 1.6$ au, $e<0.4$) may be attributed to tidal disruption of larger ($D\gtrsim50$ m) objects during planetary encounters. These models suggest that up to 30\% of small NEOs on these orbits are fragments of tidal disruption. If this is true, it is not obvious what the implications would be for the observed taxonomic distribution (Figure \ref{fig:taxH}). On the one hand, the canonical expectation of higher density, lower porosity, and higher cohesive strength of S complex objects relative to C types \citep{carry12,chang15,carbognani17} suggests that S types are less sensitive to tidal disruption and thus contribute less to the observed excess of small NEOs. This would be consistent with the apparent decrease in S+Q types with increasing $H$. On the other hand, recent results suggest opposite strength and porosity trends for small ($D<200$ m) S type asteroids in the Main Belt \citep{vavra26}. Furthermore, we see no indication of an increase in C type asteroids at small sizes, which would be expected if they were preferentially disrupting under tidal encounters. Additionally, no clear evidence is seen for a decrease in abundance of the V and A spectral types which should have similar density to S complex objects and thus should be similarly resistant to tidal disruption. As such, our preferred interpretation is that tidal disruptions do not adequately explain the  ensemble of taxonomic trends presented here, though additional work is clearly needed to better understand the behavior of different compositional types during tidal encounters as well as the structural (density, porosity, cohesion) properties of small NEOs.

Given these arguments against source region, thermal modification, discovery bias, and tidal resurfacing as mechanisms for producing the S+Q trend with $H$, we are left with size-dependent differences in surface properties as the most plausible explanation. These properties may include regolith grain size, degree of impact shock, and/or composition. It is well established that smaller asteroids have coarser regolith \citep[e.g.][]{gundlach13,maclennan22}. The spectral implications of coarser regolith are less clear. Different meteorites, even within the same type, can display either shallower or deeper absorption bands with increasing grain size \citep[e.g.][]{reddy15,bowen23,maclennan24,ridenhour24}, and such changes can have implications for taxonomic assignment. Thus, it is not unreasonable to suggest that the observed S+Q trend might be tied to changing regolith grain size. In this scenario, underlying compositions across the NEO population would be invariant with object size, but a change in spectral properties would be attributable to differences in regolith particle size. This is a possibility that can not be fully ruled out with the data presented here. However, it seems unlikely that increasing grain size and an associated diversity of spectral changes could systematically alter the observed abundance of S+Q type objects by a factor of two across the $H$ range considered here. Furthermore, if the observed taxonomic trend were solely due to grain size effects, we would expect smaller bodies with coarser regolith to have lower albedo due to increased shadowing and greater particle surface area available to scatter incident light. In fact, the opposite trend is observed: small NEOs tend to have higher albedo \citep{nesvorny24b}. As such, changing grain size is not our preferred explanation. 

Meteorites contain evidence of high pressure impacts in the form of various shock melted features \citep{rubin85}. Generally shock melt results in dark lithologies that make up a small volume fraction of most meteorite types, including ordinary chondrites \citep{stoeffler91}. Shock darkening of ordinary chondrites has been demonstrated to reduce the depth of absorption bands and can cause a shift in taxonomic assignments from S types to C or X types \citep{reddy14,kohout20,battle22,sanchez25}. While shock darkening of small NEOs could explain both the decrease in S+Q class objects and the increase in the X class, it is unclear how to reconcile these large scale trends with the small volume fraction of shocked material seen in meteorites. In fact, significant portions of the NEO population would have to consist entirely of shock darkened objects to produce such a dramatic shift in taxonomic assignments. Fully shock darkened samples are exceedingly rare amongst meteorites, suggesting this is unlikely to be a common state in the NEO population. An observed increase in albedo with increasing $H$ \citep{nesvorny24b} also indicates that impact shock is not a dominant process for small NEOs. Furthermore, the effects of shock darkening are similar for HED meteorites. If shock darkening were a size-dependent process affecting large numbers of NEOs, this would suggest that the fraction of V type asteroids should also decrease with $H$. However, no appreciable change in the abundance of V types is seen from $H=14$ to $H=28$ (Figure \ref{fig:taxH}). So while shock processes can help to explain the spectral properties of individual objects \cite[e.g][]{reddy14,battle22}, they are unlikely to be affecting significant portions of the NEO population.

Compositional trends are a final possibility for the observed size-dependent change in taxonomic type and is our preferred explanation. In this scenario the observed change in spectral types is a consequence of the underlying composition of the population changing with object size. Previous studies have suggested size-dependent compositional changes in the NEO population. For example, \citet{popescu18} suggested that an apparent increase in A type asteroids at sizes $<300$ meters was indicative of a collisionally comminuted population of mantle fragments from differentiated parent bodies. The results of our color sample do not support this result, however it is difficult to unambiguously identify olivine-rich A type objects with visible colors alone. As another example, \citet{vernazza08} showed that the typical composition of km-scale S type NEOs was analogous to LL ordinary chondrites. This was unexpected because over 60\% of large NEOs have this composition whereas only 8\% of meteorites are LL chondrites. Recent work \citep{broz24,broz24b,marsset24} seems to have resolved this issue. In short, these works show that the meteorite flux on Earth may be dominated by a few young asteroid families in the Main Belt. Newly formed collisional families have steep size frequency distributions with an excess of meter-scale bodies (i.e. the direct parent bodies of meteorites) and thus can dominate the meteorite flux for periods of up to a few tens of millions of years. If the youngest asteroid families in the Main Belt are dominating the small NEO and meteorite flux, then this would imply S class NEOs with LL-like compositions should become increasingly less common at small sizes \citep{sanchez24}. Furthermore, the overall distribution of taxonomic types at high $H$ should more closely resemble the youngest and largest Main Belt families. The decrease in S+Q type objects is thus likely an indication of the increasing contribution at small sizes of objects from primitive families like those of Veritas, Polana, and Eos \citep{broz24b}. This implies that the preference for high $H$ objects coming from the $\nu_6$ resonance (Figure \ref{fig:source}) is less important for the observed taxonomic distribution than the families across the Main Belt that are young enough to still have significant numbers of small ($<<100$ meter) members. Finally, our results suggest that the ``top of the atmosphere" population of common impactors, i.e. the direct parent bodies of meteorites, is dominated by non-S+Q types, and that atmospheric filtering thus imposes a strong bias against non-ordinary chondrite meteorite types \citep{broz24b,shober25}.

This relationship between spectral properties and NEO size can be better understood with additional data. Specifically collecting spectral data for objects with $H>27$ will improve statistics at the smallest sizes where our analysis only includes 13 S type and 8 X type objects (Figure \ref{fig:fractions}). Observations that can constrain the composition of S+Q class objects in this size regime could test for the predicted decreasing frequency of LL-like objects amongst meter and decameter scale NEOs \citep{vernazza08,sanchez24}. Filling out these statistics can help to refine the size(s) at which compositional transitions occur. Aside from improving understanding of the connections between meteorites and their direct parent bodies in space, these data would have important implications for planetary defense. Impact risk assessment models for individual objects or probabilistic treatment of the population as a whole must account for size dependent properties like composition and density \citep[e.g.][]{mathias17,dotson24}. Though challenging due to intrinsic faintness, future observations of the most likely Earth impactors, namely decameter-scale NEOs on Earth-like orbits, would be highly valuable additions to the state of knowledge for multiple research areas.

\begin{acknowledgments}

We are grateful to Steven R. Chesley and an anonymous reviewer for their careful reading and thoughtful suggestions that helped to significantly improve this manuscript. This work was enabled by NASA grants 80NSSC21K1328 and NNX17AH06G and NNX14AN82G awarded in support of the Mission Accessible Near-Earth Object Survey (MANOS). We are grateful to all of telescope operators, engineers, instrument scientists, and support astronomers who helped make these observations possible. We thank the generous allocations of telescope time recommended by the NOAO/NOIRLab and Lowell Observatory Telescope Allocation Committees. In addition to the immensely valuable resources provided by Bill Gray on his Project Pluto website (\url{projectpluto.com}), we would also like to thank Bill for assisting with ephemeris calculations for our attempted recovery of 2022 BX5 in archival images. A number of undergraduate and graduate students were involved with the observations presented in this work. We thank Gonzalo Muniz, Peter Fatka, Mary Hinkle, and Dan Avner for their involvement. NM is grateful to KM for fruitful discussions about subtrahends and minuends.

Lowell Observatory sits at the base of the San Francisco Peaks, mountains sacred to tribes throughout the region. We honor their past, present, and future generations, who have lived here for millennia and will forever call this place home. This work in this paper made use of the Lowell Discovery Telescope (LDT) at Lowell Observatory. Lowell is a private, non-profit institution dedicated to astrophysical research and public appreciation of astronomy and operates the LDT in partnership with Boston University, the University of Maryland, the University of Toledo, Northern Arizona University and Yale University. The Large Monolithic Imager was built by Lowell Observatory using funds provided by the National Science Foundation (AST-1005313). The upgrade of the DeVeny optical spectrograph for installation at the LDT was funded by a generous grant from John and Ginger Giovale and by a grant from the Mt. Cuba Astronomical Foundation. This work was based in part on observations obtained at the Southern Astrophysical Research (SOAR) telescope, which is a joint project of the Ministério da Ciência, Tecnologia e Inovações do Brasil (MCTI/LNA), the US National Science Foundation’s NOIRLab, the University of North Carolina at Chapel Hill (UNC), and Michigan State University (MSU). This work is based in part on observations at NSF Kitt Peak National Observatory, NSF NOIRLab, which is managed by the Association of Universities for Research in Astronomy (AURA) under a cooperative agreement with the U.S. National Science Foundation. The authors are honored to be permitted to conduct astronomical research on I'oligam Du’ag (Kitt Peak), a mountain with particular significance to the Tohono O’odham.

\end{acknowledgments}

%

\vspace{5mm}
\facilities{LDT (LMI), LDT (DeVeny), SOAR (Goodman), Mayall (MOSAIC-1 wide-field camera, MOSAIC-3), IRTF (SpeX), Gemini:South (GMOS)}


\software{\texttt{astropy} \citep{astropy13,astropy18},  
          \texttt{Source Extractor} \citep{Bertin96},
          \texttt{SCAMP} \citep{Bertin06},
          \texttt{Photometry Pipeline} \citep{Mommert17},
          \texttt{astroquery} \citep{Ginsburg19}
          }

\appendix
\section{Lightcurve Plots} \label{appendix}
\restartappendixnumbering

In Section \ref{sec:LC} we presented 49 newly measured lightcurves. Here we show 3-panel plots for each of these lightcurves. The format of each plot is the same. The titles contain the object's primary provisional designation, the best-fit period $P$, and the best fit amplitude $A$. The top two panels in each plot show the Lomb-Scargle (LS) periodogram (calculated with the \texttt{astropy} LombScargle class) and the $\chi^2$ values associated with Fourier fits to a range of possible rotation periods. Generally the best-fit period is derived from the minimum in the $\chi^2$ plot, which often appears at 2$\times$ the maximum in the LS periodogram. More details on this procedure are provided in the main text and in Table \ref{tab:LC}. The bottom panel in each plot shows the data phase folded to the best-fit period.

\begin{figure}
    \centering
    \includegraphics[width=0.49\textwidth]{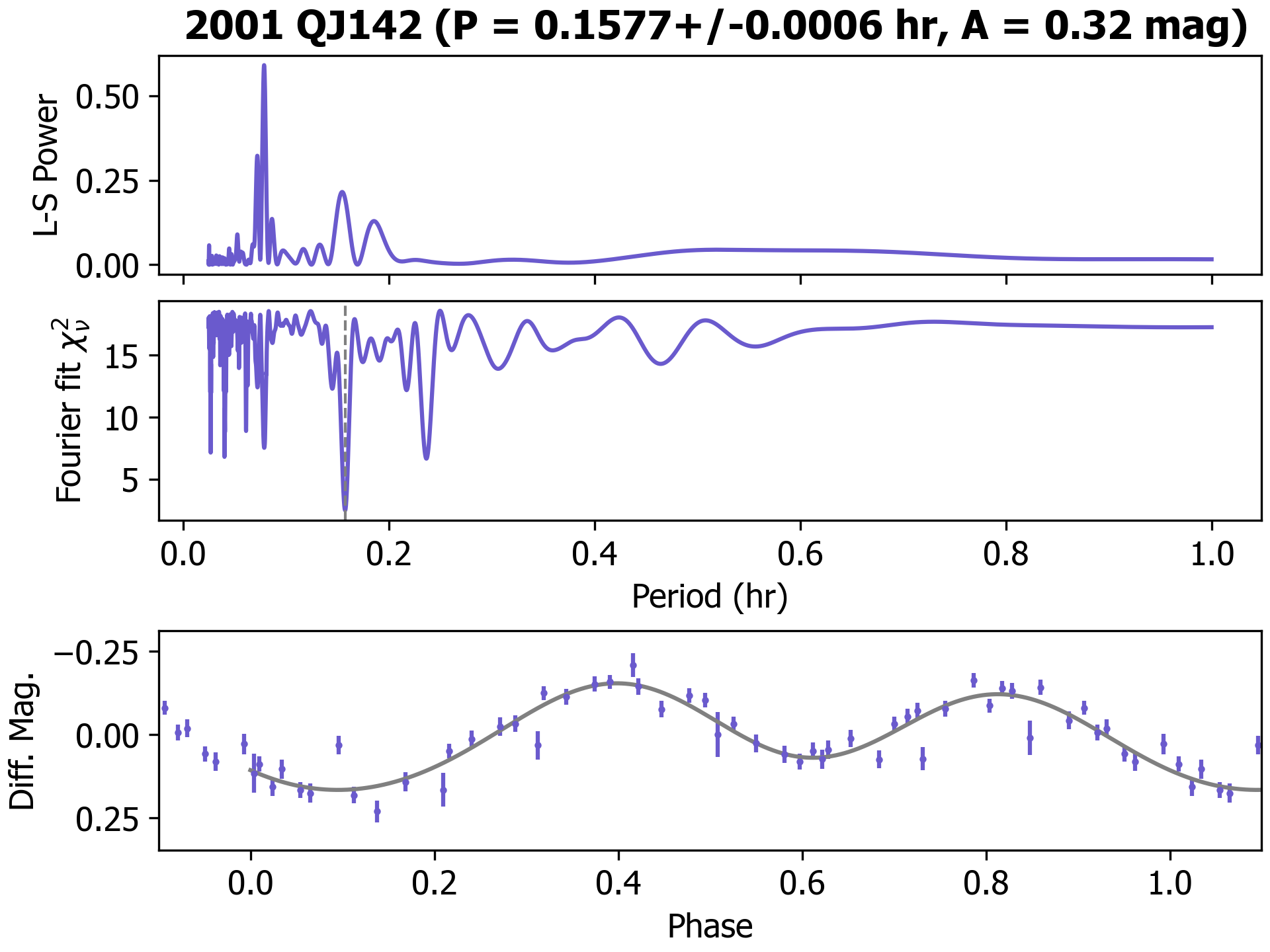}
    \includegraphics[width=0.49\textwidth]{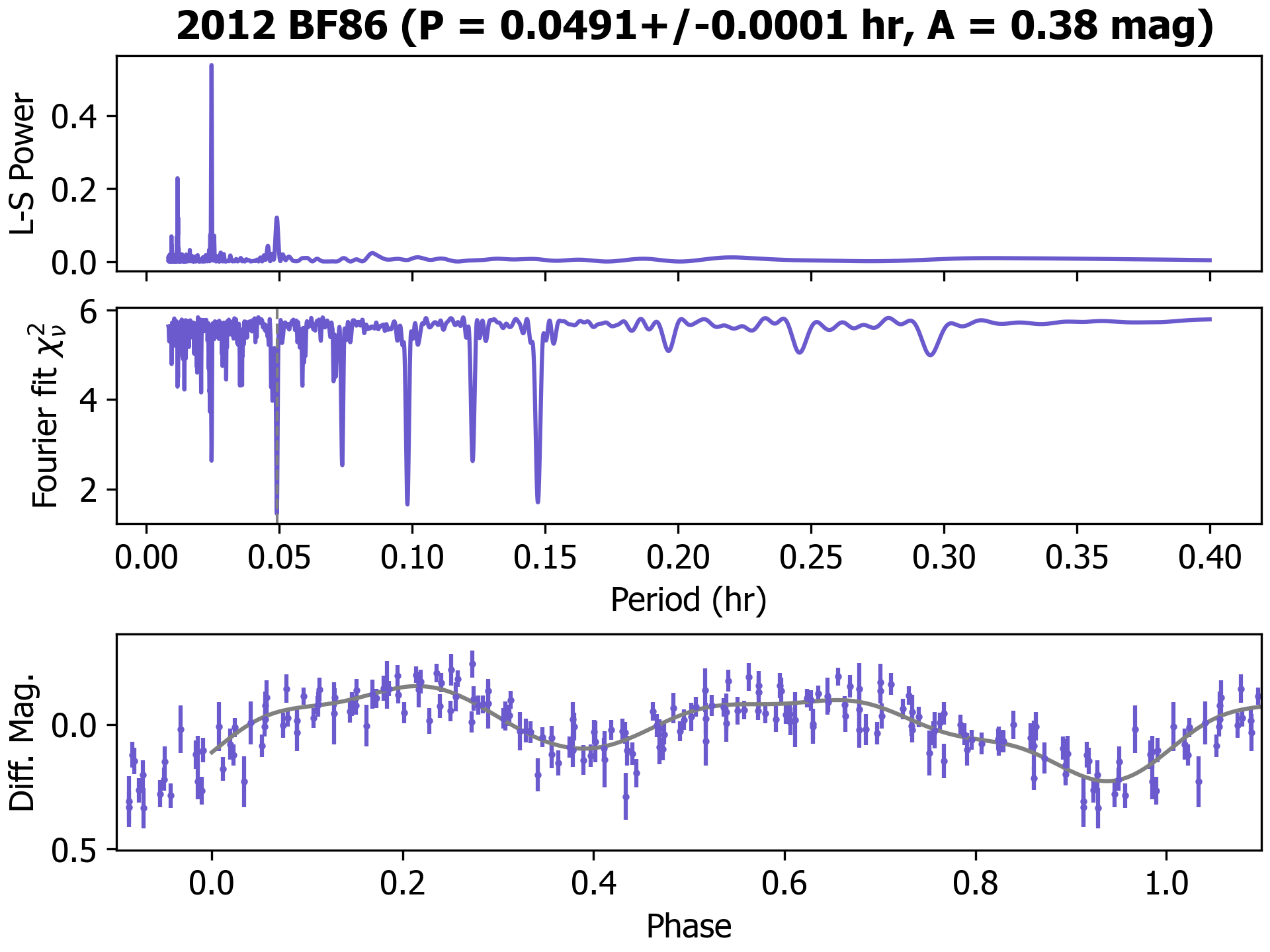}
    \includegraphics[width=0.49\textwidth]{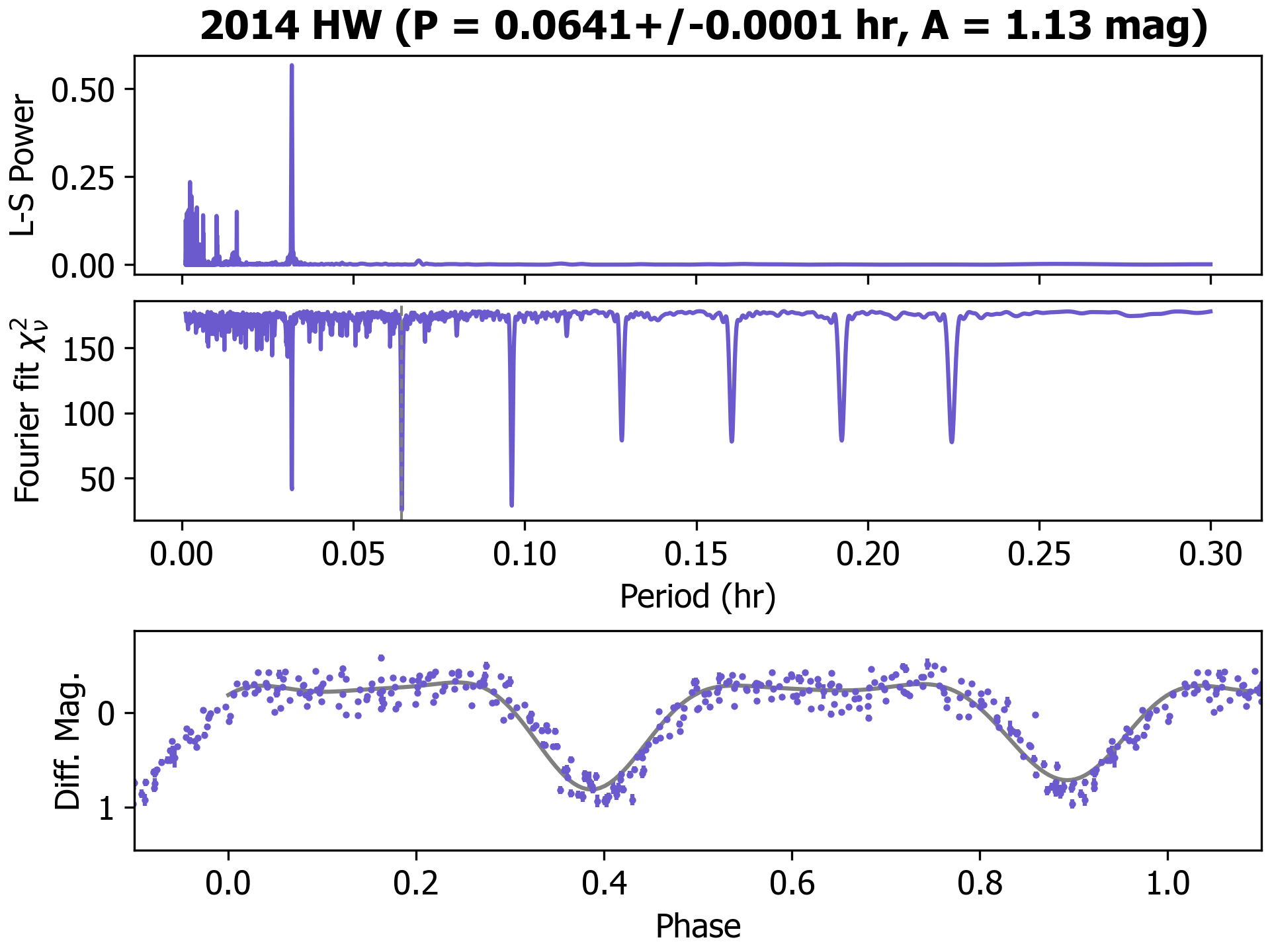}
    \includegraphics[width=0.49\textwidth]{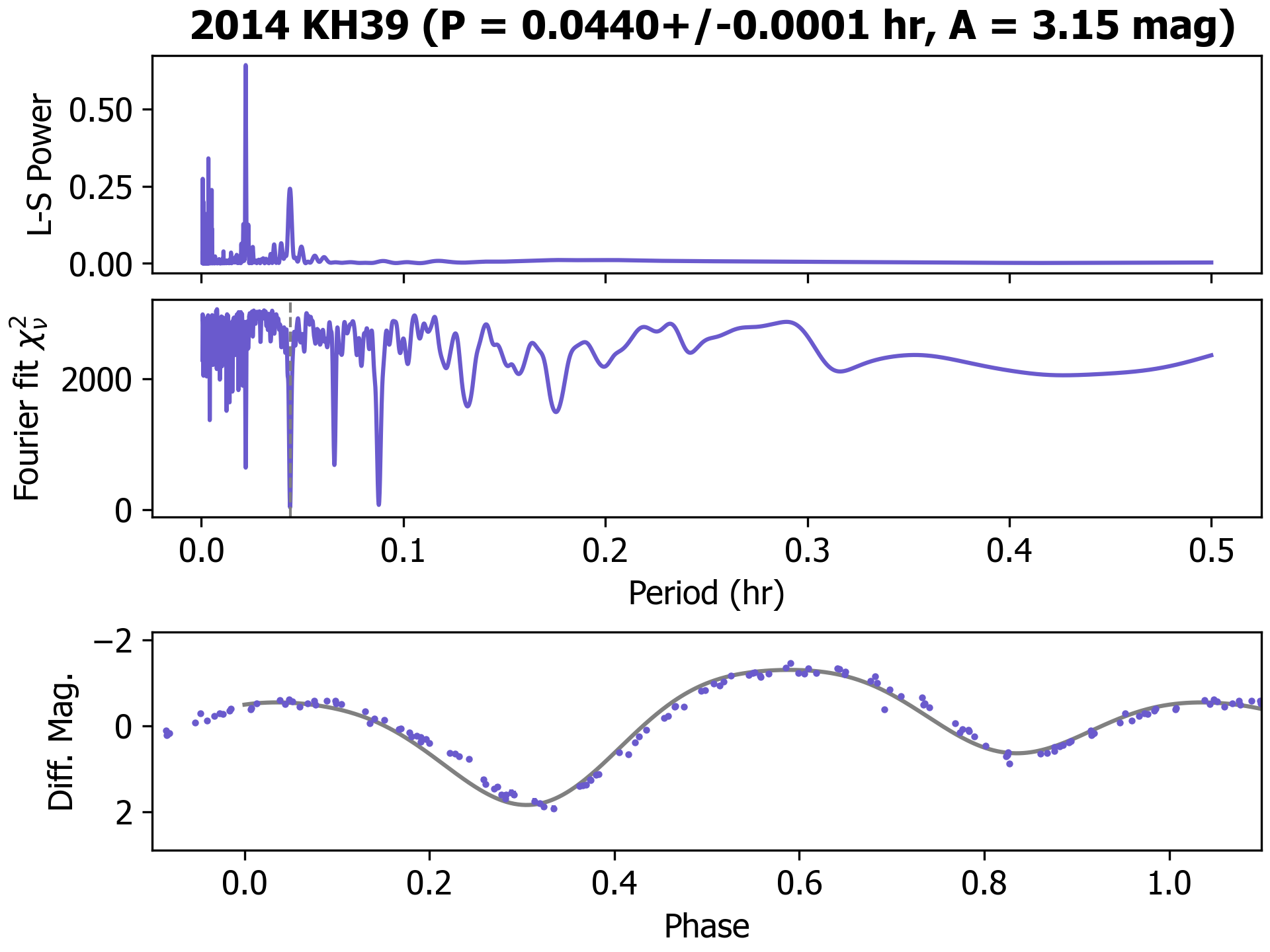}
    \includegraphics[width=0.49\textwidth]{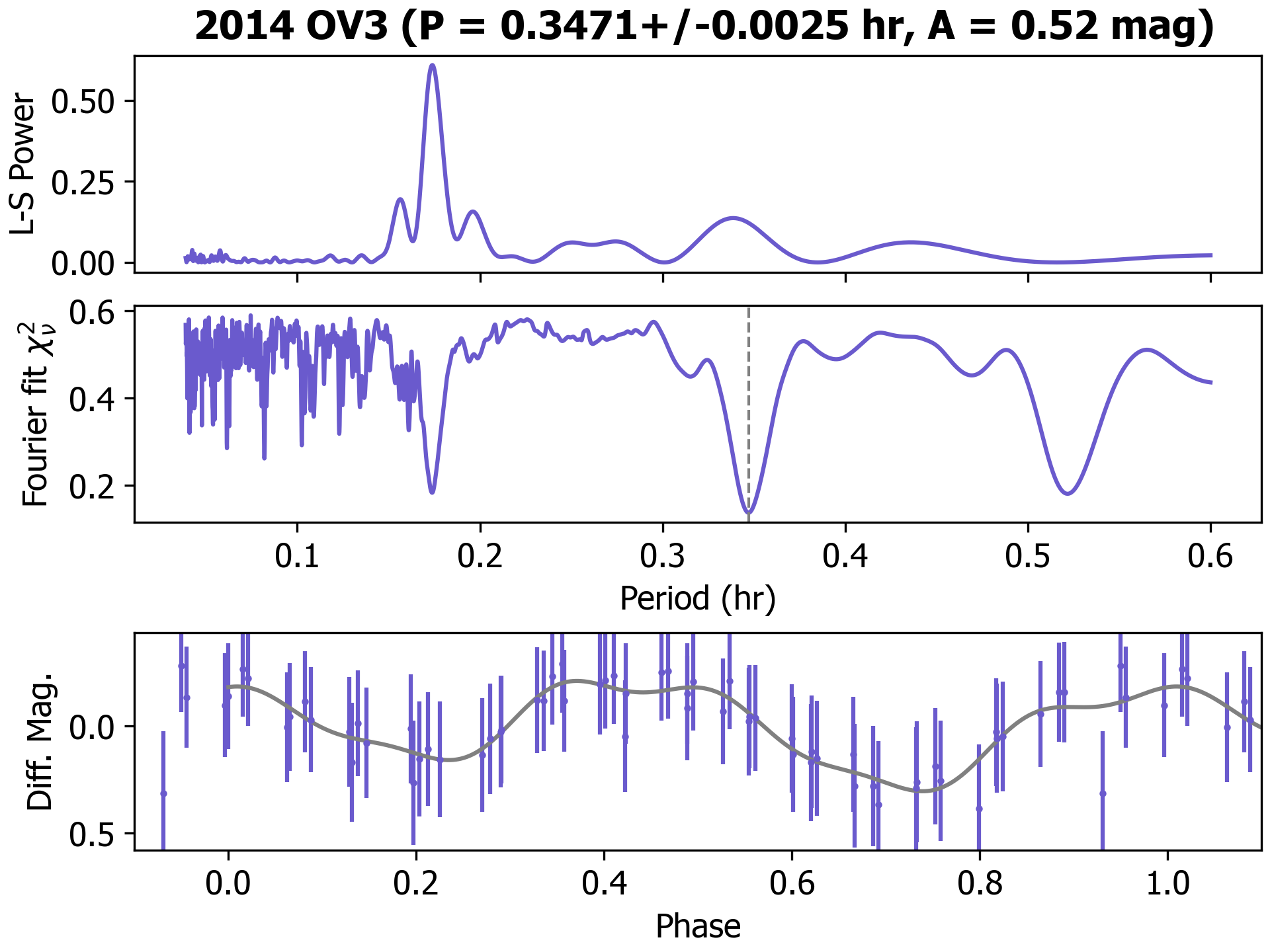}
    \includegraphics[width=0.49\textwidth]{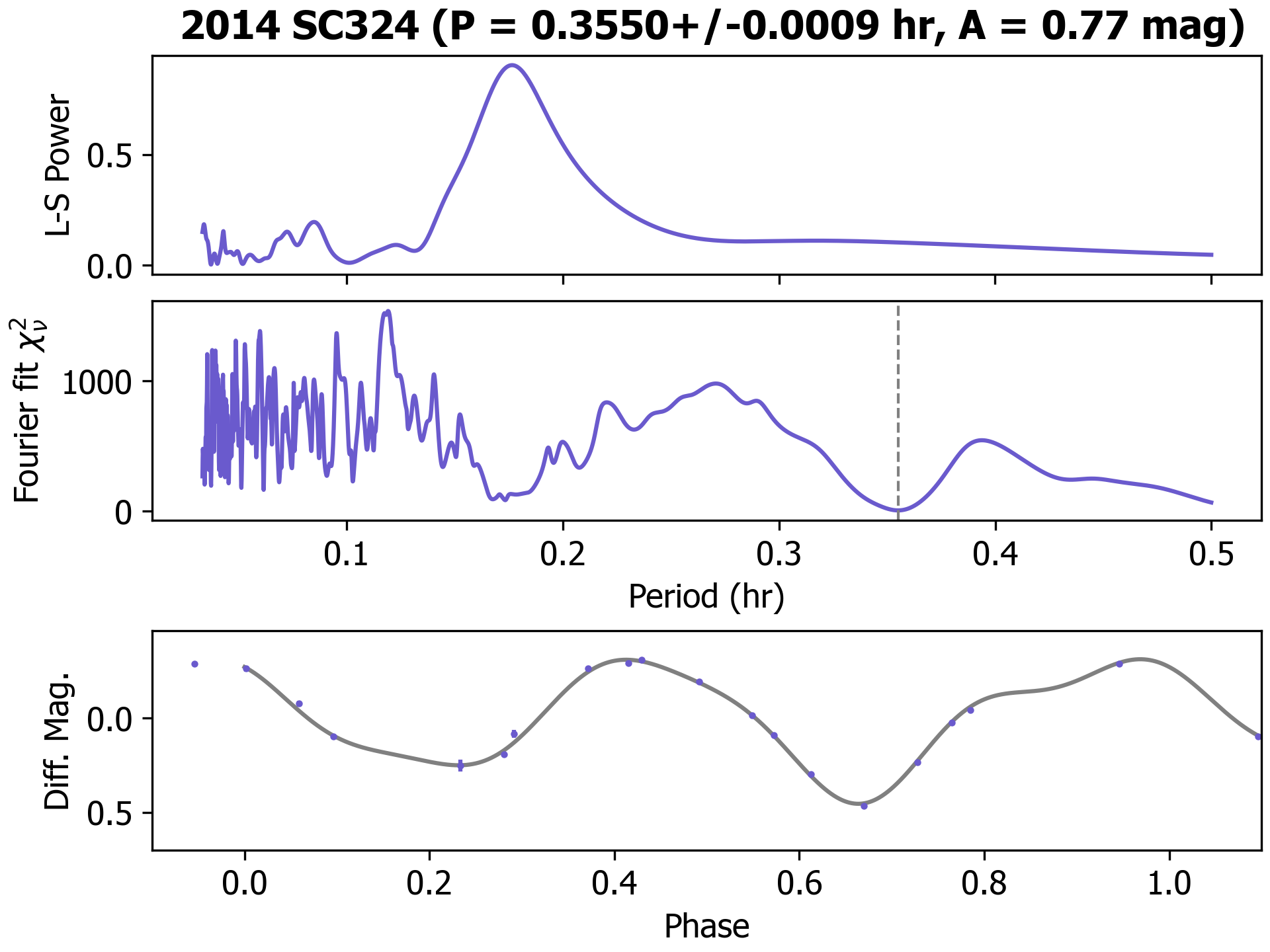}
    \caption{Three-panel plots of rotational lightcurves. The top panel is a Lomb-Scargle (LS) periodogram, the middle panel is the $\chi^2$ statistic associated with Fourier series fits to the data as a function of rotation period, and the bottom panel shows the photometric data points phase folded to the best fit period, which is indicated in the title of each subplot and by the vertical gray dashed line in the middle panel.}
    \label{fig:A1}
\end{figure}

\begin{figure}
    \centering
    \includegraphics[width=0.49\textwidth]{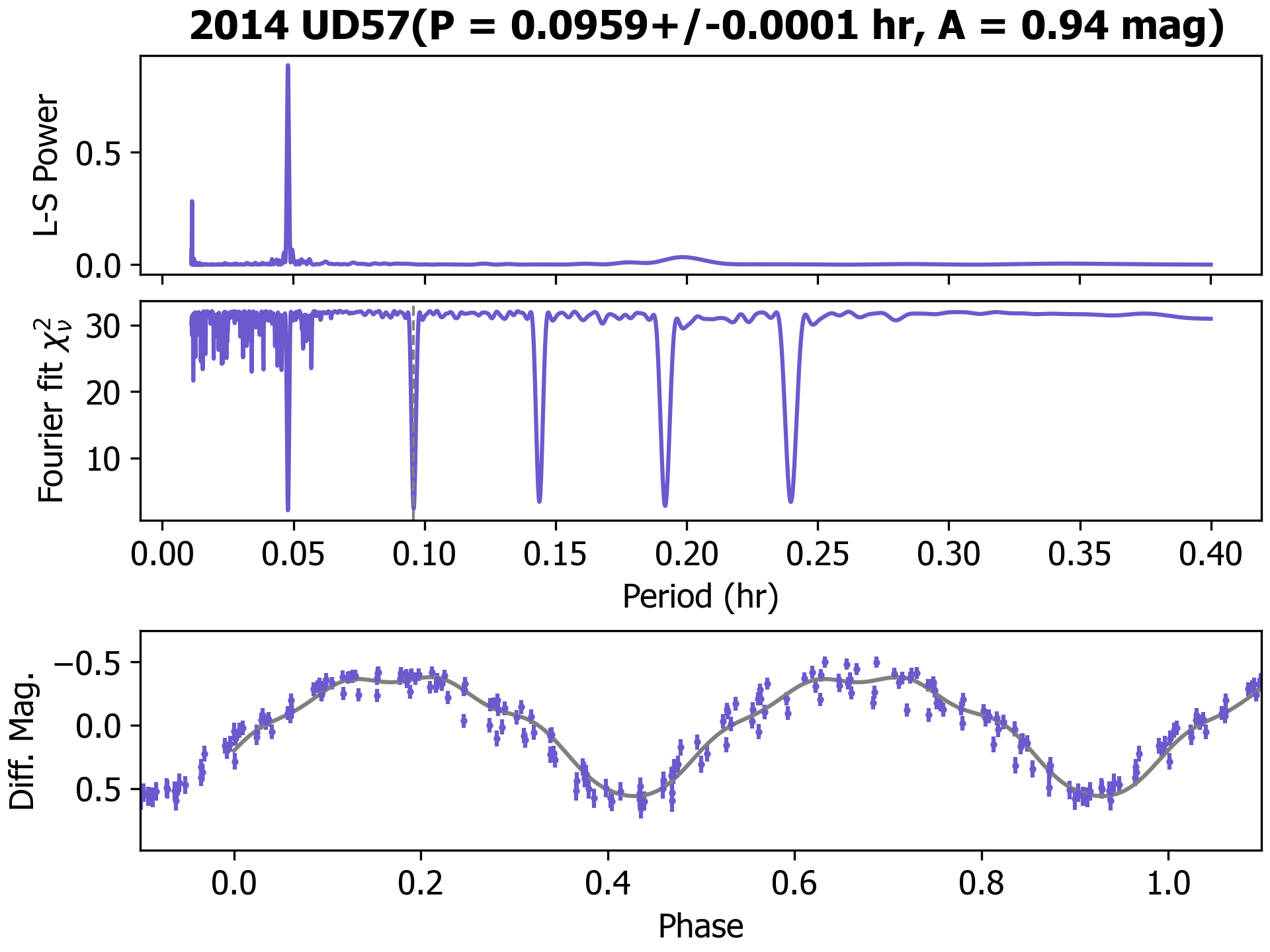}
    \includegraphics[width=0.49\textwidth]{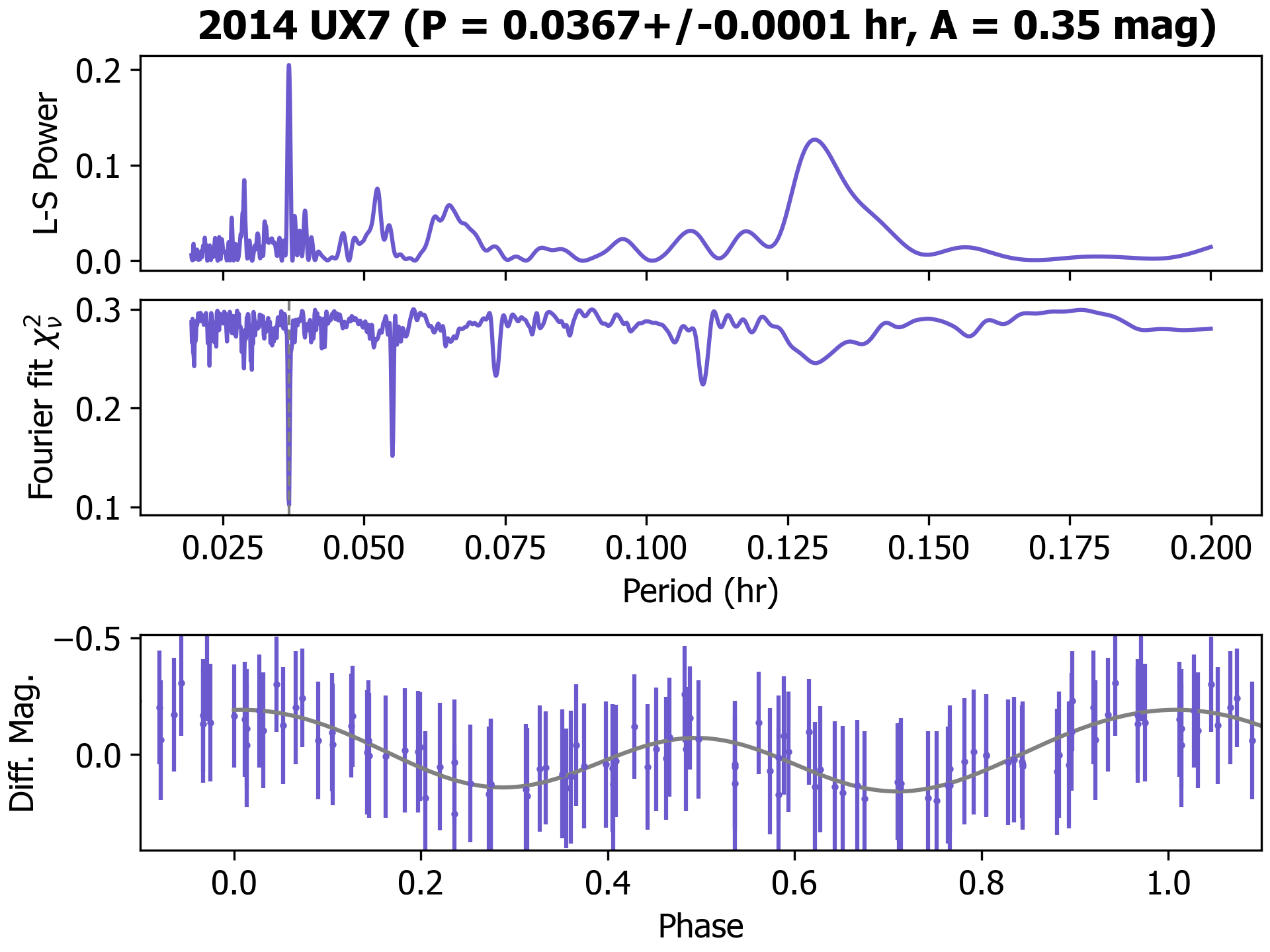}
    \includegraphics[width=0.49\textwidth]{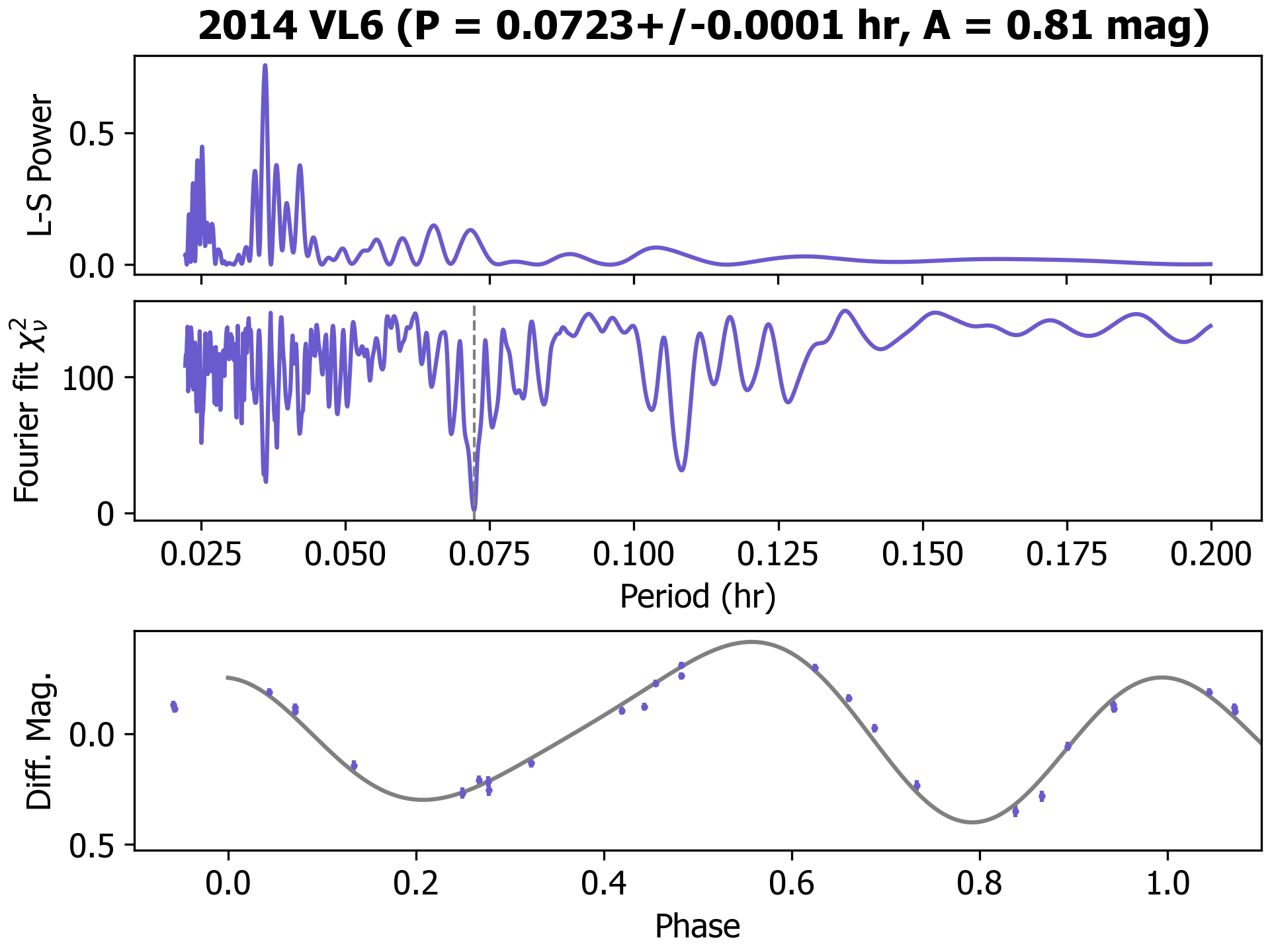}
    \includegraphics[width=0.49\textwidth]{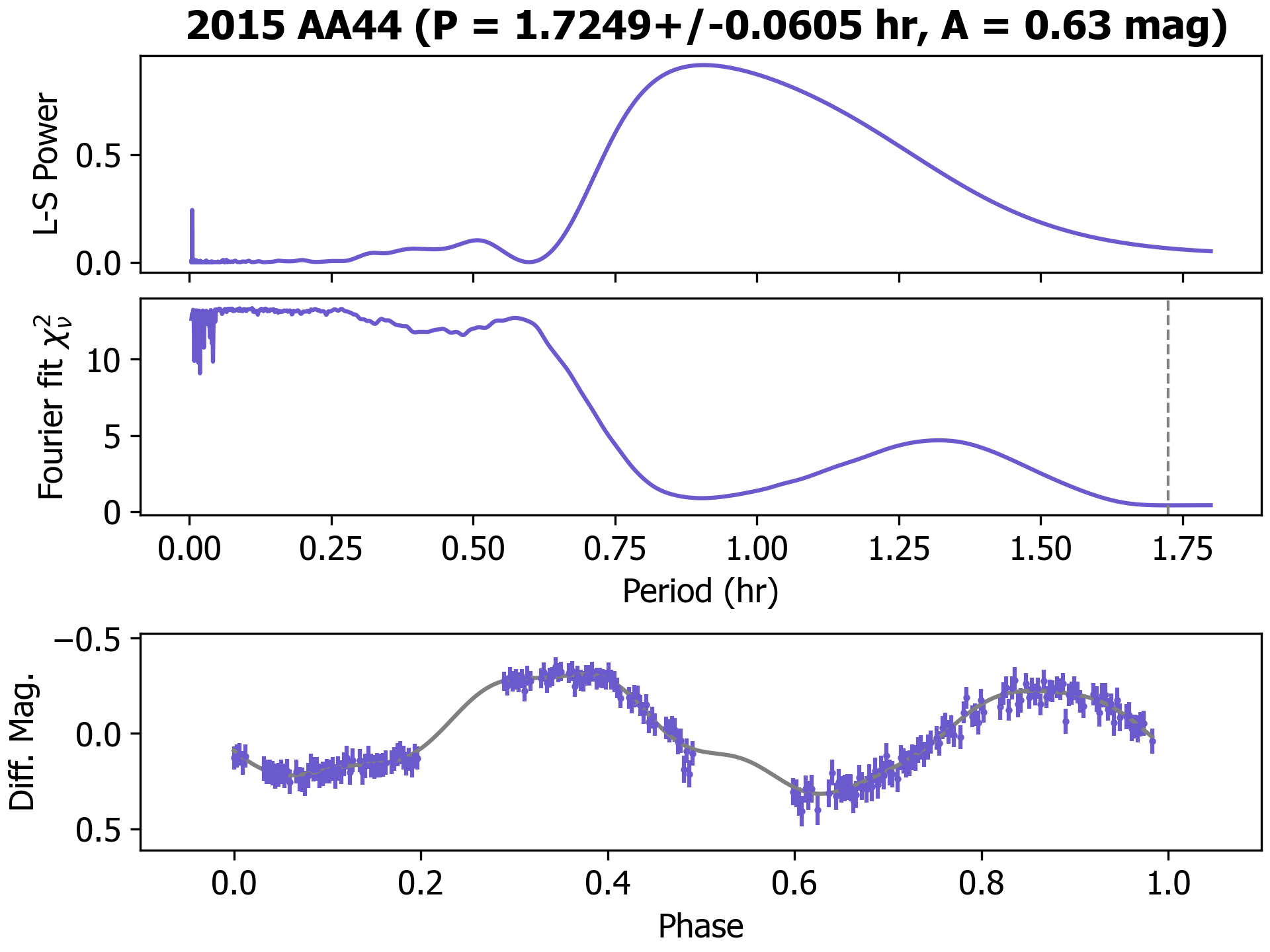}
    \includegraphics[width=0.49\textwidth]{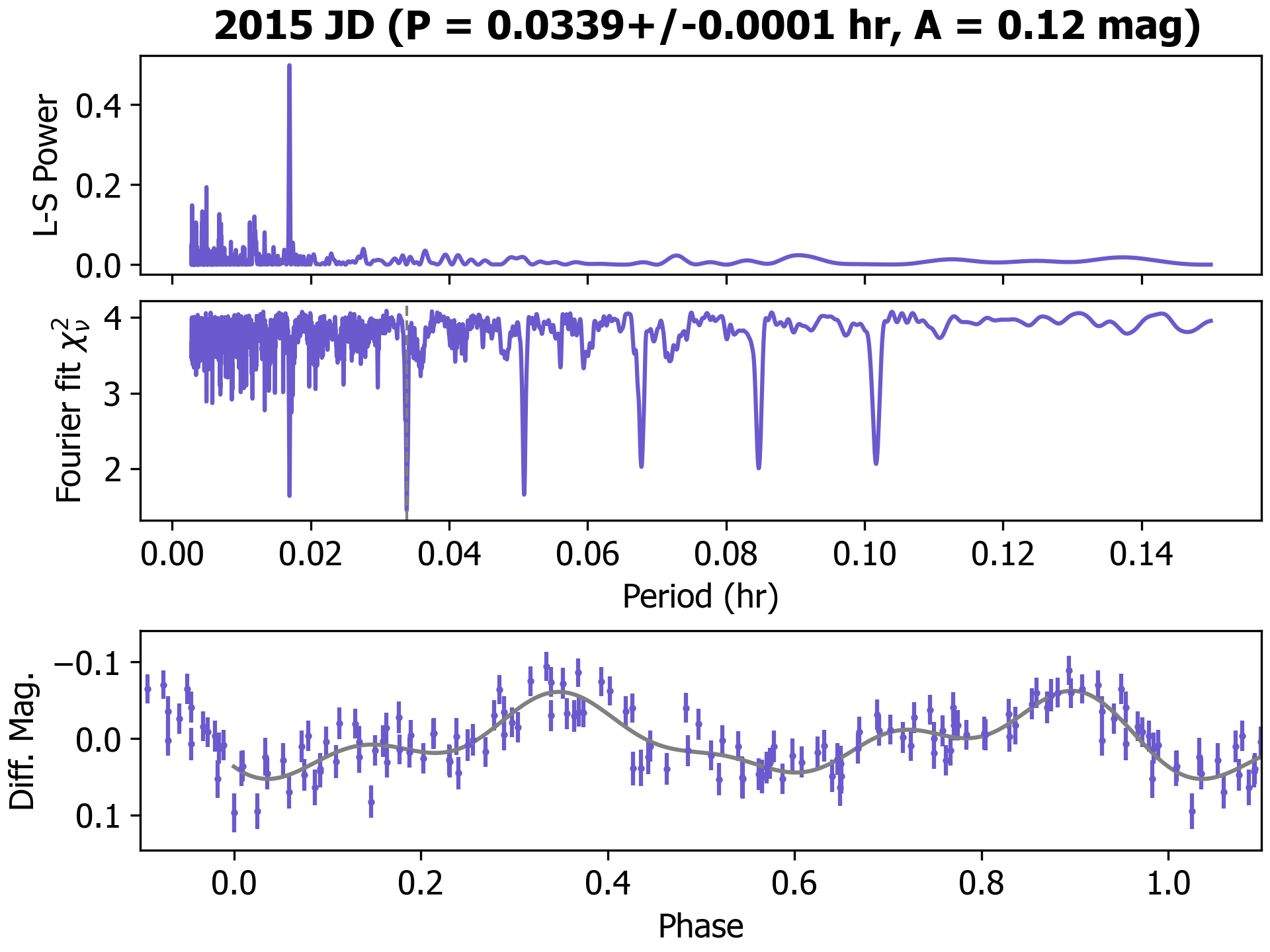}
    \includegraphics[width=0.49\textwidth]{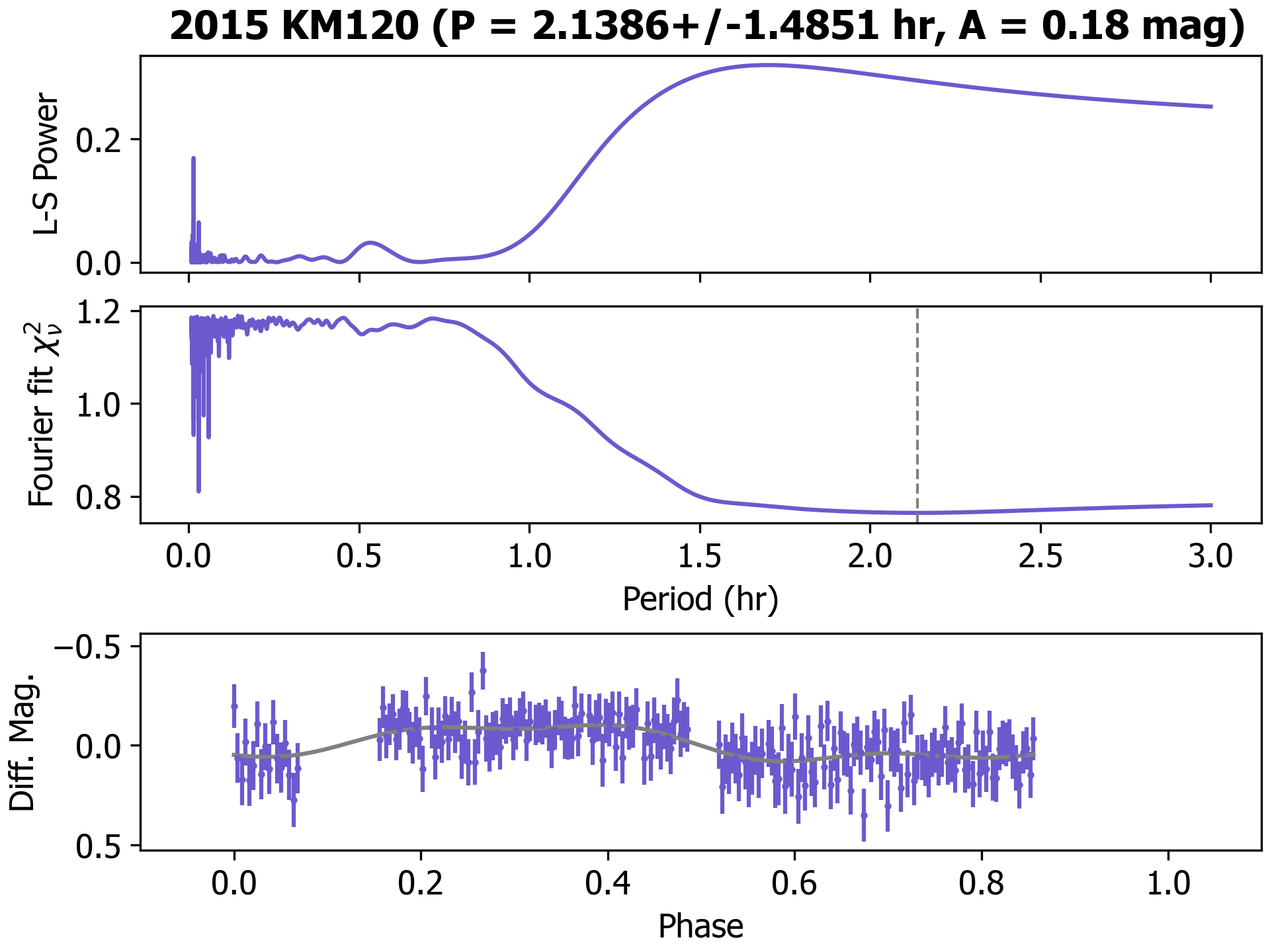}
    \caption{Continuation of Figure \ref{fig:A1}.}
\end{figure}

\begin{figure}
    \centering
    \includegraphics[width=0.49\textwidth]{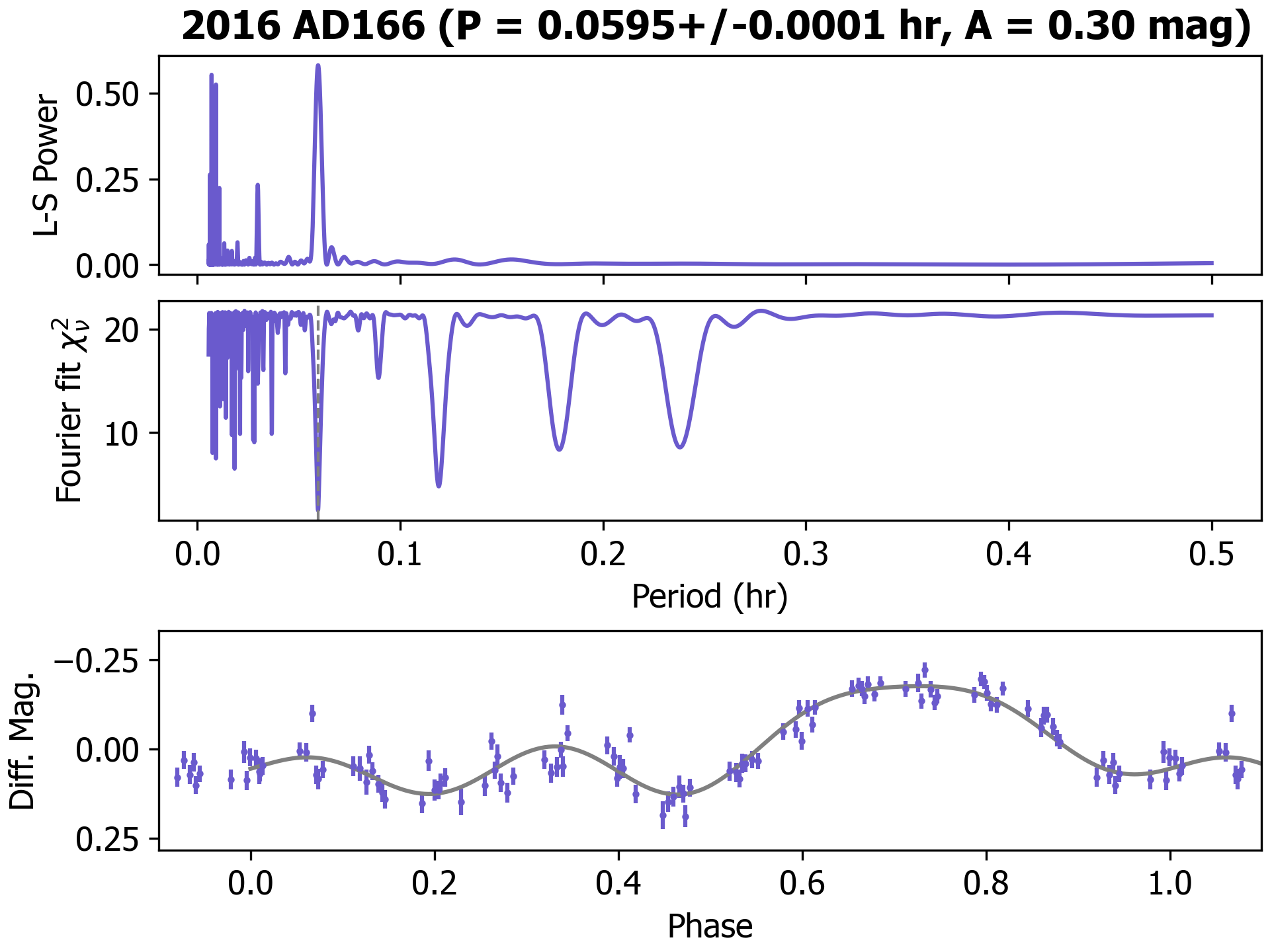}
    \includegraphics[width=0.49\textwidth]{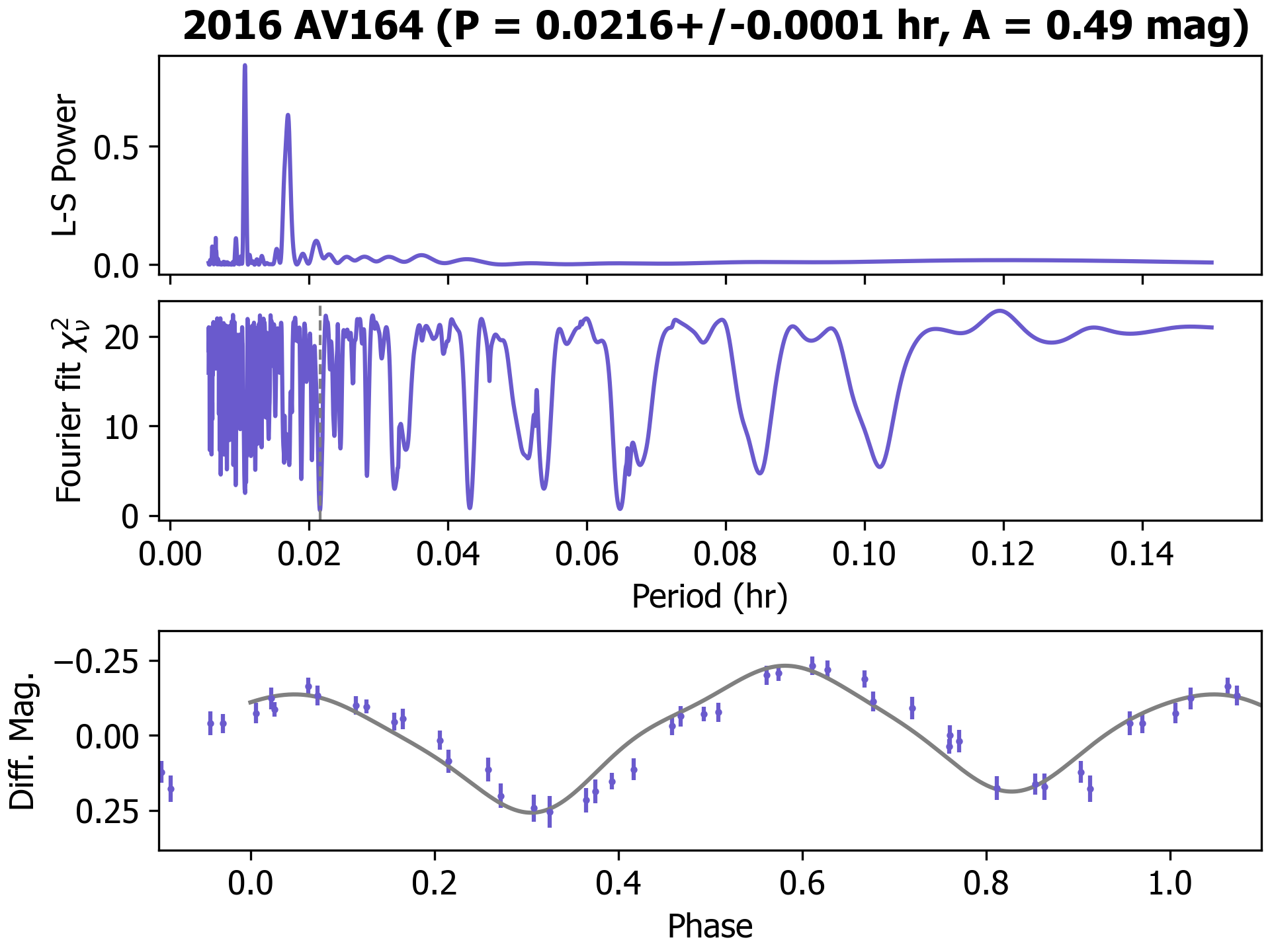}
    \includegraphics[width=0.49\textwidth]{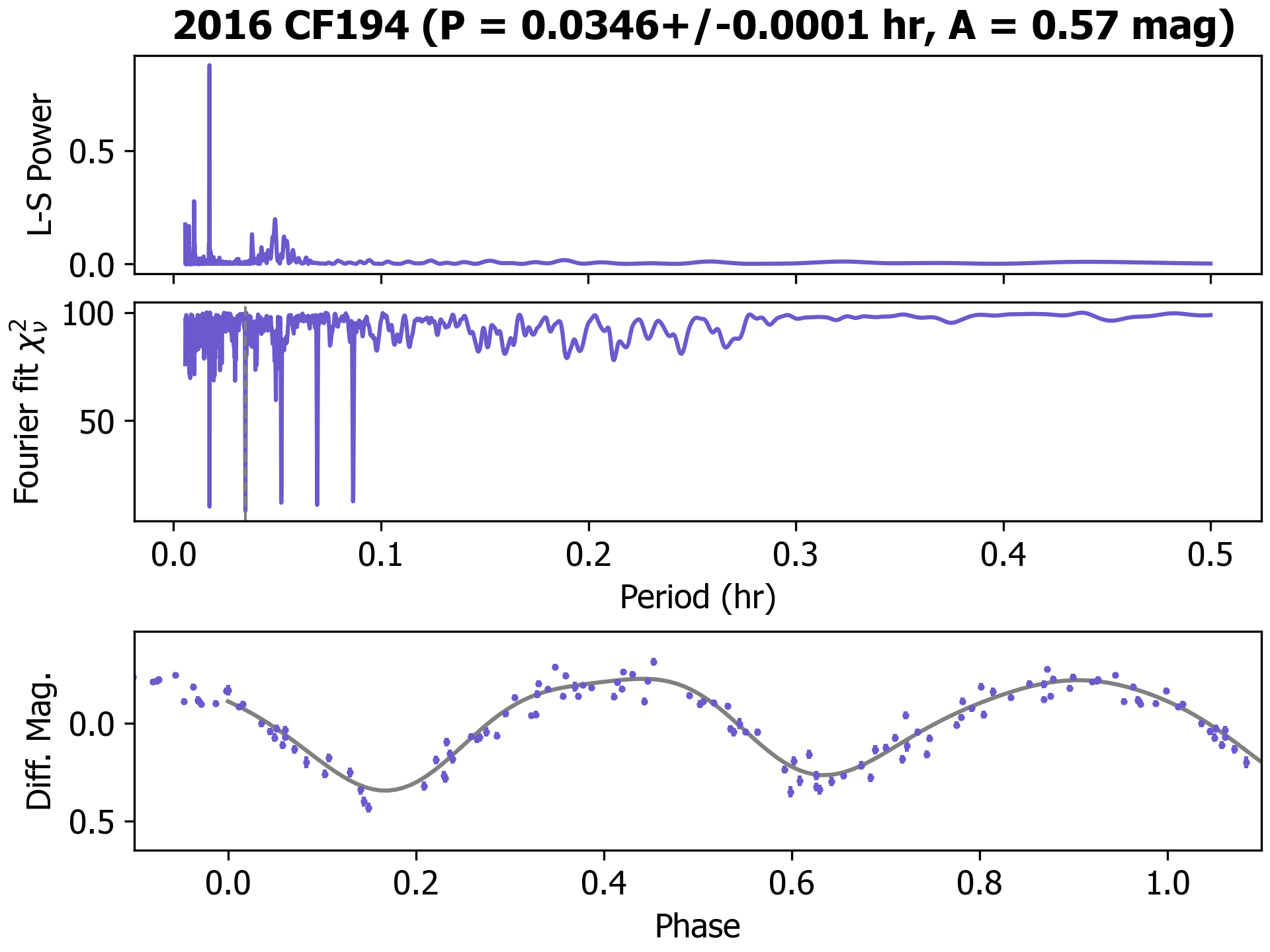}
    \includegraphics[width=0.49\textwidth]{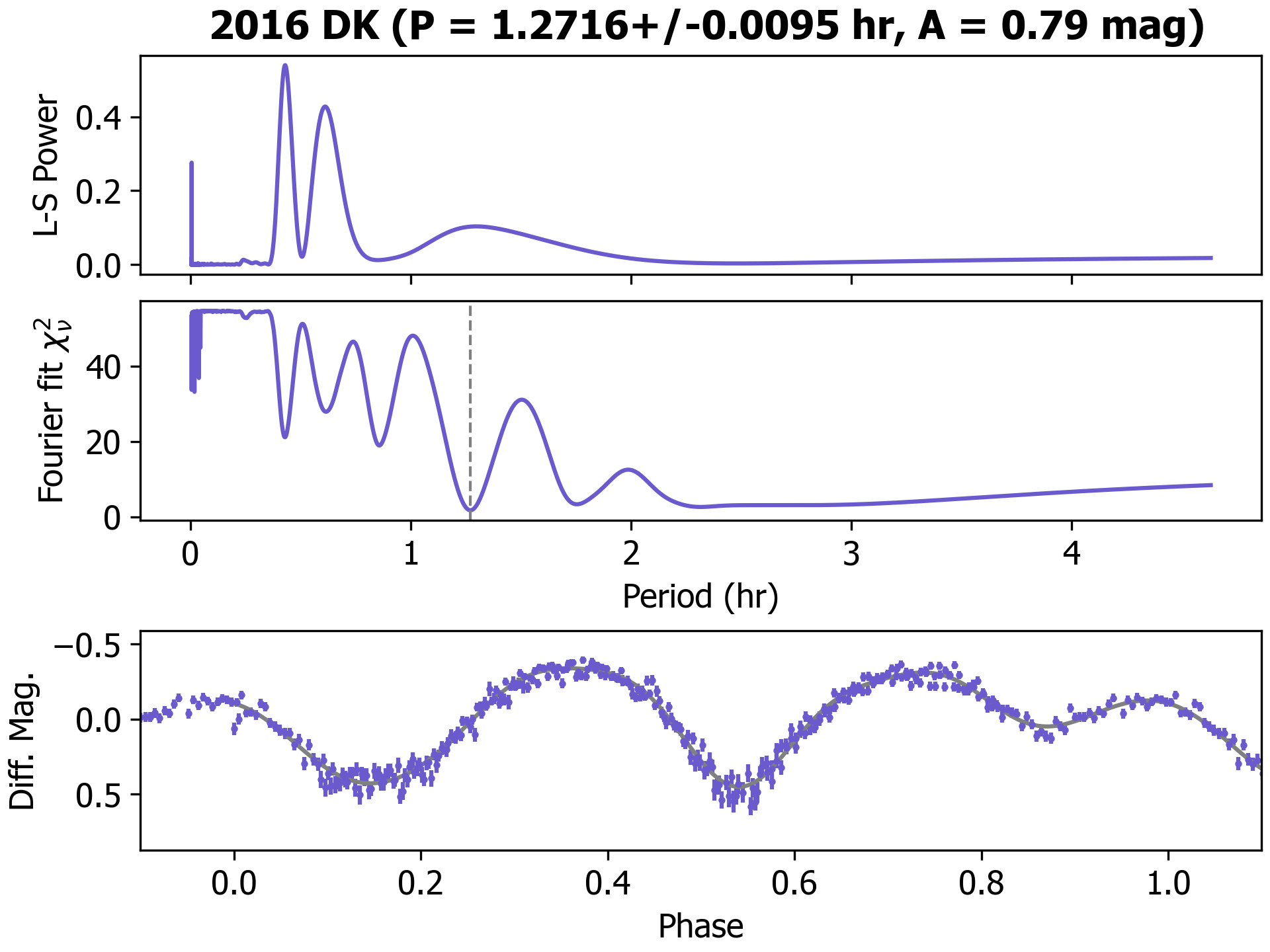}
    \includegraphics[width=0.49\textwidth]{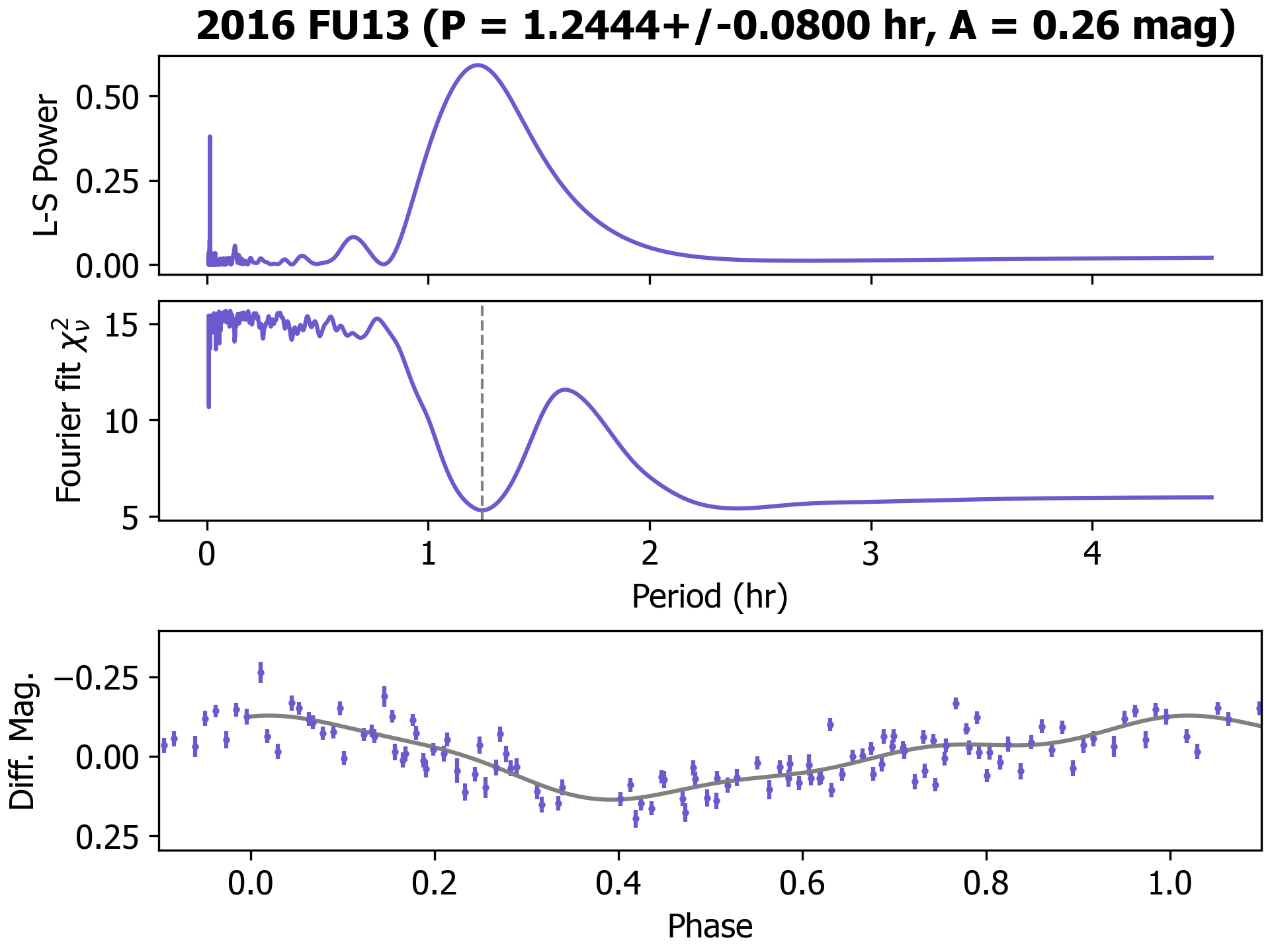}
    \includegraphics[width=0.49\textwidth]{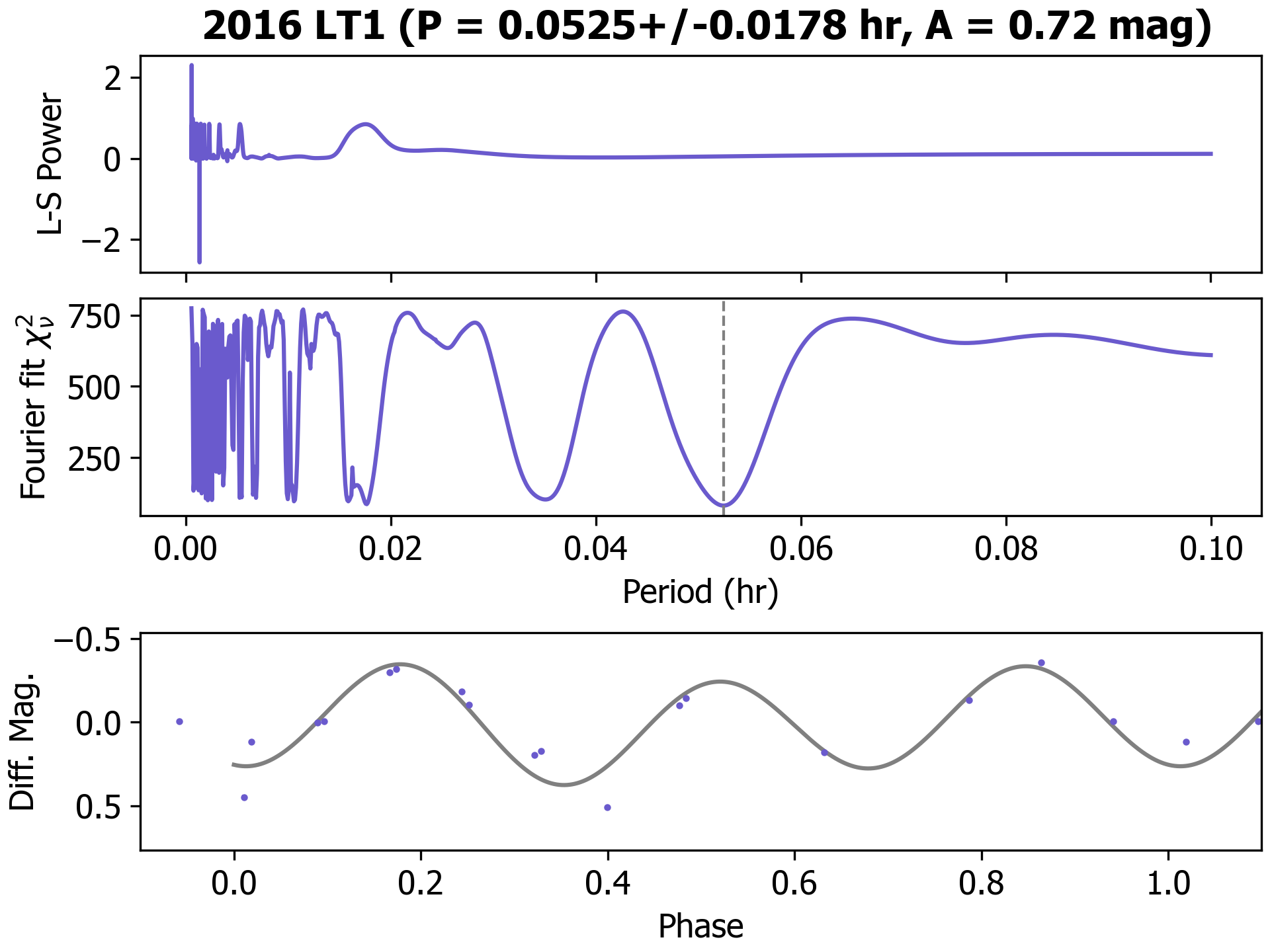}
    \caption{Continuation of Figure \ref{fig:A1}.}
\end{figure}

\begin{figure}
    \centering
    \includegraphics[width=0.49\textwidth]{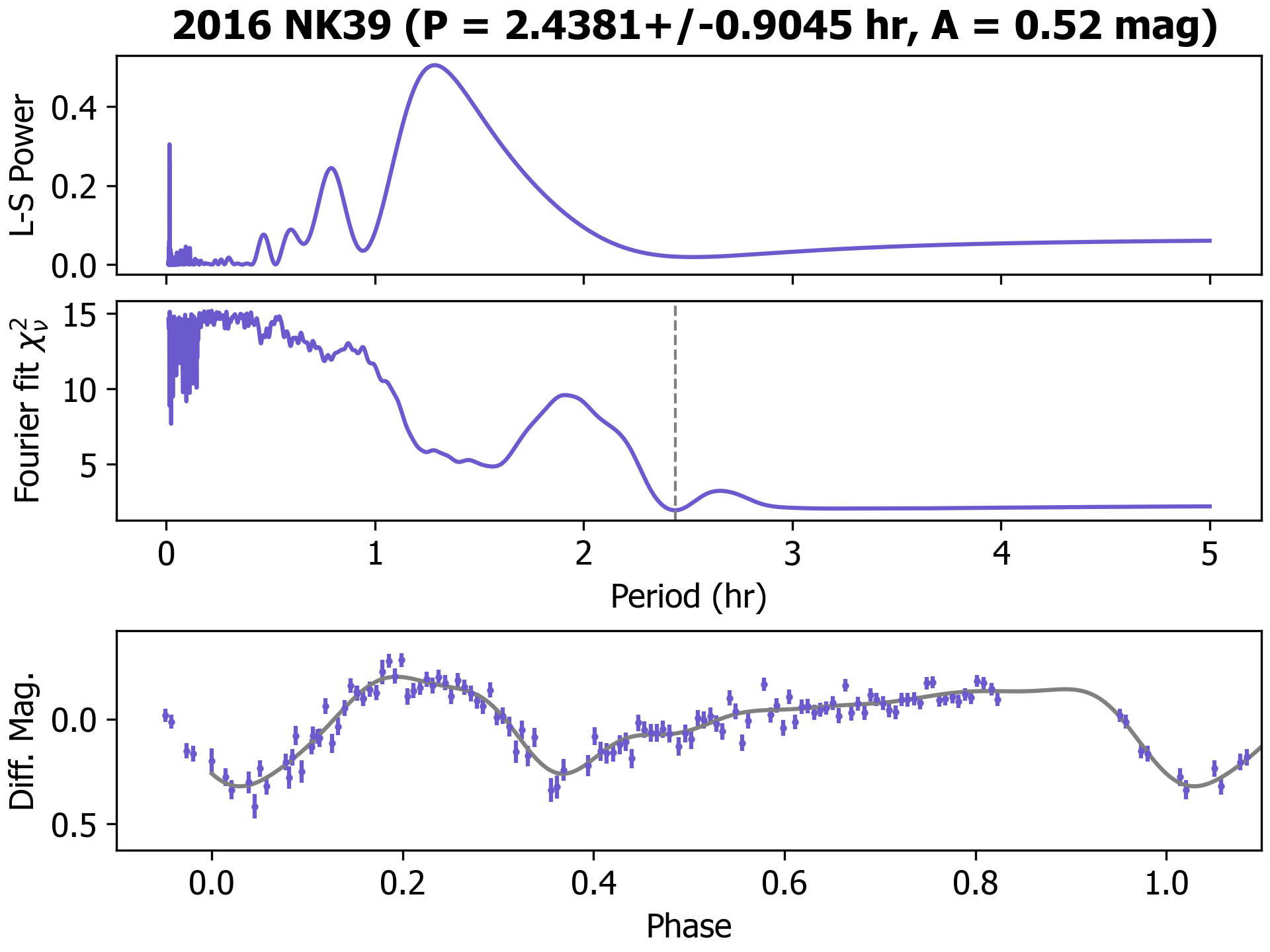}
    \includegraphics[width=0.49\textwidth]{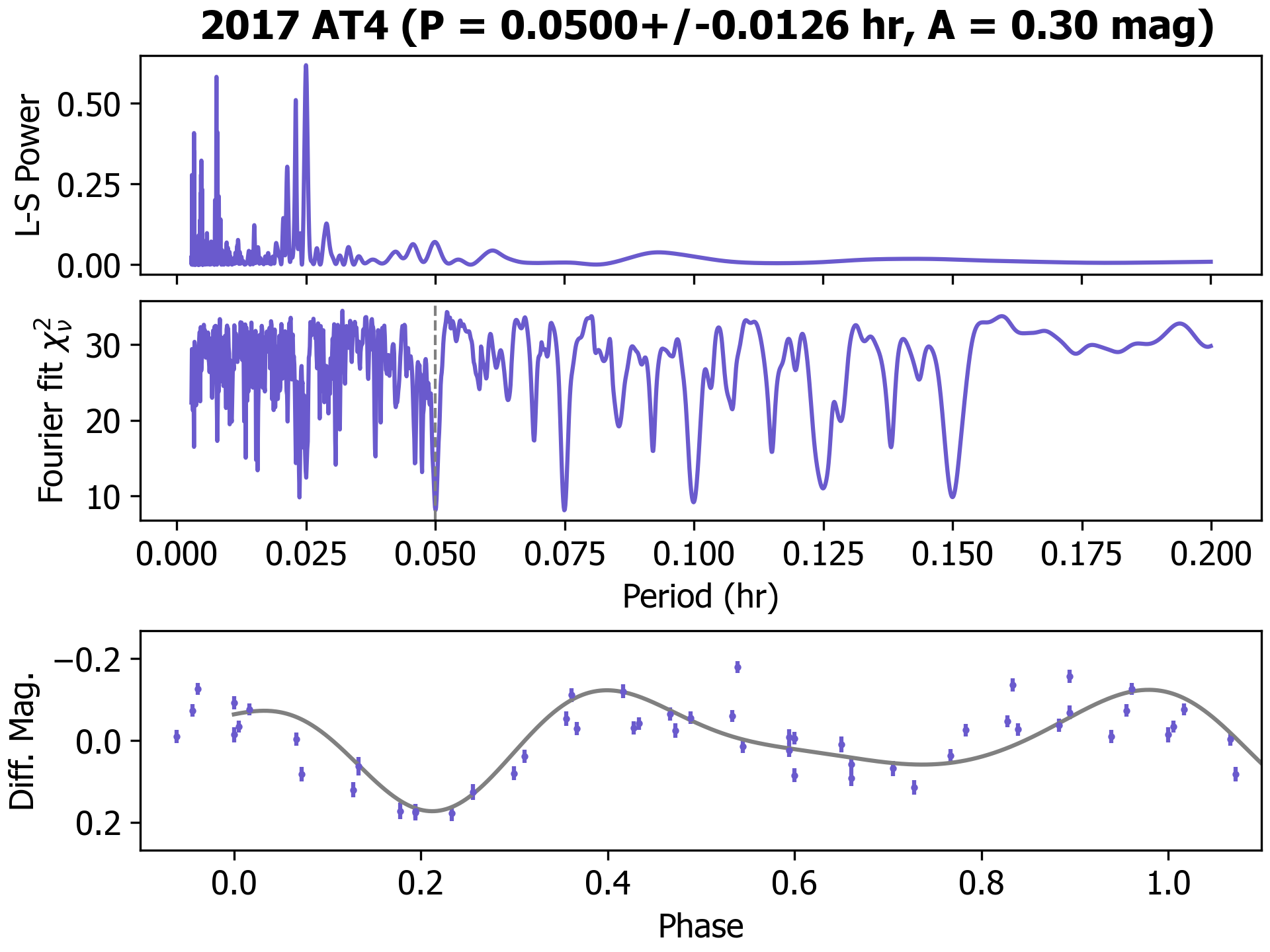}
    \includegraphics[width=0.49\textwidth]{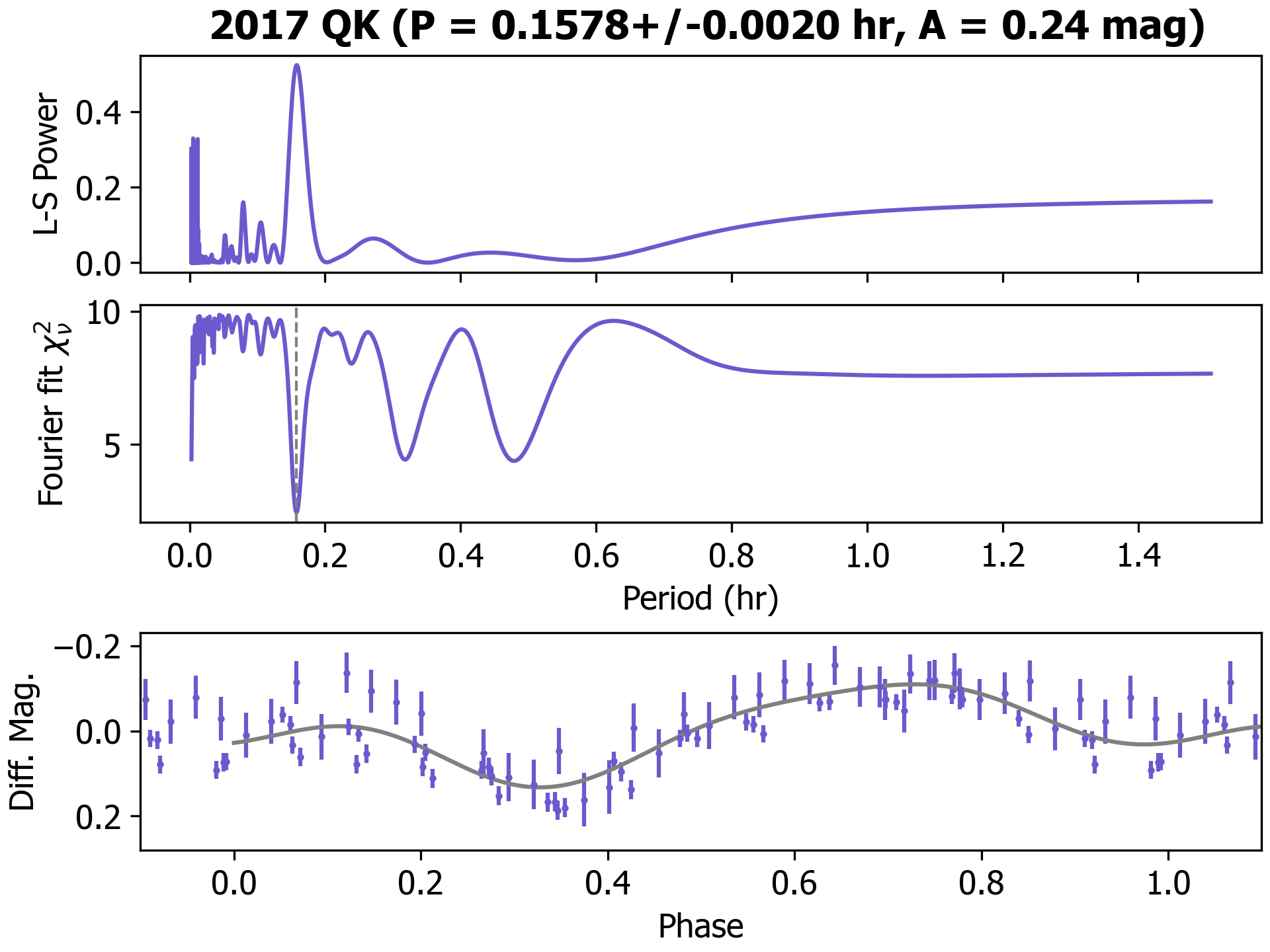}
    \includegraphics[width=0.49\textwidth]{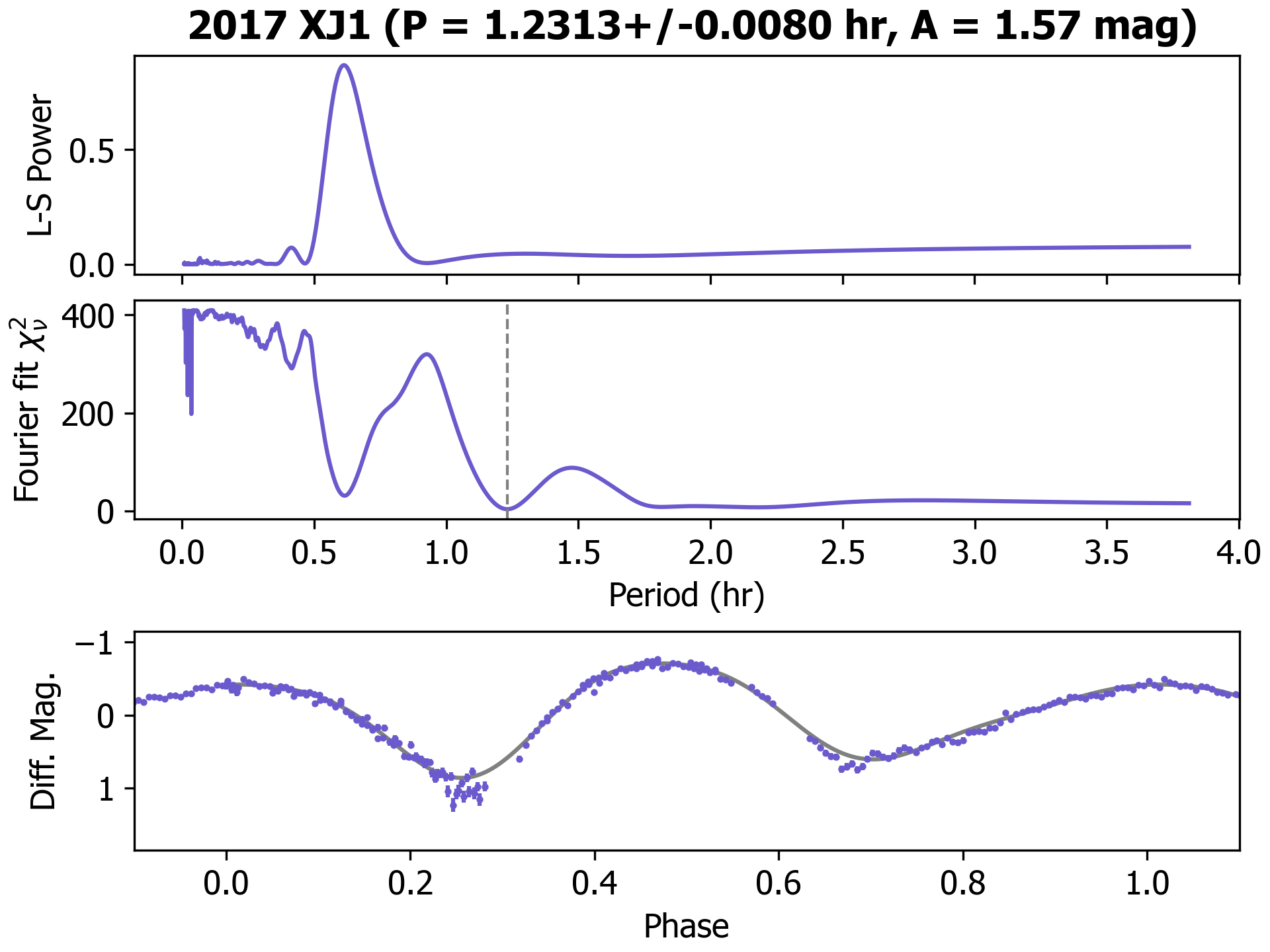}
    \includegraphics[width=0.49\textwidth]{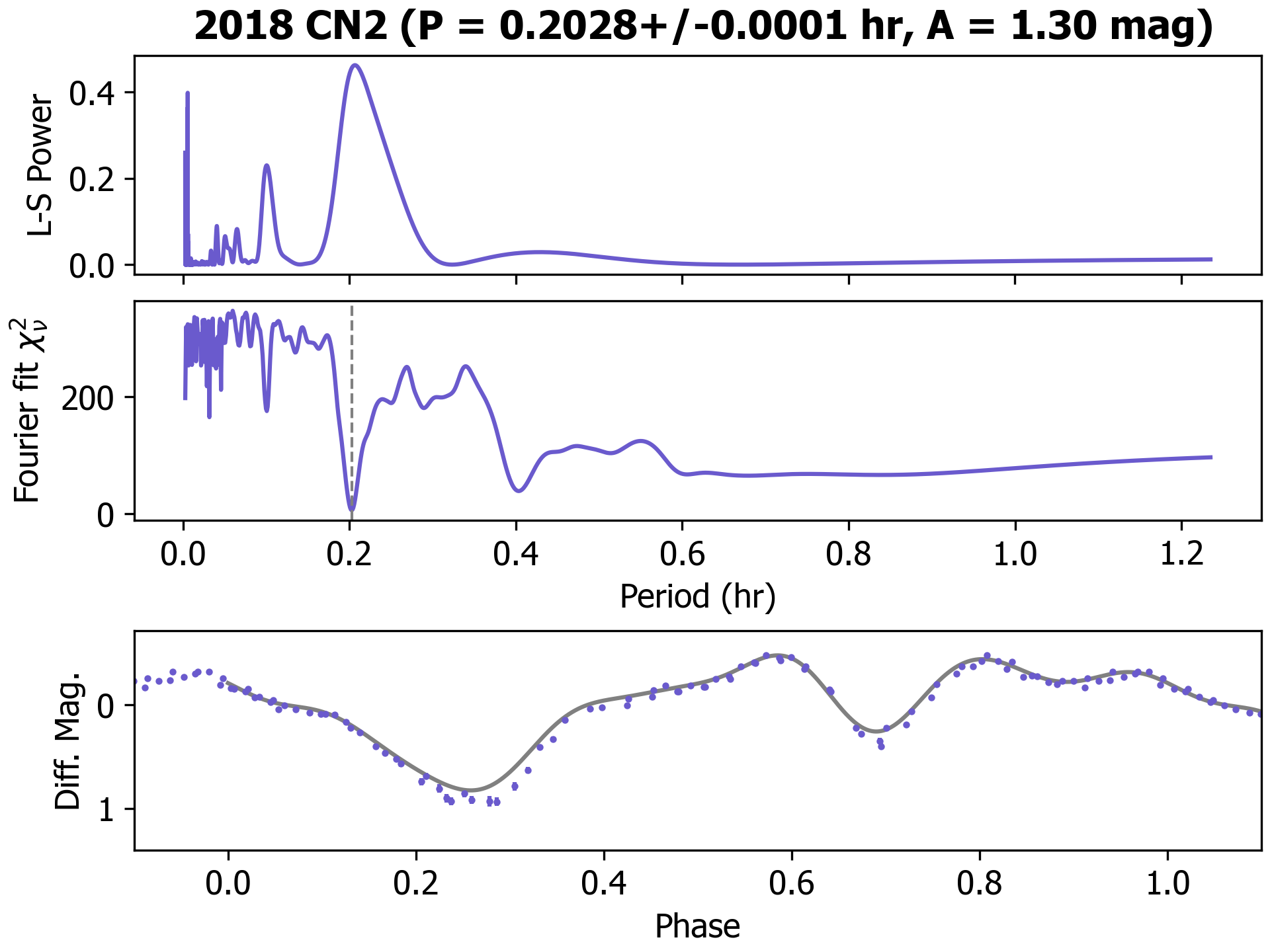}
    \includegraphics[width=0.49\textwidth]{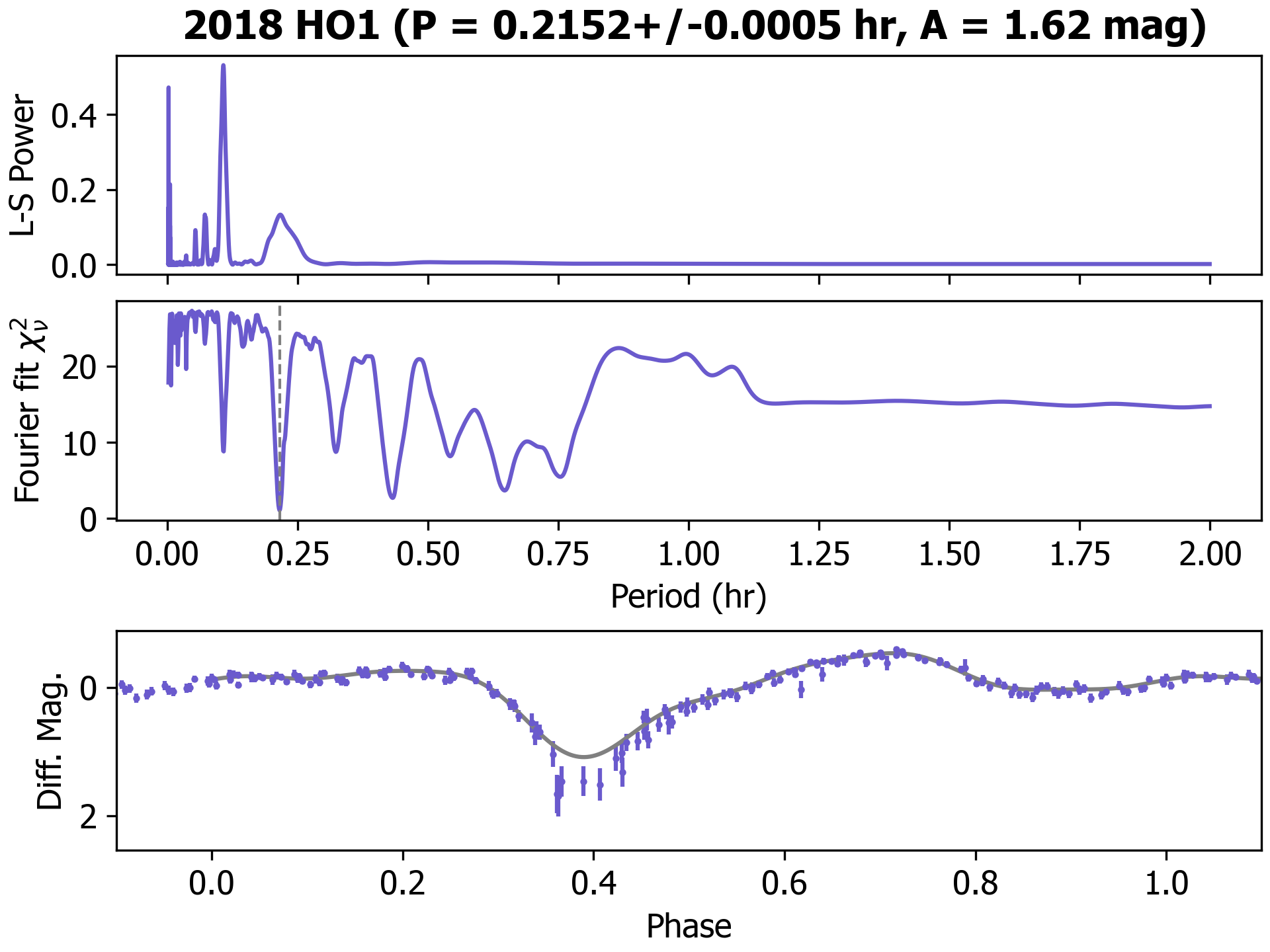}
    \caption{Continuation of Figure \ref{fig:A1}.}
\end{figure}

\begin{figure}
    \centering
    \includegraphics[width=0.49\textwidth]{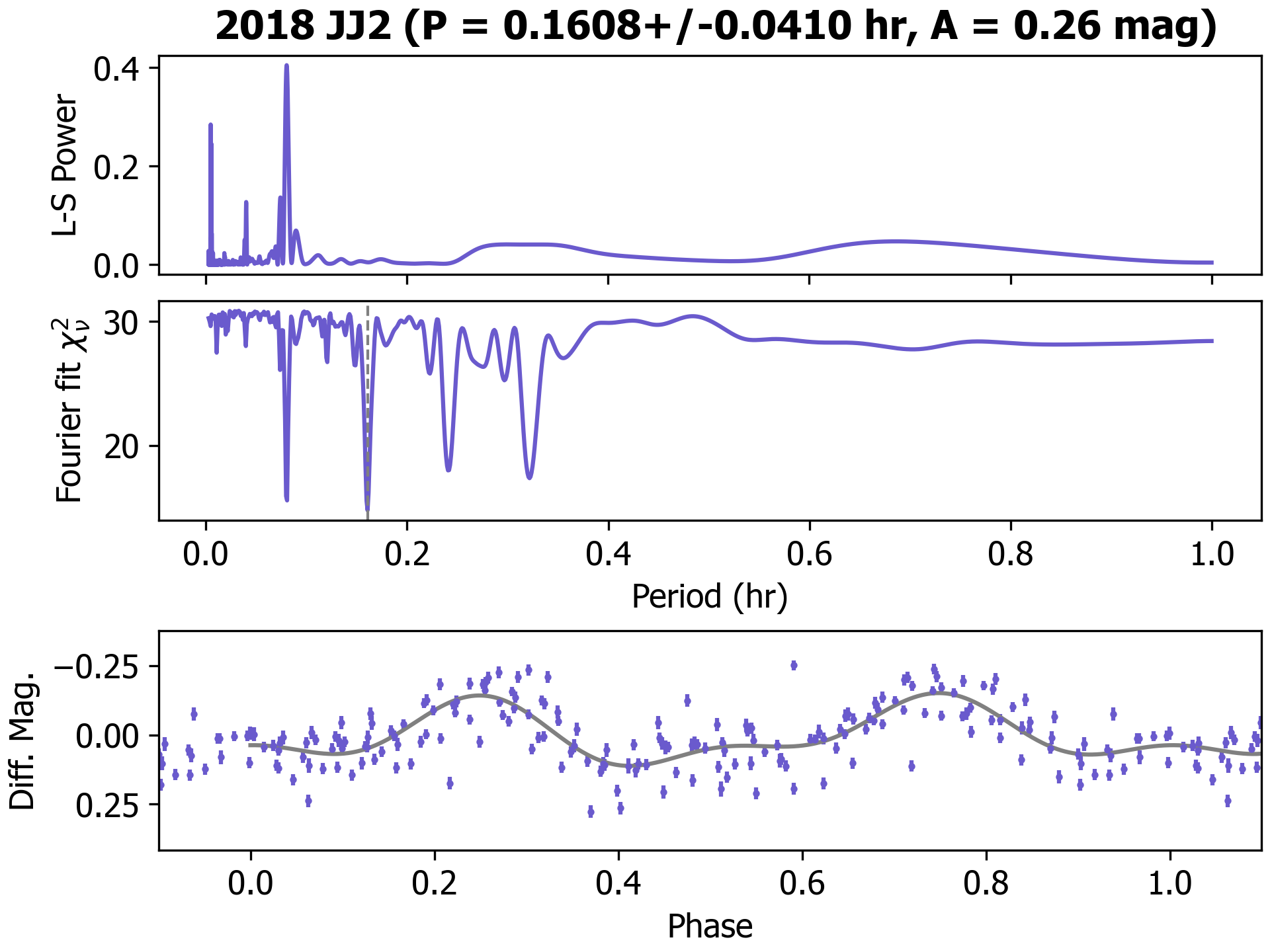}
    \includegraphics[width=0.49\textwidth]{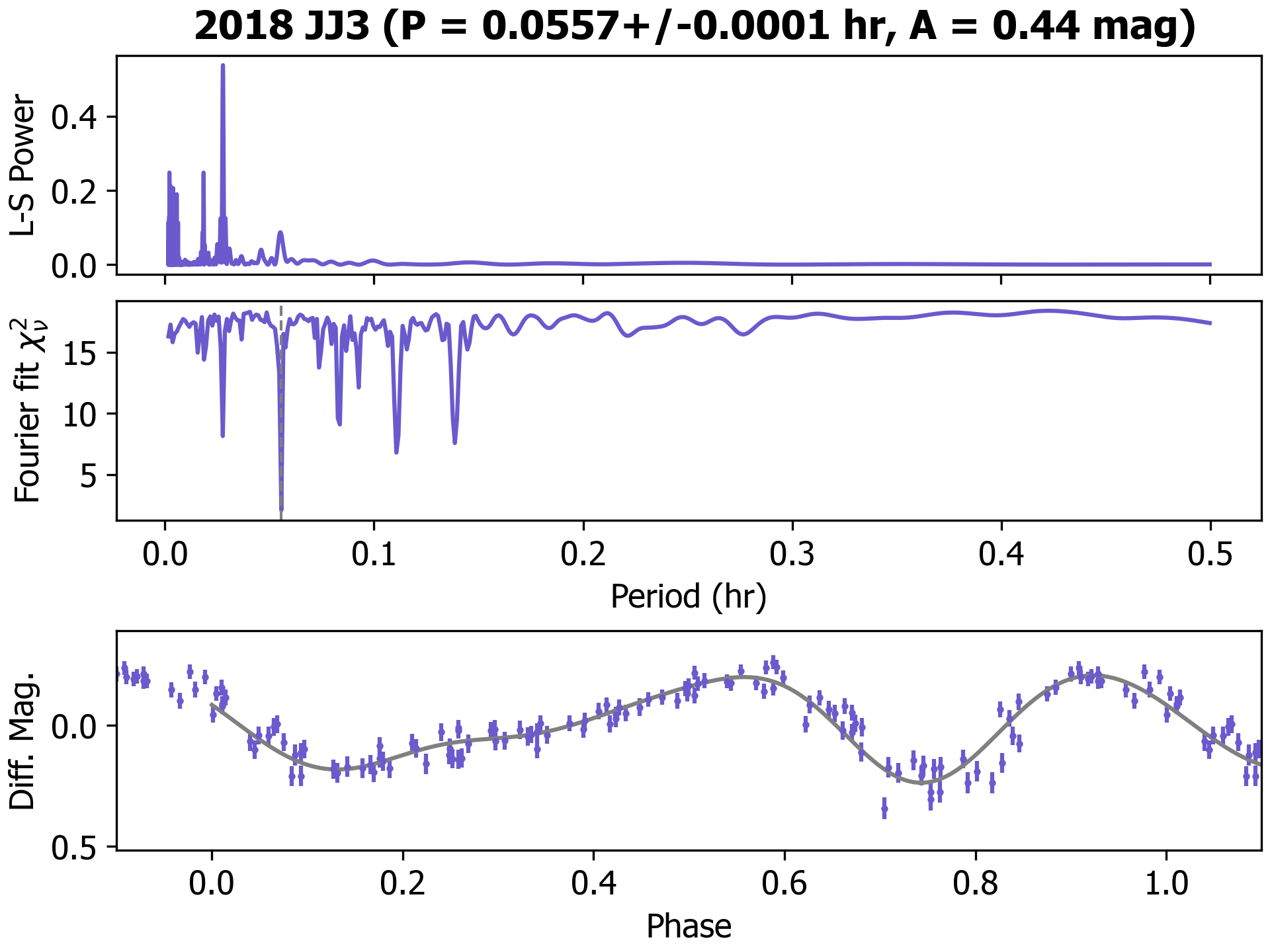}
    \includegraphics[width=0.49\textwidth]{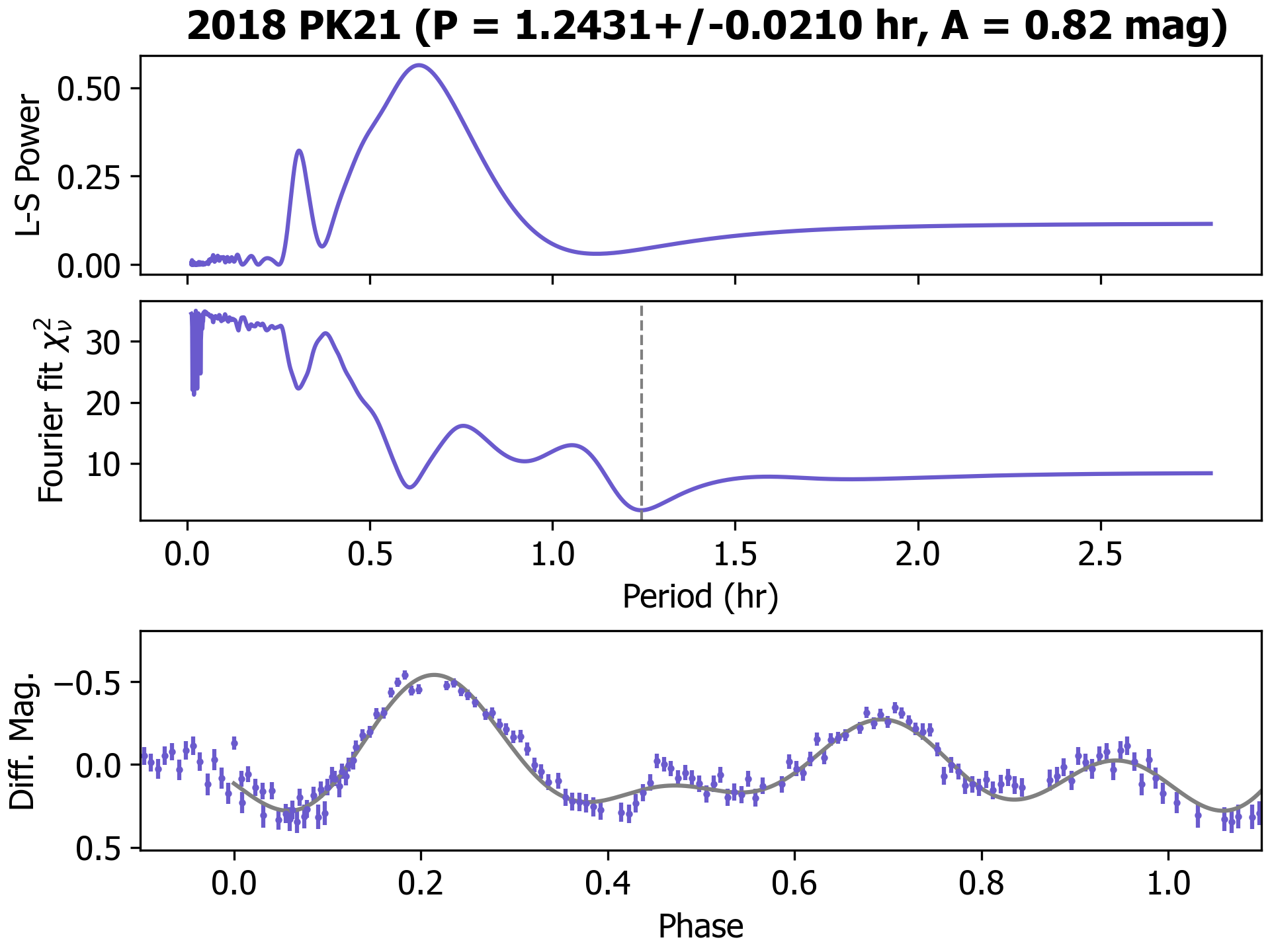}
    \includegraphics[width=0.49\textwidth]{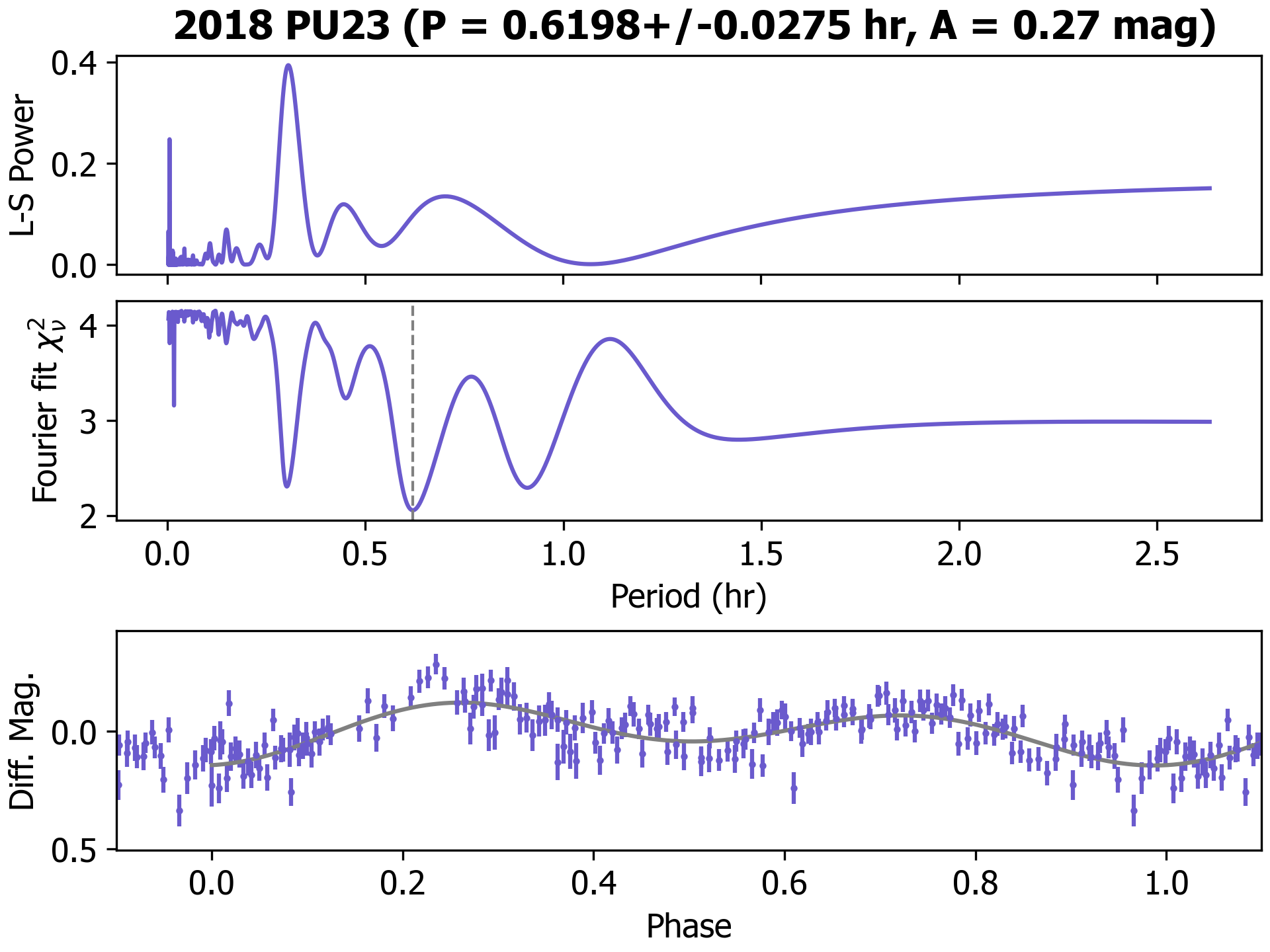}
    \includegraphics[width=0.49\textwidth]{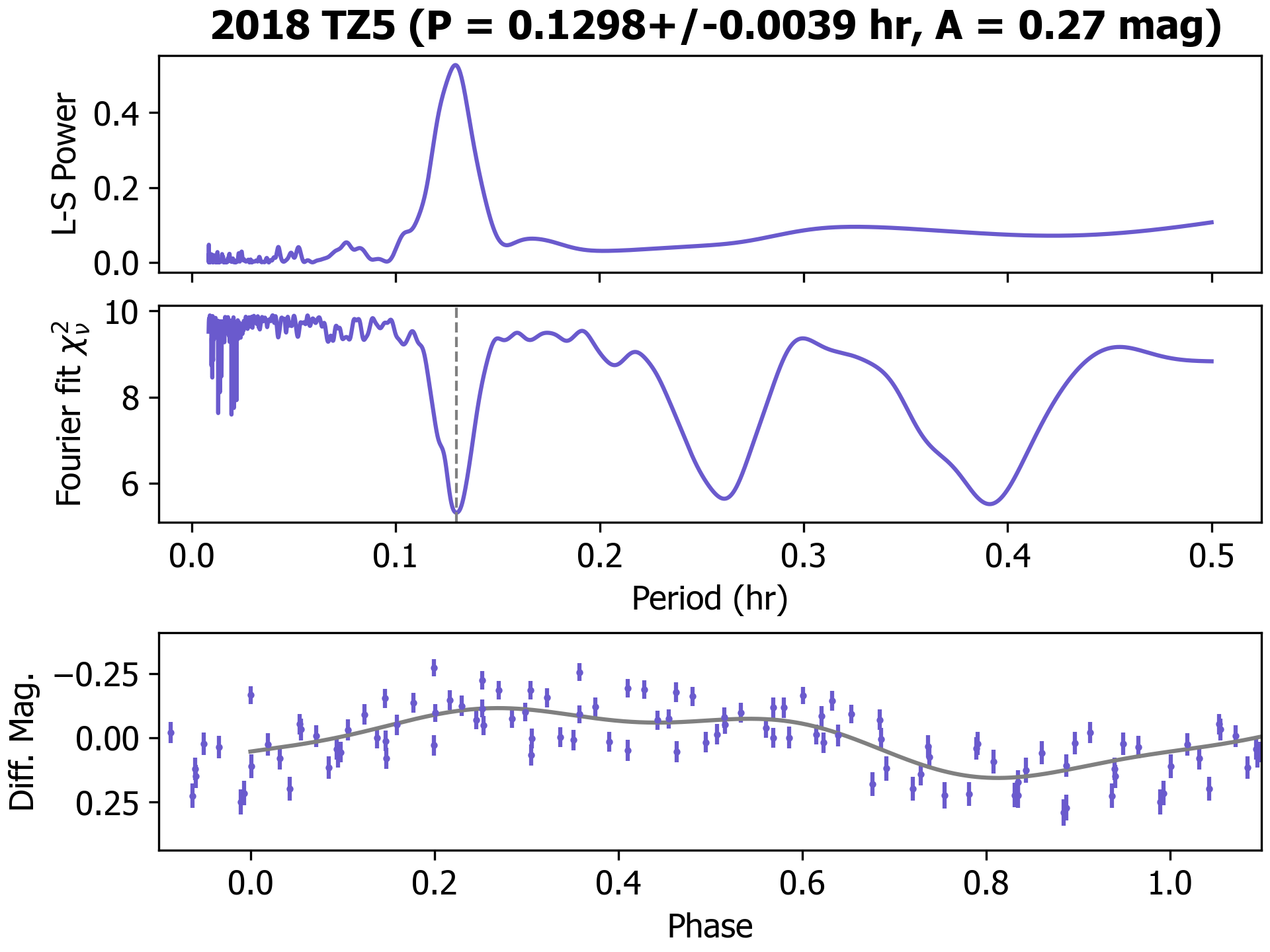}
    \includegraphics[width=0.49\textwidth]{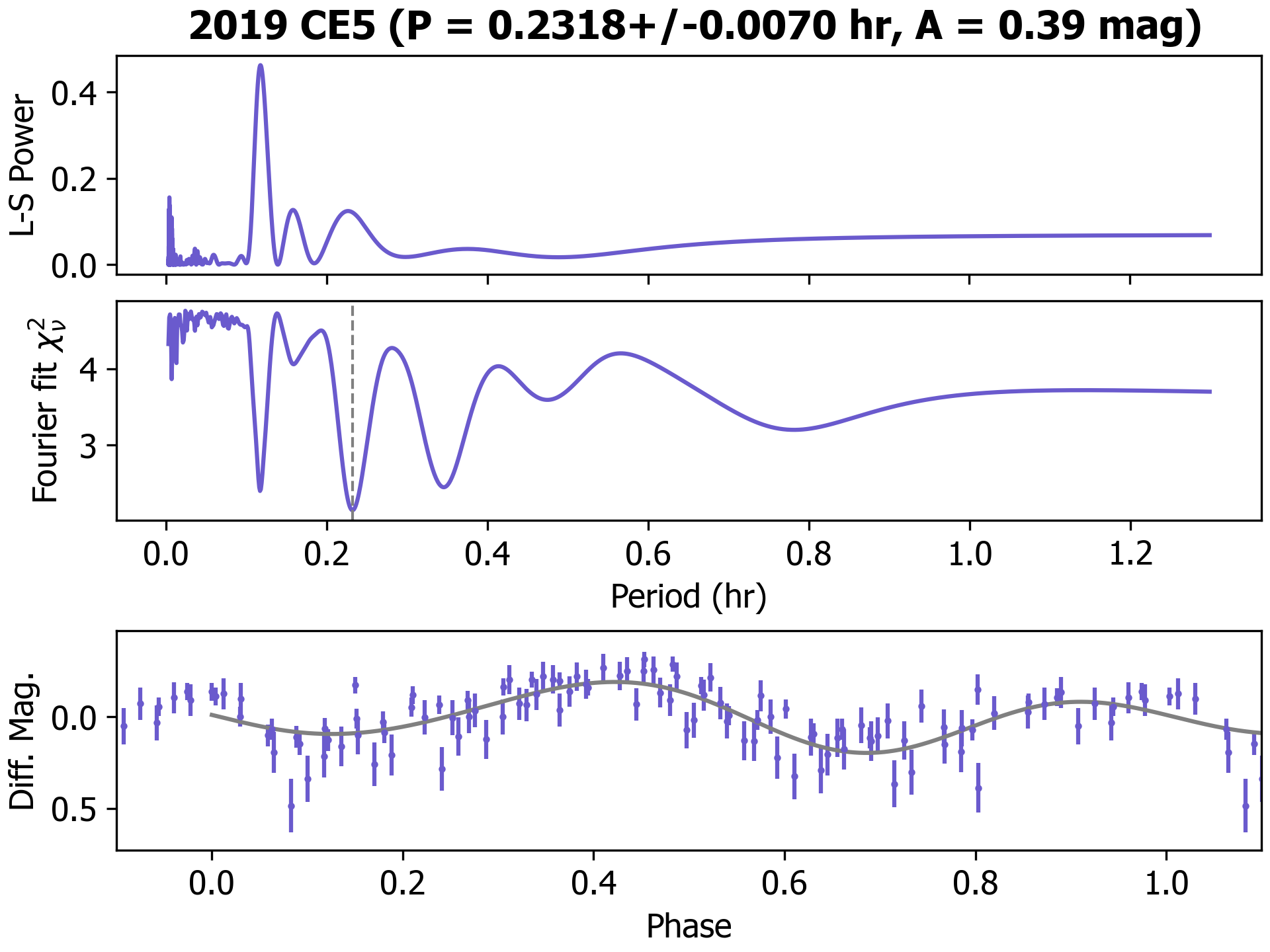}
    \caption{Continuation of Figure \ref{fig:A1}.}
\end{figure}

\begin{figure}
    \centering
    \includegraphics[width=0.49\textwidth]{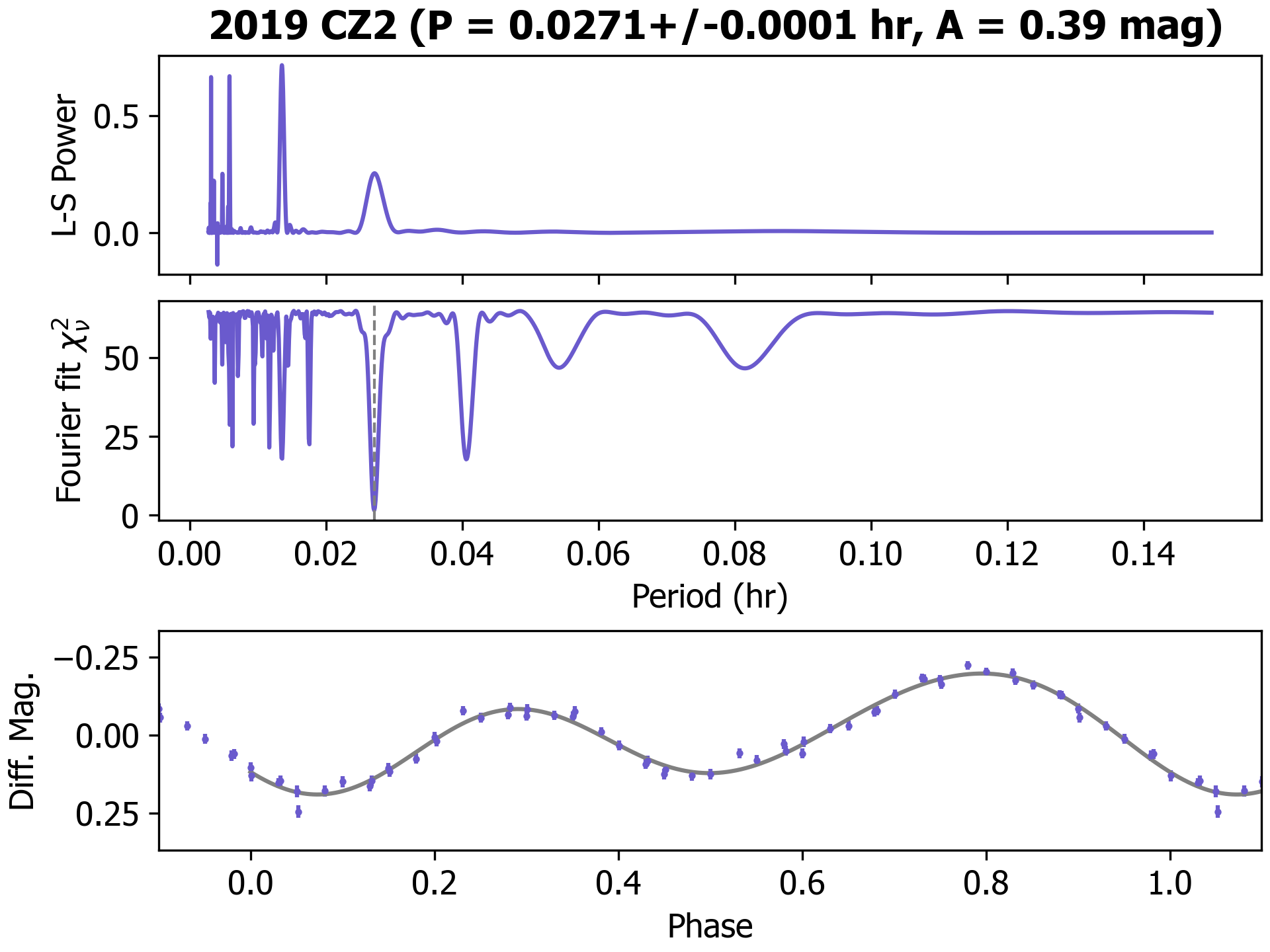}
    \includegraphics[width=0.49\textwidth]{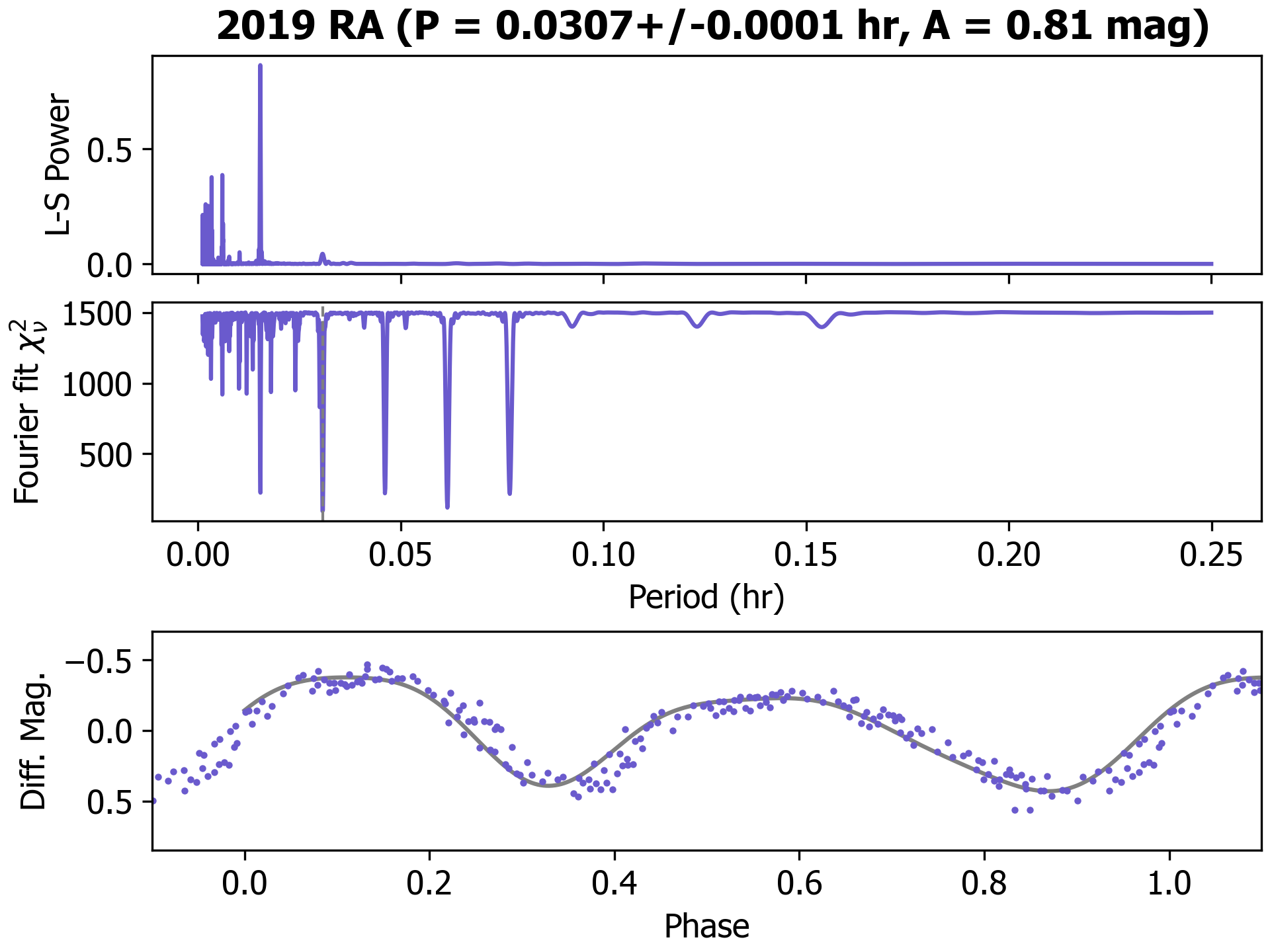}
    \includegraphics[width=0.49\textwidth]{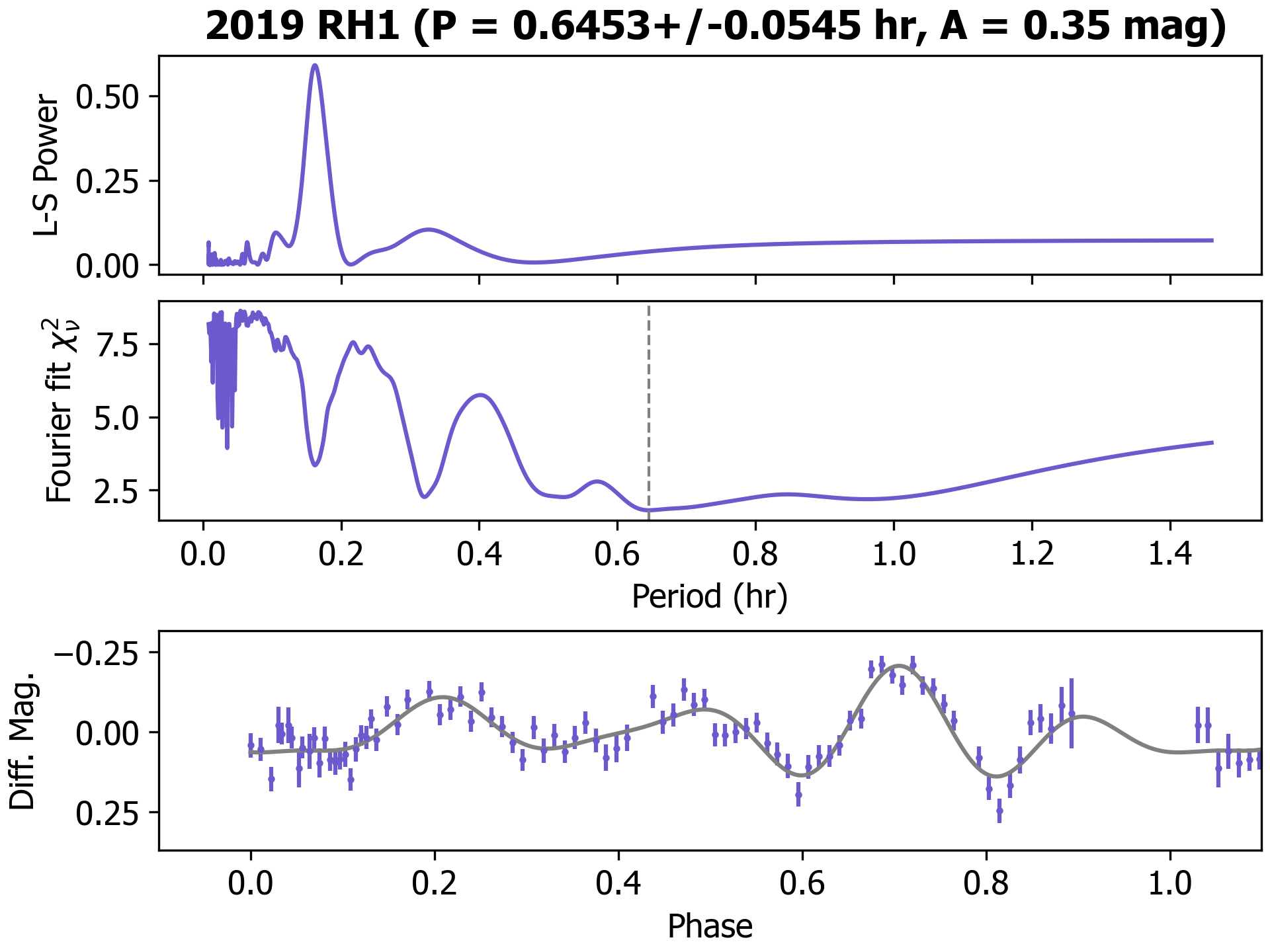}
    \includegraphics[width=0.49\textwidth]{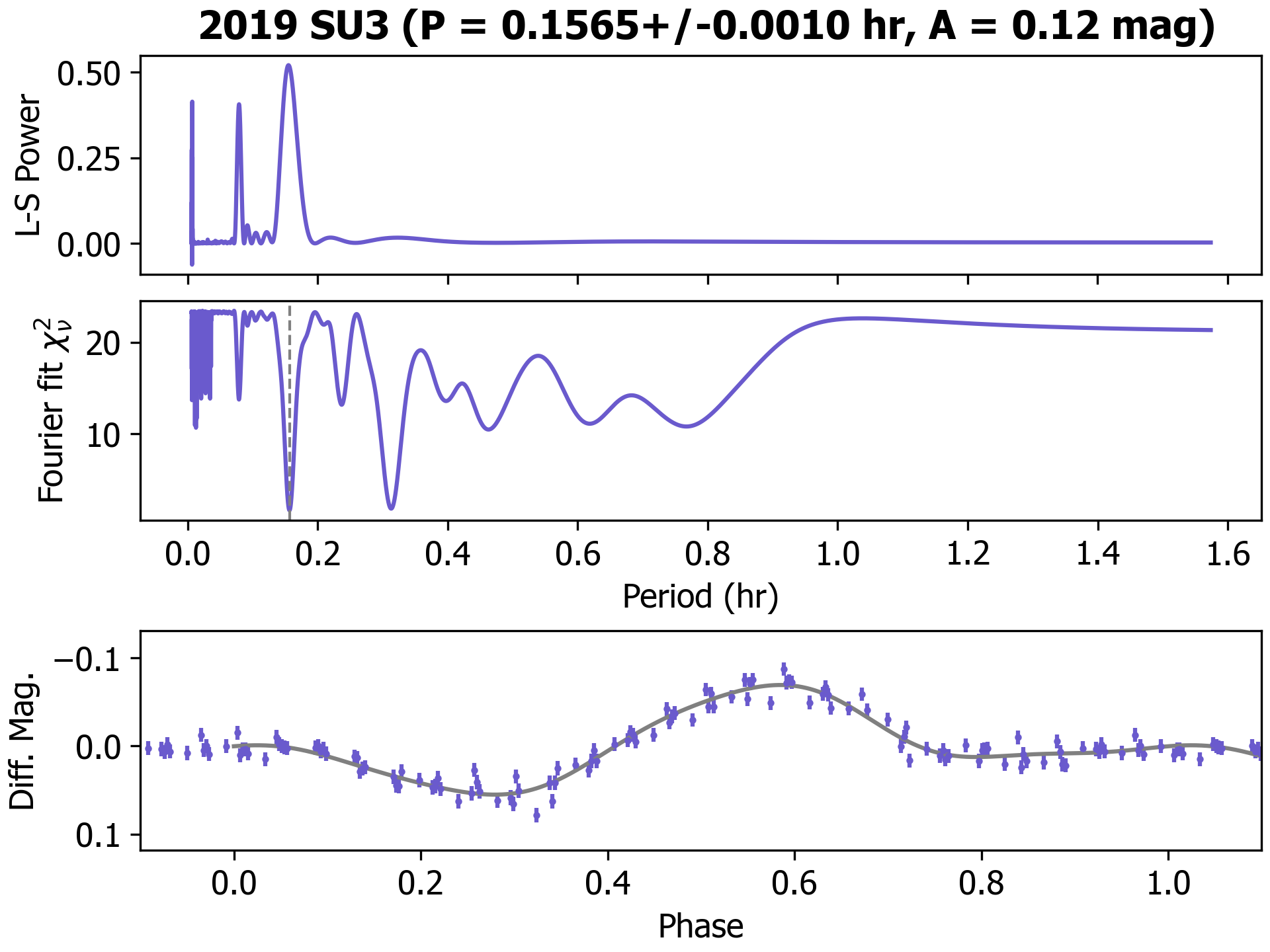}
    \includegraphics[width=0.49\textwidth]{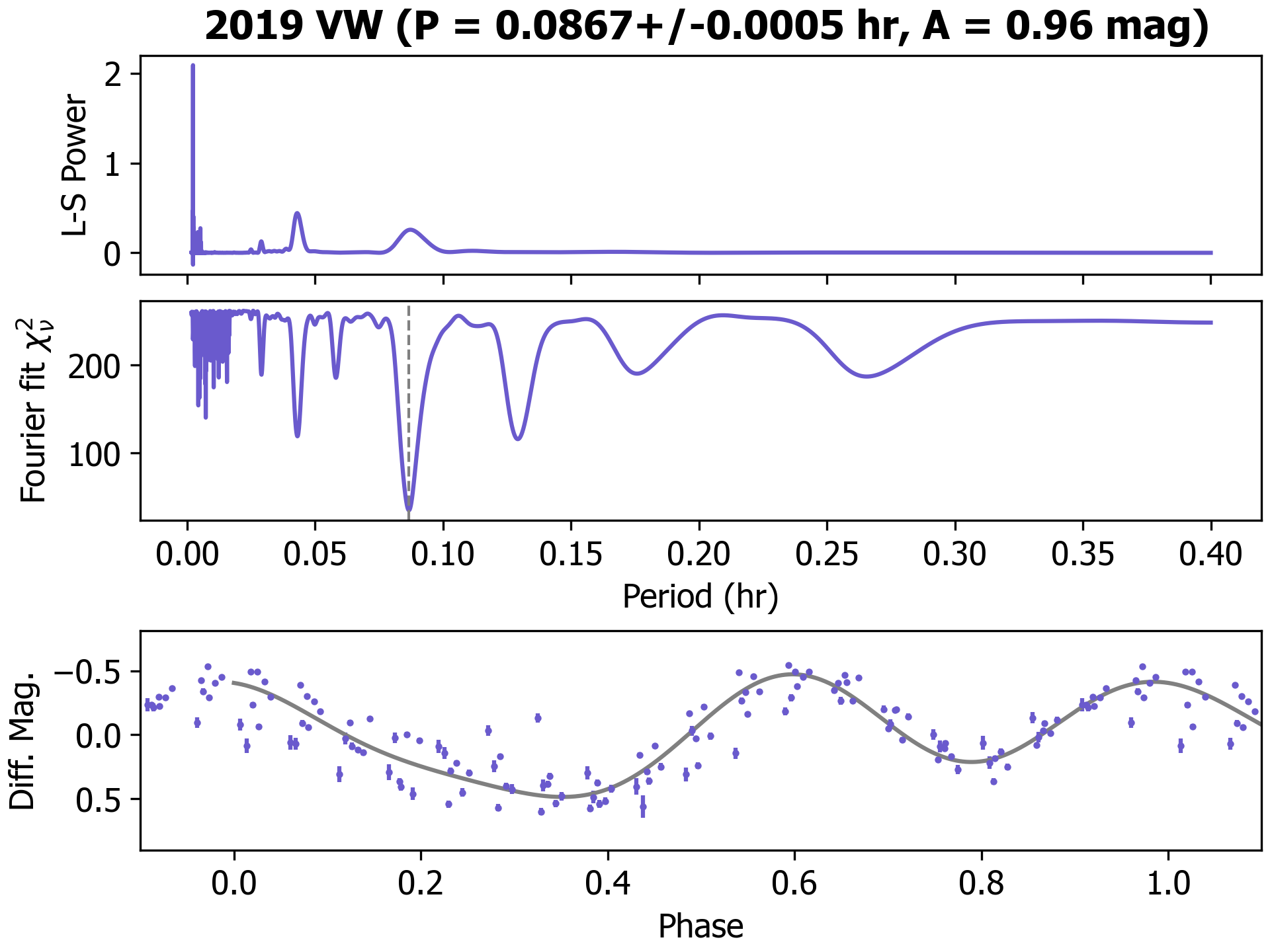}
    \includegraphics[width=0.49\textwidth]{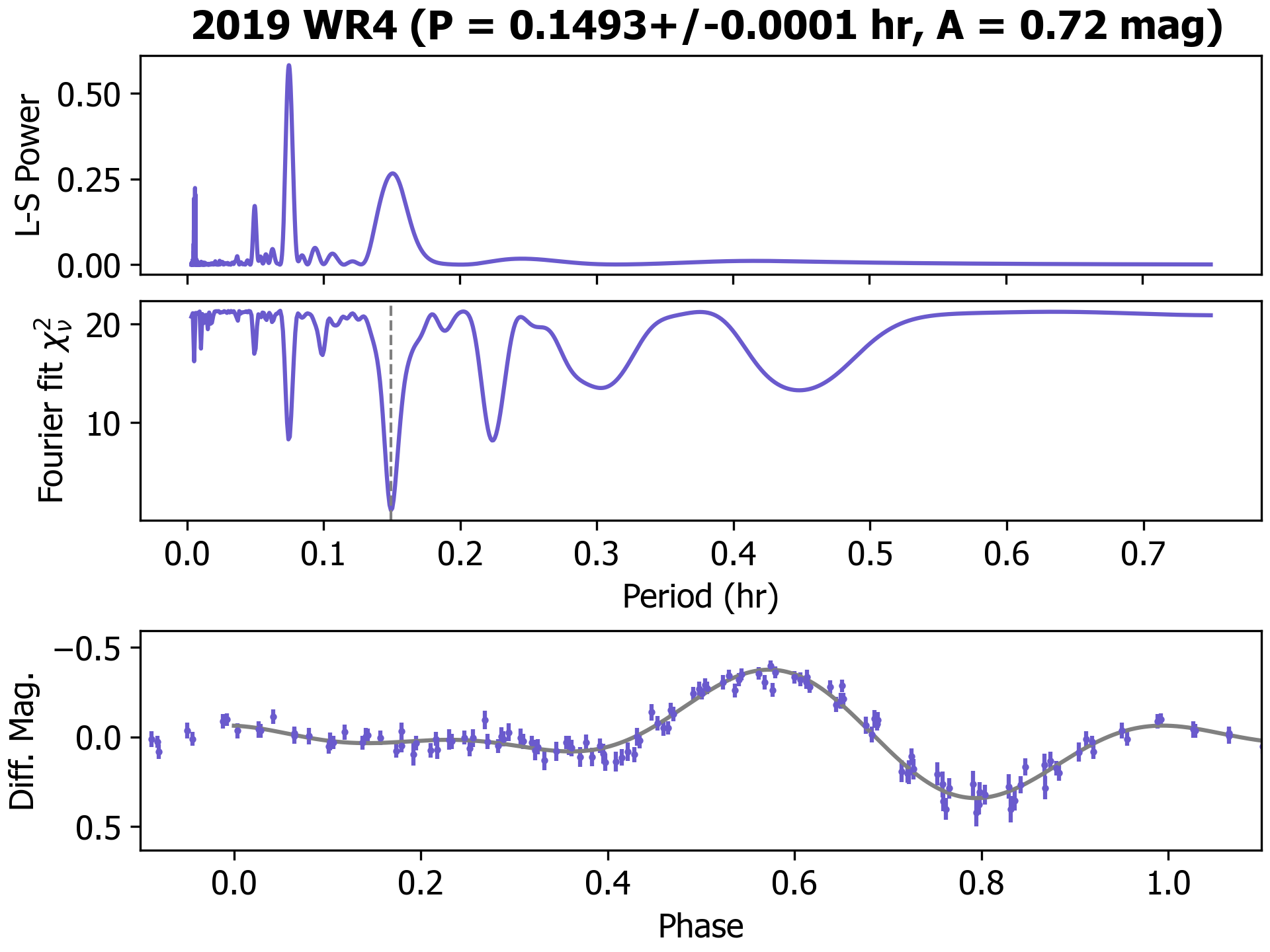}
    \caption{Continuation of Figure \ref{fig:A1}.}
\end{figure}

\begin{figure}
    \centering
    \includegraphics[width=0.49\textwidth]{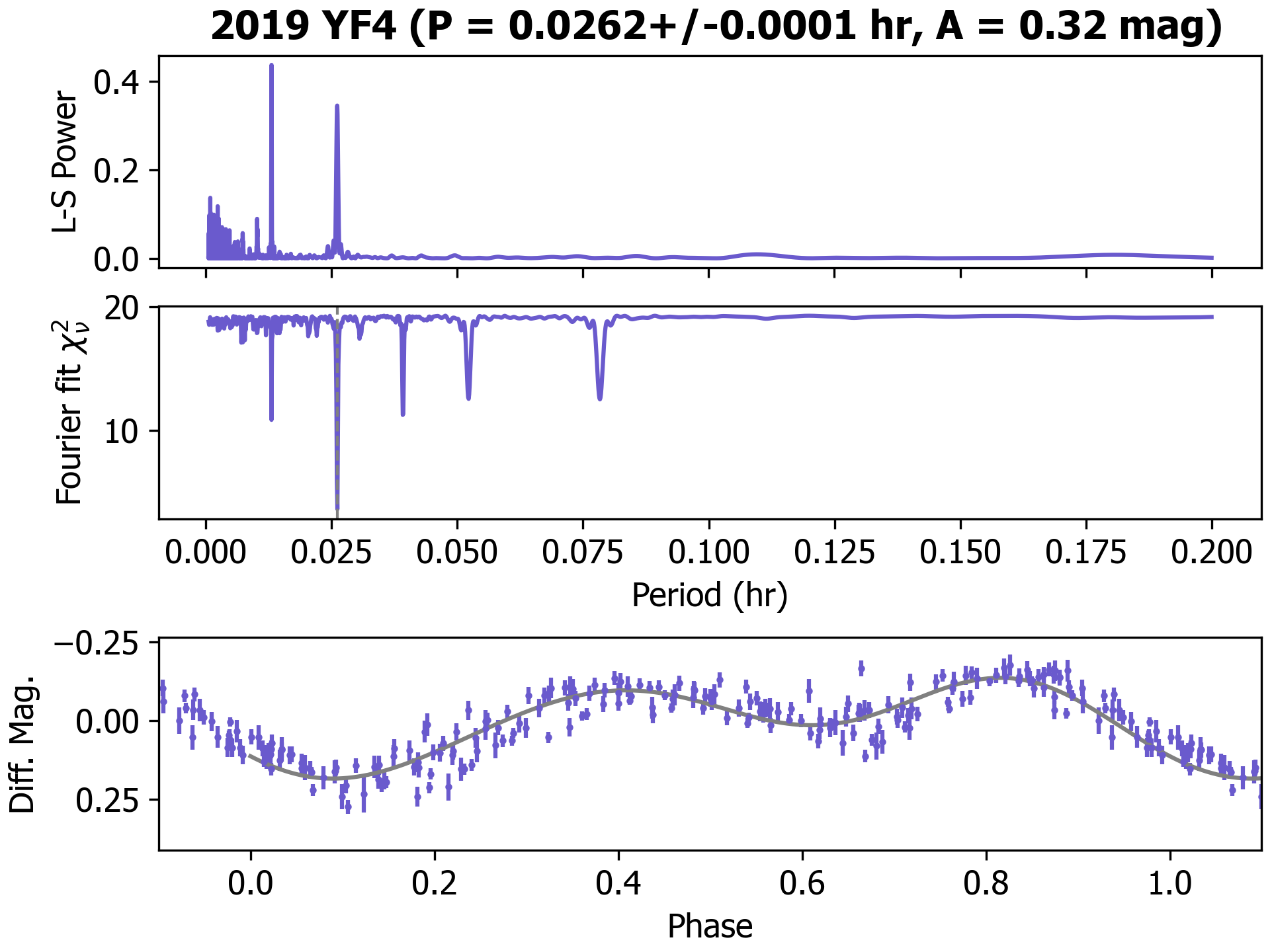}
    \includegraphics[width=0.49\textwidth]{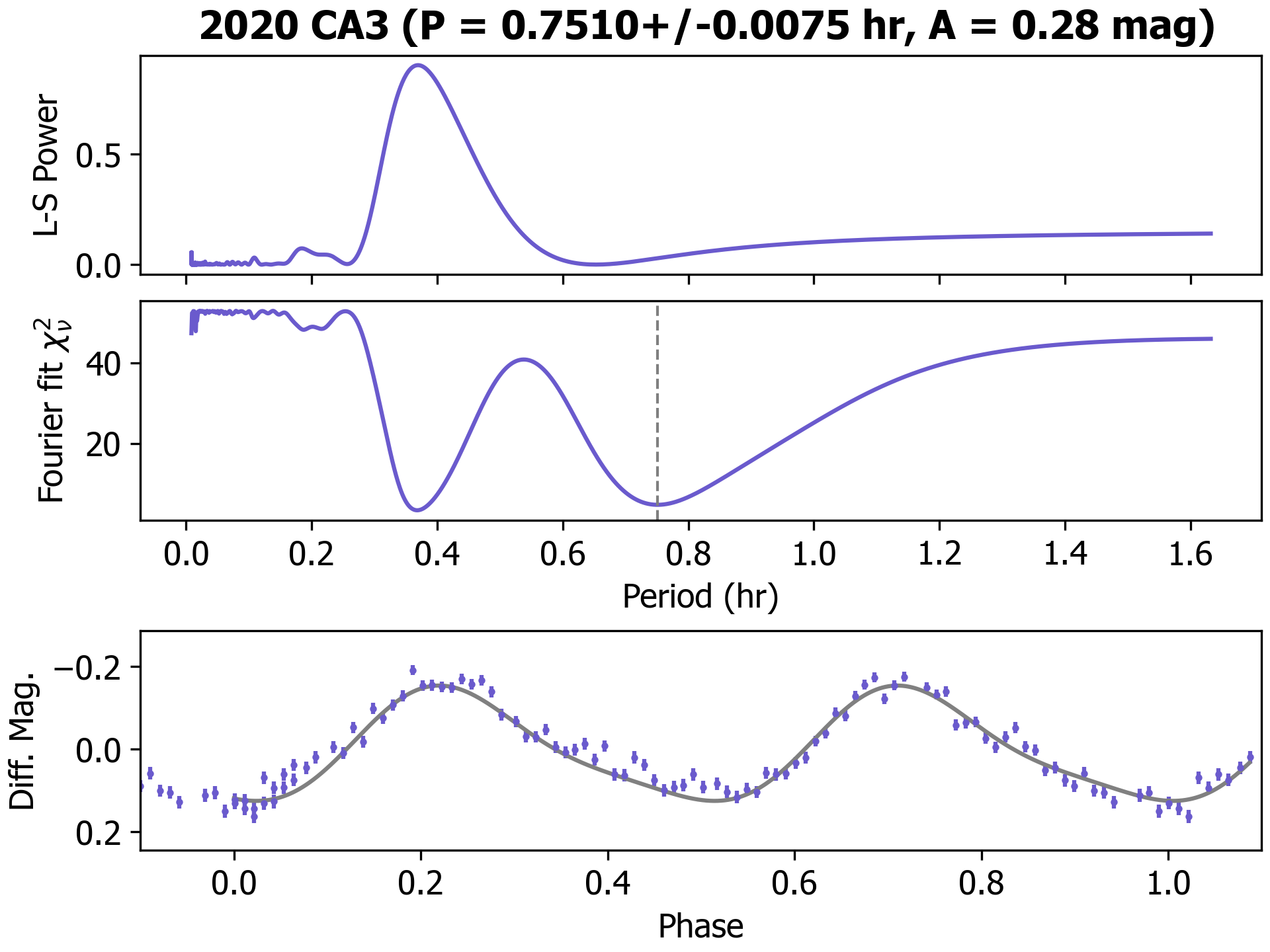}
    \includegraphics[width=0.49\textwidth]{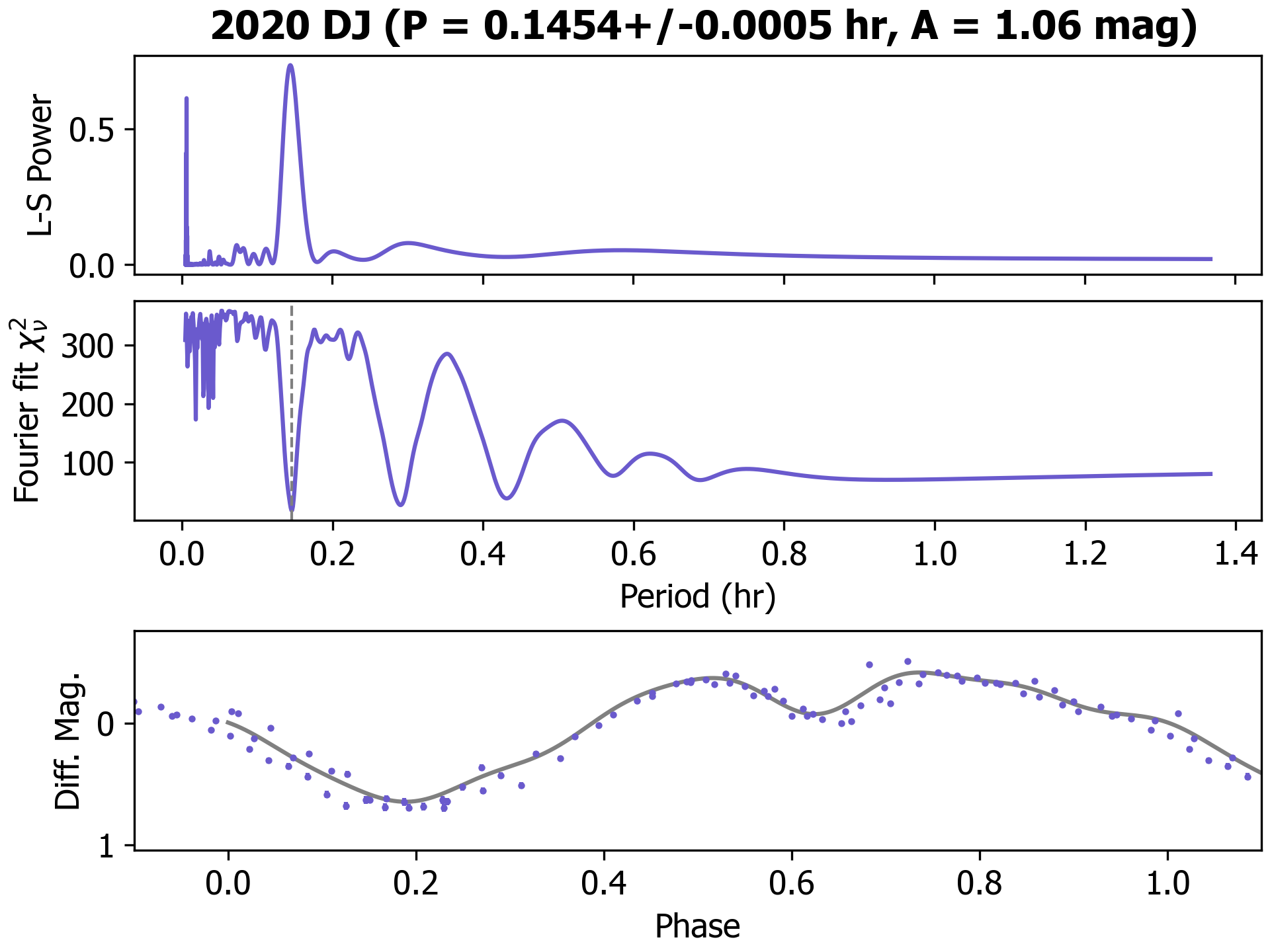}
    \includegraphics[width=0.49\textwidth]{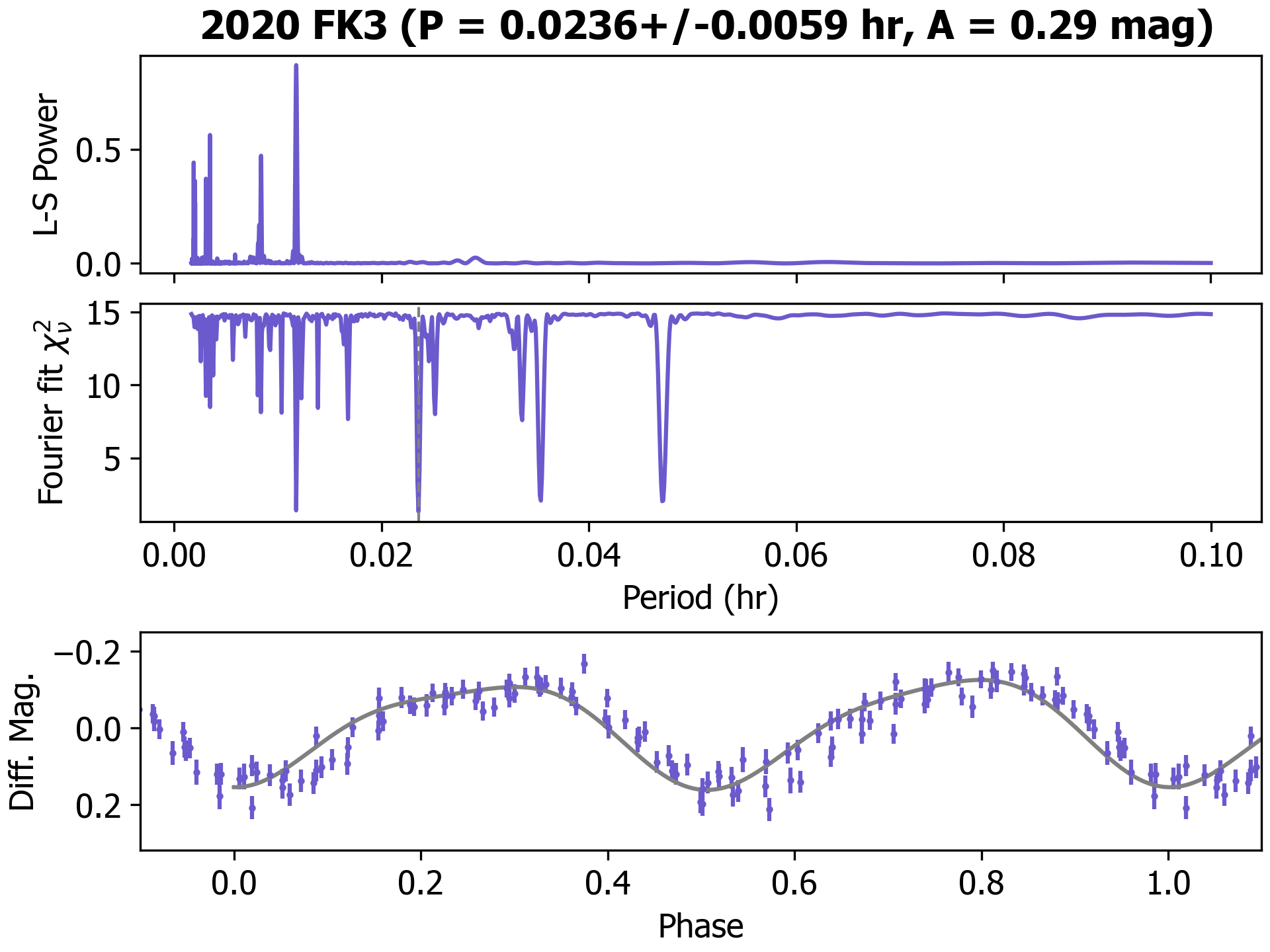}
    \includegraphics[width=0.49\textwidth]{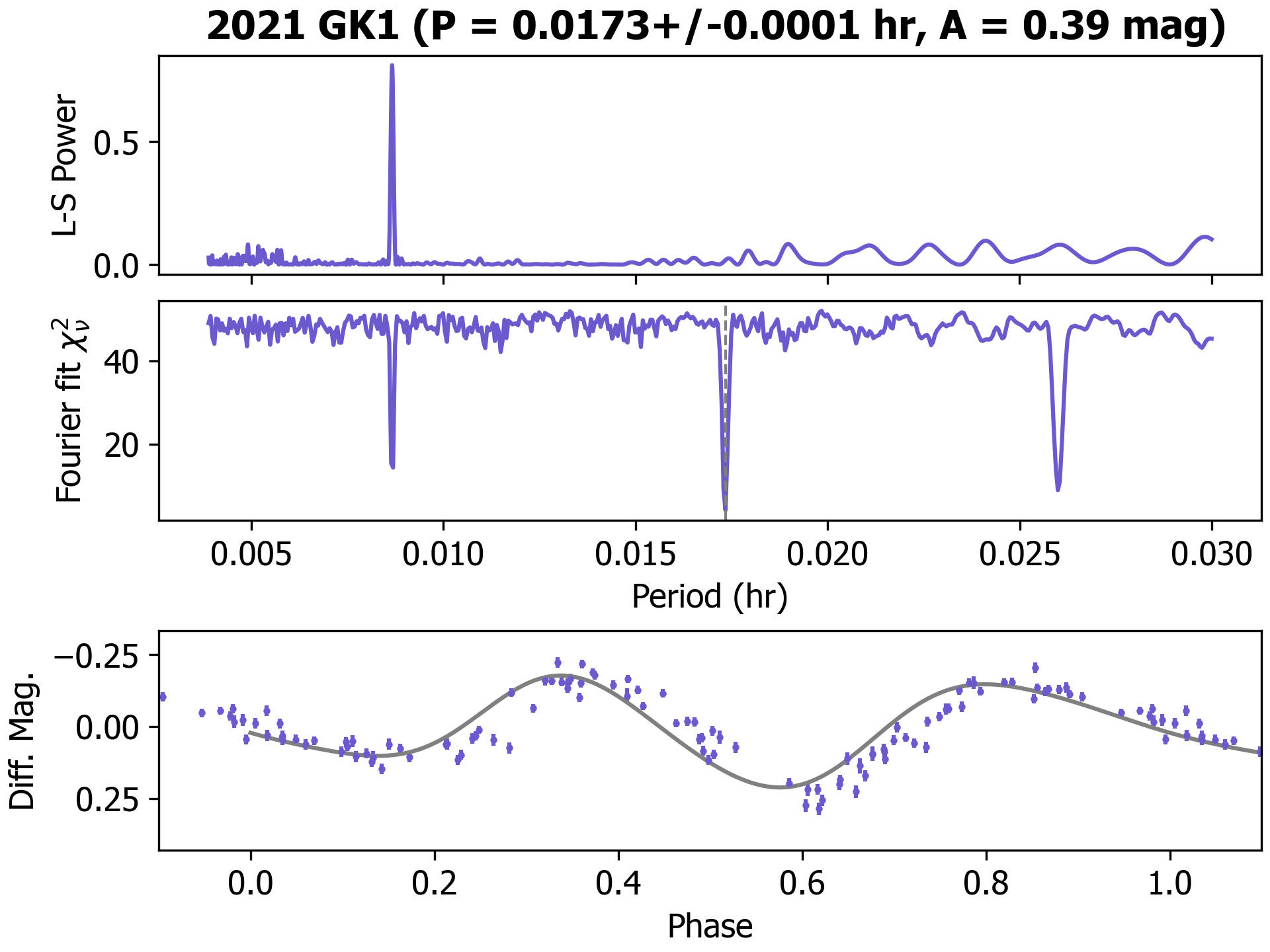}
    \includegraphics[width=0.49\textwidth]{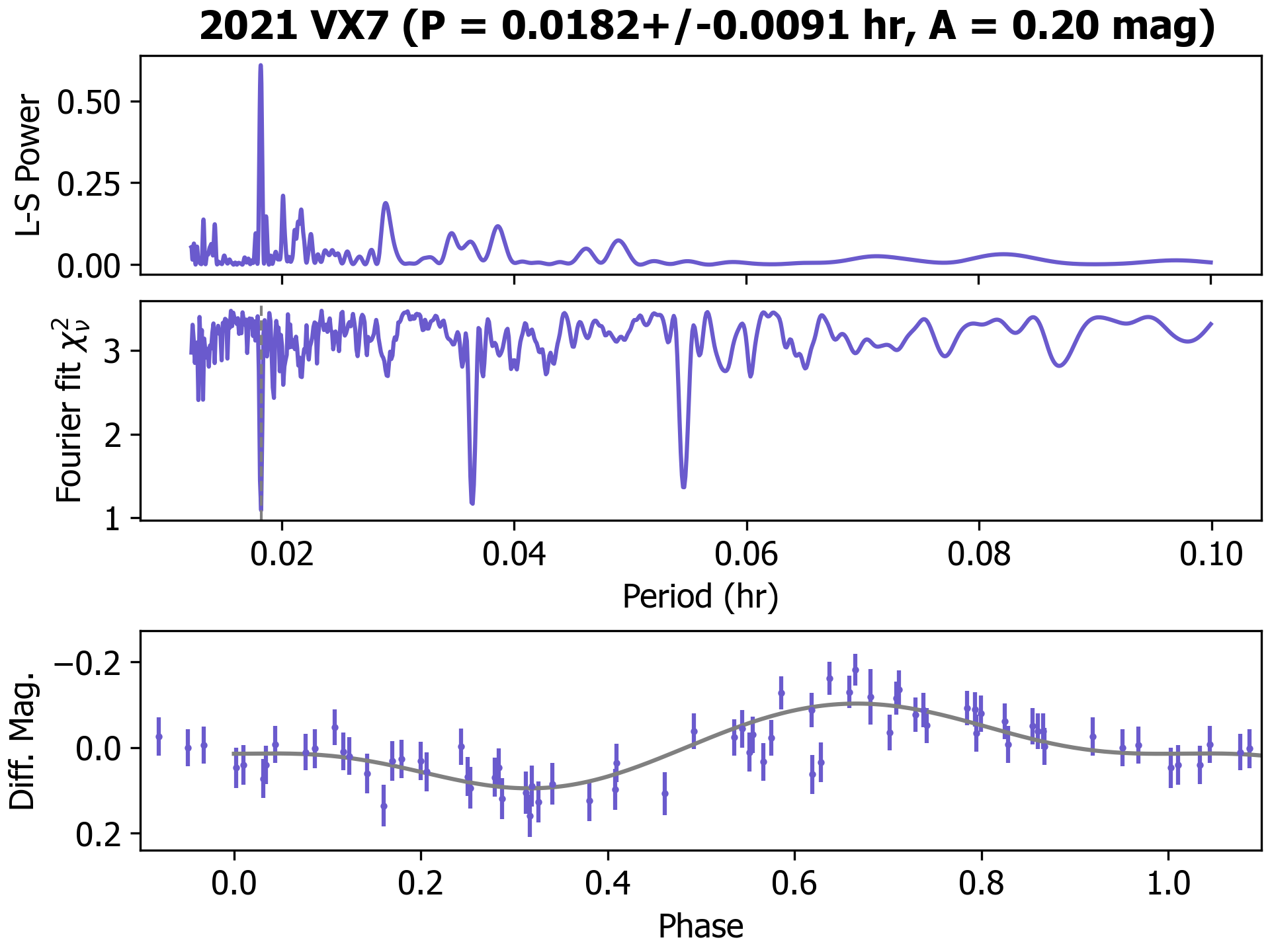}
    \caption{Continuation of Figure \ref{fig:A1}.}
\end{figure}

\begin{figure}
    \centering
    \includegraphics[width=0.49\textwidth]{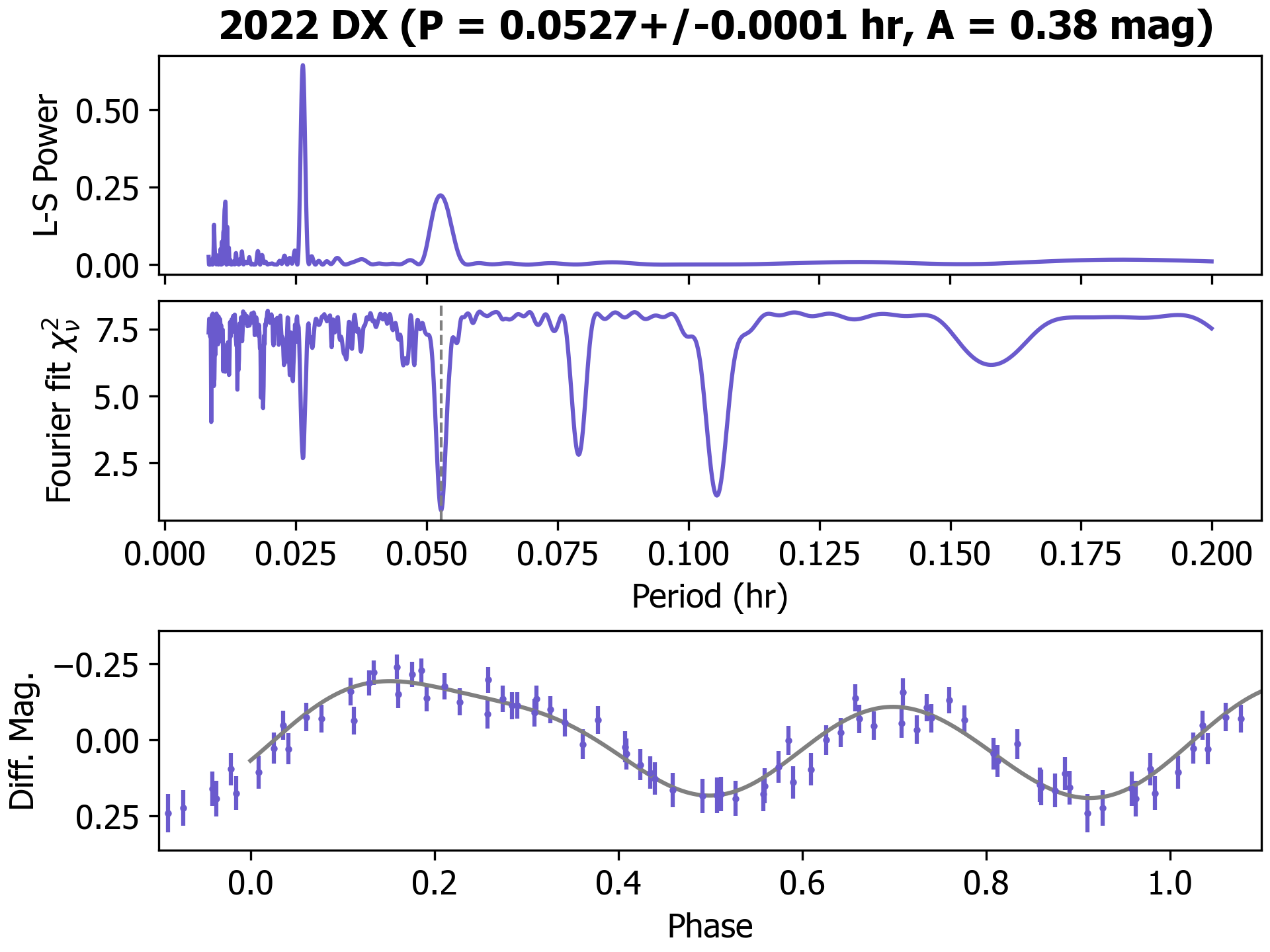}
    \includegraphics[width=0.49\textwidth]{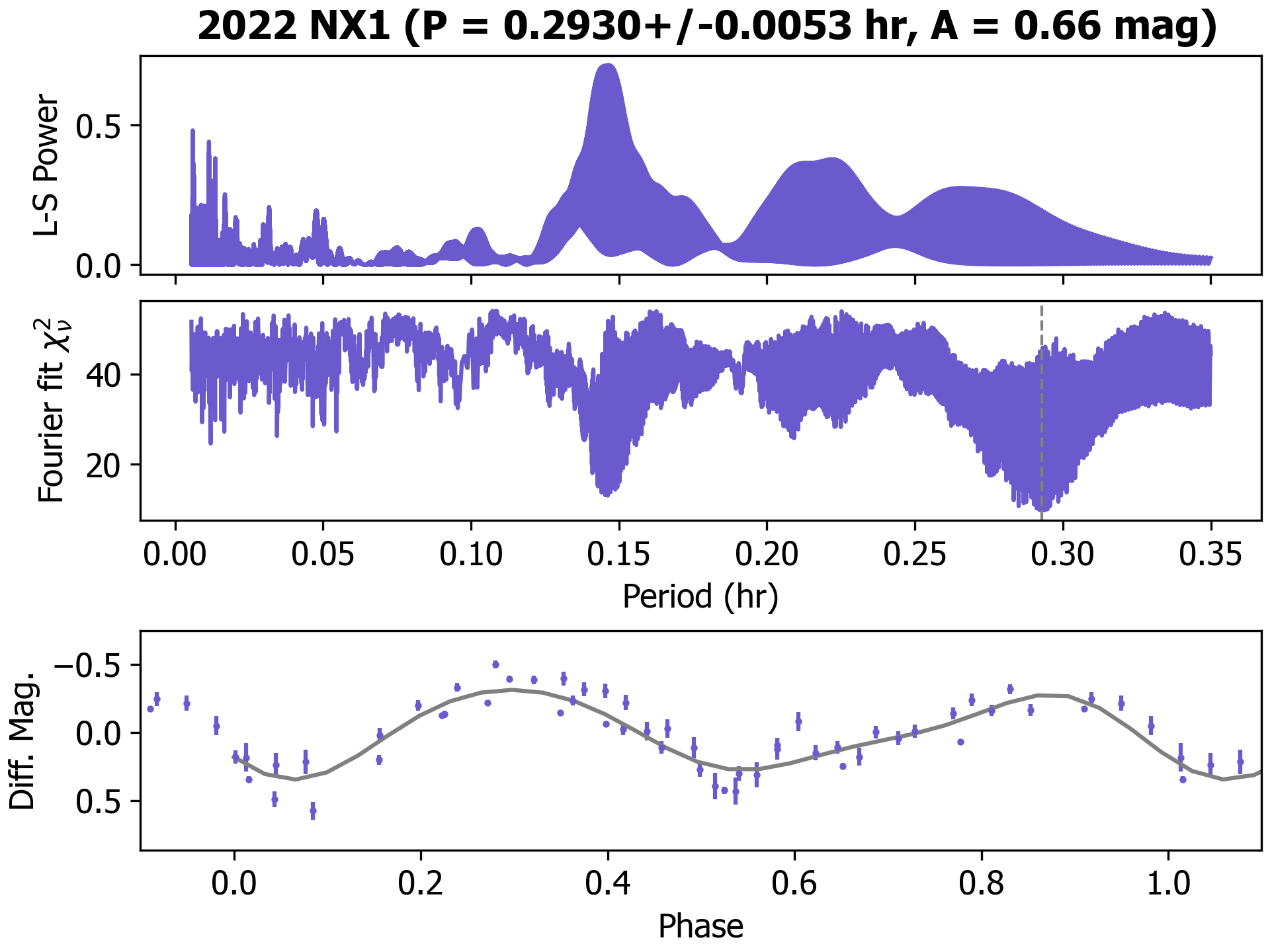}
    \includegraphics[width=0.49\textwidth]{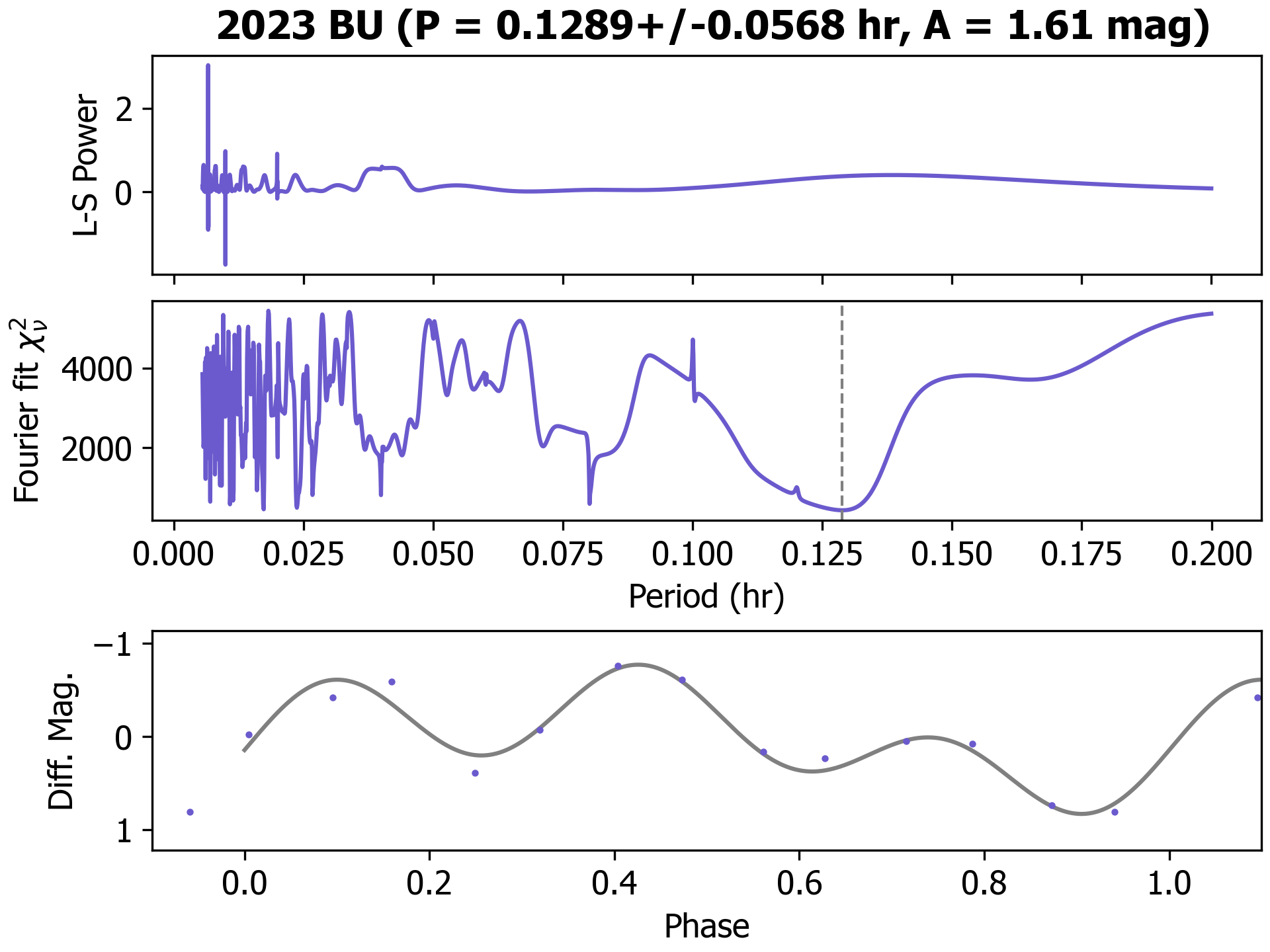}
    \includegraphics[width=0.49\textwidth]{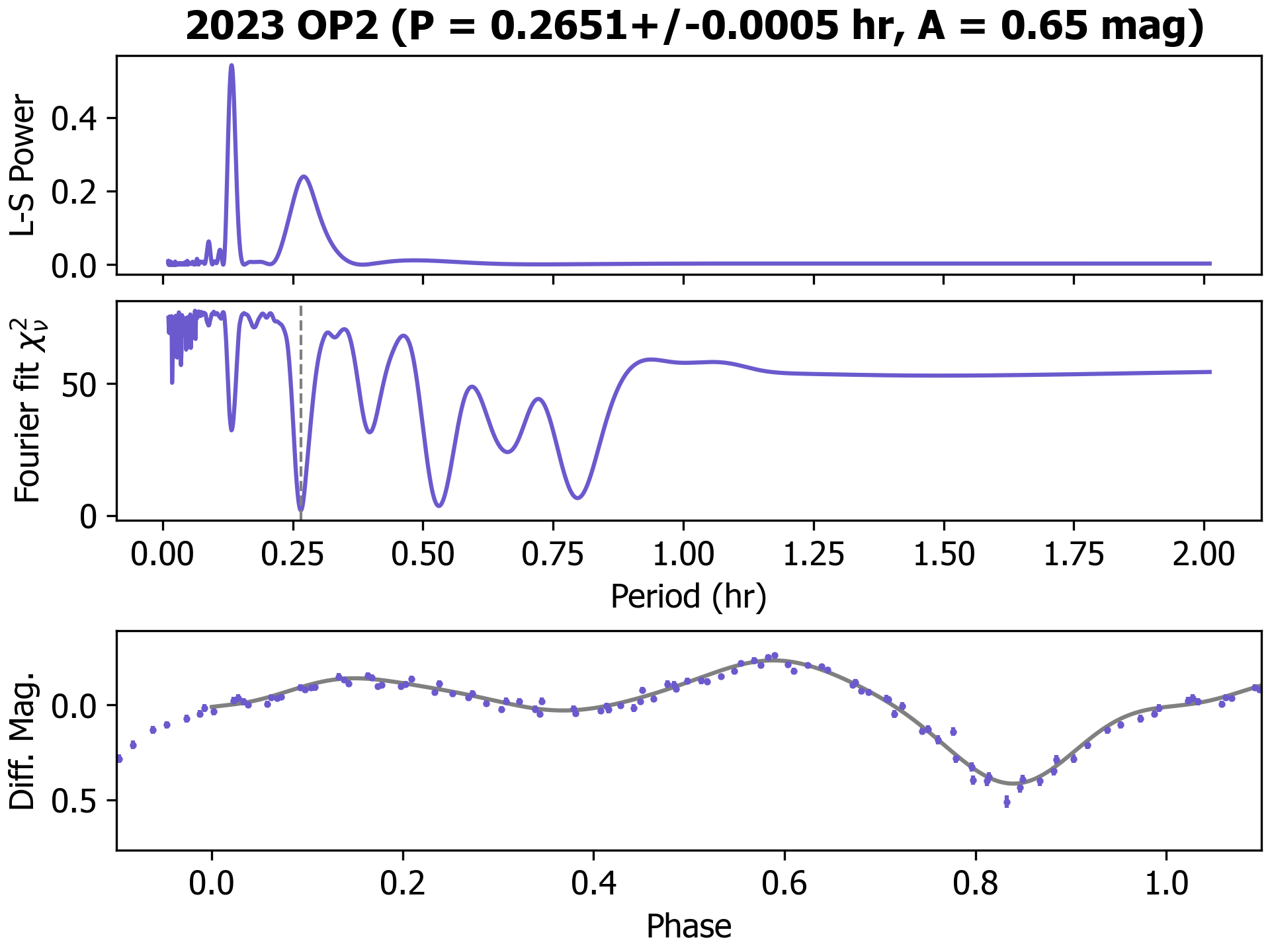}
    \includegraphics[width=0.49\textwidth]{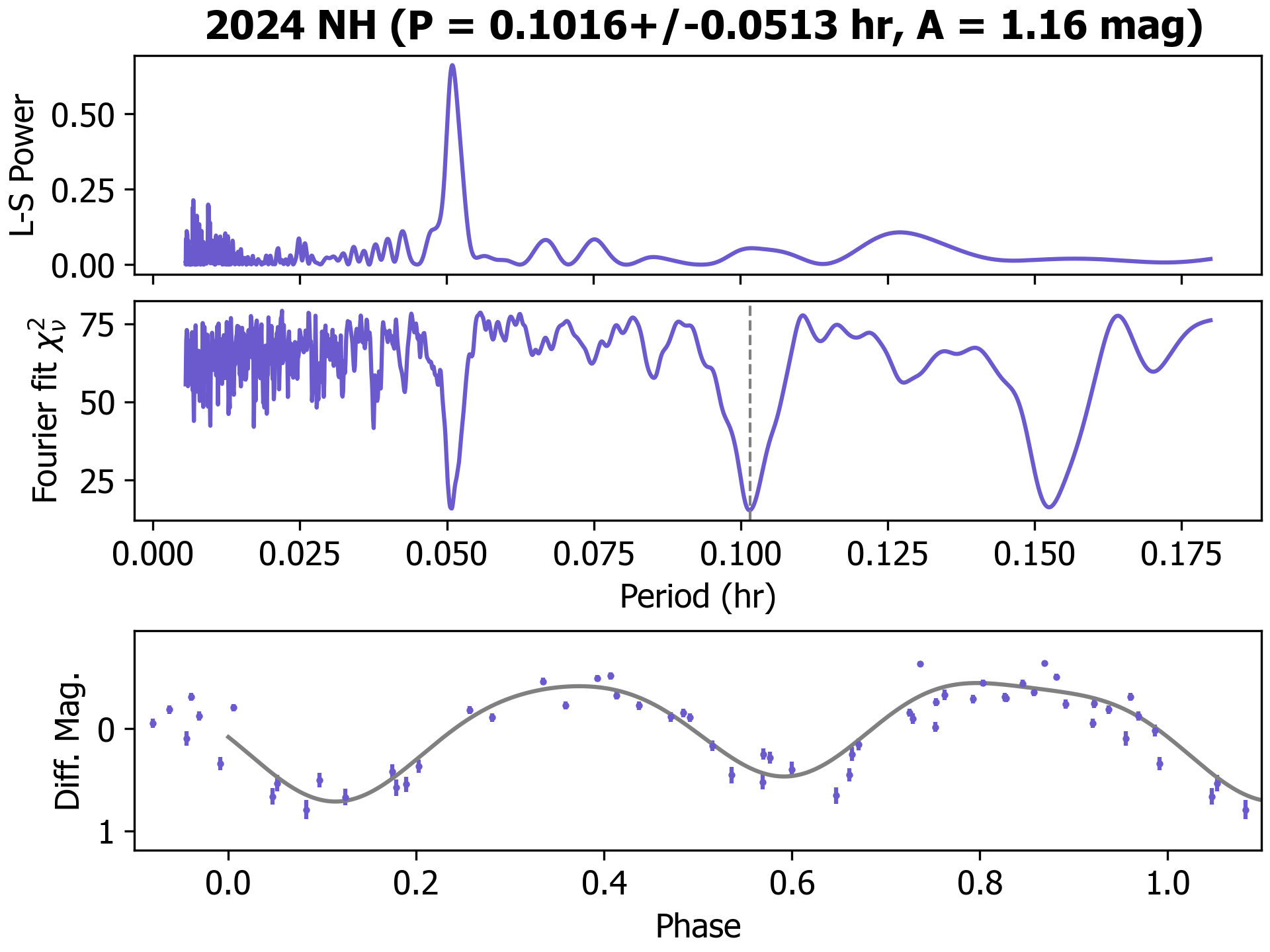}
    \includegraphics[width=0.49\textwidth]{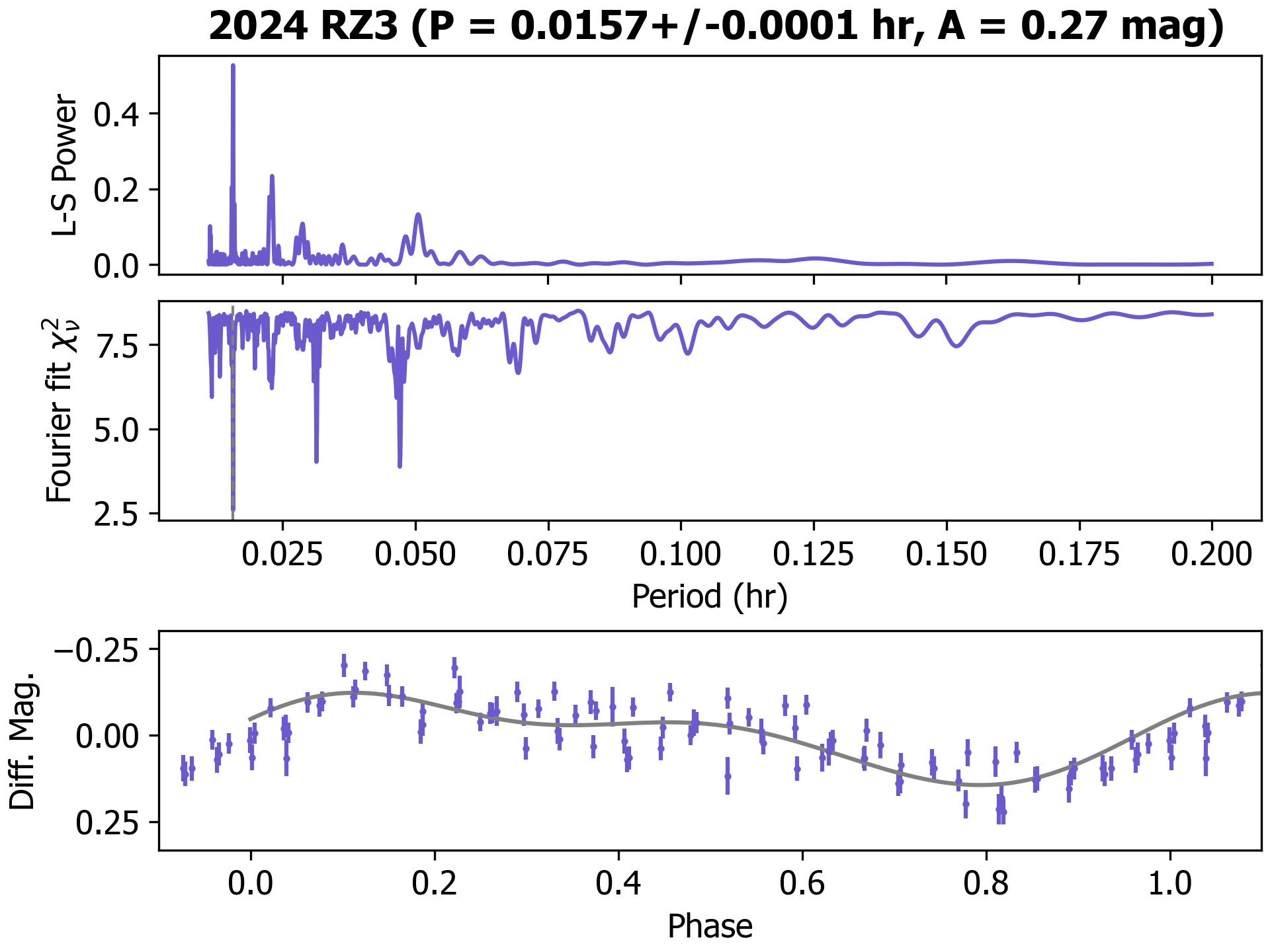}
    \caption{Continuation of Figure \ref{fig:A1}.}
\end{figure}

\begin{figure}
    \centering
    \includegraphics[width=0.49\textwidth]{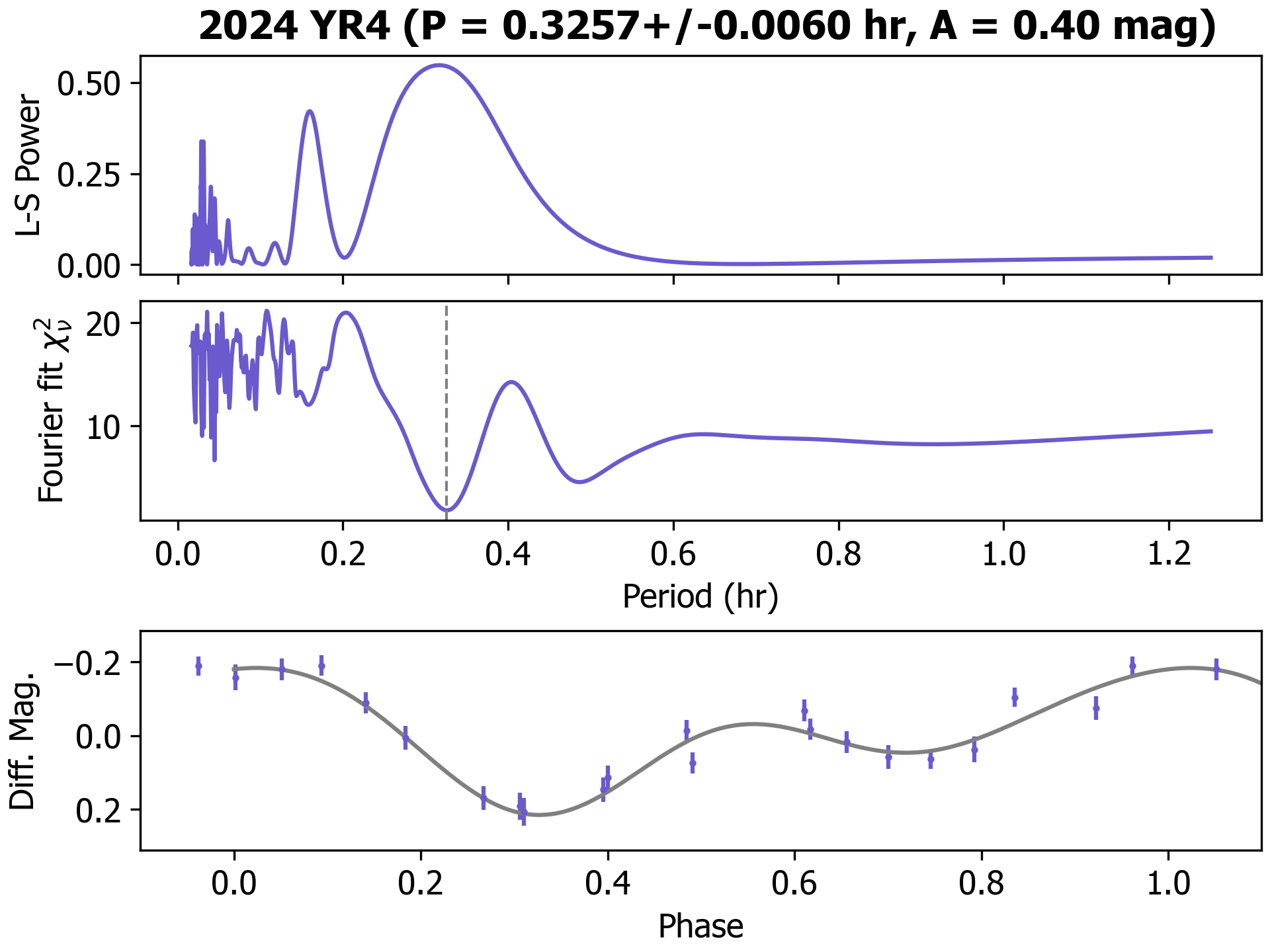}
    \caption{Continuation of Figure \ref{fig:A1}.}
\end{figure}


\bibliography{manos_colors.bib}{}
\bibliographystyle{aasjournal}



\end{document}